\documentclass[11pt]{article}
\usepackage[margin=1.1in]{geometry}
\usepackage[T1]{fontenc}
\usepackage{lmodern}
\usepackage{amsmath,amssymb,amsthm}
\usepackage{microtype}
\usepackage{xcolor}
\usepackage{enumitem}
\usepackage{array}
\usepackage{tikz}
\usepackage{placeins}
\makeatletter
\def\@part[#1]#2{%
  \ifnum\c@secnumdepth>\m@ne
    \refstepcounter{part}\addcontentsline{toc}{part}{\thepart\hspace{1em}#1}%
  \else
    \addcontentsline{toc}{part}{#1}%
  \fi
  \addvspace{2ex}%
  {\centering\interlinepenalty\@M\normalfont
   \ifnum\c@secnumdepth>\m@ne{\large\bfseries\partname\nobreakspace\thepart\par}\nobreak\vskip1ex\fi
   {\Large\bfseries #2\par}\markboth{}{}}%
  \nobreak\vskip3ex\@afterheading}
\def\@spart#1{%
  \addvspace{2ex}%
  {\centering\interlinepenalty\@M\normalfont\Large\bfseries #1\par}%
  \nobreak\vskip3ex\@afterheading}
\makeatother
\usepackage{hyperref}
\hypersetup{colorlinks=true,linkcolor=blue!50!black,citecolor=blue!50!black,urlcolor=blue!50!black}

\newtheorem{theorem}{Theorem}[section]
\newtheorem{lemma}[theorem]{Lemma}
\newtheorem{corollary}[theorem]{Corollary}
\newtheorem{proposition}[theorem]{Proposition}

\newtheorem*{theoremA}{Theorem A}
\newtheorem*{theoremB}{Theorem B}
\newtheorem*{theoremC}{Theorem C}
\newtheorem*{corollaryD}{Corollary D}
\newtheorem*{theoremE}{Theorem E}
\newtheorem*{theoremF}{Theorem F}
\newtheorem*{theoremG}{Theorem G}
\newtheorem*{theoremH}{Theorem H}
\theoremstyle{definition}
\newtheorem{definition}[theorem]{Definition}
\newtheorem{problem}[theorem]{Open problem}
\theoremstyle{remark}
\newtheorem{remark}[theorem]{Remark}

\newcommand{\Z}{\mathbb{Z}}
\newcommand{\F}{\mathbb{F}_2}
\newcommand{\Fp}{\mathbb{F}_p}
\newcommand{\Q}{\mathbb{Q}}
\newcommand{\C}{\mathbb{C}}
\newcommand{\Prob}{\mathrm{Prob}}
\newcommand{\Hh}{\mathrm{H}}

\newcommand{\Leaves}{\mathrm{Leaves}}
\newcommand{\Mult}{\textnormal{\textsc{Mult}}}

\newcommand{\Full}{\textnormal{\textsc{Full}}}
\newcommand{\Nz}{N_0}
\newcommand{\Qg}{\mathcal{G}}
\newcommand{\Kg}{K_{\mathrm g}}
\newcommand{\es}{\varepsilon^*_5}
\newcommand{\et}{\varepsilon^*_{10}}
\newcommand{\Lop}{\mathrm{Lop}\text{-}\allowbreak\mathrm{AE}\text{-}\allowbreak\mathrm{SparseTri}}
\newcommand{\eps}{\varepsilon}
\DeclareMathOperator{\polyop}{poly}
\DeclareMathOperator{\polylog}{polylog}
\DeclareMathOperator{\Tr}{Tr}
\DeclareMathOperator{\lift}{lift}
\DeclareMathOperator{\cl}{cl}
\makeatletter
\newcommand{\poly}{\@ifnextchar\log{\poly@log}{\polyop}}
\def\poly@log\log{\polylog}
\makeatother

\newcommand{\cnt}{\mathrm{cnt}}
\newcommand{\Pq}{\mathcal{P}}
\newcommand{\Iq}{\mathcal{I}}
\newcommand{\Hyp}{\mathsf{H}}

\title{Thin matrix products up to \texorpdfstring{$D\le N^{0.172}$}{D <= N\^{}0.172}\texorpdfstring{\\}{} and lossless reductions for 3SUM, Exact Triangle and APSP\texorpdfstring{\\}{, }with 3SUM over abelian groups and a dichotomy for coordinatewise problems}
\author{Ruhollah Majdoddin\\ \normalsize\texttt{r.majdodin@gmail.com}}
\date{}

\begin{document}
\maketitle

\begin{abstract}
Alman and Vassilevska Williams (arXiv:2610.06783) gave the first polynomial improvements over the textbook algorithms for 3SUM and APSP, with deterministic exponents $1.9992$ for 3SUM, $2.9983$ for Exact Triangle and $2.99942$ for APSP. We improve both steps of their algorithm and obtain deterministic algorithms for 3SUM in $n^{1.9945+o(1)}$ time and for Exact Triangle, $(\min,+)$-product and APSP in $n^{2.9945+o(1)}$ time. We also improve on their proof sketch for 3XOR and solve it deterministically in $O(n^{1.99451})$ time.

Their first step is a matrix theorem: for an $N\times D$ and a $D\times N$ integer matrix with $D\le N^\varepsilon$, $\varepsilon<0.1204$, it computes $N^2/\sqrt D$ prescribed entries of the product in time $O(N^2/D^\gamma)$, for a constant $\gamma>0$. With Schönhage's five-term partial-product tensor in place of his ten-term identity, we extend it to every $\varepsilon<0.172$; this also extends their refutation of the hinted Online Matrix--Vector (OMv) conjectures to hints of size $n^{0.172}$.

Their second step reduces the three problems to the lopsided sparse triangle problem, keeping a quarter of the matrix theorem's saving for 3SUM and a half for Exact Triangle and APSP. Our reductions keep all of it, by calling the counting version of that problem adaptively and building composite moduli one prime at a time; over $\mathbb F_2$ the same method, with linear hashes whose rows are chosen one at a time from small-bias sets in place of the moduli, reduces 3XOR directly to the triangle problem and also keeps all of it.

Finally we answer, for two families, which other problems on three lists the method makes truly subquadratic. It solves 3SUM over $\mathbb F_p^d$ for every fixed prime $p$ with the exponents of 3XOR, and over every finite abelian group of exponent $p^2$, such as $\mathbb Z_4^d$, in $O(n^{1.99726})$. For coordinatewise problems with a fixed relation $P$ on every coordinate triple, an exhaustive computation with certified verdicts shows that every Boolean $P$, and every functional $P$ over three letters, either is the solution set of an equation over $\mathbb Z$, $\mathbb F_2$ or $\mathbb F_3$, and the problem is then truly subquadratic, or has a gadget for Orthogonal Vectors, and the problem is then hard under the Orthogonal Vectors hypothesis. The triangle versions show the same dichotomy one level up, between truly subcubic and cubic under the Orthogonal Vectors hypothesis. Over four letters, just four functional relations, up to symmetry, satisfy neither condition.
\end{abstract}
\begin{quote}\small\itshape
\textbf{Use of AI.} This paper was written with Anthropic's Claude models, Claude Opus 5.5 and Claude Fable 5.1, through Claude Code, in October 2026. Under the author's direction they searched the literature, searched for and developed the proofs, wrote and ran the ancillary scripts, drafted and revised the text, and checked it against the cited sources; the research questions, the choice among the directions proposed and the decision on every statement are the author's. The computations can be re-run from the ancillary files (Section~\ref{a:sec:anc}). The Lean formalisation described in Section~\ref{m:sec:lean} was written in the same way. The author takes full responsibility for the content.
\end{quote}
\section{Introduction}\label{m:sec:intro}

Alman and Vassilevska Williams~\cite{ZT}, below ``the preprint'' (we refer to its version v1 of 5 October 2026), give the first polynomial improvements over the textbook algorithms for 3SUM and APSP: deterministic algorithms with exponents $1.9992$ for 3SUM, $2.9983$ for Exact Triangle and $2.99942$ for APSP. The engine of their result is a matrix theorem: a prescribed set of at most $N^2/\sqrt D$ entries of the product of an $N\times D$ and a $D\times N$ integer matrix is computed deterministically in time $O(N^2/D^\gamma)$ for a constant $\gamma>0$, polynomially below $N^2$, whenever $D\le N^\varepsilon$ for some $\varepsilon$ below a limit $\et=0.1204\ldots$. Deterministic reductions from 3SUM, Exact Triangle and APSP to the lopsided sparse triangle problem, which the matrix theorem solves, turn this into the exponents. Its concluding section names two places where the exponents can be improved: the matrix theorem, for which it asks for other base identities ``to handle $D>N^{0.12}$'', and the reductions, of which ``many \ldots\ lose a constant fraction of the saving in the exponent'': ``Which of them can be improved?''~\cite[Section~6]{ZT}. This paper takes up both questions, and a third: which other problems the method reaches.

The first subject of this paper is the matrix theorem. Part~\hyperref[m:partI]{I} replaces the preprint's ten-term identity by Schönhage's five-term partial-product tensor and extends the range to every $\varepsilon<\es=0.172\ldots$, Coppersmith's 1982 exponent from that tensor, and at the preprint's threshold $N\ge D^{18}$ replaces preprocessing $O(N^2/D^{0.063})$ and query time $O(D^{0.437})$ by $O(N^2/D^{0.092})$ and $O(D^{0.403})$ (Theorem~\hyperref[m:introA]{A}). As the preprint observes, any improvement to the matrix theorem ``carries over to all the other speedups'' it obtains~\cite[Section~6]{ZT}, so Part~\hyperref[m:partI]{I} improves each of them. We work out Online Set Disjointness (Theorem~\hyperref[m:introA]{A}: the matrix theorem's data structure answers it, with $0/1$ matrices), the hinted OMv conjectures (Section~\ref{m:sec:omv}), and 3SUM, Exact Triangle and APSP (Section~\ref{x:sec:consequences}). The real-valued versions of 3SUM, Exact Triangle, the $(\min,+)$-product and APSP (the preprint's Theorem~35) follow by inserting the new matrix theorem into the preprint's argument unchanged, and we do not restate them. The other consequences gain from both parts and are listed below. Part~\hyperref[m:partI]{I} also answers in part the preprint's question whether other base identities could serve: within a natural class, most cannot (partly proved, partly numerical; Section~\ref{m:sec:barriers}).

Part~\hyperref[m:partII]{II} concerns the second step. Throughout, the \emph{saving} of a bound $n^{2-\delta}$ or $n^{3-\delta}$ is $\delta$; the preprint's reductions keep a quarter, a half and a sixth of what the matrix theorem saves (the preamble of Part~\hyperref[m:partII]{II} recalls where these fractions come from), and ours keep all of it, at the price of calling the counting version of the lopsided sparse triangle problem adaptively (Theorems~\hyperref[m:introB]{B} and~\hyperref[m:introC]{C}).

Each part improves the preprint's exponents on its own (Part~\hyperref[m:partII]{II} only for very large $n$, Section~\ref{m:sec:intro:reductions}); together they give deterministic 3SUM in time $n^{1.99450+o(1)}$, and Exact Triangle, $(\min,+)$-product and APSP in time $n^{2.99450+o(1)}$ (Corollary~\hyperref[m:introD]{D} in Section~\ref{m:sec:intro:comb}). In terms of the saving this is about seven times the preprint's $0.0008$ for 3SUM, more than three times its $0.0017$ for Exact Triangle, and more than nine times its $0.00058$ for APSP (more than three times the half that its footnote~10 sketches for APSP).

Through known reductions, each part on its own also improves the remaining consequences that the preprint draws, and we do not restate them: those it proves in its Section~5 (Directed Unweighted APSP, through square $(\min,+)$-products in the proof of its Theorem~34; Zero-, Min- and Max-Weight $k$-Clique, through Exact Triangle in its Corollary~39), and those it lists in its Section~1.2 as consequences of known reductions (Tree Edit Distance, the $(\min,+)$-convolution class, Knapsack, the counting versions). Each of them calls an integer algorithm for 3SUM, Exact Triangle, the $(\min,+)$-product or APSP as a black box, and both parts give faster algorithms for these four (Part~\hyperref[m:partII]{II} directly, Theorems~\hyperref[m:introB]{B} and~\hyperref[m:introC]{C}). For the $k$-Clique problems the saving is $\lfloor k/3\rfloor$ times that of Exact Triangle, as in the preprint's Corollary~39, and Tree Edit Distance gets the exponent of APSP, as in the preprint; for the others the preprint states the consequence only qualitatively, as a truly subcubic or truly subquadratic algorithm, and so do we. Part~\hyperref[m:partII]{II} does not apply to the real-valued versions as it stands, since its reductions hash integers modulo primes; whether reductions for real inputs can also keep the whole saving we leave open.

Over $\mathbb F_2$ the same method improves on the preprint's Remark 23 on 3XOR: our reduction keeps the whole saving, where the route that the remark sketches keeps a quarter, and solves 3XOR deterministically in time $O(n^{1.99451})$, against the $O(n^{1.9992})$ that the remark's route would give\footnote{What the remark's route gives when carried out; not proved here (Section~\ref{x:sec:xor}).} (Theorem~\hyperref[m:introE]{E}). Section~\ref{x:sec:novelty} lists what is new and what we take from the preprint and from earlier work.

The preprint's method reaches beyond the three problems in its title, and Sections~\ref{g:sec:groups} and~\ref{g:sec:coord} ask how far. Before the preprint, 3SUM and 3XOR were conjectured to need quadratic time, and lower bounds for other problems on three lists of vectors were proved under these conjectures; Orthogonal Vectors, which is quadratic-hard under the Strong Exponential Time Hypothesis~\cite{Wil05}, is listed by the preprint among the conjectures it leaves unaffected (its Figure~1). Which problems of this kind does the method make truly subquadratic, with reductions of the kind the preprint introduced, and which stay hard?

We show that for 3SUM over a finite abelian group the answer depends on the group: the method solves 3SUM over every elementary abelian group $\Fp^d$, with the exponents of 3XOR for fixed $p$, and, with hashes that adapt to the instance, over every group of exponent $p^2$, such as $\Z_4^d$, with half of the saving (Theorems~\hyperref[m:introF]{F} and~\hyperref[m:introG]{G}); higher exponents remain open. For the coordinatewise problems, where each coordinate triple of the three vectors must lie in a fixed relation $P$, we prove a dichotomy (Theorem~\hyperref[m:introH]{H}): for every Boolean $P$ and every functional $P$ over three letters, either $P$ is the solution set of an equation over $\Z$, $\F$ or $\mathbb F_3$ and the problem is truly subquadratic, unconditionally, or it is quadratic-hard under the Orthogonal Vectors hypothesis, which the preprint does not touch. The dichotomy stops at four letters; the triangle versions satisfy it one level up, between truly subcubic and cubic, and Exact Triangle over $\Z_4^d$ is open (Section~\ref{m:sec:intro:beyond}).

\begin{table}[t]
\centering\small
\renewcommand{\arraystretch}{1.15}\setlength{\tabcolsep}{4pt}
\begin{tabular}{@{}lllll@{}}
 & \cite{ZT} & Part~\hyperref[m:partI]{I} alone & Part~\hyperref[m:partII]{II} alone & Parts~\hyperref[m:partI]{I} and~\hyperref[m:partII]{II}\\\hline
3SUM (\S\ref{s:sec:statement}) & $O(n^{1.9992})$ & $O(n^{1.99863})$ & $n^{1.99650+o(1)}$ & $n^{1.99450+o(1)}$\\
Exact Triangle (\S\ref{e:sec:statement}) & $O(n^{2.9983})$ & $O(n^{2.9973})$ & $n^{2.99650+o(1)}$ & $n^{2.99450+o(1)}$\\
APSP (\S\ref{e:sec:apsp}) & $O(n^{2.99942})$ & $O(n^{2.99909})$ & $n^{2.99650+o(1)}$ & $n^{2.99450+o(1)}$\\
3XOR (\S\ref{x:sec:xor}) & $O(n^{1.9992})^\dagger$ & $O(n^{1.99863})^\dagger$ & $O(n^{1.99651})$ & $O(n^{1.99451})$\\
3SUM over $\Fp^d$, fixed $p$ (\S\ref{g:sec:elementary}) & -- & -- & $O(n^{1.99651})$ & $O(n^{1.99451})$\\
\begin{tabular}[t]{@{}l@{}}Exact Triangle over $\Fp^d$, fixed $p$\\\quad(\S\ref{g:sec:triangle:el})\end{tabular} & -- & -- & $O(n^{2.99651})$ & $O(n^{2.99451})$\\
\begin{tabular}[t]{@{}l@{}}3SUM over groups of exponent $p^2$\\\quad(\S\ref{g:sec:G}; half of $\varepsilon\gamma$)\end{tabular} & -- & -- & $O(n^{1.99826})$ & $O(n^{1.99726})$\\\hline
$\varepsilon\gamma$ & $0.0035$ & $0.0055$ & $0.0035$ & $0.0055$\\
saving kept & $1/4,1/2,1/6$ & $1/4,1/2,1/6$ & all & all
\end{tabular}
\caption{Deterministic exponents, from Corollaries~\ref{x:cor:fg} and~\ref{x:cor:xor}; ``saving kept'' is the fraction of $\varepsilon\gamma$ kept for 3SUM, Exact Triangle and APSP, and Section~\ref{x:sec:xor} keeps all of it for 3XOR as well. For 3XOR the preprint's Remark 23 sketches a route in one sentence, without an exponent; the entries marked $\dagger$ are what that route would give when carried out (Section~\ref{x:sec:xor}). The three rows on groups are Corollaries~\ref{g:cor:explicitA}, \ref{g:cor:explicitT}(a) and~\ref{g:cor:explicitG} and the remarks after them, with counting calls for $\Fp^d$ (3SUM and Exact Triangle) and one detection call per sub-instance for 3SUM over groups of exponent $p^2$, which keeps half of $\varepsilon\gamma$; the preprint has no reduction for these problems, so their first two columns are empty (--); for 3SUM over $\Fp^d$ one algorithm serves every prime in $n^{1.99450+o(1)}$ (Theorem~\ref{g:thm:App}(b)). The section after each problem is where its reductions are proved.}
\label{tab:summary}
\end{table}

\subsection{The matrix theorem}\label{m:sec:intro:matrix}

The matrix theorem of the preprint computes a prescribed set $W$ of entries of a thin product $XY$, $X\in\Z^{N\times D}$, $Y\in\Z^{D\times N}$, in time polynomially below $N^2$. The preprint states it in a data-structure form (preprocess, then query single entries; its Theorem 24), in a form for prescribed entries (its Theorem 25), and in an explicit instance at $N\ge D^{18}$ (its Corollary 26), which is the one its reductions use; the form for prescribed entries is proved through the data-structure form, with one query per wanted entry (the preprint's proof of Theorem 25), in which the query time enters through $|W|D^{q}\le N^2/D^{1/2-q}$, and the data-structure form is needed in its own right only for the hinted OMv conjectures (Section~\ref{m:sec:omv}); Section~\ref{c:sec:prelim} quotes all three. Both theorems hold for every $\varepsilon<\et=\ln4/(5\ln10)=0.1204\ldots$, where $D\le N^\varepsilon$.\footnote{The preprint writes $\varepsilon^*$ for $\et$; we add the index to set it apart from our $\es$.} Part~\hyperref[m:partI]{I} extends the range to Coppersmith's exponent:
\phantomsection\label{m:introA}\begin{theoremA}[Theorem~\ref{m:thm:general} and Corollary~\ref{m:cor:explicit}]
\cite[Theorems 24 and 25]{ZT} hold for every $\varepsilon<\es=\ln4/(5\ln5)=0.17227\ldots$ (we state them with the logarithmic factors absorbed into $\gamma$, as the preprint notes they can be). Explicitly, \cite[Corollary 26]{ZT} holds with preprocessing $O(N^2/D^{0.092})$ and queries $O(D^{0.403})$; for $N\ge D^{14}$ it holds with preprocessing $O(N^2/D^{0.077})$ and queries $O(D^{0.421})$.
\end{theoremA}
As in the preprint, the saving $\gamma$ tends to $0$ as $\varepsilon$ approaches the limit; Table~\ref{m:tab:rates} gives values. Table~\ref{tab:matrix} sets the two matrix theorems side by side. At $N\ge D^{14}$ the preprint states no corollary; the entry of the table is the best value of the preprint's inequality at this threshold, $\gamma\approx0.045$ at $q=0.43$ (Table~\ref{m:tab:rates}; its own Table 2 lists $0.035$ at $c=15$), a \emph{rate value}, that is, a numerical optimum of a closed-form bound rather than a theorem (Remark~\ref{m:rem:exact}).

With Corollaries~\ref{m:cor:explicit} and~\ref{m:cor:tradeoff} in place of the preprint's Corollaries 26 and 31, its refutation of the hinted OMv conjectures 5.2 and 5.7 of van den Brand, Nanongkai and Saranurak~\cite{BNS19} extends from $\tau<0.1204$ to every $\tau<\es\approx0.17227$ (Corollary~\ref{m:cor:omv}). For their Conjecture 5.12, with two hints of sizes $n^{\tau_1}$ and $n^{\tau_2}$, the preprint's refutation needs the first hint to be polynomially smaller than the second, $\tau_1<\et\tau_2$. By using a phase that the preprint's proof leaves idle, Conjecture 5.12 fails for every $\tau_1<\es$, whatever $\tau_2$, and, with the blocks of the preprint's proof taken on the other side, for every $\tau_2<\es\tau_1$ (Proposition~\ref{m:rem:omv512}). The main lower bounds of~\cite{BNS19}, at $\tau\approx0.53$ and $\tau_1+\tau_2\approx1.41$, remain out of reach.

\begin{table}[ht]
\centering\small
\renewcommand{\arraystretch}{1.15}\setlength{\tabcolsep}{4pt}
\begin{tabular}{@{}>{\raggedright}p{0.24\textwidth}>{\raggedright}p{0.335\textwidth}>{\raggedright\arraybackslash}p{0.375\textwidth}@{}}
 & \cite{ZT}, Schönhage's identity & Part~\hyperref[m:partI]{I}, partial-product tensor\\\hline
terms per level (private) & $10$ ($1$) & $5$ ($1$)\\
ratio $c=L/m$ needed & $c>10$ & $c>5$ (Corollary~\ref{m:cor:tradeoff})\\
limit of $D\le N^{\varepsilon}$ & $\et=\ln4/(5\ln10)=0.1204$ & $\es=\ln4/(5\ln5)\approx0.17227$\\
$N\ge D^{18}$: prep., query & $N^2/D^{0.063}$, $D^{0.437}$ (Cor.~26) & $N^2/D^{0.092}$, $D^{0.403}$ (Cor.~\ref{m:cor:explicit}(a))\\
$N\ge D^{14}$: prep., query & $N^2/D^{0.045}$, $D^{0.43}$ (optimised; its Tab.~2 at $c=15$: $0.035$) & $N^2/D^{0.077}$, $D^{0.421}$ (Cor.~\ref{m:cor:explicit}(b))\\
Conj.~5.2, 5.7 refuted for & $\tau<0.1204$ (Cor.~40) & $\tau<0.17227$ (Cor.~\ref{m:cor:omv})\\
Conj.~5.12 refuted for & $\tau_1<0.1204\,\tau_2$ (Cor.~40) & $\tau_1<0.172$ or $\tau_2<0.172\,\tau_1$ (Prop.~\ref{m:rem:omv512})
\end{tabular}
\caption{The matrix theorem: the preprint's numbers and those of Part~\hyperref[m:partI]{I}, which runs the tensor at $2L$ levels in two halves of $L$ (Section~\ref{m:sec:squared}); $c=L/m$ per half. The conjectures (hinted OMv) are those of~\cite{BNS19}; Cor.~26, Cor.~40 and Tab.~2 in the middle column are the preprint's; prep.\ is the preprocessing time.}
\label{tab:matrix}
\end{table}

The preprint's base is Schönhage's identity for $\langle3,1,3\rangle\oplus\langle1,4,1\rangle$ in ten terms (its Lemma 6), and the preprint asks in its Section 6 whether other identities could serve: ``to handle $D>N^{0.12}$, one could look for new base identities that compute longer inner products per multiplication than Schönhage's \ldots\ It is not immediately clear how to make use of other tensors, since our analysis uses specific properties of Schönhage's identity, such as its sparsity and the way the leaves of its recursion are shared among the output entries \ldots\ If other tensors cannot be used in this way, it would be interesting to understand why.''

Part~\hyperref[m:partI]{I} answers with one tensor that can be used: Schönhage's five-term partial-product tensor, the construction behind Coppersmith's first proof of $\alpha>0.172$ \cite[p.~467]{Cop82}. Per level, read with its first factor as the output, it computes an outer product $\langle2,1,2\rangle$ and an inner product of length $2$ in five terms. It has the property that the preprint's pruning (computing only the terms that the wanted entries need) uses: one term $P_0$ that only the inner output needs, its \emph{private} term, and one term $P_{ij}$ for each outer output.

It also has a defect that Schönhage's ten-term identity does not have: one variable belongs to both the inner and the outer product. In consequence the products of one run of the recursion overlap and, unlike in the preprint (the paragraph after its Lemma 9: ``The $K$ products can be completely independent''), they cannot have independent right factors. We restore independent factors in two steps. We re-index the inputs through $\F$-linear maps that are bijective where it matters, and we tensor the run with a copy in which left and right inputs are exchanged, which is Coppersmith's last step. The result has the shape the preprint's analysis needs: many independent products of blocks of $X$ and $Y$, each left factor depending only on its block of rows and each right factor only on its block of columns, with every wanted entry the sum of a fixed number of outputs ($2m+1$, one per run of the recursion). From there on the preprint's Sections 2.4 and 4 apply with changed constants.

Within a natural class, most other identities cannot be used (Section~\ref{m:sec:barriers}, partly proved in Appendix~\ref{a:sec:barriers}, partly numerical). In particular, an exact identity of the same shape, with a private term for the inner output and one term for each outer output, whose inner variables are disjoint from the outer ones and whose error does not reach the outputs the preprint reads (Appendix~\ref{a:sec:exact}), has inner length at most $1$ (Proposition~\ref{a:prop:exact}); the shared variable is the price of inner length $2$. Measured by the rate $\varepsilon\gamma$ it gives, our tensor is the best known base identity for the preprint's method: no decomposition of the shape of any of the $145$ full products found by computer search gives a saving, over any field and of any rank (Corollary~\ref{a:cor:145}), Vrana's decomposition gives none (Proposition~\ref{a:prop:vrana}), and the other identities examined score lower or give none (numerical; Appendix~\ref{a:sec:overview}). It is not shown to be optimal: the thin full products that Appendix~\ref{a:sec:full} does not exclude, the smallest being $\langle34,2,35\rangle$, are open, and identities of other forms were not examined.

The preprint mentions the algorithm this tensor comes from. It says that \cite{Cop82} contains two algorithms, the first of which ``achieves an $n^2\,\mathrm{poly}\log n$ running time for $n\times n^{0.172}\times n$ matrix multiplication'' and is not used there, and that the preprint builds on the second, ``in a way which ultimately achieves the worse bound $\alpha>0.12$'' \cite[Section 1.1]{ZT}. Part~\hyperref[m:partI]{I} brings the preprint's pruned method to the first algorithm's tensor, and the limit of the method becomes exactly that algorithm's exponent $\ln4/(5\ln5)=0.17227\ldots$ (Proposition~\ref{m:prop:ceiling}); at $L=5m$ the construction reproves Coppersmith's bound with exact multiplications and without his constant matrices (Section~\ref{m:sec:ceiling}).

\subsection{Reductions that keep the whole saving}\label{m:sec:intro:reductions}

Part~\hyperref[m:partII]{II} assumes the statement $\Hyp(\varepsilon,\gamma)$ (Definition~\ref{c:hyp}): the counting version $\#\Lop(N,D)$ of Lopsided All-Edges Sparse Triangle, with at most $N^2/\sqrt D$ query pairs, can be solved deterministically in time $O(N^2/D^\gamma)$ whenever $D\le N^\varepsilon$. The preprint's Corollary 26 gives $\Hyp(1/18,0.063)$, and Corollary~\ref{m:cor:explicit}(b) gives $\Hyp(1/14,0.077)$. So $\Hyp(\varepsilon,\gamma)$ is not a hypothesis but a name for what a matrix theorem provides: every theorem below that assumes it holds unconditionally with these parameters, and the explicit exponents stated with it are unconditional. $\Hyp^{\det}(\varepsilon,\gamma)$ is the same statement for the detection problem (Section~\ref{c:sec:prelim}). The preprint's reductions keep a quarter (3SUM), a half (Exact Triangle) and a sixth (APSP) of the saving $\varepsilon\gamma$. The reductions of Part~\hyperref[m:partII]{II} keep all of it, up to a factor $n^{o(1)}$.

They work as follows. Where the preprint's reductions use a single prime modulus, and the reduction of Kopelowitz, Pettie and Porat~\cite{KPP16} uses random hash functions, ours run $O(\log n)$ stages, each with a composite modulus. Each modulus is built one prime at a time, and each prime is chosen to minimise the number of false positives among the candidates that survived the previous stages. These restricted counts are exactly what the counting algorithm of $\Hyp(\varepsilon,\gamma)$ returns. For the all-edges version of Exact Triangle, the last round of each stage certifies the surviving incidences (pairs of an edge $(a,b)$ and a piece, one of the blocks into which the third part is cut) as true or false from the counts under all candidate primes. Building a modulus one prime at a time goes back to Chan and Lewenstein~\cite{CL15} and Chan and He~\cite{CH20}, choosing each prime to minimise the false positives to Fischer, Kaliciak and Polak~\cite{FKP24}, and rounds that build fresh moduli on the survivors of earlier rounds appear in~\cite{CL15} and in Fischer, Jin and Xu~\cite{FJX25}, whose counts range over the whole instance and are computed by FFT. What is new is that the counts are taken only over the survivors of the earlier moduli and are supplied by the counting algorithm, without a quadratic or cubic counting cost (Section~\ref{x:sec:novelty}).

The preprint points to two of the ingredients. Its footnote~9 (in the proof of its Theorem~17) observes that its Exact Triangle reduction can be made combinatorial by counting the false positives with the algorithm for $\#\Lop$ itself, ``for instance, by choosing the modulus one prime factor at a time as in [FKP24, Lemma~3.2]''; with a single modulus that reduction still keeps half of the saving. Its footnote~10 (in its Section~6) observes that APSP keeps all of the saving of Exact Triangle when it goes through the all-edges version of Exact Triangle, and that the reduction of~\cite{KPP16} to offline Set Disjointness keeps all of 3SUM's saving but is randomized, while its derandomization~\cite{FKP24} keeps far less. What this paper adds is the 3SUM reduction, the reductions from Exact Triangle and from its all-edges version that keep the whole saving instead of half, and their composition with Part~\hyperref[m:partI]{I}. The price is threefold: the called algorithm must count; the calls are adaptive, while the preprint's reduction is not; and the time carries a factor $(\log n)^{r+O(1)}$ against a loss $n^{\varepsilon/r}$, where $r$ is the number of primes in a modulus. With the preprint's Corollary 26 as the counting algorithm, the gain over the preprint therefore appears only once $n$ has at least about $200{,}000$ decimal digits for 3SUM and $210{,}000$ for Exact Triangle, whatever $r$ is chosen (Sections~\ref{s:sec:numbers} and~\ref{e:sec:numbers}); in the same way, the combination (Corollary~\hyperref[m:introD]{D}) improves on Part~\hyperref[m:partI]{I} alone only for very large $n$. This caveat concerns Part~\hyperref[m:partII]{II} only: Part~\hyperref[m:partI]{I} alone gives 3SUM in $O(n^{1.99863})$ with no growing polylogarithmic factor.

\phantomsection\label{m:introB}\begin{theoremB}[Theorem~\ref{s:thm:main3sum}, Corollaries~\ref{s:cor:uniform} and~\ref{s:cor:variants}]
Assume $\Hyp(\varepsilon,\gamma)$ with $0<\varepsilon,\gamma\le\frac12$, and let $\nu\ge1$ be a constant. Then 3SUM on a set of at most $n$ integers in $[1,n^\nu]$ can be solved deterministically in time $n^{2-\varepsilon\gamma+o(1)}$, and in time $O(n^{2-\varepsilon\gamma+\varepsilon/r}(\log n)^{r+4})$ for every fixed integer $r\ge1$; the standard variants take the same time up to a factor $O(\log^2n)$.
\end{theoremB}

\phantomsection\label{m:introC}\begin{theoremC}[Theorems~\ref{e:thm:A} and~\ref{e:thm:B}, Corollaries~\ref{e:cor:uniform} and~\ref{e:cor:C}]
Assume $\Hyp(\varepsilon,\gamma)$ with $0<\varepsilon,\gamma\le\frac12$, and let $\nu\ge1$ be a constant. Then Exact Triangle and its all-edges version, on $n$ vertices per part with integer weights of absolute value at most $n^\nu$, can be solved deterministically in time $n^{3-\varepsilon\gamma+o(1)}$. The same holds for the $(\min,+)$-product of two $n\times n$ integer matrices and for APSP on directed $n$-vertex graphs with no negative cycles, with entries and weights of absolute value at most $n^\nu$. For every fixed integer $r\ge1$ the time is at most $O(n^{3-\varepsilon\gamma+\varepsilon/(2r)}(\log n)^{r+5})$ in all four cases.
\end{theoremC}

The 3SUM reduction is also a deterministic version, for the counting problem and with many adaptive calls, of the reduction of Kopelowitz, Pettie and Porat~\cite{KPP16} from 3SUM to offline Set Disjointness (Theorem~\ref{s:thm:kpp}): an algorithm that counts the intersections of offline Set Disjointness with the parameters of Fischer, Kaliciak and Polak~\cite{FKP24} deterministically in time $n^{2-\eta}$ yields deterministic 3SUM in time $n^{2-\eta+\delta}$ for every $\delta>0$, whereas the derandomization of Fischer, Kaliciak and Polak~\cite{FKP24}, which needs only detection, keeps about $\eta/25$ of the saving by their proof as written.

\subsection{The combination}\label{m:sec:intro:comb}

\phantomsection\label{m:introD}\begin{corollaryD}[Corollary~\ref{x:cor:fg}]
Let $\nu\ge1$ be a constant. 3SUM on $n$ integers of absolute value at most $n^\nu$ can be solved deterministically in time $n^{1.99450+o(1)}$, and Exact Triangle (also in its all-edges version), the $(\min,+)$-product and APSP with integer weights of absolute value at most $n^\nu$ in time $n^{2.99450+o(1)}$.
\end{corollaryD}

The preprint states $O(n^{1.9992})$ for 3SUM, $O(n^{2.9983})$ for Exact Triangle and $O(n^{2.99942})$ for APSP (its Theorems 19 and 22); its abstract rounds the last to $O(n^{2.9995})$. The preprint presents its algorithms in their simplest form rather than with the best exponents it had: it reports algorithms with slightly better exponents that are significantly more complicated and that it did not include (the paragraph ``Even faster algorithms'' at the end of its Section~1.2), and its footnote~10 sketches two such improvements, adding that ``substantially better improvements than these would be more exciting''. We compare with the exponents it states. Table~\ref{tab:summary} separates the contributions of the two parts. ``Part~\hyperref[m:partI]{I} alone'' is instance (b) of Corollary~\ref{m:cor:explicit} ($N\ge D^{14}$, $\varepsilon\gamma=0.077/14=0.0055$, the largest product $\varepsilon\gamma$ among our explicit instances) with the preprint's reductions; ``Part~\hyperref[m:partII]{II} alone'' is the preprint's Corollary 26 ($N\ge D^{18}$, $\varepsilon\gamma=0.063/18=0.0035$) with the reductions of Part~\hyperref[m:partII]{II}. The preprint's reductions keep a quarter, a half and a sixth of $\varepsilon\gamma$ for 3SUM, Exact Triangle and APSP; its footnote~10 sketches a route to APSP that keeps a half, which would give $n^{3-\varepsilon'+o(1)}=n^{2.99825+o(1)}$, and against it the APSP gain of Part~\hyperref[m:partII]{II} equals its Exact Triangle gain. The factor $n^{o(1)}$ in the exponents of Part~\hyperref[m:partII]{II} is the large power of $\log n$ discussed in Section~\ref{m:sec:intro:reductions}. The preprint's own rounding, the choice of its explicit instance and the numerical optima behind these exponents are discussed at the end of Section~\ref{x:sec:consequences}.

\subsection{3XOR, 3SUM over groups and coordinatewise problems}\label{m:sec:intro:beyond}

\paragraph{3XOR.} We prove the preprint's Remark 23 and improve on it. The remark's route keeps a quarter of the saving; our reduction keeps all of it up to a factor $(\log n)^8$, and a second one keeps half of it with a single call to the detection algorithm (Theorem~\hyperref[m:introE]{E}). Over $\mathbb F_2$ the reduction of Part~\hyperref[m:partII]{II} becomes simpler: the modulus becomes a linear hash whose rows are chosen one at a time from a small-bias set, as the remark proposes, and each row halves the false positives up to a factor $1+o(1)$, so the loss $n^{o(1)}$ becomes a fixed power of $\log n$. The details are in Section~\ref{x:sec:xor}.

\phantomsection\label{m:introE}\begin{theoremE}[Theorems~\ref{x:thm:xorcount}, \ref{x:thm:xordet} and Corollary~\ref{x:cor:xor}]
Assume $\Hyp(\varepsilon,\gamma)$ with $0<\varepsilon,\gamma\le\frac12$. Then 3XOR on three lists of at most $n$ vectors of $\mathbb F_2^d$, $d=O(\log n)$, is solved deterministically in time $O(n^{2-\varepsilon\gamma}(\log n)^8)$, and with one call to the detection algorithm in time $O(n^{2-\varepsilon\gamma/2}\log n)$. With Corollary~\ref{m:cor:explicit}(b) this is $O(n^{1.99451})$ and $O(n^{1.99726})$; with the preprint's Corollary 26 it is $O(n^{1.99651})$ and $O(n^{1.99826})$.
\end{theoremE}

\paragraph{3SUM over groups and coordinatewise problems.} Sections~\ref{g:sec:groups} and~\ref{g:sec:coord} ask how far the method of Section~\ref{x:sec:xor} reaches. 3SUM over a finite abelian group $G$ asks for $x_1+x_2+x_3=0$ with $x_i$ from three lists of $n$ elements of $G$, $|G|\le n^{O(1)}$; 3XOR is the case $G=\F^d$. Over the elementary groups $\Fp^d$ the method carries over for every prime $p$ (Theorem~\hyperref[m:introF]{F}); for $p=2$ these are the bounds of Theorem~\hyperref[m:introE]{E}, so Section~\ref{g:sec:groups} generalises Section~\ref{x:sec:xor} and does not improve its exponents. Groups with a large non-elementary part have no family of homomorphic hashes of the kind the method uses (Proposition~\ref{g:prop:C}), but hashes that adapt to the instance reach every group of exponent $p^2$ (Theorem~\hyperref[m:introG]{G}); groups of higher exponent and of mixed primes remain open (Section~\ref{g:sec:beyond}).

The coordinatewise problem $\Prob(P)$, for a fixed relation $P\subseteq\Sigma^3$, asks for $x_i$ from three lists of $n$ vectors of $\Sigma^d$, $d\le c\log n$, with $(x_1[j],x_2[j],x_3[j])\in P$ for every coordinate $j$; 3XOR and Orthogonal Vectors are examples. For two-letter alphabets, and for functional relations over three letters, Section~\ref{g:sec:coord} proves a dichotomy at the quadratic level (Theorem~\hyperref[m:introH]{H}). For Boolean relations it refines, in the dense model, the cubic-level dichotomy of Bringmann, Fischer and Künnemann~\cite{BFK19}: of the $21$ orbits that their Theorem~3 proves subcubic, $10$ are truly subquadratic and $11$ are quadratic-hard under the Orthogonal Vectors hypothesis, and their quadratic lower bound for the parity relations rested on a 3-XOR hypothesis, which the preprint's Remark~23 refutes in its form for $n$ vectors of dimension $O(\log n)$ and Section~\ref{x:sec:xor} proves with an explicit exponent. A relation has a gadget for Orthogonal Vectors of some size exactly when it has one of size~$1$, and exactly when one of its restrictions to two letters per position is a non-representable Boolean relation (Proposition~\ref{g:prop:F}(d)). At four letters the three groups no longer suffice: the Latin squares isotopic to $\Z_4$ are representable only over groups that contain $\Z_4$, and Theorem~\hyperref[m:introG]{G} solves them. There the dichotomy also stops: $4$ of the $167342$ functional relations over four letters up to symmetry lie in neither class, and for relations that are not functional this happens already at three letters, with $4$ orbits (Proposition~\ref{g:prop:F}); these eight orbits all reduce to one problem, whose complexity is open (Proposition~\ref{g:prop:thirdT}).

Section~\ref{g:sec:triangle} treats the triangle versions, Exact Triangle over $G$ and the coordinatewise triangle problem $\mathrm{Tri}(P)$, in which the group elements or the letters sit on the edges of a complete tripartite graph and the baseline is cubic. Exact Triangle over $\Fp^d$ keeps the whole saving, uniformly in $p$, which gives $O(n^{2.99451})$ unconditionally for fixed $p$ (Theorems~\ref{g:thm:triFp} and~\ref{g:thm:triuniform}, Corollary~\ref{g:cor:explicitT}). We could not carry the method of Theorem~\hyperref[m:introG]{G} over to triangles (Remark~\ref{g:rem:trinogo} says why), and Exact Triangle over $\Z_4^d$ is open (Open problem~\ref{g:prob:triexp}); unless the Nondeterministic Strong Exponential Time Hypothesis (NSETH) of Carmosino et al.~\cite{CGIMPS16} fails, its cubic hardness cannot be proved by a deterministic reduction from CNF-SAT (Proposition~\ref{g:prop:tricert}). $\mathrm{Tri}(P)$ has the dichotomy of Theorem~\hyperref[m:introH]{H} at the cubic level, for the same relations and with the Orthogonal Vectors hypothesis on the hard side (Theorem~\ref{g:thm:triD}), and the Latin squares isotopic to $\Z_4$ fall into the open case (Proposition~\ref{g:prop:triF}).

\phantomsection\label{m:introF}\begin{theoremF}[Theorems~\ref{g:thm:A}, \ref{g:thm:B}, \ref{g:thm:App} and Corollary~\ref{g:cor:explicitA}]
Let $p$ be a fixed prime, $c>0$ and $0<\varepsilon,\gamma\le\frac12$. If $\Hyp^{\det}(\varepsilon,\gamma)$ holds, 3SUM over $\Fp^d$ with $p^d\le n^c$ is solved deterministically with one detection call in time $O(n^{2-\varepsilon\gamma/2}\log n)$, and if $\Hyp(\varepsilon,\gamma)$ holds, with adaptive counting calls in time $O(n^{2-\varepsilon\gamma}(\log n)^8)$. If $\Hyp(\varepsilon,\gamma)$ holds, one algorithm, uniform in $p$, solves 3SUM over $\Fp^d$ for every prime $p$ with adaptive counting calls in time $n^{2-\varepsilon\gamma+o(1)}$. With Corollary~\ref{m:cor:explicit}(b) these bounds are $O(n^{1.99726})$, $O(n^{1.99451})$ and $n^{1.99450+o(1)}$; the uniform algorithm of Theorem~\ref{g:thm:App}(a), which makes one detection call for the small primes, runs in time $O(n^{1.99863})$.
\end{theoremF}

\phantomsection\label{m:introG}\begin{theoremG}[Theorem~\ref{g:thm:G} and Corollary~\ref{g:cor:explicitG}]
Assume $\Hyp^{\det}(\varepsilon,\gamma)$ with $0<\varepsilon\le\frac14$ and $0<\gamma\le\frac12$, and let $p$ be a fixed prime and $c>0$. 3SUM over every finite abelian group $G$ of exponent dividing $p^2$, $|G|\le n^c$, for example $\Z_4^d$, is solved deterministically in time $O(n^{2-\varepsilon\gamma/2}(\log n)^4)$. With Corollary~\ref{m:cor:explicit}(b) this is $O(n^{1.99726})$.
\end{theoremG}

\phantomsection\label{m:introH}\begin{theoremH}[Theorems~\ref{g:thm:D} and~\ref{g:thm:E}]
Let $P\subseteq\{0,1\}^3$, or $P=\{(a,b,g(a,b))\}$ for a map $g\colon\{0,1,2\}^2\to\{0,1,2\}$. Exactly one of the following holds. Either $P$ is the solution set of an equation $\varphi_1(a)+\varphi_2(b)+\varphi_3(c)=t$ over $\Z$, $\F$ or $\mathbb F_3$, and $\Prob(P)$ is solved deterministically in time $O(n^{1.99451})$ for every $c$; or $P$ has a one-coordinate gadget for Orthogonal Vectors, and, unless the Orthogonal Vectors hypothesis fails, for every $\delta>0$ there is a $c$ for which $\Prob(P)$ has no $O(n^{2-\delta})$ algorithm. The classification rests on an exhaustive computation with certified verdicts (Section~\ref{g:sec:computer}).
\end{theoremH}

\subsection{How to read the paper}\label{m:sec:intro:guide}

Parts~\hyperref[m:partI]{I} and~\hyperref[m:partII]{II} are independent of each other: they meet only in the statement $\Hyp(\varepsilon,\gamma)$ of Section~\ref{c:sec:prelim}, which Part~\hyperref[m:partI]{I} proves for new parameters and Part~\hyperref[m:partII]{II} assumes. For the matrix theorem, read Section~\ref{c:sec:prelim}, the preamble of Part~\hyperref[m:partI]{I}, which recalls the preprint's framework, and Sections~\ref{m:sec:tensor}--\ref{m:sec:explicit}; Section~\ref{m:sec:omv} (hinted OMv) and Section~\ref{m:sec:barriers} with Appendix~\ref{a:sec:barriers} (other identities) can be read separately. For the reductions, read Section~\ref{c:sec:prelim} and Part~\hyperref[m:partII]{II}, whose preamble gives the idea in one paragraph; Section~\ref{s:sec:kpp} is used nowhere else. Section~\ref{x:sec:xor} follows the idea of Part~\hyperref[m:partII]{II} and, for its triangle version, Section~\ref{e:sec:statement}; Section~\ref{g:sec:groups} builds on Section~\ref{x:sec:xor}, Section~\ref{g:sec:coord} on Sections~\ref{x:sec:xor} and~\ref{g:sec:groups}, and Section~\ref{g:sec:triangle} on Sections~\ref{e:sec:statement} and~\ref{x:sec:xor}--\ref{g:sec:coord}.

\paragraph{Status.} Every theorem, lemma, proposition and corollary is proved, some with computer assistance: the classifications of Section~\ref{g:sec:coord} rest on exhaustive computations with certified verdicts, and a few rank computations of Part~\hyperref[m:partI]{I} on exact computer calculations (Section~\ref{a:sec:anc}). Rate values, the scores of Section~\ref{m:sec:barriers} and statements marked numerical are computed and are not theorems. The mathematics outside Sections~\ref{x:sec:xor}--\ref{g:sec:triangle} is also formalised in Lean~4, with the exceptions that Section~\ref{m:sec:lean} lists; of Section~\ref{x:sec:xor} the correctness of the algorithms, their counts and the exponents are formalised, and of Sections~\ref{g:sec:groups}--\ref{g:sec:triangle} the correctness and the counts of the main algorithms and the dichotomy of Theorem~\hyperref[m:introH]{H} are formalised (Section~\ref{m:sec:lean} lists what is not).

\paragraph{Organisation and notation.} Section~\ref{c:sec:prelim} quotes the preprint's definitions and its matrix theorem and defines the statement $\Hyp(\varepsilon,\gamma)$. Part~\hyperref[m:partI]{I} (Sections~\ref{m:sec:tensor}--\ref{m:sec:barriers}) proves the matrix theorem with Schönhage's partial-product tensor and extends the preprint's refutation of the hinted OMv conjectures (Sections~\ref{m:sec:tensor}--\ref{m:sec:omv}), and Section~\ref{m:sec:barriers} explains why most other base identities cannot replace the tensor; Appendix~\ref{a:sec:barriers} gives the details and proves those barrier statements that are proved rather than numerical. Part~\hyperref[m:partII]{II} (Sections~\ref{c:sec:primes}--\ref{e:sec:statement}) proves the reductions given $\Hyp(\varepsilon,\gamma)$. Section~\ref{x:sec:consequences} combines the two parts, Section~\ref{x:sec:novelty} says what is known and what is new, Section~\ref{x:sec:xor} gives a fourth reduction that keeps the whole saving, for 3XOR, a problem outside the title's three, for which the preprint claims a deterministic truly subquadratic algorithm in its Remark~23 and in its introduction; the section has its own notation. Section~\ref{g:sec:groups} extends it to 3SUM over finite abelian groups (Theorems~\hyperref[m:introF]{F} and~\hyperref[m:introG]{G}), Section~\ref{g:sec:coord} to the coordinatewise problems $\Prob(P)$ (Theorem~\hyperref[m:introH]{H}), and Section~\ref{g:sec:triangle} to the triangle versions of both, Exact Triangle over groups and $\mathrm{Tri}(P)$; these three sections use the notation of Section~\ref{x:sec:xor} where they overlap with it and define their own otherwise. Section~\ref{m:sec:status} describes the machine checks: the ancillary scripts of the arXiv submission and the Lean formalisation. We use the preprint's notation and its v1 numbering throughout: its results are named with its plain numbers, as in ``the preprint's Lemma 9'', while the results of this paper carry section numbers; the citation~\cite{ZT} is written out where the preprint's text is quoted, where a statement or proof is taken over from it, and in the tables. The two parts keep their own notation, each symbol being defined where it is used; in particular the letters $m$, $L$, $c$, $r$, $\Lambda$, $P$, $Q$, $T$, $\Pi$, $M$, $s$, $t$, $\mu$ and $\rho$ denote different quantities in Part~\hyperref[m:partI]{I} and in Part~\hyperref[m:partII]{II}.
\section{The preprint's problems and the statement \texorpdfstring{$\Hyp(\varepsilon,\gamma)$}{H(epsilon,gamma)}}\label{c:sec:prelim}

\paragraph{The problems.} The preprint~\cite{ZT} defines the following problems (quoted from its Section 3.1, Definitions 13 and 14).
\begin{quote}
\textbf{Definition} (Lopsided All-Edges Sparse Triangle, $\Lop(n,D)$). We are given an unweighted undirected tripartite graph with two parts $A$ and $B$ of $n$ vertices each and a middle part $M$ of at most $D$ vertices, with arbitrary edges in $M\times A$ and $M\times B$. Let $W\subseteq A\times B$ be the set of edges between $A$ and $B$. Decide, for every edge $(a,b)\in W$, whether it lies in a triangle with some vertex of $M$, that is, whether $a$ and $b$ have a common neighbor in $M$.

{\sloppy\textbf{Definition} (Lopsided All-Edges Sparse Triangle Counting, $\#\Lop(n,D)$). This is the same problem as $\Lop(n,D)$, except that for every edge $(a,b)\in W$ we ask for the number of triangles it lies in, that is, the number of common neighbors of $a$ and $b$ in $M$.\par}
\end{quote}
In matrix language (also from the preprint): with the biadjacency matrices $X\in\{0,1\}^{n\times D}$ of $A\times M$ and $Y\in\{0,1\}^{D\times n}$ of $M\times B$, $\#\Lop(n,D)$ asks for the entries $(XY)[a,b]$, $(a,b)\in W$, of the integer product. Below we write $\#\Lop(N,D)$, with $N$ for the number of vertices per side as in the matrix theorem.

\begin{definition}[$\Hyp(\varepsilon,\gamma)$]\label{c:hyp}
For $0<\varepsilon,\gamma\le\frac12$, $\Hyp(\varepsilon,\gamma)$ is the statement that there are a deterministic algorithm $\mathcal O$ and a constant $C_{\mathcal O}$ such that for all integers $N\ge1$ and $D\ge2$ with $D\le N^\varepsilon$, the algorithm $\mathcal O$ solves every instance of $\#\Lop(N,D)$ with $|W|\le N^2/\sqrt D$ in at most $C_{\mathcal O}N^2/D^\gamma$ steps of a word RAM with $O(\log N)$-bit words. We call $\mathcal O$ the \emph{counting algorithm}.
\end{definition}

The two bounds $\gamma\le\frac12$ and $\varepsilon\le\frac12$ in the definition restrict nothing that matters. For $\gamma>\frac12$ the statement would be false, since the algorithm must output $|W|=\lfloor N^2/\sqrt D\rfloor$ numbers. The bound $\varepsilon\le\frac12$ is used only to bound the input and output of each call (Lemmas~\ref{s:lem:shape} and~\ref{e:lem:shape}), and the statement with the same words and some $\varepsilon>\frac12$ implies $\Hyp(\frac12,\gamma)$.

We write $\Hyp^{\det}(\varepsilon,\gamma)$ for the statement of Definition~\ref{c:hyp} with the detection problem $\Lop(N,D)$ in place of $\#\Lop(N,D)$, and call its algorithm the \emph{detection algorithm}. It is weaker than $\Hyp(\varepsilon,\gamma)$, since a count answers detection; Part~\hyperref[m:partII]{II} assumes $\Hyp(\varepsilon,\gamma)$, and Sections~\ref{x:sec:xor}--\ref{g:sec:triangle} assume $\Hyp^{\det}(\varepsilon,\gamma)$ where one detection call suffices.

The statement $\Hyp(\varepsilon,\gamma)$ is not a conjecture but the interface between the parts of the paper, and it is proved for the parameters we use: the preprint's Corollary 26 gives $\Hyp(1/18,\,0.063)$ (Lemma~\ref{c:lem:hzt} below), and Part~\hyperref[m:partI]{I} gives $\Hyp(1/18,\,0.092)$ and $\Hyp(1/14,\,0.077)$ (Corollary~\ref{m:cor:explicit} and the paragraph after its proof). The reductions of Part~\hyperref[m:partII]{II} and of Sections~\ref{x:sec:xor}--\ref{g:sec:triangle} assume only $\Hyp(\varepsilon,\gamma)$, or $\Hyp^{\det}(\varepsilon,\gamma)$, so each of them improves automatically whenever the matrix theorem does; how much of the saving $\varepsilon\gamma$ each keeps is stated with it. The unconditional exponents of the paper come from $\Hyp(1/14,\,0.077)$, with $\varepsilon\gamma=0.0055$.

\paragraph{The matrix theorem.} With $\et:=\ln4/(5\ln10)=0.1204\ldots$ as in Section~\ref{m:sec:intro:matrix}, the preprint's data-structure form of its matrix theorem, its form for prescribed entries, and their explicit instance read:
\begin{quote}
\textbf{Theorem} (24 of~\cite{ZT}). For every $\varepsilon<\et$ and every $q>0$, there is a $\gamma>0$ such that the following holds. Given as input matrices $X\in\Z^{N\times D}$ and $Y\in\Z^{D\times N}$, where $2\le D\le N^\varepsilon$, whose entries are integers of absolute value at most $N^{O(1)}$, we can preprocess them deterministically in $O(N^2\log^2D/D^\gamma)$ time and space, after which any single entry $(XY)[I,J]$ can be computed deterministically in $O(D^q\log D)$ time.

\textbf{Theorem} (25 of~\cite{ZT}). For every $\varepsilon<\et$ and every $\kappa>0$ there is a $\gamma>0$ such that the following holds. Given as input matrices $X\in\Z^{N\times D}$ and $Y\in\Z^{D\times N}$, where $2\le D\le N^\varepsilon$, whose entries are integers of absolute value at most $N^{O(1)}$, as well as a set $W$ of at most $N^2/D^\kappa$ positions of an $N\times N$ matrix, the entries $(XY)[I,J]$, $(I,J)\in W$, can be computed deterministically in time $O(N^2\log^2D/D^\gamma)$.

\textbf{Corollary} (26 of~\cite{ZT}). Let $N\ge D^{18}$, and let $X\in\Z^{N\times D}$ and $Y\in\Z^{D\times N}$ have entries of absolute value at most $N^{O(1)}$. We can preprocess them deterministically in $O(N^2/D^{0.063})$ time and space, after which any single entry $(XY)[I,J]$ can be computed deterministically in $O(D^{0.437})$ time. Hence, for every set $W$ of positions of an $N\times N$ matrix, the entries $(XY)[I,J]$, $(I,J)\in W$, can be computed deterministically in $O(|W|\,D^{0.437}+N^2/D^{0.063})$ time, which is $O(N^2/D^{0.063})$ whenever $|W|\le N^2/\sqrt D$.
\end{quote}
\begin{lemma}\label{c:lem:hzt}
$\Hyp(1/18,\,0.063)$ holds.
\end{lemma}

\begin{proof}
Apply \cite[Corollary~26]{ZT} to the biadjacency matrices $X\in\{0,1\}^{N\times D}$ of $A\times M$ and $Y\in\{0,1\}^{D\times N}$ of $M\times B$, whose product has the numbers of common neighbours as entries. Its condition $N\ge D^{18}$ is the same as $D\le N^{1/18}$, and its last sentence computes the entries at $W$ in time $O(N^2/D^{0.063})$ whenever $|W|\le N^2/\sqrt D$.
\end{proof}

The definition speaks of $0/1$ matrices only, although the preprint's Corollary 26 and Corollary~\ref{m:cor:explicit} take integer matrices: Part~\hyperref[m:partII]{II} reduces its integer instances to $0/1$ ones by binary splitting (Lemma~\ref{s:lem:split}), so that nothing beyond the $0/1$ counting problem is asked of the counting algorithm.

\part{The matrix theorem with Schönhage's partial-product tensor}\label{m:partI}

Part~\hyperref[m:partI]{I} is written as a set of changes to the preprint's Sections 2 and 4, which the reader should have at hand. It keeps the preprint's letters and v1 numbering and proves only what changes; whatever is not mentioned is taken over unchanged. The paragraph ``The preprint's framework in brief'' recalls the notions of those two sections that Part~\hyperref[m:partI]{I} uses, the paragraph after it gives the route through Sections~\ref{m:sec:tensor}--\ref{m:sec:barriers}, and Table~\ref{c:tab:objects} lists the objects whose values or definitions change.

\paragraph{The preprint's framework in brief.} These are the notions of the preprint's Sections 2 and 4 that Part~\hyperref[m:partI]{I} uses, in its words where they are short, with the preprint's subsection for each.

\emph{The identity as a trilinear polynomial} (Section 2.2). An identity is written $\sum_\lambda\varphi_\lambda\psi_\lambda\chi_\lambda=G+E$, a sum over its \emph{terms} $\lambda$, where $\varphi_\lambda$ is a linear form in the \emph{left} variables, $\psi_\lambda$ one in the \emph{right} variables and $\chi_\lambda$ one in the \emph{output} variables; $G$ is the polynomial whose coefficients are the wanted products, an outer product $\sum_{i,j}x_iy_jz_{ij}$ and an inner product with output $z_0$, and $E$ is an error polynomial. The preprint's identity is Schönhage's for $\langle3,1,3\rangle\oplus\langle1,4,1\rangle$ (Lemma 6), with ten terms $P_{ij}$ ($i,j\le3$) and $P_0$. The variables of the inner product are \emph{inner}, those of the outer product \emph{outer}. A term $\lambda$ \emph{contributes to} an output variable $z$ if $z$ appears in $\chi_\lambda$: every term contributes to $z_0$, and only $P_{ij}$ contributes to $z_{ij}$.

\emph{Strings, arrays and the recursion} (Sections 2.3.1--2.3.2). The identity is applied at $L$ levels of recursion. A \emph{left string} of length $L$ is a string of $L$ left variables, its \emph{variable at level $\ell$} being the $\ell$-th; right strings and output strings likewise. An \emph{array} on the left strings assigns an integer to each of them, and its \emph{slice} $a_s$ at a left variable $s$ is the array $a_s[u']:=a[su']$ on the strings of length $L-1$. The recursion $\Full(a,b)$ takes a left array $a$ and a right array $b$, forms for each term $\lambda$ the arrays $A_\lambda=\sum_s\varphi_\lambda(s)\,a_s$ and $B_\lambda=\sum_t\psi_\lambda(t)\,b_t$ (this step is the \emph{encoding}), where $\varphi_\lambda(s)$ is the coefficient of $s$ in $\varphi_\lambda$, recurses on each pair $(A_\lambda,B_\lambda)$, and assembles the output array from the results as the forms $\chi_\lambda$ prescribe (the \emph{decoding}). Its calls form a recursion tree whose \emph{leaves} are the strings $\tau=\tau_1\cdots\tau_L$ of one term per level; at the leaf $\tau$ the recursion multiplies two numbers $\Phi_\tau(a)$ and $\Psi_\tau(b)$, which depend on $a$ alone and on $b$ alone. A leaf \emph{contributes to} an output string $w$ if $\tau_\ell$ contributes to $w_\ell$ at every level, and the recursion returns the array $\Mult(a,b)[w]=\sum_{\tau\text{ contributing to }w}\Phi_\tau(a)\Psi_\tau(b)$ (Lemma 7), which is also $\sum_{u,v}\prod_{\ell}\gamma(u_\ell,v_\ell,w_\ell)\,a[u]\,b[v]$ over all left strings $u$ and right strings $v$, with $\gamma(s,t,z)$ the coefficient of $stz$ in $G+E$ (Lemma 8). One execution of $\Full$ on a pair of arrays is a \emph{run}.

\emph{Products from strings, tiles and encodings} (Sections 2.3.3--2.3.4 and 2.4.1). The \emph{inner set} of a string is the set of levels at which it has an inner variable. With $m$ inner levels, a left string with inner set $Q$, $|Q|=m$, indexes an entry of an $\Nz\times D$ matrix $X_Q$, its outer part the row and its inner part the column; a right string with inner set $Q$ indexes an entry of a $D\times\Nz$ matrix $Y_Q$, and an output string with inner set $Q$ an entry of $X_QY_Q$. If $a$ and $b$ hold the entries of $K=\binom Lm$ matrices $X_Q$, $Y_Q$ side by side, one run computes all $K$ products at once: $\Mult(a,b)[w]=(X_QY_Q)[w]$ (Lemma 9). Only strings whose inner sets have exactly $m$ elements are used: the input arrays vanish on the other strings, and only the output strings whose inner sets have exactly $m$ elements are read; we call this the preprint's \emph{selection}. Since every monomial of $E$ contains an outer output and an inner input (the first observation of Section 2.2), a product of level monomials that contains a monomial of $E$ cannot lead from selected input strings to a selected output string, which is why $E$ does no harm (the proof of Lemma 9); an error with this property is called \emph{harmless} below (Appendix~\ref{a:sec:exact} makes this a condition on the directions of the error monomials). To multiply $X\in\Z^{N\times D}$ by $Y\in\Z^{D\times N}$, $X$ is cut into \emph{row blocks} of $\Nz$ rows and $Y$ into \emph{column blocks} of $\Nz$ columns, $K_0=\lfloor\sqrt K\rfloor$ consecutive blocks form a \emph{band}, a row band together with a column band is a \emph{tile}, and one run per tile computes its $K_0^2$ block products, each with its own subset $Q$; $M=K\Nz^2$ counts the output entries of all $K$ products of a run, which the preprint calls the output entries of a tile (only $K_0^2\Nz^2$ of them are entries of $XY$). A tile spans $K_0\Nz$ rows and columns; Theorem 30 assumes $N\ge\sqrt K\Nz$, the \emph{tile fit}, so that padding $N$ to a multiple of $K_0\Nz$ at most doubles it. The \emph{encoding} of a band, the result of all encoding steps on its input array, is the array of the $10^L$ numbers $\Phi_\tau(a)$, or $\Psi_\tau(b)$; it depends on the band alone, is computed once in $O(10^L)$ operations and is shared by all tiles of the band.

\emph{Pruning and the counts of leaves} (Sections 2.4.2--2.4.3). Given the set $U$ of the output strings wanted in a tile, the pruned recursion visits exactly the leaves of $\Leaves(U)$, those contributing to some string of $U$, and the sets passed to its calls have total size at most $(L+1)|\Leaves(U)|$ (Lemma 10); a call on a set $S$ takes $O(L|S|)$ operations, so the cost is $O(L^2)$ per leaf visited (Section 2.4.4). The leaves contributing to $w$ choose $P_{ij}$ wherever $w$ has $z_{ij}$ and any term at the $m$ levels of its inner set; the one that chooses $P_0$ at all of them is the \emph{private leaf} of $w$, which contributes to no other output string, and the \emph{order} of a leaf is $m$ minus the number of levels at which it chooses $P_0$. Exactly $\alpha_d$ leaves of order $d$ contribute to each output string, and a tile has $\beta_d$ leaves of order $d$ in all. Bounding the leaves of order $d$ in $\Leaves(U)$ both by $|U|\alpha_d$ and by $\beta_d$ gives $|\Leaves(U)|\le\sum_{d=0}^m\min\{|U|\alpha_d,\beta_d\}$ (Lemma 11). The $\beta_d$ decay: $\beta_d\le\rho^d\beta_0$ for the decay rate $\rho=9m/(L-m+1)$ of Section 4.3, which is below $1$ for $L\ge10m$ (eq.~(7); Section 2.4.3 has the case $L=19m$, eq.~(5), with rate $\frac12$).

\emph{Cubes, boxes and the data structure} (Sections 4.2--4.4). A \emph{cube} is a string of $L$ symbols, each a term or a star $\ast$; its leaves are obtained by replacing every star by any term, and its \emph{value} is the sum of the products $\Phi_\tau(a)\Psi_\tau(b)$ at its leaves. The cube of $w$, with stars on its inner set, has value $(X_QY_Q)[w]$. For a \emph{switching order} $t$, a \emph{box} is a cube in which (i)~at most $m-t$ symbols are $P_0$ or stars and (ii)~every star is at a lower level than every $P_0$, and the $\alpha_t$ boxes of $w$ partition its leaves of order at least $t$ (Lemmas 27 and 28). The data structure stores the encodings of all bands and, for every tile, the values of all its boxes, computed by a dynamic program from the boxes with fewer stars (Lemma 29); a query adds the products at the leaves of order below $t$ in the cube of $w$ to the values of the boxes of $w$. Theorem 30 bounds the preprocessing by the boxes, $O(Lm\,\rho^t/(1-\rho))$ per entry of $XY$, plus the encodings, $N\cdot10^L/(\sqrt K\Nz)$, and a query by $L\sum_{d\le t}\alpha_d$. Corollary 31 sets $L=cm$ and $t=\theta m$, rounded up, which turns this into the exponents $\gamma$ (preprocessing $O(N^2\log^2D/D^\gamma)$) and $q$ (query $O(D^q\log D)$) and the range $\varepsilon<R_c(\gamma)$ for $D\le N^\varepsilon$, the range coming from the condition $D^\gamma10^L/(\sqrt K\Nz)\le N$ (eq.~(10)) that keeps the encodings within the preprocessing time.

\paragraph{The route.} Part~\hyperref[m:partI]{I} changes the identity, the strings, the products of a run and the counts of leaves, and keeps the rest; the first line of each section names the part of the preprint it replaces. Section~\ref{m:sec:tensor} replaces the preprint's Section 2.2 and Lemma 6: it writes Schönhage's partial-product tensor in the preprint's notation and explains its defect, the shared variable $y_1$, without which five terms cannot give an inner product of length $2$. Section~\ref{m:sec:hash} is new: it defines the $\F$-linear re-indexing maps and the good sets on which they are bijective. Section~\ref{m:sec:squared} replaces the preprint's Sections 2.3.3 and 2.3.4, with Lemma 9 and the tiling there, and its Section 2.4.1 for the encodings: it tensors each run with its left/right-exchanged copy, the \emph{squared construction}, so that the products of a run have independent factors again, and tiles $X$ and $Y$ with it. Section~\ref{m:sec:leaves} replaces the preprint's Section 2.4.3 and Lemma 11: it counts the leaves of a squared run by their order, and the counts $\alpha'_d$, $\beta'_d$ replace $\alpha_d$, $\beta_d$. Section~\ref{m:sec:ds} replaces the preprint's Sections 4.2--4.4, that is, Lemmas 27--29, Theorem 30, Corollary 31 and Theorems 24 and 25: it runs the data structure on the squared construction and proves Theorem~\hyperref[m:introA]{A}. Section~\ref{m:sec:ceiling}, the counterpart of the paragraph ``Where $\varepsilon^*$ comes from'' in the preprint's Section 4.4, explains where the limit $\es$ comes from. Section~\ref{m:sec:explicit}, the counterpart of the preprint's Corollary 26, computes the explicit constants of Table~\ref{tab:matrix}; Section~\ref{m:sec:omv} extends its Corollary 40, the consequence for hinted OMv; and Section~\ref{m:sec:barriers} answers the question of its Section 6, why most other base identities cannot replace the tensor. Table~\ref{c:tab:objects} is for reference while reading these sections and need not be understood first.

\begin{table}[ht]
\centering\small
\renewcommand{\arraystretch}{1.12}\setlength{\tabcolsep}{4pt}
\begin{tabular}{@{}>{\raggedright}p{0.22\textwidth}>{\raggedright}p{0.33\textwidth}>{\raggedright\arraybackslash}p{0.37\textwidth}@{}}
object & \cite{ZT} & Part~\hyperref[m:partI]{I}\\\hline
base identity & Schönhage's $\langle3,1,3\rangle\oplus\langle1,4,1\rangle$, $10$ terms, disjoint variables (Lemma 6) & partial-product tensor, $5$ terms, one shared variable $y_1$ (Lemma~\ref{m:lem:identity})\\
levels, inner levels & $L$, $m$ (one run) & $2L$, $2m$: two halves of $L$, $m$; left and right exchanged in the second\\
block size $\Nz$ & $3^{L-m}$ & $4^{L-m}$\\
products per run & $K=\binom Lm$ independent $X_QY_Q$ & $\Kg^2$ block products $X_{Q_k}Y_{Q_{k'}}$ over $2m+1$ runs (Theorem~\ref{m:thm:correct})\\
tile & $K_0\times K_0$ grid, $K_0=\lfloor\sqrt K\rfloor$ & $\Kg\times\Kg$ grid, $\Kg\ge\mu_\infty^2K$ good sets\\
$M$ (entries of a tile) & $K\Nz^2$ & $\Kg^2\Nz^2$\\
encodings & $10^L$ per band & $25^L$ per band and per run\\
tile fit & $N\ge\sqrt K\Nz$ & $N\ge K\Nz$\\
leaves of order $d$ per entry & $\alpha_d=\binom md9^d$ & $\alpha'_d=\binom{2m}d4^d$\\
leaves of order $d$ in a tile & $\beta_d=\binom L{m-d}9^{L-m+d}$ & $\beta'_d=\sum\beta_{d_1}\beta_{d_2}$, $\beta_e=\binom L{m-e}4^{L-m+e}$\\
decay rate & $\rho=9m/(L-m+1)$, $L\ge10m$ & $\rho=4m/(L-m+1)$, $L\ge5m$\\
boxes & conditions (i), (ii) (Section 4.2) & (i), (ii) with $2m$, and (iii) at most $m$ per half\\
switching order & $t=\lceil\theta m\rceil$, $0<\theta<0.9$ & $t=\lceil2\theta m\rceil$, $0<\theta<0.8$\\
base and range & $\Lambda$, $R_c(\gamma)$ (eq.~(11)) & $\Lambda_c(\gamma)$, $R'_c(\gamma)$ \eqref{m:eq:Rc}\\
limit & $\varepsilon^*=\ln4/(5\ln10)$, here $\et$ & $\es=\ln4/(5\ln5)$\\
Corollary 26 instance & $c=21$, $\theta=\frac19$: $0.063$ / $0.437$ & $c=14.5$, $\theta=\frac1{19}$: $0.092$ / $0.403$; at $D^{14}$: $c=11.5$, $\theta=\frac1{18}$
\end{tabular}
\caption{Objects of the preprint whose values or definitions change in Part~\hyperref[m:partI]{I}, with the preprint's letters.}
\label{c:tab:objects}
\end{table}

\section{The tensor}\label{m:sec:tensor}

This section replaces the preprint's Section 2.2 and Lemma 6.

\paragraph{Schönhage's construction in Coppersmith's form.} Coppersmith~\cite[p.~467]{Cop82} attributes the following construction to Schönhage~\cite{Sch81}. In Coppersmith's variables it is the identity
\begin{multline*}
(a_{11}+x^2a_{12})(b_{21}+x^2b_{11})c_{11}+(a_{11}+x^2a_{13})\,b_{31}\,(c_{11}-xc_{21})+(a_{11}+x^2a_{22})(b_{21}-xb_{12})\,c_{12}\\
+(a_{11}+x^2a_{23})(b_{31}+xb_{12})(c_{12}+xc_{21})-a_{11}(b_{21}+b_{31})(c_{11}+c_{12})\\
=x^2(a_{11}b_{11}c_{11}+a_{11}b_{12}c_{21}+a_{12}b_{21}c_{11}+a_{13}b_{31}c_{11}+a_{22}b_{21}c_{12}+a_{23}b_{31}c_{12})+x^3P(a,b,c,x),
\end{multline*}
with a polynomial $P$. With the matrices
\[
A=\begin{pmatrix}a_{11}&a_{12}&a_{13}\\0&a_{22}&a_{23}\end{pmatrix},\qquad B=\begin{pmatrix}b_{11}&b_{12}\\b_{21}&0\\b_{31}&0\end{pmatrix},\qquad C=\begin{pmatrix}c_{11}&c_{12}\\c_{21}&c_{22}\end{pmatrix},
\]
the coefficient of $x^2$ is the partial matrix product $\mathrm{trace}\,(ABC)$, so the identity is an approximate evaluation of this partial product with five multiplications, and Coppersmith uses it to prove $\mathrm{Rank}(\langle N,N,N^\alpha\rangle)=O(N^2(\log N)^2)$ with $\alpha=2\log2/(5\log5)=0.17227$~\cite[p.~467]{Cop82}. His remarks on p.~469 place it in Schönhage's family with two parameters $k$ and $n$, whose exponent $\alpha=2\log((k-1)(n-1)+1)/((kn+1)\log(kn+1))$ is largest at $k=n=2$. (The ancillary file \texttt{anc/identity\_check.py} verifies the identity symbolically.) We call the tensor \emph{Schönhage's partial-product tensor}.

\paragraph{The identity in the preprint's notation.} We read the $a$-variables as the outputs and rename, so that the identity takes the shape of the preprint's Lemma 6: the left variables $x_1:=b_{21}$, $x_2:=b_{31}$, $p_1:=b_{11}$, $p_2:=b_{12}$, the right variables $y_1:=c_{11}$, $y_2:=c_{12}$, $q:=c_{21}$, and the output variables $z_{11}:=a_{12}$, $z_{21}:=a_{13}$, $z_{12}:=a_{22}$, $z_{22}:=a_{23}$, $z_0:=a_{11}$. Let
\[
G:=\sum_{i,j=1}^2x_iy_j\,z_{ij}+(p_1y_1+p_2q)\,z_0,\qquad
\hat p:=\begin{pmatrix}p_1&-p_2\\0&p_2\end{pmatrix},\qquad \hat q:=\begin{pmatrix}0&0\\-q&q\end{pmatrix}.
\]
Every row of $\hat q$ sums to zero; the second column of $\hat p$ sums to zero and the first to $p_1$. The five terms $P_{ij}$ ($i,j\in\{1,2\}$) and $P_0$ have the linear forms
\begin{alignat*}{3}
\varphi_{P_{ij}}&=x_i+\hat p_{ij}, &\qquad \psi_{P_{ij}}&=y_j+\hat q_{ij}, &\qquad \chi_{P_{ij}}&=z_{ij}+z_0,\\
\varphi_{P_0}&=-(x_1+x_2), & \psi_{P_0}&=y_1+y_2, & \chi_{P_0}&=z_0,
\end{alignat*}
which are Coppersmith's five terms at $x=1$, in the order $P_{11},P_{21},P_{12},P_{22},P_0$.

\begin{lemma}\label{m:lem:identity}
Summing over the five terms, $\sum_\lambda\varphi_\lambda\psi_\lambda\chi_\lambda=G+E$, where
\[
E:=\sum_{i,j=1}^2\big(x_i\hat q_{ij}+\hat p_{ij}y_j+\hat p_{ij}\hat q_{ij}\big)z_{ij}
=p_1y_1z_{11}-p_2y_2z_{12}+p_2y_2z_{22}+p_2q\,z_{22}-x_2q\,z_{21}+x_2q\,z_{22}.
\]
\end{lemma}

\begin{proof}
The proof of~\cite[Lemma 6]{ZT} applies verbatim. The coefficient of $z_{ij}$ comes only from $P_{ij}$ and is $(x_i+\hat p_{ij})(y_j+\hat q_{ij})=x_iy_j+(x_i\hat q_{ij}+\hat p_{ij}y_j+\hat p_{ij}\hat q_{ij})$. The coefficient of $z_0$ is $\sum_{i,j}(x_i+\hat p_{ij})(y_j+\hat q_{ij})-(x_1+x_2)(y_1+y_2)$; the products $x_iy_j$ cancel, the cross terms are $\sum_ix_i\sum_j\hat q_{ij}+\sum_jy_j\sum_i\hat p_{ij}=0+p_1y_1$, and $\sum_{i,j}\hat p_{ij}\hat q_{ij}=p_2q$.
\end{proof}

The two observations of the preprint's Section 2.2 hold, the second with one change. First, every monomial of $E$ contains an outer output $z_{ij}$ and an inner input $p_1$, $p_2$ or $q$, so $E$ is harmless for the preprint's selection, as there. Second, every term contributes to $z_0$ (with coefficient $1$), and only $P_{ij}$ contributes to $z_{ij}$; the four terms $P_{ij}$ are distinct, and $P_0$ contributes to $z_0$ alone. As in the preprint, the sign of the private term sits in $\varphi_{P_0}$, so every $\chi_\lambda$ is a sum of output variables without coefficients. The change is that $y_1$ is both an outer variable (in $x_iy_1z_{i1}$) and an inner one (in $p_1y_1z_0$): it is \emph{shared}.

The shared variable is what buys the inner length $2$. An exact identity with this outer product and five terms, whose inner variables are disjoint from the outer ones and whose error is harmless, has inner length at most $1$ (Proposition~\ref{a:prop:exact}), and a border identity cannot reach $2$ with disjoint inner variables either (Proposition~\ref{m:rem:border} below); so at five terms a shared variable is necessary for inner length $2$. The rest of Part~\hyperref[m:partI]{I} shows that the price can be paid.

\paragraph{Why the shared variable matters.} We explain why one run of the tensor cannot compute independent products, why Coppersmith's remedy does not fit the preprint's pruning, and what we do instead.

In the preprint's framework the identity is tensored over $L$ levels, with the inner product chosen at a set $Q$ of $m$ levels, and Lemma~\ref{m:lem:identity} then gives for each $Q$ a matrix product of shape $2^{L-m}\times2^m\times2^{L-m}$. A left string (one left variable per level) still determines $Q$, as in the preprint: it has $p$'s exactly on $Q$. A right string does not. The string $y_1y_1\cdots y_1$ pairs, for every $Q$, with the left string that has $p_1$ on $Q$ and $x_1$ elsewhere, and so contributes to the entry of the $Q$-th product in the first row and column with the inner index $(1,\dots,1)$. One number would have to be an entry of every right factor $Y_Q$ at once, so the $\binom Lm$ products cannot have independent right factors.

This is a property of the support of the tensor, not of its decomposition: the coefficients $\gamma_\ell$ of the preprint's Lemma 8 are the coefficients of the tensor $G+E$, so $\Mult$ is the same for every decomposition of it. Coppersmith's remedy~\cite[p.~468]{Cop82} writes the wanted inputs as $A'A$ and $BB'$ with constant matrices $A'$, $B'$ whose maximal minors are all nonzero, so that the product computed is $A'(AB)B'$; in our orientation this handles the shared variable, but every entry of the wanted product becomes a linear combination of many entries of the power's product (Section~\ref{x:sec:novelty} quotes Coppersmith). In the preprint's tiling, however, each entry is ``a single output of a single run'' \cite[Section 2.3.4]{ZT}, and with $N^2/\sqrt D$ wanted entries spread over all rows we expect Coppersmith's remedy to need almost every output, so that the pruning would save nothing (an argument, not a theorem). We instead let all products of a run share their right factor, through a re-indexing that is injective where it matters (Section~\ref{m:sec:hash}), and then make the shared factor independent again by a transposed copy of the run (Section~\ref{m:sec:squared}).

\begin{proposition}[the disjoint tensor]\label{m:rem:border}
The tensor $G(2,2,2)=\langle2,1,2\rangle\oplus\langle1,2,1\rangle$ has border rank $6$. Hence no border identity in five terms computes $\langle2,1,2\rangle$ together with an inner product of length $2$ whose variables are disjoint from the outer ones.
\end{proposition}

\begin{proof}
The $4+2$ products of $G(2,2,2)$ give rank at most $6$. The lower bound $\underline R(G(2,2,2))\ge6$ is a finite computation by weak border apolarity, given in Appendix~\ref{a:sec:border} after Remark~\ref{a:rem:koszul}, where the notions it uses are introduced. For the consequence: by Lemma~\ref{a:lem:normal}, an inner form of length $2$ on variables disjoint from the outer ones is $G(2,2,2)$ after a change of basis, and a five-term border identity for it would give $\underline R(G(2,2,2))\le5$.
\end{proof}

\section{Re-indexing maps and good sets}\label{m:sec:hash}

This section is new; it has no counterpart in the preprint. Section~\ref{m:sec:tensor} showed that a right string of a run does not determine the inner set $Q$, so the right strings cannot be read as entries of $\binom Lm$ independent factors $Y_Q$. We therefore read the indices of every string through two fixed linear maps over $\F$ that do not depend on $Q$: the levels at which a string has $p_2$ or $q$ are mapped to an inner index in $\F^m$, and the levels at which it has $y_2$ or $z_{i2}$ to a row or column index in $\F^{L-m}$, in both cases by summing the columns of a fixed matrix at those levels. What a good set $Q$ guarantees is that both sums are bijections, so that no index is lost or counted twice. Section~\ref{m:sec:squared} says which letters are read through which map.

Let $L>m\ge1$. A \emph{$\Gamma$-map} is an $m\times L$ matrix $\Gamma$ over $\F$ and a \emph{$\Delta$-map} an $(L-m)\times L$ matrix $\Delta$ over $\F$; we call both \emph{re-indexing maps}. They are not hash functions in the sense of the preprint's Section 3.2 and Part~\hyperref[m:partII]{II}, which reduce modulo a prime (in Part~\hyperref[m:partII]{II}, a product of primes) and have false positives: on the sets that matter, re-indexing maps are bijective. For a set $T\subseteq[L]$ we write $\Gamma[T]:=\sum_{\ell\in T}\Gamma_{\cdot\ell}\in\F^m$ for the sum of the columns of $\Gamma$ at $T$, and $\Delta[S]\in\F^{L-m}$ likewise. A set $Q\subseteq[L]$ of $m$ levels is \emph{good} if the columns of $\Gamma$ at $Q$ and the columns of $\Delta$ at $[L]\setminus Q$ are both bases. Then $T\mapsto\Gamma[T]$ is a bijection from the subsets of $Q$ onto $\F^m$, and $S\mapsto\Delta[S]$ a bijection from the subsets of $[L]\setminus Q$ onto $\F^{L-m}$, since every vector is the sum of a unique subset of a basis.

The next lemma finds maps with many good sets. Let $\mu_k:=\prod_{i=1}^k(1-2^{-i})$, the probability that a uniform $k\times k$ matrix over $\F$ is invertible, and $\mu_\infty:=\lim_k\mu_k=0.2887\ldots$, so $\mu_\infty^2=0.08339\ldots$; note $\mu_k>\mu_\infty$ for all $k$.

\begin{lemma}\label{m:lem:hashes}
There are a $\Gamma$-map and a $\Delta$-map with at least $\mu_m\mu_{L-m}\binom Lm\ge\mu_\infty^2\binom Lm$ good sets, and a deterministic algorithm finds them, depending only on $L$ and $m$, in $\binom Lm(2^m+2^{L-m})L^{O(1)}$ operations on $O(\log N)$-bit words whenever $L\le\log_2N$.
\end{lemma}

\begin{proof}
For uniform $\Gamma$ the columns at a fixed $Q$ form a uniform $m\times m$ matrix, so they are a basis with probability $\mu_m$; the same holds for $\Delta$ at $[L]\setminus Q$ with $\mu_{L-m}$, independently. Hence the expected number of good sets is $\mu_m\mu_{L-m}\binom Lm$, and we derandomize by the method of conditional expectations. Fix the columns of $\Gamma$ one level at a time. When the columns at the levels $1,\dots,\ell$ are fixed and $r$ of them lie in $Q$, the columns of $\Gamma$ at $Q$ are a basis with conditional probability $\mu_{m-r}$ if these $r$ fixed columns are linearly independent, and $0$ otherwise, since the remaining $m-r$ uniform columns extend $r$ independent vectors to a basis with probability $\prod_{i=r}^{m-1}(1-2^{i-m})=\mu_{m-r}$. So the conditional expectation of the number of sets $Q$ with $\Gamma$ invertible at $Q$ is a sum of $\binom Lm$ such terms, each found by one rank computation. Choosing at each level the column, among the $2^m$ values, that maximises it never decreases it, so the final $\Gamma$ is invertible at at least $\mu_m\binom Lm$ sets. Then fix $\Delta$ in the same way, maximising the sum of $\Pr[\Delta$ invertible at $[L]\setminus Q]$ over the sets $Q$ at which $\Gamma$ is invertible; at the end at least $\mu_{L-m}\mu_m\binom Lm$ sets are good. With all probabilities multiplied by $2^{L^2}$ the arithmetic is on integers of $O(L^2)$ bits, that is $O(L)$ words, and each evaluation costs $\binom Lm L^{O(1)}$ operations; there are $L(2^m+2^{L-m})$ evaluations.
\end{proof}

\section{The squared run}\label{m:sec:squared}

This section replaces the preprint's Sections 2.3.3 and 2.3.4, with Lemma 9 and the tiling there, and its Section 2.4.1 for the encodings. We call the algorithm of Sections~\ref{m:sec:squared}--\ref{m:sec:ds} the \emph{squared construction} and each of its $2m+1$ runs a \emph{squared run}.

\paragraph{The idea.} A squared run has $2L$ levels. The first $L$ use the identity of Lemma~\ref{m:lem:identity} as it stands and the last $L$ use it with left and right exchanged, so a left string is an $x$-string followed by a $y$-string, and a right string a $y$-string followed by an $x$-string. The $x$-half of a string determines an inner set $Q$, the levels with $p_1$ or $p_2$. The $x$-half of the left string selects the block $X_Q$ and the $x$-half of the right string the block $Y_Q$, and the two sets are independent of each other, so one squared run addresses every block product $X_{Q_k}Y_{Q_{k'}}$ of the grid of good sets. The $y$-half of a string does not determine an inner set, and this is why the shared variable does no harm: an array never has to know the inner set of the other side. The inner index of a product is read through the re-indexing maps of Section~\ref{m:sec:hash}, from the levels with $p_2$ in the $x$-half and with $q$ in the $y$-half, and the $2m+1$ runs, indexed by the difference $s$ between the numbers of these letters in the two halves, make sure that only monomials of $G$ contribute (Theorem~\ref{m:thm:correct}; Remark~\ref{m:rem:degree} reads the runs as a degree split). The rest of the section tiles $X$ and $Y$ into bands and tiles as in the preprint, with a $\Kg\times\Kg$ grid of good sets in place of the preprint's $K_0\times K_0$ grid, and counts the encodings.

\paragraph{Per-level identities.} The proofs of the preprint's Lemmas 7 and 8 treat one level at a time and take the slices at a level over that level's alphabet (four left letters in the first half below, three in the second), which is all they need. So these lemmas with their proofs, the recursion $\Full$ of the preprint's Section 2.3.1, the leaves, the encodings and $\Mult$ are unchanged if the identity used at level $\ell$ depends on $\ell$. We run $2L$ levels. At the levels $1,\dots,L$ (the \emph{first half}) we use Lemma~\ref{m:lem:identity} as it stands. At the levels $L+1,\dots,2L$ (the \emph{second half}) we use it with left and right variables exchanged: the left variables there are $y_1,y_2,q$ and the left form of a term $\lambda$ is $\psi_\lambda$, the right variables are $x_1,x_2,p_1,p_2$ and the right form is $\varphi_\lambda$. The terms and their output forms $\chi_\lambda$ are the same at every level, so the relation ``contributes to'' of the preprint's Section 2.3.2, and with it every notion of the preprint that concerns leaves, cubes and boxes, is the same at all $2L$ levels. In the preprint's Lemma 8, the factor at level $\ell$ is then $\gamma_\ell(u_\ell,v_\ell,w_\ell)$, the coefficient in $G+E$ of the monomial $u_\ell v_\ell w_\ell$, at every level.

\paragraph{Strings.} An \emph{$x$-string} is a string over $\{x_1,x_2,p_1,p_2\}$, a \emph{$y$-string} one over $\{y_1,y_2,q\}$, and a \emph{$z$-string} one over the five output variables, all of length $L$. For an $x$-string $\xi$, $Q(\xi)$ is the set of levels where it has $p_1$ or $p_2$ (its inner set), $T(\xi)$ the set where it has $p_2$, and $\iota(\xi)\in\{1,2\}^{L-|Q(\xi)|}$ the indices $i$ of its letters $x_i$ in increasing order of the levels. For a $y$-string $\eta$, $T(\eta)$ is the set of levels where it has $q$ and $S(\eta)$ the set where it has $y_2$. For a $z$-string $\zeta$, $Q(\zeta)$ is the set of levels with $z_0$, $\iota(\zeta)$ the first indices $i$ of its letters $z_{ij}$ in increasing order of the levels, and $S(\zeta)$ the set of levels with a letter $z_{i2}$. A left string of the run is $u=u'u''$ with an $x$-string $u'$ (first half) and a $y$-string $u''$ (second half); a right string is $v=v'v''$ with a $y$-string $v'$ and an $x$-string $v''$; an output string is $w=w'w''$ with two $z$-strings. Only the output strings with $|Q(w')|=|Q(w'')|=m$ are read.

\paragraph{The products.} Fix re-indexing maps $\Gamma,\Delta$ as in Lemma~\ref{m:lem:hashes} and a list $\Qg=(Q_1,\dots,Q_{\Kg})$ of good sets, with $\Kg\ge\mu_\infty^2\binom Lm$; both halves use the same $\Gamma$, $\Delta$ and $\Qg$. Let $\Nz:=4^{L-m}$ and $D:=4^m$. We index the rows of an $\Nz\times D$ matrix by $\{1,2\}^{L-m}\times\F^{L-m}$, its columns by $\F^m\times\F^m$, and the columns of a $D\times\Nz$ matrix by $\F^{L-m}\times\{1,2\}^{L-m}$. For every $Q\in\Qg$ let $X_Q\in\Z^{\Nz\times D}$ and $Y_Q\in\Z^{D\times\Nz}$ be arbitrary. For $s\in\{-m,\dots,m\}$ define the input arrays of the \emph{run $s$}:
\begin{align*}
a_s[u'u'']&:=X_{Q(u')}\big[(\iota(u'),\Delta[S(u'')]),\,(\Gamma[T(u')],\Gamma[T(u'')])\big]&&\text{if }Q(u')\in\Qg,\ |T(u'')|-|T(u')|=s,\\
b_s[v'v'']&:=Y_{Q(v'')}\big[(\Gamma[T(v')],\Gamma[T(v'')]),\,(\Delta[S(v')],\iota(v''))\big]&&\text{if }Q(v'')\in\Qg,\ |T(v'')|-|T(v')|=s,
\end{align*}
and $a_s[u]:=0$, $b_s[v]:=0$ otherwise (the example after Theorem~\ref{m:thm:correct} and Figure~\ref{fig:example} show these arrays in the case $L=3$, $m=1$). The left array depends on the matrices $X_Q$ and on the global re-indexing maps, never on the inner set of the second half, and the right array depends only on the $Y_Q$; this is what makes the shared variable of Section~\ref{m:sec:tensor} harmless.

In words, the next theorem says that the $2m+1$ runs together compute, for every pair $(Q_k,Q_{k'})$ of good sets, every entry of the block product $X_{Q_k}Y_{Q_{k'}}$ as a single output, which is what the preprint's Lemma 9 gives for one run of Schönhage's identity with $k=k'$.

\begin{theorem}\label{m:thm:correct}
For every output string $w=w'w''$ with $Q(w'),Q(w'')\in\Qg$,
\[
\sum_{s=-m}^{m}\Mult(a_s,b_s)[w]\;=\;\big(X_{Q(w')}Y_{Q(w'')}\big)\big[(\iota(w'),\Delta[S(w'')]),\,(\Delta[S(w')],\iota(w''))\big],
\]
and for fixed $Q_k,Q_{k'}\in\Qg$ the map from the output strings $w$ with $Q(w')=Q_k$, $Q(w'')=Q_{k'}$ to the positions on the right is a bijection onto the $\Nz^2$ entries of $X_{Q_k}Y_{Q_{k'}}$.
\end{theorem}

\begin{proof}
By the preprint's Lemma 8, $\Mult(a_s,b_s)[w]$ is the sum over left strings $u$ and right strings $v$ of $\prod_\ell\gamma_\ell(u_\ell,v_\ell,w_\ell)\,a_s[u]b_s[v]$. We determine which summands can be nonzero, as in the proof of the preprint's Lemma 9. Let $Q':=Q(w')$ and $Q'':=Q(w'')$.

\emph{First half.} The only monomials of $G+E$ with $z_0$ are $p_1y_1z_0$ and $p_2qz_0$, with coefficient $1$. So at every level of $Q'$ the left letter $u'_\ell$ is $p_1$ or $p_2$, and since $a_s[u]\ne0$ requires $|Q(u')|=m$, we get $Q(u')=Q'$: the letters of $u'$ outside $Q'$ are $x$'s. At a level of $Q'$ the pair is $(p_1,y_1)$ or $(p_2,q)$, so $v'$ has a $q$ at every level of $T(u')$. At a level $\ell\notin Q'$ with $w'_\ell=z_{ij}$, the monomials of $G+E$ with $z_{ij}$ and an $x$-letter are $x_iy_jz_{ij}$ and, for $i=2$, $x_2qz_{2j}$ (Lemma~\ref{m:lem:identity}). Hence $|T(v')|\ge|T(u')|$, with equality exactly when $v'$ has no $q$ outside $Q'$.

\emph{Second half.} Symmetrically, with the $x$-letters now on the right: $b_s[v]\ne0$ forces $Q(v'')=Q''$, the pairs at $Q''$ are $(y_1,p_1)$ or $(q,p_2)$ (left letter first), and $|T(u'')|\ge|T(v'')|$, with equality exactly when $u''$ has no $q$ outside $Q''$.

\emph{The run.} A nonzero summand of run $s$ has $|T(u'')|-|T(u')|=s=|T(v'')|-|T(v')|$, hence
\[
\big(|T(v')|-|T(u')|\big)+\big(|T(u'')|-|T(v'')|\big)=s-s=0 .
\]
Both brackets are nonnegative, so both vanish, and no $q$ occurs at an outer level. So at every level the monomial is one of $G$ (coefficient $1$): $x_iy_jz_{ij}$ at the outer levels and $p_1y_1z_0$ or $p_2qz_0$ at the inner ones. The summands that remain are therefore indexed by a pair of sets $T'\subseteq Q'$, $T''\subseteq Q''$ (the levels with $p_2$, in $u'$ and in $v''$): $u'$ has inner set $Q'$, $\iota(u')=\iota(w')$ and $T(u')=T'$; $v'$ has $T(v')=T'$ and $S(v')=S(w')$; $v''$ has inner set $Q''$, $\iota(v'')=\iota(w'')$ and $T(v'')=T''$; $u''$ has $T(u'')=T''$ and $S(u'')=S(w'')$. Each such pair lies in exactly one run, $s=|T''|-|T'|$. Summing over $s$, with $I_w:=(\iota(w'),\Delta[S(w'')])$ and $J_w:=(\Delta[S(w')],\iota(w''))$,
\[
\sum_s\Mult(a_s,b_s)[w]=\sum_{T'\subseteq Q',\,T''\subseteq Q''}X_{Q'}\big[I_w,(\Gamma[T'],\Gamma[T''])\big]\,Y_{Q''}\big[(\Gamma[T'],\Gamma[T'']),J_w\big].
\]
Since $Q'$ and $Q''$ are good, $(T',T'')\mapsto(\Gamma[T'],\Gamma[T''])$ is a bijection onto $\F^m\times\F^m$, and the sum is the claimed entry. For the bijection: $w'$ with $Q(w')=Q'$ is determined by $\iota(w')$ and $S(w')\subseteq[L]\setminus Q'$, and $w''$ likewise; $\iota$ ranges over $\{1,2\}^{L-m}$, and $S\mapsto \Delta[S]$ is a bijection onto $\F^{L-m}$ because the sets are good.
\end{proof}

\paragraph{Example.} Let $L=3$ and $m=1$, and take $\Gamma=(1\ 1\ 1)$ and $\Delta=\begin{pmatrix}0&1&1\\1&0&1\end{pmatrix}$, for which all three sets $\{1\},\{2\},\{3\}$ are good. The output string $w=w'w''$ with $w'=z_{21}z_0z_{12}$ and $w''=z_0z_{11}z_{22}$ has $Q(w')=\{2\}$, $\iota(w')=(2,1)$, $S(w')=\{3\}$ and $Q(w'')=\{1\}$, $\iota(w'')=(1,2)$, $S(w'')=\{3\}$. By Theorem~\ref{m:thm:correct} the runs compute at $w$ the entry of $X_{\{2\}}Y_{\{1\}}$ in the row $(\iota(w'),\Delta[S(w'')])=((2,1),(1,1))$ and the column $(\Delta[S(w')],\iota(w''))=((1,1),(1,2))$, as the sum of four summands, one for each pair $T'\subseteq\{2\}$, $T''\subseteq\{1\}$. For $T'=\{2\}$, $T''=\{1\}$ the left string is $u=u'u''$ with $u'=x_2p_2x_1$ ($Q(u')=\{2\}$, $T(u')=\{2\}$, $\iota(u')=(2,1)$) and $u''=q\,y_1y_2$ ($T(u'')=\{1\}$, $S(u'')=\{3\}$), and the right string is $v=v'v''$ with $v'=y_1q\,y_2$ ($T(v')=\{2\}$, $S(v')=\{3\}$) and $v''=p_2x_1x_2$ ($Q(v'')=\{1\}$, $T(v'')=\{1\}$, $\iota(v'')=(1,2)$). Both lie in the run $s=|T(u'')|-|T(u')|=|T(v'')|-|T(v')|=0$, and at every level the monomial is one of $G$ with coefficient $1$ ($x_2y_1z_{21}$, $p_2qz_0$, $x_1y_2z_{12}$ in the first half; $p_2qz_0$, $x_1y_1z_{11}$, $x_2y_2z_{22}$ in the second, read with left and right exchanged). So the summand is
\[
a_0[u]\,b_0[v]=X_{\{2\}}\big[((2,1),(1,1)),\,((1),(1))\big]\cdot Y_{\{1\}}\big[((1),(1)),\,((1,1),(1,2))\big],
\]
the inner index being $(\Gamma[T(u')],\Gamma[T(u'')])=(\Gamma[\{2\}],\Gamma[\{1\}])=((1),(1))$, an element of $\F^1\times\F^1$. The other three summands come from $u=x_2p_1x_1\,y_1y_1y_2$, $v=y_1y_1y_2\,p_1x_1x_2$ (run $0$, inner index $((0),(0))$), from $u=x_2p_1x_1\,q\,y_1y_2$, $v=y_1y_1y_2\,p_2x_1x_2$ (run $1$) and from $u=x_2p_2x_1\,y_1y_1y_2$, $v=y_1q\,y_2\,p_1x_1x_2$ (run $-1$). The left array reads its second half $u''$ only through $|T(u'')|$, $\Gamma[T(u'')]$ and $\Delta[S(u'')]$, and a $y$-string does not determine an inner set, so $a_s[u]$ does not depend on the inner set $Q(w'')$ of the output it serves. The ancillary file \texttt{anc/example\_squared.py} generates every number of this example (the maps are the first pair, in lexicographic order, for which all three sets are good) and checks Theorem~\ref{m:thm:correct} by brute force at $L=3$, $m=1$ on all $2304$ read output strings. Figure~\ref{fig:example} shows the summand with $T'=\{2\}$, $T''=\{1\}$.

\begin{figure}[t]
\centering
\begin{tikzpicture}[x=1.05cm,y=0.72cm,font=\small]
\fill[gray!20] (1.5,-0.5) rectangle (2.5,2.5);
\fill[gray!20] (4.0,-0.5) rectangle (5.0,2.5);
\draw[gray] (3.75,-0.6) -- (3.75,3.3);
\node at (2,3.05) {first half};
\node at (5.5,3.05) {second half};
\foreach \x/\l in {1/1,2/2,3/3,4.5/4,5.5/5,6.5/6} \node[gray] at (\x,2.65) {\scriptsize level \l};
\node[anchor=east] at (0.45,2) {output $w=w'w''$};
\node[anchor=east] at (0.45,1) {left string $u=u'u''$};
\node[anchor=east] at (0.45,0) {right string $v=v'v''$};
\foreach \x/\t in {1/z_{21},2/z_0,3/z_{12},4.5/z_0,5.5/z_{11},6.5/z_{22}} \node at (\x,2) {$\t$};
\foreach \x/\t in {1/x_2,2/p_2,3/x_1,4.5/q,5.5/y_1,6.5/y_2} \node at (\x,1) {$\t$};
\foreach \x/\t in {1/y_1,2/q,3/y_2,4.5/p_2,5.5/x_1,6.5/x_2} \node at (\x,0) {$\t$};
\node[align=center] at (2,-1.3) {$\iota(w')=(2,1)$\\ $S(w')=\{3\}$};
\node[align=center] at (5.5,-1.3) {$\iota(w'')=(1,2)$\\ $S(w'')=\{3\}$};
\node[align=left,anchor=north] at (3.75,-2.2) {row of $X_{\{2\}}$: $(\iota(w'),\Delta[S(w'')])=((2,1),(1,1))$\\ column of $Y_{\{1\}}$: $(\Delta[S(w')],\iota(w''))=((1,1),(1,2))$\\ inner index: $(\Gamma[T(u')],\Gamma[T(u'')])=((1),(1))$};
\end{tikzpicture}
\caption{The example of Section~\ref{m:sec:squared} ($L=3$, $m=1$): the output string $w$, and the left and right strings of its summand with $T'=\{2\}$, $T''=\{1\}$ (run $s=0$). Shaded: the inner sets $Q(w')=\{2\}$ and $Q(w'')=\{1\}$ (level $4$). The row of the block product takes $\iota$ from the first half and $\Delta[S]$ from the second, the column the other way round; the inner index is read off the levels with $p_2$ in $u'$ and with $q$ in $u''$.}
\label{fig:example}
\end{figure}

\begin{remark}[the runs are a degree split]\label{m:rem:degree}
Give $p_1$ and every $z_{ij}$ weight $2$, $p_2$ and $q$ weight $1$, and the other variables weight $0$. Every monomial of $G$ has weight $2$ and every monomial of $E$ weight $3$ or $4$. In run $s$ every nonzero left string has weight $2m+s$, every nonzero right string weight $2m-s$, and every read output string weight $4(L-m)$; the total $4L$ forces weight $2$ at every level, which is the proof above in other words. It is also Coppersmith's construction at $x=1$: the terms printed in Section~\ref{m:sec:tensor} are the terms of Lemma~\ref{m:lem:identity} with every variable multiplied by $x$ to its weight. Fixing the two half-weights separately would need $(m+1)^2$ runs; their difference $s$ suffices.
\end{remark}

\paragraph{The tiling.} Let $X\in\Z^{N\times D}$ and $Y\in\Z^{D\times N}$, and index the columns of $X$ and the rows of $Y$ by $\F^m\times\F^m$ through a fixed bijection. Cut $X$ into row blocks of $\Nz$ rows, each indexed by $\{1,2\}^{L-m}\times\F^{L-m}$, and group $\Kg$ consecutive row blocks into a \emph{row band}, the $k$-th block of a band playing the role of $X_{Q_k}$. Cut $Y$ in the same way into column blocks and column bands, the $k$-th block of a column band playing $Y_{Q_k}$. So, as in the preprint's Section 2.3.4, the matrices $X_{Q_k}$ and $Y_{Q_k}$ of Theorem~\ref{m:thm:correct} are blocks of $X$ and $Y$. A \emph{tile} is a row band together with a column band; its $2m+1$ runs compute all $\Kg^2$ block products of the tile by Theorem~\ref{m:thm:correct}, each entry of $XY$ in the tile being the sum of one output of each run, at a position fixed in advance. As in the preprint's Section 2.3.4, we first pad $N$ to a multiple of $\Kg\Nz$, which at most doubles it when $N\ge\Kg\Nz$. A tile has $M:=\Kg^2\Nz^2$ output entries. The difference from the preprint is that the grid is $\Kg\times\Kg$, with all good sets on both sides, where the preprint's single run must spend $K_0^2\le\binom Lm$ of its subsets on a $K_0\times K_0$ grid, $K_0=\lfloor\sqrt K\rfloor$; this is where the gain in the range of $D$ comes from (Section~\ref{m:sec:ceiling}).

\paragraph{Encodings per band.} The left arrays $a_s$ of a tile depend only on its row band, and the right arrays only on its column band, as the definitions show. So, as in the preprint's Section 2.4.1, we encode each band once per run and share the encodings among all tiles of the band. An encoding has $5^{2L}=25^L$ numbers, and computing it takes $O(25^L)$ operations by the recursion: in the first half a left array has at most $4^{L-k}3^L$ entries per vertex at depth $k$, and $\sum_k5^k4^{L-k}3^L=O(15^L)$; in the second half $\sum_k5^{L+k}3^{L-k}=O(25^L)$, and symmetrically for right arrays, the series being geometric because $5>4>3$. Forming an array takes $O(L)$ operations for each of its at most $\binom Lm2^L3^L\le12^L$ nonzero entries, which is also $O(25^L)$.

\begin{remark}[one half alone]\label{m:rem:half}
Without the transposed copy, one run of $L$ levels with the right array $c[v]:=Y[\Gamma[T(v)],\Delta[S(v)]]$ computes $X_QY$ for all good $Q$ against one shared $2^m\times2^{L-m}$ matrix $Y$, that is, one stacked product $\langle\Kg2^{L-m},2^m,2^{L-m}\rangle$. A tile is then $\Kg$ row blocks by a single column block, every column block needs its own encoding, and the range of $D$ shrinks; its rate is lower ($\varepsilon\gamma=0.0049$ in the exact leaf count of Remark~\ref{m:rem:exact}, numerical; ancillary file \texttt{anc/shared\_variable\_variants.py}, output line A), which is why we square.
\end{remark}

\section{Leaves}\label{m:sec:leaves}

This section replaces the preprint's Section 2.4.3 and Lemma 11.

\paragraph{What leaves are for.} The preprint's analysis rests on two counts of leaves, recalled in the brief before Section~\ref{m:sec:tensor}: the number $\alpha_d$ of leaves of order $d$ that contribute to an output string, the same for every string, and the number $\beta_d$ of leaves of order $d$ in a tile, which decays in $d$. This section carries both counts over to the squared run. The only new point is that a leaf has two halves, so that the counts are convolutions of the counts for one half.

\paragraph{Products at leaves and the counts.} A leaf $\tau$ is a string of $2L$ terms. Let
\[
\Pi(\tau):=\sum_{s=-m}^{m}\Phi_\tau(a_s)\,\Psi_\tau(b_s),
\]
the \emph{product at} $\tau$, which takes $2m+1$ multiplications. By the preprint's Lemma 7 for each run and Theorem~\ref{m:thm:correct}, every read entry of a tile is $\sum_{\tau}\Pi(\tau)$ over the leaves $\tau$ contributing to its output string $w$; this is the only change in the meaning of ``the product at a leaf'' in the preprint's Sections 2.4 and 4. The leaves contributing to $w$ form the \emph{cube} of $w$: any term at the $2m$ levels of $Q(w):=Q(w')\cup(L+Q(w''))$, and $P_{ij}$ at every other level where $w$ has $z_{ij}$. The \emph{private leaf} of $w$ has $P_0$ at all of $Q(w)$, and the \emph{order} of a leaf is $2m$ minus its number of $P_0$'s. For $0\le e\le m$ let $\beta_e:=\binom L{m-e}4^{L-m+e}$, the number of strings of $L$ terms with exactly $m-e$ symbols $P_0$, and let $K:=\binom Lm$, so $\beta_0=K\Nz$.

\begin{lemma}[analogue of~{\cite[Lemma 11]{ZT}}]\label{m:lem:counts}
Let $\rho:=4m/(L-m+1)$.
\begin{enumerate}[label=(\alph*),nosep]
\item For every read output string $w$ and every $0\le d\le2m$, exactly $\alpha'_d:=\binom{2m}d4^d$ leaves of order $d$ contribute to $w$.
\item The leaves with at most $m$ symbols $P_0$ in each half and order $d$ number $\beta'_d:=\sum_{d_1+d_2=d}\beta_{d_1}\beta_{d_2}$ (with $0\le d_1,d_2\le m$), and $\beta'_d\le(d+1)\rho^d\beta_0^2\le(d+1)\rho^dM/\mu_\infty^4$.
\item For every set $U$ of read output strings, $|\Leaves(U)|\le\sum_{d=0}^{2m}\min\{|U|\alpha'_d,\,\beta'_d\}$.
\end{enumerate}
\end{lemma}

\begin{proof}
(a) A leaf of order $d$ in the cube of $w$ replaces $P_0$ by one of the four other terms at $d$ of the $2m$ levels of $Q(w)$. This number does not depend on $w$; for the half-counts it is Vandermonde's identity $\sum_{d_1+d_2=d}\binom m{d_1}4^{d_1}\binom m{d_2}4^{d_2}=\binom{2m}d4^d$. (b) Choose $m-d_1$ levels with $P_0$ in the first half and $m-d_2$ in the second, and one of four terms elsewhere. As in the preprint's eq.~(7), $\beta_e/\beta_{e-1}=4(m-e+1)/(L-m+e)\le\rho$, so $\beta_e\le\rho^e\beta_0$ and $\beta'_d\le(d+1)\rho^d\beta_0^2$; finally $\beta_0^2=K^2\Nz^2\le M/\mu_\infty^4$ since $\Kg\ge\mu_\infty^2 K$. (c) A leaf of $\Leaves(U)$ contributes to some $w\in U$, so it has $P_0$ only at $Q(w)$, hence at most $m$ in each half, and its order is some $d\in[0,2m]$. The leaves of order $d$ in $\Leaves(U)$ number at most $|U|\alpha'_d$ by (a) and at most $\beta'_d$ by (b), which is the argument of the preprint's Lemma 11.
\end{proof}

The bound of (c) is linear in $|U|$ on each piece and does not depend on which output strings are in $U$, because $\alpha'_d$ is the same for every output string; so it holds for every $W$, not only on average, and sums over the tiles as in the preprint's Section 2.4.4. The pruned recursion of the preprint's Section 2.4.2 and its Lemma 10 apply verbatim on the $2L$ levels, with $\Pi(\tau)$ returned at a leaf.

\begin{remark}[the exact leaf count]\label{m:rem:exact}
For a set $W$ known in advance, the pruned recursion can use the inequality of Lemma~\ref{m:lem:counts}(c) unsplit; we call this bound the \emph{exact leaf count}. The explicit corollaries use instead the \emph{split form}, the bound of Corollary~\ref{m:cor:tradeoff}, which splits the leaves at the switching order $t$. A \emph{rate value} is the supremum of the saving $\gamma$ (or of $\varepsilon\gamma$) over the parameters $c$ and $\theta$ of such a closed-form bound as $m\to\infty$, evaluated and maximised numerically, with no explicit threshold on $m$; the preprint's Table 2 likewise evaluates its Corollaries 31 and 32 numerically, at fixed $c$, rounded down. Rate values are numerical, not theorems. The exact leaf count gives the largest rate: the product of the saving and of the range, maximised over the ratio $c=L/m$, is $\varepsilon\gamma=0.005825$ at $c\approx10.2$, against $0.003935$ for Schönhage's identity in the same model (ancillary files \texttt{anc/shared\_variable\_variants.py}, output line H, and \texttt{anc/profile\_model.py}). The split form gives $0.005652$ at $c\approx10.6$, against $0.003710$ at $c\approx22.2$ for Schönhage's identity (\texttt{anc/explicit\_constants.py}, output lines E3).
\end{remark}

\section{The data structure}\label{m:sec:ds}

This section replaces the preprint's Sections 4.2--4.4, that is, Lemmas 27--29, Theorem 30, Corollary 31 and Theorems 24 and 25 there. The preprint's Section 4 uses Schönhage's identity only through the cube structure of an output entry, the uniform counts $\alpha_d$, the decay of the $\beta_d$, the encodings per band, and the word size. Each has a counterpart above.

\paragraph{What the data structure does.} The data structure is the preprint's (recalled in the brief before Section~\ref{m:sec:tensor}): a query adds the products $\Pi(\tau)$ at the leaves of order below the switching order $t$ in the cube of $w$ to the precomputed values of the boxes of $w$, and the preprocessing computes all box values by a dynamic program. In the squared construction the query time is governed by $\sum_{d<t}\alpha'_d$ and the preprocessing by the number of boxes, which decays like $\rho^t$ by Lemma~\ref{m:lem:counts}(b); the choice of $t$ trades the one against the other (Corollary~\ref{m:cor:tradeoff}).

\paragraph{Balanced boxes.} Fix the switching order $t$ with $0\le t\le2m$. A \emph{box} is a cube $\pi$ on the $2L$ levels (symbols: the five terms and $\ast$) such that (i)~at most $2m-t$ of its symbols are $P_0$ or stars, (ii)~every star is at a lower level than every $P_0$, and (iii)~at most $m$ of its symbols in each half are $P_0$ or stars. Conditions (i) and (ii) are those of the preprint's Section 4.2 with $m$ replaced by $2m$; (iii), which balances the stars and symbols $P_0$ between the two halves, is new, and it removes a factor: without it the count of the preprint's Lemma 29 would run over all leaves of $2L$ levels with at most $2m-t$ symbols $P_0$, about $\binom{2L}{2m}/\binom Lm^2=\Theta(\sqrt m)$ times more than the leaves that serve read outputs. The boxes of an output string $w$ and the sets $\mathcal B_V$, $V\subseteq Q(w)$, $|V|=2m-t$, are defined as in the preprint's Section 4.2 (the sets defined before its Lemma 27 and used in its Lemma 28) with $Q:=Q(w)$, a set of $2m$ levels ordered by their index.

\begin{lemma}\label{m:lem:boxes}
\begin{enumerate}[label=(\alph*),nosep]
\item Lemmas 27 and 28 of~\cite{ZT} hold verbatim for every read output string $w$, with $Q=Q(w)$, $m$ replaced by $2m$, nine by four, and the products $\Pi(\tau)$: the entry of $w$ is the sum of the $\sum_{d<t}\alpha'_d$ products $\Pi(\tau)$ at the leaves of order below $t$ in its cube and of the values of its $\alpha'_t$ boxes, every leaf of order at least $t$ of the cube lying in exactly one of them.
\item There are at most $(2m+1)\sum_{d=t}^{2m}\beta'_d$ boxes, and the dynamic program of~\cite[Lemma 29]{ZT} computes the values of all of them, and stores them in a trie, in $O(L)$ time and space per box.
\end{enumerate}
\end{lemma}

\begin{proof}
(a) Lemmas 27 and 28 of~\cite{ZT} are statements about subsets $Z\subseteq V\subseteq Q$ of an ordered set of levels and about the terms other than $P_0$ at the remaining levels of $Q$; of the identity they use only that the leaves contributing to $w$ form the cube of $w$. The boxes of $w$ satisfy (iii), since their symbols $P_0$ and stars lie in $Q(w)$, which has $m$ levels in each half. (b) A box with $f$ symbols $P_0$ or stars arises from the leaf obtained by replacing its stars with $P_0$, which has at most $m$ symbols $P_0$ in each half by (iii) and order $2m-f\ge t$ by (i), by turning its $e\le f\le2m$ lowest symbols $P_0$ into stars; with Lemma~\ref{m:lem:counts}(b) this gives the count. In the dynamic program, replacing the highest star of a box by a term keeps (i) and (ii) by the argument of the preprint's Lemma 29, and keeps (iii), since it replaces a star by $P_0$ or lowers the count of one half by one. So the program never leaves the boxes, and each value is the sum of five values computed before. The base case reads $\Pi(\tau)$, that is $2m+1$ products, in $O(L)$ operations, since the encodings of the $2m+1$ runs of a band are stored side by side, indexed by the leaf, so that one search of $O(L)$ steps finds all of them.
\end{proof}

\paragraph{What else changes.} The list of changes against the preprint's Sections 2.4 and 4 is the following.
\begin{enumerate}[label=(\arabic*),nosep]
\item Terms per level $10\to5$, non-private terms $9\to4$, so $\alpha_d=\binom md9^d\to\alpha'_d=\binom{2m}d4^d$, $\beta_d\to\beta'_d$, $\rho=9m/(L-m+1)\to4m/(L-m+1)$, and $L\ge10m\to L\ge5m$; the cube of an output has $2m$ free levels.
\item The sign of $P_0$ is in $\varphi_{P_0}$, as in the preprint, so nothing changes there.
\item The product at a leaf is $\Pi(\tau)$, summed over the $2m+1$ runs; the encodings are per band and per run, and the box values are stored already summed over the runs, since the dynamic program is linear in the leaf products.
\item The tile is $\Kg\times\Kg$ block products with $\Kg\ge\mu_\infty^2 K$, so the box count per output entry carries the constant $\mu_\infty^{-4}<144$.
\item A query must invert the re-indexing maps: the row of $I$ in its block gives $(\iota(w'),\Delta[S(w'')])$ and the column of $J$ gives $(\Delta[S(w')],\iota(w''))$, so with the inverses of the columns of $\Delta$ at $[L]\setminus Q$, stored for each $Q\in\Qg$, the output string is found in $O(L^2)$ operations.
\item The re-indexing maps are found once by Lemma~\ref{m:lem:hashes}.
\item Word size: at $x=1$ all coefficients of the forms are $0,\pm1$, an encoded number is a signed sum of at most $12^L$ input entries, and a value is a sum of at most $(2m+1)25^L$ products; since $N\ge\Nz=4^{L-m}\ge4^{4L/5}$ we have $25^L\le N^{2.91}$, so all integers have $O(\log N)$ bits, and $L\le\log_2 N$.
\item $\Nz=4^{L-m}$ instead of $3^{L-m}$, and $M=\Kg^2\Nz^2$ counts only the entries of the $\Kg^2$ block products.
\item The tile fit is $N\ge K\Nz$ instead of $N\ge\sqrt K\Nz$, and the encoding term has $K$ in place of $\sqrt K$, because the grid is $\Kg\times\Kg$.
\item The levels are ordered $1,\dots,2L$, the first half below the second, and \emph{lower} in the preprint's Lemmas 27--29 refers to this order.
\end{enumerate}
Items (1), (8), (9) are also in the table at the start of Part~\hyperref[m:partI]{I}.

\begin{theorem}[analogue of~{\cite[Theorem 30]{ZT}}]\label{m:thm:tech}
Let $m\ge1$, $D=4^m$, $L\ge5m$, $0\le t\le2m$, $K=\binom Lm$, $\Nz=4^{L-m}$, $\rho=4m/(L-m+1)$, and $N\ge K\Nz$. Given $X\in\Z^{N\times D}$ and $Y\in\Z^{D\times N}$ with entries of absolute value at most $N^{O(1)}$, we can preprocess them deterministically in time and space
\begin{equation}\label{m:eq:prep}
O\Big(L\,m^2\,\frac{\rho^t}{1-\rho}\,N^2+\frac{m\,N\cdot25^L}{K\,\Nz}+N\,L^{O(1)}\Big),
\end{equation}
after which any single entry $(XY)[I,J]$ can be computed deterministically in $O\big(L\sum_{d=0}^{t}\alpha'_d+L^2\big)$ time. For a set $W$ of positions, the entries at $W$ take $|W|$ times the query time plus~\eqref{m:eq:prep}.
\end{theorem}

\begin{proof}
The proof of~\cite[Theorem 30]{ZT}, with the changes listed above. \emph{Preprocessing}: choose the re-indexing maps and the list $\Qg$ by Lemma~\ref{m:lem:hashes}, in $K(2^m+2^{L-m})L^{O(1)}\le N L^{O(1)}$ operations since $K2^{L-m}\le K\Nz\le N$ (and Lemma~\ref{m:lem:hashes} applies, since $N\ge\Nz=4^{L-m}\ge2^L$ gives $L\le\log_2N$), and store the inverses of $\Delta$ at $[L]\setminus Q$, $Q\in\Qg$, in $O(KL^3)$; tile, padding $N$ to a multiple of $\Kg\Nz$, which at most doubles it; form and encode the arrays of all bands, at most $4N/(\Kg\Nz)\le4N/(\mu_\infty^2 K\Nz)$ of them, $2m+1$ arrays each at $O(25^L)$ (Section~\ref{m:sec:squared}), which is the second term; compute the boxes of each of the at most $4N^2/M$ tiles by Lemma~\ref{m:lem:boxes}(b), at most $(2m+1)\sum_{d\ge t}\beta'_d\le(2m+1)^2\rho^tM/(\mu_\infty^4(1-\rho))$ per tile by Lemma~\ref{m:lem:counts}(b), at $O(L)$ each, which is the first term. \emph{Query}: locate the tile, the pair $(Q_k,Q_{k'})$ and the output string with change (5), in $O(L^2)$; then add the $\sum_{d<t}\alpha'_d$ products $\Pi(\tau)$ and the $\alpha'_t$ box values of Lemma~\ref{m:lem:boxes}(a), each found in $O(L)$. Correctness is Lemma~\ref{m:lem:boxes}(a) with Theorem~\ref{m:thm:correct}, and the word size is change (7).
\end{proof}

Let $\Hh(x):=-x\ln x-(1-x)\ln(1-x)$, and for $c>5$ and $\gamma\ge0$ let
\begin{equation}\label{m:eq:Rc}
\Lambda_c(\gamma):=\Big(\frac{25}{4}\Big)^{c}e^{-c\Hh(1/c)}\,4^{1+\gamma},\qquad
R'_c(\gamma):=\frac{\ln4}{\ln\Lambda_c(\gamma)}=\frac{\ln2}{c\ln5-\frac c2\Hh(1/c)-(c-1)\ln2+\gamma\ln2},
\end{equation}
the counterparts of the base $\Lambda$ and of $R_c(\gamma)$ in the preprint's eq.~(11).

\begin{corollary}[analogue of~{\cite[Corollary 31]{ZT}}]\label{m:cor:tradeoff}
Let $c>5$ and $0<\theta<0.8$, where $\theta$ is the switching order as a fraction of the $2m$ inner levels, $t=\lceil2\theta m\rceil$ (in the preprint's Corollary 31, $t=\lceil\theta m\rceil$ of $m$ levels); let $\rho_c:=4/(c-1)$, and let
\[
\gamma:=\theta\log_2\frac{c-1}4>0\qquad\text{and}\qquad q:=\frac{\Hh(\theta)+\theta\ln4}{\ln2}.
\]
Given $X\in\Z^{N\times D}$ and $Y\in\Z^{D\times N}$ with entries of absolute value at most $N^{O(1)}$, where $2\le D\le N^\varepsilon$ and $\varepsilon<R'_c(\gamma)$, we can preprocess them deterministically in $O(N^2\log^3\!D/D^\gamma)$ time and space, after which any single entry $(XY)[I,J]$ can be computed deterministically in $O(D^q\log D)$ time. The constants depend on $c$, $\theta$ and $\varepsilon$. Moreover $R'_c(0)$ increases to $\es:=\ln4/(5\ln5)=0.17227\ldots$ as $c$ decreases to $5$.
\end{corollary}

\begin{proof}
We repeat the proof of~\cite[Corollary 31]{ZT} (that is, of its Corollary 26 with general parameters), applying Theorem~\ref{m:thm:tech} with $m:=\lceil\log_4D\rceil$ after padding $D$ to $4^m$, $L:=\lceil cm\rceil$ and $t:=\lceil2\theta m\rceil$; the assumption $D\le N^\varepsilon$ now reads $N\ge4^{(m-1)/\varepsilon}$. \emph{Boxes}: $\rho<\rho_c$ gives $\rho^t\le\rho_c^{2\theta m}=D^{-\gamma}$, and $1/(1-\rho)<1/(1-\rho_c)$. \emph{Queries}: with $x:=(1-\theta)/\theta$, where $4x>1$ because $\theta<0.8$, we have $4^d\le(4x)^tx^{-d}$ for $d\le t$, so $\sum_{d\le t}\binom{2m}d4^d\le(4x)^t(1+1/x)^{2m}\le4x\big((4x)^\theta/(1-\theta)\big)^{2m}=4x\,D^q$, using $(4x)^\theta/(1-\theta)=4^\theta e^{\Hh(\theta)}$. \emph{Encodings}: both the second term of~\eqref{m:eq:prep} being $O(mN^2/D^\gamma)$ and the condition $N\ge K\Nz$ follow from
\begin{equation}\label{m:eq:encrange}
\frac{D^\gamma\cdot25^L}{\Kg\,\Nz}\le N,
\end{equation}
the first by rearranging, since $\Kg\le K$, and the second because $25^L=(1+4)^{2L}\ge\binom{2L}{2m}16^{L-m}\ge K^2\Nz^2$. With $\Kg\ge\mu_\infty^2 K$, $K\ge e^{L\Hh(m/L)}/(L+1)$, and the fact that $L\ln\frac{25}4-L\Hh(m/L)$ increases in $L$ with derivative $\ln\frac{25}4+\ln(1-m/L)<\ln\frac{25}4$, so that rounding $L$ up from $cm$ costs at most a factor $\frac{25}4$, the left-hand side of~\eqref{m:eq:encrange} is at most
\begin{equation}\label{m:eq:lhs}
\frac{25/4}{\mu_\infty^2}\,(cm+2)\,\Lambda_c(\gamma)^m<75\,(cm+2)\,\Lambda_c(\gamma)^m .
\end{equation}
As $\varepsilon<R'_c(\gamma)$ says $\Lambda_c(\gamma)<4^{1/\varepsilon}$, \eqref{m:eq:encrange} holds once $m$ exceeds a constant depending on $c,\theta,\varepsilon$, and for smaller $m$ the corollary holds trivially. The third term of~\eqref{m:eq:prep} is $O(N^2/D^\gamma)$ since $N\ge D^{1/\varepsilon}$. Theorem~\ref{m:thm:tech} gives preprocessing $O(Lm^2N^2/D^\gamma)=O(N^2\log^3\!D/D^\gamma)$ and query $O(L\,D^q+L^2)=O(D^q\log D)$. Finally, $\frac{d}{dc}\big(2c\ln5-c\Hh(1/c)-(c-1)\ln4\big)=\ln\frac{25(c-1)}{4c}>0$, and at $c=5$ the identity $\Hh(\frac15)+\frac45\ln4=\ln5$ gives $10\ln5-5\Hh(\frac15)-4\ln4=5\ln5$, so $R'_c(0)$ increases to $\ln4/(5\ln5)$ as $c\downarrow5$.
\end{proof}

The following theorem is the first sentence of Theorem~\hyperref[m:introA]{A}; Corollary~\ref{m:cor:explicit} gives the second.

\begin{theorem}[analogue of~{\cite[Theorems 24 and 25]{ZT}}]\label{m:thm:general}
For every $\varepsilon<\es=\ln4/(5\ln5)=0.17227\ldots$ and every $q>0$ there is a $\gamma>0$ such that, for $X\in\Z^{N\times D}$ and $Y\in\Z^{D\times N}$ with $2\le D\le N^\varepsilon$ and entries of absolute value $N^{O(1)}$, a deterministic preprocessing in $O(N^2/D^\gamma)$ time and space allows any single entry of $XY$ to be computed deterministically in $O(D^q)$ time. For every $\varepsilon<\es$ and $\kappa>0$ there is a $\gamma>0$ such that the entries of $XY$ at any set $W$ of at most $N^2/D^{\kappa}$ positions can be computed deterministically in $O(N^2/D^\gamma)$ time.
\end{theorem}

\begin{proof}
As the proofs of~\cite[Theorems 24 and 25]{ZT}: choose $c>5$ with $R'_c(0)>\varepsilon$, then $\theta$ small enough that $q(\theta)<q/2$ (so that $D^{q(\theta)}\log D=O(D^q)$) and $R'_c(\gamma(\theta))>\varepsilon$, apply Corollary~\ref{m:cor:tradeoff}, and absorb the logarithms by halving $\gamma$; for $W$ take $q:=\kappa/2$ and ask $|W|$ queries.
\end{proof}

\section{Where the limit comes from}\label{m:sec:ceiling}

This section is the counterpart of the paragraph ``Where $\varepsilon^*$ comes from'' in the preprint's Section 4.4, after the proof of its Corollary 31. Two constraints bound the range of $D$ in Corollary~\ref{m:cor:tradeoff}: the leaf count must decay, which needs $c>5$, and the encodings of the bands must cost less than $N^2$, which at $c=5$ coincides with the tile fit. The limit $\es$ is where the second constraint becomes tight as $c$ decreases to $5$. The paragraph after the proof explains why it is Coppersmith's exponent: at $c=5$ one tile of the squared construction is Coppersmith's product.

\begin{proposition}\label{m:prop:ceiling}
The range of Corollary~\ref{m:cor:tradeoff} is $\varepsilon<\es=\ln4/(5\ln5)=0.17227$, against $\et=\ln4/(5\ln10)=0.12041$ for the preprint's Corollary 31. Both limits have the form $\varepsilon^*_R=2\ln q_0/(R\ln R)$, with $R$ terms and inner length $q_0$ per level ($R=10$, $q_0=4$; $R=5$, $q_0=2$ per half), and both arise in the same way: the leaf count needs $c>R$, and at $c=R$ the tile fit and the encoding constraint coincide.
\end{proposition}

\begin{proof}
The box bound of Theorem~\ref{m:thm:tech} saves only if $\rho<1$, and $\rho\to\rho_c=4/(c-1)$, so $c>5$ is needed; for every $c>5$ and $\theta\to0$ the saving is positive. The range $\varepsilon<R'_c(\gamma)\le R'_c(0)$ is decreasing in $c$ (Corollary~\ref{m:cor:tradeoff}), so its supremum is the value $\es$ at $c=5$. At $c=5$, the identity $\Hh(\frac15)+\frac45\ln4=\ln5$ says that $\binom{5m}m4^{4m}$, the largest term of $(1+4)^{5m}$, is $5^{5m}$ up to polynomial factors. So one half selects $K\Nz\approx5^L$ rows, a tile has $M\approx25^L$ entries, as many as the run has leaves, and the encodings cost $N\cdot25^L/(K\Nz)\approx N\cdot5^L$, which is below $N^2$ exactly when $N\ge5^L=D^{1/\es}$, which is also the tile fit $N\ge K\Nz$. For Schönhage's identity the same computation with $\Hh(\frac1{10})+\frac9{10}\ln9=\ln10$ is the preprint's own explanation of $\et$ in the paragraph ``Where $\varepsilon^*$ comes from'' of its Section 4.4, after the proof of its Corollary 31.
\end{proof}

\paragraph{The limit is Coppersmith's exponent.} The value $\es=2\log2/(5\log5)$ is the $\alpha$ of Coppersmith's theorem quoted in Section~\ref{m:sec:tensor}, and this is not a coincidence. At $c=5$ one tile of the squared construction is a product of shape $\langle\Kg\Nz,D,\Kg\Nz\rangle=\langle\Theta(5^L/\sqrt L),\,2^{2L/5},\,\Theta(5^L/\sqrt L)\rangle$, computed by $2m+1$ runs of $25^L$ multiplications each. It takes $2m+1$ runs of $25^L=\Theta(N'^2L)$ bilinear multiplications, $O(N'^2\log^2N')$ in all, for $N'=\Kg\Nz$, and $D=2^{2L/5}=5^{\alpha L}\ge N'^\alpha$ with $\alpha=\es$: Coppersmith's $\mathrm{Rank}(\langle N,N,N^\alpha\rangle)=O(N^2(\log N)^2)$, after permuting the dimensions, for $N=N'$ in this sequence; so Theorem~\ref{m:thm:correct} at $L=5m$ reproves Coppersmith's bound along a sequence of sizes, with exact multiplications and without constant matrices. Up to a permutation of the dimensions (his thin dimension is the third, ours the middle one) and up to constant and polynomial factors, the tile is Coppersmith's final shape $\langle K_C5^{M_C}M_C^{-1/2},\,K_C5^{M_C}M_C^{-1/2},\,2^{2M_C/5}\rangle$ with his border-rank bound $5^{2M_C}$~\cite[p.~469]{Cop82}, his power $M_C$ being our $L$ and his $K_C$ a constant; our $25^L$ counts exact multiplications, and~\cite{LG17} states the shape in our orientation. Coppersmith's construction is the full product at its critical ratio.

So the $\Kg\times\Kg$ grid loses nothing at the limit. The reason is that the products of one half share their right factor and therefore stack along one dimension (one half computes $\langle\Kg2^{L-m},2^m,2^{L-m}\rangle$), so each half puts the whole factor $\Kg$ on one side of the tile. Schönhage's identity instead gives $K$ disjoint products, which a tiling can arrange only as a $K_0\times K_0$ grid, $K_0\approx\sqrt K$, while Coppersmith's Proof version 2 converts disjoint copies into one product by the direct-sum theorem. This is why the preprint's tiling stops at $0.1204$, below the $0.1402$ of Proof version 2,\footnote{The preprint cites Coppersmith's Proof version 2 for this exponent \cite[Sections 2.3 and 2.3.4]{ZT}. Nissim and Schwartz show that this proof is incorrect as written (it implicitly assumes that border rank is additive over disjoint copies) and reprove the same bound by an iterated stacking argument: ``We revisit this work, show that the second proof is incorrect, and propose an alternative one'' \cite{NS24}. The preprint uses the number only for comparison, so nothing in it depends on this.} as the preprint says: ``Tiling enlarges only the outer dimensions, so it handles a smaller inner dimension than Coppersmith's algorithm from the same identity (at best $D\approx N^{0.1204}$ rather than $N^{0.1402}$ [Cop82])'' \cite[Section 2.3.4]{ZT}.

With $W$ known in advance and the exact leaf count of Remark~\ref{m:rem:exact}, the range extends to about $0.1784$ (Schönhage: $0.1292$); these are numerical and concern only the non-adaptive theorem (ancillary file \texttt{anc/exact\_ceiling.py}).

\section{Explicit constants}\label{m:sec:explicit}

This section is the counterpart of the preprint's Corollary 26: it applies Theorem~\ref{m:thm:tech} with $c=14.5$, $\theta=\frac1{19}$ at the threshold $N\ge D^{18}$ and with $c=11.5$, $\theta=\frac1{18}$ at $N\ge D^{14}$, in the notation of Corollary~\ref{m:cor:tradeoff}, and carries the constants through, which gives the exponents of Table~\ref{tab:matrix}. The steps of the proof follow the terms of the preprocessing bound~\eqref{m:eq:prep} and the query time of Theorem~\ref{m:thm:tech}: the box count gives the saving $\gamma$, the number of leaves of order at most $t$ gives the query exponent $q$, and the encoding condition~\eqref{m:eq:encrange}, which is where the threshold enters, gives both the bound on the second term and the condition $N\ge K\Nz$ of Theorem~\ref{m:thm:tech}, and decides from which $m$ on the parameters are valid. The paragraph after the proof says how the two instances were chosen, and Table~\ref{m:tab:rates} compares the two identities at other thresholds and query times.

\begin{corollary}[analogue of~{\cite[Corollary 26]{ZT}}]\label{m:cor:explicit}
Let $X\in\Z^{N\times D}$ and $Y\in\Z^{D\times N}$ have entries of absolute value at most $N^{O(1)}$.
\begin{enumerate}[label=(\alph*),nosep]
\item If $N\ge D^{18}$, we can preprocess $X,Y$ deterministically in $O(N^2/D^{0.092})$ time and space, after which any single entry $(XY)[I,J]$ can be computed deterministically in $O(D^{0.403})$ time. Hence for every set $W$ of positions the entries at $W$ take $O(|W|\,D^{0.403}+N^2/D^{0.092})$ time, which is $O(N^2/D^{0.092})$ whenever $|W|\le N^2/\sqrt D$.
\item If $N\ge D^{14}$, the same holds with preprocessing $O(N^2/D^{0.077})$, queries $O(D^{0.421})$, and $O(|W|\,D^{0.421}+N^2/D^{0.077})$, which is $O(N^2/D^{0.077})$ whenever $|W|\le N^2/\sqrt D$.
\end{enumerate}
\end{corollary}

\begin{proof}
(a) \emph{Setting up.} Let $m:=\lceil\log_4D\rceil$ and pad the inner dimension to $4^m<4D$ with zero columns of $X$ and zero rows of $Y$; this changes no entry of $XY$ and $D$ by a factor below $4$. So from now on $D=4^m$, and $N\ge D^{18}$ now reads $N\ge4^{18(m-1)}$. Let $L:=\lceil14.5\,m\rceil$ and $t:=\lceil2m/19\rceil$, and apply Theorem~\ref{m:thm:tech}; its condition $N\ge K\Nz$ is verified with the encodings.

\emph{Boxes.} $\rho=4m/(L-m+1)<4/13.5=8/27$, so $1/(1-\rho)<27/19$, and since $t\ge2m/19$, $\rho^t\le(8/27)^{2m/19}=D^{-\gamma}$ with $\gamma:=\log_2(27/8)/19=0.09236\ldots$. The first term of~\eqref{m:eq:prep} is $O(m^3N^2/D^\gamma)$.

\emph{Queries.} With $x=18$: $4^d\le72^t18^{-d}$ for $d\le t$, and $t<2m/19+1$, so
\[
\sum_{d=0}^{t}\binom{2m}d4^d\le72^t\Big(\frac{19}{18}\Big)^{2m}<72\Big(72^{1/19}\cdot\frac{19}{18}\Big)^{2m}=72\,D^{q},
\]
where $q:=\log_2\big(72^{1/19}\cdot\frac{19}{18}\big)=0.40273\ldots$,
so a query takes $O(m\,D^q+m^2)$ time.

\emph{Encodings.} By~\eqref{m:eq:encrange} and~\eqref{m:eq:lhs}, it suffices that $75(14.5m+2)\Lambda^m\le N$ with
\[
\Lambda:=\Lambda_{14.5}(\gamma)=\Big(\frac{25}4\Big)^{14.5}e^{-14.5\,\Hh(1/14.5)}\cdot4\cdot\Big(\frac{27}8\Big)^{2/19}=4.1456\ldots\cdot10^{10}.
\]
The base $4^{18}=6.8719\ldots\cdot10^{10}$ of the assumption $N\ge4^{18(m-1)}$ is larger than $\Lambda$ by a factor of more than $1.657$, so the inequality holds whenever $1.657^m\ge4^{18}\cdot75\,(14.5m+2)$, which is the case for all $m\ge72$. (For $m<72$, $D$ is bounded by a constant and the corollary holds trivially. With exact binomial coefficients and $\Kg\ge\mu_\infty^2 K$, a computation shows that~\eqref{m:eq:encrange} itself holds from $m=62$ on; we do not need this.) The third term of~\eqref{m:eq:prep} is $O(NL^{O(1)})\le O(N^2/D)$.

\emph{Conclusion.} For $m\ge72$ Theorem~\ref{m:thm:tech} applies: preprocessing $O(m^3N^2/D^{0.09236})$ and queries $O(mD^{q}+m^2)$ with $q=0.40273\ldots$. Since $m=O(\log D)$ and the exponents $0.09236-0.092$ and $0.403-q$ are positive, these are $O(N^2/D^{0.092})$ and $O(D^{0.403})$. The bound for $W$ follows by asking $|W|$ queries, and $0.403+0.092<\frac12$.

(b) The same with $L:=\lceil11.5\,m\rceil$, $t:=\lceil2m/18\rceil$, $\rho<4/10.5=8/21$, $\gamma:=\log_2(21/8)/18=0.07735\ldots$; $x=17$, $\sum_{d\le t}\binom{2m}d4^d<68\,D^{q}$ with $q:=\log_2(68^{1/18}\cdot\frac{18}{17})=0.42065\ldots$; $\Lambda:=\Lambda_{11.5}(\gamma)=2.1169\ldots\cdot10^8$, against $4^{14}=2.6843\ldots\cdot10^8$, a factor of more than $1.268$, so~\eqref{m:eq:encrange} holds whenever $1.268^m\ge4^{14}\cdot75\,(11.5m+2)$, that is for all $m\ge131$ (from $m=110$ with exact binomials). Rounding gives $0.077<0.07735$ and $0.421>0.42065$, and $0.421+0.077<\frac12$.
\end{proof}

\paragraph{The statements $\Hyp$.} Applied to the biadjacency matrices of $A\times M$ and $M\times B$, as the preprint's Corollary 26 is in Lemma~\ref{c:lem:hzt}, part (a) gives $\Hyp(1/18,\,0.092)$ and part (b) gives $\Hyp(1/14,\,0.077)$ (Definition~\ref{c:hyp}): the conditions $N\ge D^{18}$ and $N\ge D^{14}$ are $D\le N^{1/18}$ and $D\le N^{1/14}$, and the bounds for $|W|\le N^2/\sqrt D$ are the running times that the definition asks for.

\paragraph{How the instances were chosen.} Instance (b) is the best point of a grid ($c$ in steps of $0.05$, $\theta=1/k$ for integers $k$, and a validity threshold $m_0\le150$ for~\eqref{m:eq:encrange}) at $N\ge D^{14}$. Instance (a) is not the grid's best point at $N\ge D^{18}$: $c=14.6$, $\theta=\frac1{19}$ gives $\gamma=0.0929$ (from $m=112$) and the same rounded exponents, and we keep $c=14.5$. Its switching order $t=\lceil2m/19\rceil$ is about $2/19\approx0.105$ times the $m$ inner levels of one half, close to the preprint's $t=m/9$. The preprint's own inequality at $c=21$, $\theta=\frac19$ reproduces its numbers ($\gamma=0.0640$, $q=0.4277$, $\Lambda=4.1986\cdot10^{10}$, a factor $1.6367$, valid from $m=58$; the $m\ge60$ of the preprint's proof of its Corollary 26 is valid, not tight).

Among thresholds $N\ge D^{1/\varepsilon}$ with $1/\varepsilon\in\{10,\dots,16,18\}$, instance (b) maximises the rounded product $\varepsilon\gamma$ ($0.0055$, against $0.00511$ for (a)); this is why we use (b) for the reductions of Part~\hyperref[m:partII]{II} (Section~\ref{x:sec:consequences}).

\begin{table}[t]
\centering\small
\setlength{\tabcolsep}{4pt}
\begin{tabular}{c|ccc|cc}
 & \multicolumn{3}{c|}{data structure: $\gamma$ for query time $D^{q}$} & \multicolumn{2}{c}{wanted entries: $\gamma$ for $|W|\le N^2/D^{\kappa}$}\\
threshold & $q=0.25$ & $q=0.43$ & $q=0.50$ & $\kappa=0.50$ & $\kappa=1.00$\\\hline
$N\ge D^{18}$ & $0.0340\,/\,0.0525$ & $0.0660\,/\,0.1018$ & $0.0797\,/\,0.1229$ & $0.0667\,/\,0.0946$ & $0.1570\,/\,0.2177$\\
$N\ge D^{14}$ & $0.0232\,/\,0.0415$ & $0.0450\,/\,0.0805$ & $0.0543\,/\,0.0971$ & $0.0479\,/\,0.0785$ & $0.1142\,/\,0.1820$\\
$N\ge D^{12}$ & $0.0165\,/\,0.0347$ & $0.0319\,/\,0.0672$ & $0.0385\,/\,0.0810$ & $0.0352\,/\,0.0676$ & $0.0846\,/\,0.1577$\\
$N\ge D^{10}$ & $0.0084\,/\,0.0264$ & $0.0162\,/\,0.0511$ & $0.0196\,/\,0.0617$ & $0.0187\,/\,0.0536$ & $0.0455\,/\,0.1257$\\
$N\ge D^{8}$ & $-\,/\,0.0160$ & $-\,/\,0.0309$ & $-\,/\,0.0372$ & $-\,/\,0.0341$ & $-\,/\,0.0808$\\
$N\ge D^{6}$ & $-\,/\,0.0017$ & $-\,/\,0.0033$ & $-\,/\,0.0040$ & $-\,/\,0.0039$ & $-\,/\,0.0094$
\end{tabular}
\caption{Rate values (not explicit theorems) of Corollaries 31, 32 of~\cite{ZT} and of Corollary~\ref{m:cor:tradeoff} (with $|W|$ queries, as in its Corollary 32), in the form of its Table 2; each cell is Schönhage's identity / the squared construction. Unlike the preprint's Table 2, whose rows are values of $c$, the rows here are thresholds, since $c$ means different things for the two identities: each cell is the supremum of $\gamma$ over $c$ at that threshold, rounded down; ``$-$'' means no saving ($D>N^{0.1204}$). For $W$ the best $\theta$ is used, where $\gamma=\kappa-q$. Each value holds for every smaller $\gamma$ with a threshold $m_0$ that depends on the margin; the only explicit entries of Part~\hyperref[m:partI]{I} are those of Corollary~\ref{m:cor:explicit}. The values are printed by \texttt{anc/explicit\_constants.py}.}
\label{m:tab:rates}
\end{table}

\paragraph{Beyond the two instances.} Table~\ref{m:tab:rates} compares the two identities over thresholds, query times and sizes of $W$, by rate values (Remark~\ref{m:rem:exact}), not explicit theorems. The squared construction wins in every cell, by a factor of about $1.4$--$1.5$ at $N\ge D^{18}$, about $1.6$--$1.8$ at $N\ge D^{14}$, about $2$ at $N\ge D^{12}$ and about $3$ at $N\ge D^{10}$; for $D$ between $N^{0.1204}$ and $N^{\es}$ only the squared construction gives a saving.

\FloatBarrier
\section{Hinted OMv}\label{m:sec:omv}

In the three hinted variants of Online Matrix--Vector multiplication of van den Brand, Nanongkai and Saranurak~\cite{BNS19}, which the preprint recalls in its Section 5.4, the input arrives in phases; the numbers 5.2, 5.7 and 5.12 are the conjecture numbers of~\cite{BNS19}, which the preprint keeps, and all three conjectures are over the Boolean semiring. For Conjecture 5.2, Phase 1 receives $M\in\{0,1\}^{n\times t}$, Phase 2 $V\in\{0,1\}^{t\times n}$, and Phase 3 an index $i$, after which $MV_{[n],i}$, the product of $M$ and column $i$ of $V$, is output. For Conjecture 5.7, Phase 1 receives $N\in\{0,1\}^{n\times n}$ and $V\in\{0,1\}^{t\times n}$, Phase 2 a list $I\in[n]^t$ of column indices, and Phase 3 an index $j$, after which $N_{[n],I}V_{[t],j}$ is output, $N_{[n],I}$ being the $n\times t$ matrix whose $k$-th column is column $I_k$ of $N$. For Conjecture 5.12, Phase 1 receives $U\in\{0,1\}^{n\times t_1}$, $N\in\{0,1\}^{n\times n}$ and $V\in\{0,1\}^{t_2\times n}$, Phase 2 a list $I\in[n]^{t_1}$, Phase 3 a list $J\in[n]^{t_2}$, and Phase 4 a pair $(i,j)$, after which the Boolean entry $(UN_{I,J}V)_{i,j}$ is output, $N_{I,J}$ being the $t_1\times t_2$ matrix with entries $(N_{I,J})_{a,b}=N_{I_a,J_b}$, in which, as~\cite{BNS19} note, rows and columns of $N$ may repeat. Here $t=n^\tau$ and $t_k=n^{\tau_k}$, and the letter $N$ for the $n\times n$ matrix, which is the preprint's, is not the dimension $N$ of the matrix theorem. Each conjecture says that no algorithm is polynomially faster, at the same time, than the obvious ones, which multiply everything out when a hint arrives or wait for the query (two for Conjectures 5.2 and 5.7, three for Conjecture 5.12). For Conjectures 5.2 and 5.7 the product to be formed is an $n\times t$ by a $t\times n$ matrix, so $N=n$ and $D=t$, and for $n\ge t^{18}$ the data structure of Corollary~\ref{m:cor:explicit}(a) beats both obvious algorithms, whatever $\omega$, since $\omega(1,1,\tau)\ge2$: preprocessing in $O(n^{2-0.092\tau})$ when the hint arrives, and a column of the product, $n$ queries, in $O(n^{1+0.403\tau})$. For Conjecture 5.12 the same holds blockwise, with $N=t_2$ and $D=t_1$, as in the preprint's proof of its Corollary 40.

This section extends the preprint's refutation, its Corollary 40, from $\tau<\et$ to $\tau<\es$, with the same proof and better exponents; Proposition~\ref{m:rem:omv512} then refutes Conjecture 5.12 on a larger region, by using a phase that the proof leaves idle. The preprint's Corollary 40 uses the data structure only through its Corollary 26 (for $\tau<1/18$) and through its Corollary 31 with $q=\frac12$ (for general $\tau<\et$); the reductions themselves (column queries, blocks of $t_2$ rows, $0/1$ inputs computed over $\Z$) do not touch the identity.

\begin{corollary}[analogue of~{\cite[Corollary 40]{ZT}}]\label{m:cor:omv}
Conjectures 5.2 and 5.7 of~\cite{BNS19} fail for every $0<\tau<1/18$: Phase 2 takes $O(n^{2-0.092\tau})$ time and Phase 3 takes $O(n^{1+0.403\tau})$ time. Conjecture 5.12 fails for every $0<\tau_1<\tau_2/18$: Phase 3 takes $O(n^{1+\tau_2-0.092\tau_1})$ time and Phase 4 takes $O(n^{\tau_2+0.403\tau_1})$ time, with polynomial Phase 1 and a Phase 2 that only stores $I$. With $0.077$ and $0.421$ in place of $0.092$ and $0.403$ the same holds for every $\tau<1/14$ and every $\tau_1<\tau_2/14$. More generally, Conjectures 5.2 and 5.7 fail for every $0<\tau<\es=0.17227\ldots$, and Conjecture 5.12 fails for every $0<\tau_1<\es\tau_2$.
\end{corollary}

\begin{proof}
The proof of~\cite[Corollary 40]{ZT}, with Corollary~\ref{m:cor:explicit}(a), respectively (b), in place of its Corollary 26 and Corollary~\ref{m:cor:tradeoff}, with $c,\theta$ chosen as in Theorem~\ref{m:thm:general} for $q=\frac12$, in place of its Corollary 31. For Conjecture 5.12 the blocks of $t_2$ rows of the preprint's Corollary 40, each a product of a $t_2\times t_1$ by a $t_1\times t_2$ matrix, still apply, since $t_2=n^{\tau_2}\ge n^{18\tau_1}=D^{18}$ (respectively $D^{14}$) when $\tau_1<\tau_2/18$ (respectively $\tau_2/14$).
\end{proof}

The discussion after the preprint's Corollary 40 moves with the threshold. For $\tau<\es$, where $\omega(1,1,\tau)=2$~\cite{VXXZ24}, the end of the update--query trade-offs of~\cite{BNS19} with the fastest queries loses its support. For the distance oracles of Haeupler, Long and Saranurak~\cite{HLS24}, new evidence of hardness would be needed for $\delta\le\es$, and the remaining cases of the conjecture still rule out update time $O(n^{1.82})$ rather than $O(n^{1.87})$, since $2-\es=1.8277\ldots$ (rounded down as the preprint rounds $2-0.1204=1.8796$ to $1.87$). The lower bound of van den Brand, Song and Zhou~\cite{BSZ24} for dynamic attention needs a new hypothesis for $\tau<\es$. Online Set Disjointness, in the list of consequences of the preprint's introduction, has a data structure for $D\le N^{0.17}$ (was $N^{0.12}$). The main bounds of~\cite{BNS19} are unaffected.

\begin{remark}[checks for Sections~\ref{m:sec:tensor}--\ref{m:sec:explicit}]\label{m:rem:brute}
The ancillary files of this arXiv submission contain the checks. \texttt{anc/identity\_check.py} verifies Lemma~\ref{m:lem:identity} and Remark~\ref{m:rem:degree} symbolically. The script \texttt{anc/squared\_check.py} implements Sections~\ref{m:sec:hash}--\ref{m:sec:ds} as written, re-indexing maps included: $0$ wrong outputs at $2+2$, $3+3$ and $4+4$ levels ($m=1,2$), while one run without the split by $s$ fails; exact counts in Lemma~\ref{m:lem:counts}; every query right at every switching order, with box counts within Lemma~\ref{m:lem:boxes}(b). \texttt{anc/explicit\_constants.py} computes the constants of Corollary~\ref{m:cor:explicit}, the grid of instances and Table~\ref{m:tab:rates}. Table~\ref{tab:anc} in Section~\ref{a:sec:anc} lists all ancillary files.
\end{remark}

\begin{proposition}[Conjecture 5.12 on a larger region]\label{m:rem:omv512}
Conjecture 5.12 of~\cite{BNS19} fails for every $0<\tau_1<1/18$ and every $0<\tau_2<1$: Phase 2 takes $O(n^{2-0.092\tau_1})$ time, Phase 3 only stores $J$, and Phase 4 takes $O(n^{\tau_2+0.403\tau_1})$ time. It also fails for every $0<\tau_2<\tau_1/18$: Phase 2 only stores $I$, Phase 3 takes $O(n^{1+\tau_1-0.092\tau_2})$ time, and Phase 4 takes $O(n^{\tau_1+0.403\tau_2})$ time. More generally, it fails for every $\tau_1<\es$ and for every $\tau_2<\es\tau_1$. The region $\tau_1<\es\tau_2$ of Corollary~\ref{m:cor:omv} lies inside the first of these. The point $\tau_1+\tau_2\approx1.41$ of the main bound of~\cite{BNS19} lies in neither, since it would need $\tau_2>1$ or $\tau_1>1$.
\end{proposition}

\begin{proof}
We compute over $\Z$ with $0/1$ matrices as in the proof of Corollary~\ref{m:cor:omv}. \emph{First region.} In Phase 2 the matrices $X:=U$ and $Y:=N_{I,[n]}$, the $t_1\times n$ matrix whose $a$-th row is row $I_a$ of $N$, are known. Preprocess them by Corollary~\ref{m:cor:explicit}(a) with $n$ in the role of its $N$ and inner dimension $D:=t_1$; it applies since $n\ge t_1^{18}$ for $\tau_1<1/18$, and it takes $O(n^2/D^{0.092})=O(n^{2-0.092\tau_1})$ time, below $n^{\omega(1,\tau_1,1)}$ since $\omega(1,\tau_1,1)\ge2$. Column $b$ of $UN_{I,J}$ is column $J_b$ of $XY$, so $(UN_{I,J}V)_{i,j}$ is $1$ if and only if $(XY)[i,J_b]>0$ and $V_{b,j}=1$ for some $b\le t_2$: these are $t_2$ queries, in $O(t_2D^{0.403})=O(n^{\tau_2+0.403\tau_1})$ time, below $n^{\tau_1+\tau_2}$. \emph{Second region.} In Phase 3 the matrices $X:=N_{I,J}$, of size $t_1\times t_2$, and $Y:=V$ are known; the inner dimension is $D:=t_2$. Cut the $n$ columns of $V$ into $n/t_1$ blocks of $t_1$ columns and preprocess the product of $X$ with each block, a $t_1\times t_2$ by $t_2\times t_1$ product, by Corollary~\ref{m:cor:explicit}(a) with $t_1$ in the role of its $N$; it applies since $t_1\ge t_2^{18}$ for $\tau_2<\tau_1/18$. This takes $(n/t_1)\cdot O(t_1^2/D^{0.092})=O(n^{1+\tau_1-0.092\tau_2})$ time, below $n^{\omega(\tau_2,\tau_1,1)}$, which is at least $n^{1+\tau_1}$ because the product $N_{I,J}V$ has $n^{1+\tau_1}$ entries. Row $a$ of $N_{I,J}V$ is row $a$ of $XY$, so $(UN_{I,J}V)_{i,j}$ is $1$ if and only if $U_{i,a}=1$ and $(XY)[a,j]>0$ for some $a\le t_1$: these are $t_1$ queries, in $O(n^{\tau_1+0.403\tau_2})$ time, below $n^{\tau_1+\tau_2}$. \emph{General thresholds.} For $\tau_1<\es$, respectively $\tau_2<\es\tau_1$, use Corollary~\ref{m:cor:tradeoff} as in the proof of Corollary~\ref{m:cor:omv}, with $q=\frac12<1$. Both index identities, with repeated indices in $I$ and $J$, were also checked by brute force on random Boolean instances (\texttt{anc/omv512\_check.py}).
\end{proof}

Of the three obvious algorithms that~\cite[p.~31]{BNS19} list before Conjecture 5.12, the first multiplies $UN_{I,[n]}$ in Phase 2 and the second $N_{I,J}V$ in Phase 3; the two regions run these with the data structure. The preprint's proof and that of Corollary~\ref{m:cor:omv} multiply $UN_{I,J}$ in Phase 3 instead and leave Phase 2 idle; using Phase 2 removes the condition that the first hint be polynomially smaller than the second.

\section{Why not other identities}\label{m:sec:barriers}

This section answers the preprint's Section 6 question, quoted in Section~\ref{m:sec:intro:matrix}, for the class of identities it suggests; it supports, but is not part of, the results of Sections~\ref{m:sec:tensor}--\ref{m:sec:omv}. Appendix~\ref{a:sec:overview} gives the details and the numerical scores, and the rest of Appendix~\ref{a:sec:barriers} the proofs, with the hypotheses stated there. Consider a base identity per level for an outer product $\langle k,1,n\rangle$ with outputs $z_{ij}$ and one inner output $z_0$ whose bilinear form has rank $q_0$, the inner length, in $R$ terms, used with Bini's interpolation if it is a border identity~\cite{Bin80}.

In short, the method needs a base whose outer outputs are rank-one and singly covered and whose inner product is long, and the inner product of such an identity is capped: by $R-k-n+2$ with at most one shared variable, exact or border, and by $R-k-n$ for exact identities with disjoint inner variables and harmless error. Appendix~\ref{a:sec:overview} lists the shapes with their numerical scores, including those the caps leave open.

\paragraph{The proved barriers.} Under the hypotheses of Appendix~\ref{a:sec:barriers} the following hold.
\begin{enumerate}[label=(\arabic*),nosep]
\item \emph{Private leaves.} If $R_0$ terms are pure inner and every outer output lies in at least $R_1$ terms that cover no other outer output, a saving needs $R_0^2R_1^{2(c-1)}<q_0$ (Lemma~\ref{a:lem:private}).
\item \emph{Exact identities.} An exact identity whose inner variables are disjoint from the outer ones and whose error is harmless for the preprint's selection (both terms are recalled in the brief before Section~\ref{m:sec:tensor}) has $q_0\le R-k-n$ (Proposition~\ref{a:prop:exact}).
\item \emph{Border identities.} A border identity has $q_0\le R-k-n+1$ with disjoint inner variables and effective inner length (Appendix~\ref{a:sec:border}) at most $R-k-n+2$ with one shared variable (Propositions~\ref{a:prop:borderG} and~\ref{a:prop:borderS}), in whatever coordinates the inner form is given (Lemma~\ref{a:lem:normal}); for $(k,n)=(2,2),(2,3),(3,3)$ and $kn+1$ terms, computer-calculated flattening ranks (with the substitution recursion for $(3,3)$) sharpen this to the family's $(k-1)(n-1)+1$ (Remark~\ref{a:rem:koszul}), and for $(k,n)=(2,2)$ with disjoint inner variables border apolarity sharpens $q_0\le2$ to $q_0\le1$ (Proposition~\ref{m:rem:border}).
\item \emph{Shared variables.} A run with at least two products in which every output of every product is a correct single output (Appendix~\ref{a:sec:shared}) of a distinct entry and block has no two shared variables on one side, and, if the blocks used for one entry are pairwise disjoint (as in our tilings), at most one shared variable (Theorem~\ref{a:thm:shared}).
\item \emph{Vrana's partial product} with his decomposition gives no saving (Proposition~\ref{a:prop:vrana}).
\end{enumerate}

\emph{Full products.} With a full product $\langle o_1,b,o_2\rangle$ as base and every output string selected, the leaf count of the pruned recursion is bounded below by the sizes $S_l$ of the output supports of the terms alone: if $\sum_lS_l^t\ge o_1o_2\,b^{t/2}$ for every $t\in[0,1]$, some set of $N^2/\sqrt D$ wanted entries costs as much as the full product up to a subexponential factor (Theorem~\ref{a:thm:full}). Two bounds that hold for every decomposition over every field, one from spanning trees of the output grid and one on the rank, give this for every shape with $\kappa\le b$ or $\kappa^{2(b-1)}\ge b^{\kappa-1}$, $\kappa=\min(o_1,o_2)$ (Proposition~\ref{a:prop:criterion}), which covers every rotation of the $145$ decompositions found by computer search (Corollary~\ref{a:cor:145}). Numerically the criterion reaches every shape whose smaller side is at most about $7b+10$ and every square of side at most about $11b+12$, and for $b\le4$ the open shapes nearest to that region could not beat our rate even if they saved (Appendix~\ref{a:sec:full}).

\paragraph{What is numerical.} Every score, the cases $R_0\ge2$ not excluded by the private-leaf condition ($kn\le30$; ancillary file \texttt{anc/pure\_inner\_cases.py}), and the open shapes of the family (Appendix~\ref{a:sec:overview}) are numerical. The sharper caps for $(2,2)$, $(2,3)$, $(3,3)$ are computer-assisted: they rest on exact rank computations, the flattening ranks of Remark~\ref{a:rem:koszul} and, for $(2,2)$ with disjoint inner variables, the ranks of Proposition~\ref{m:rem:border}. The barriers above are proved in Appendix~\ref{a:sec:barriers} under the hypotheses stated there: one inner output, outer product $\langle k,1,n\rangle$, Schönhage's sparsity ($R_1=1$; Appendix~\ref{a:sec:overview}), Bini's interpolation, all outputs selected; and, for a full product as base, Theorem~\ref{a:thm:full} and Proposition~\ref{a:prop:criterion} hold for every decomposition of a shape in the criterion's region (Appendix~\ref{a:sec:full}).

\part{Reductions that keep the whole saving}\label{m:partII}

Part~\hyperref[m:partII]{II} assumes $\Hyp(\varepsilon,\gamma)$ (Definition~\ref{c:hyp}) and uses nothing else from Part~\hyperref[m:partI]{I}.

\paragraph{The idea.} Both reductions hash their integers modulo a modulus and ask the counting algorithm about the hashed instance. The algorithm counts, for every pair it is asked about, the witnesses modulo the modulus; a true witness survives every modulus, and a false one (a \emph{false positive}) survives a further prime factor with a frequency that the averaging lemma of Section~\ref{c:sec:primes} bounds. So a modulus is built one prime at a time, each prime chosen in an interval $[P,2P)$ so as to minimise the false positives that the counting algorithm counts among the current survivors. Each stage builds one modulus of $r$ primes; if the counts then exceed half the number of survivors, a true witness exists and the answer is YES, and otherwise at most half of the survivors keep a nonzero count and only they go on (the all-edges version replaces the YES exit by a certification of single incidences, Section~\ref{e:sec:alledges}). After $O(\log n)$ stages so few survivors remain that scanning them exactly costs nothing in the exponent. The preprint's reductions use one prime, scan every pair the detection algorithm accepts, and lose half or more of the saving in the balance between the two.

\paragraph{Difference to the preprint.} Table~\ref{tab:partII} sums up the differences; per problem they are these. \emph{Exact Triangle.} The preprint's Theorem 17 hashes the weights modulo one prime $p\approx\sqrt D$, chosen by fast matrix multiplication over $\Z[x]/(x^p-1)$, and scans every accepted pair in a piece of about $\sqrt D/g$ vertices; its Remark 20 balances instances against scans and keeps half of the saving, and by its footnote~9 a single modulus chosen with the counting algorithm, one prime factor at a time as in~\cite[Lemma~3.2]{FKP24}, still keeps half. Here the stages make the scans negligible, and the whole saving is kept up to a loss $n^{\varepsilon/(2r)}$. \emph{3SUM.} The preprint goes through Exact Triangle by its Theorem 21(a) and keeps a quarter of the saving. Here the probes of the self-reduction of Fischer, Kaliciak and Polak~\cite{FKP24} (Section~\ref{s:sec:instances}) are filtered by the same kind of stages, and all of the saving is kept up to a loss $n^{\varepsilon/r}$. \emph{APSP.} The preprint's Theorem 22 goes through the detection version of Exact Triangle and its Theorem 21(b) and keeps a sixth; as its footnote~10 suggests, we go through the all-edges version, with~\cite[Lemma 9.2]{CVX23} for the link from negative to exact triangles, and keep all of it.

\begin{table}[ht]
\centering\small
\renewcommand{\arraystretch}{1.15}\setlength{\tabcolsep}{4pt}
\begin{tabular}{@{}>{\raggedright}p{0.2\textwidth}>{\raggedright}p{0.36\textwidth}>{\raggedright\arraybackslash}p{0.36\textwidth}@{}}
 & \cite{ZT} & Part~\hyperref[m:partII]{II}\\\hline
modulus (Exact Triangle) & one prime $p\approx\sqrt D$, chosen by matrix multiplication over $\Z[x]/(x^p-1)$ & $O(\log n)$ stages, each a product of $r$ primes chosen with the counting algorithm\\
called algorithm & $\Lop$, detection & $\#\Lop$, counting\\
calls & non-adaptive & adaptive\\
scans & every accepted pair, in a piece of about $\sqrt D/g$ vertices & only the survivors of all stages; negligible\\
saving kept (3SUM, Exact Triangle, APSP) & $\frac14$, $\frac12$, $\frac16$ of $\varepsilon\gamma$ & all, up to $n^{\varepsilon/r}$ and $(\log n)^{r+O(1)}$
\end{tabular}
\caption{The reductions of Part~\hyperref[m:partII]{II} against those of the preprint.}
\label{tab:partII}
\end{table}

\section{Two lemmas on primes}\label{c:sec:primes}

Both reductions build their moduli from primes in an interval $[P,2P)$. The first lemma bounds the number of such primes; the second is the averaging in the proof of~\cite[Lemma~3.2]{FKP24} (p.~10), restated and proved, with the sum bound of its proof stated explicitly.

\begin{lemma}\label{c:lem:primes}
For every integer $P\ge1024$ the interval $[P,2P)$ contains at least $P/(2\ln P)$ primes.
\end{lemma}
\begin{proof}
By the explicit bounds of Rosser and Schoenfeld~\cite[Corollary~1]{RS62}, $\pi(x)>x/\ln x$ for $x\ge17$ and $\pi(x)<1.25506\,x/\ln x$ for $x>1$. Since $2P$ is not prime, the number of primes in $[P,2P)$ is at least $\pi(2P)-\pi(P)>2P/\ln(2P)-1.25506\,P/\ln P$, and $\ln(2P)\le1.1\ln P$ for $P\ge1024$, so it is at least $(2/1.1-1.25506)\,P/\ln P>P/(2\ln P)$. Only the existence of some absolute threshold is needed, which already follows from the prime number theorem.
\end{proof}

\begin{lemma}[averaging]\label{c:lem:avg}
Let $P\ge2$, $Z_{\max}\ge2$, let $m\ge1$ be an integer, and let $\Pi'$ be a nonempty set of primes $\ge P$, none dividing $m$. Let $\mathcal F$ be a finite multiset of nonzero integers $z$ with $|z|\le Z_{\max}$, each divisible by $m$. Then
\[
\sum_{p\in\Pi'}\#\{z\in\mathcal F: mp\mid z\}\le|\mathcal F|\cdot\log_PZ_{\max},
\]
so some $p\in\Pi'$ satisfies $\#\{z\in\mathcal F: mp\mid z\}\le|\mathcal F|\log_PZ_{\max}/|\Pi'|$.
\end{lemma}
\begin{proof}
A nonzero integer $z$ has at most $\log_P|z|\le\log_PZ_{\max}$ distinct prime divisors $\ge P$, since their product divides $z$ and is at least $P$ to the number of them. For $p\in\Pi'$ and $z\in\mathcal F$, $mp\mid z$ if and only if $p\mid z$, because $m\mid z$ and $\gcd(m,p)=1$. Summing over $z$ gives the inequality, and the minimum over $p$ is at most the average.
\end{proof}

\paragraph{How the lemmas are used.} In both reductions a modulus is built one prime at a time, and the two lemmas together say how much one prime buys. The reductions keep a set of \emph{survivors}, the objects that may still contain a solution: \emph{probes} for 3SUM (Section~\ref{s:sec:instances}) and \emph{incidences} for Exact Triangle (Section~\ref{e:sec:instances}). When the modulus built so far is $m$, the multiset $\mathcal F$ consists of the false positives modulo $m$ among the survivors, each recorded as the integer that vanishes modulo $m$ but not in $\Z$ ($a+b-c$ for 3SUM, the weight of the triangle for Exact Triangle), so that $Z_{\max}$ is $2U$ or $3U$, polynomial in $n$, where $U$ bounds the input integers; the pool $\Pi'$ consists of the primes in $[P,2P)$ not yet chosen. The counts of the survivors modulo $mp$ are evaluated by the counting algorithm for every $p$ in the pool; the true witnesses contribute the same to every count, so the prime with the smallest total is a prime of Lemma~\ref{c:lem:avg}, and the false positives modulo $mp$ are at most the fraction $\log_PZ_{\max}/|\Pi'|$ of those modulo $m$. Lemma~\ref{c:lem:primes} keeps the pool large: after the at most $r$ chosen primes are removed it still has $P/(4\ln P)$ members, as $r\le P/(4\ln P)$, so the fraction is at most $4\ln Z_{\max}/P$ (Lemmas~\ref{s:lem:halving} and~\ref{e:lem:step}). So every prime divides the false positives by $P$ up to a logarithm, and $r$ primes divide them by $P^r$ up to $(\log n)^r$.

\paragraph{How large the primes are.} This fixes $P$ in the parameter blocks of Sections~\ref{s:sec:instances} and~\ref{e:sec:instances}, and it is where the loss $n^{\varepsilon/r}$, respectively $n^{\varepsilon/(2r)}$, and the factor $(\log n)^{r+O(1)}$ of the theorems come from; in outline, with $\sigma$ the size of a group of the 3SUM self-reduction, respectively of a piece of the third vertex part in Exact Triangle (both defined there), the argument is as follows. A survivor has at most $\sigma^2$ false positives in 3SUM and at most $\sigma$ in Exact Triangle, and the $r$ primes of a stage must bring this below one half, so the modulus $P^r$ is about $\sigma^2$, respectively $\sigma$, up to the logarithms. $\Hyp(\varepsilon,\gamma)$ bounds the middle part of a call by $n^\varepsilon$, and the middle part is the modulus for 3SUM and the modulus times $\sigma$ for Exact Triangle; so $\sigma$ is about $n^{\varepsilon/2}$ in both, as large as $\Hyp(\varepsilon,\gamma)$ allows (the saving comes from $\sigma^{2\gamma}$), the modulus is about $n^\varepsilon$, respectively $n^{\varepsilon/2}$, and $P$ is about $n^{\varepsilon/r}$, respectively $n^{\varepsilon/(2r)}$. Each candidate prime costs one evaluation of the counts, so a step makes at most $P$ evaluations: this is the loss. The logarithm lost per prime gives the $(\log n)^r$.

\section{3SUM}\label{s:sec:statement}

\paragraph{The 3SUM variant.} Throughout, $[U]:=\{1,\dots,U\}$. The core problem is the monochromatic form used by Fischer, Kaliciak and Polak~\cite{FKP24}:

\begin{definition}[3SUM, core form]\label{s:def:3sum}
Given a set $A\subseteq[U]$ of $n$ integers, decide whether there are $a,b,c\in A$ (not necessarily distinct) with $a+b=c$.
\end{definition}

Fischer, Kaliciak and Polak state it as ``Problem 2.1 (3SUM). Decide whether in a given set $A\subseteq[U]$ there is a triple $a,b,c\in A$ with $a+b=c$'' \cite[p.~7]{FKP24}; they do not say whether $a=b$ is allowed, and we allow it. The standard variants reduce to Definition~\ref{s:def:3sum} (Lemma~\ref{s:lem:variants} in Section~\ref{s:sec:variants}): three sets $A,B,C\subseteq[-U,U]$ and $a+b=c$ or $a+b+c=0$ with $a\in A$, $b\in B$, $c\in C$; and one list $x_1,\dots,x_n\in[-U,U]$ and $x_i+x_j+x_k=0$ with distinct indices $i,j,k$.

\begin{theorem}[main]\label{s:thm:main3sum}
Assume $\Hyp(\varepsilon,\gamma)$ with $0<\varepsilon,\gamma\le\frac12$. For every constant $\nu\ge1$ and every integer $r\ge1$ there is a deterministic algorithm that, given $n$, solves 3SUM (Definition~\ref{s:def:3sum}) for every set $A\subseteq[U]$ with $|A|\le n$ and $U\le n^\nu$ in time
\[
O\big(n^{2-\varepsilon\gamma+\varepsilon/r}(\log n)^{r+4}\big),
\]
where the constant in the $O$ depends on $\nu$, $r$, $\varepsilon$ and $C_{\mathcal O}$.
\end{theorem}

\begin{corollary}[one algorithm, $r$ growing with $n$]\label{s:cor:uniform}
Under the same assumptions, one deterministic algorithm solves 3SUM on at most $n$ integers with $U\le n^\nu$ in time $n^{2-\varepsilon\gamma+O(\sqrt{\log\log n/\log n})}=n^{2-\varepsilon\gamma+o(1)}$.
\end{corollary}

\begin{corollary}[standard variants]\label{s:cor:variants}
The same bounds, with $\nu$ replaced by $\nu+1$ and an extra factor $O(\log^2 n)$ for the one-list variant, hold for the variants with three sets ($a+b=c$ or $a+b+c=0$), each of at most $n$ integers, and with one list of $n$ integers and distinct indices, the integers of absolute value at most $n^\nu$.
\end{corollary}

For $r\le 1/\gamma$ the bound of Theorem~\ref{s:thm:main3sum} is no better than $n^2$; the theorem is of interest for $r>1/\gamma$. Numbers are in Section~\ref{s:sec:numbers}.

\paragraph{What is taken from elsewhere.} Besides $\Hyp(\varepsilon,\gamma)$, the proof uses as a black box only the self-reduction of Fischer, Kaliciak and Polak~\cite[Lemma~3.3]{FKP24}, quoted in Section~\ref{s:sec:quoted}, and explicit bounds on primes (Lemma~\ref{c:lem:primes}); prior work on the method is discussed in Section~\ref{x:sec:novelty}.

\paragraph{Relation to Kopelowitz, Pettie and Porat.} The randomized reduction of~\cite{KPP16} to offline Set Disjointness keeps all of the saving, and its derandomization by Fischer, Kaliciak and Polak~\cite{FKP24} keeps about a 25th of it by their proof as written. The reduction here is deterministic and keeps all of the saving, but needs the counting version $\#\Lop$; Section~\ref{s:sec:kpp} restates it in their terms and quantifies their bound.

Section~\ref{s:sec:quoted} quotes the self-reduction of Fischer, Kaliciak and Polak that splits an instance into groups, Section~\ref{s:sec:variants} reduces the standard variants to Definition~\ref{s:def:3sum} and integer instances of the counting algorithm to $0/1$ ones, Section~\ref{s:sec:instances} defines the instances asked of it, Section~\ref{s:sec:algorithm} gives the algorithm, and Sections~\ref{s:sec:correctness} and~\ref{s:sec:cost} prove Theorem~\ref{s:thm:main3sum} and Corollary~\ref{s:cor:uniform}.

\subsection{The self-reduction of Fischer, Kaliciak and Polak}\label{s:sec:quoted}

\paragraph{The quoted lemma.} The only statement of~\cite{FKP24} used as a black box is the self-reduction, their Lemma 3.3: it splits the input set into $g$ groups, each covered by a short interval, and lists $O(g^2)$ triples of groups outside of which no solution $a+b=c$ can lie; the reduction of Section~\ref{s:sec:instances} asks the counting algorithm only about these triples. It reads (arXiv 2310.12913v2, printed pp.~8--9, verbatim except for typesetting):
\begin{quote}
\textbf{Lemma 3.3} (Deterministic Self-Reduction). Let $A\subseteq[U]$ be a set of size $n$, and let $g\ge1$. We can deterministically construct a partition of $A$ into subsets $A_1,\dots,A_g$, and a set of triples $R\subseteq[g]^3$ of size $O(g^2)$ such that:
\begin{itemize}[nosep]
\item Each set $A_i$ has size $O(n/g)$ and can be covered by an interval of length $O(U/g)$.
\item For all triples $a,b,c\in A$ with $a+b=c$, there is some $(i,j,k)\in R$ with $a\in A_i$, $b\in A_j$, $c\in A_k$.
\end{itemize}
The algorithm runs in time $\tilde O(n+g^2)$.
\end{quote}
We use \cite[Lemma~3.3]{FKP24} as stated. Its proof covers $a=b$: $R$ is the set of group triples $(i,j,k)$ with $\min(A_i)+\min(A_j)\le\max(A_k)$ and $\max(A_i)+\max(A_j)\ge\min(A_k)$, and every solution $a+b=c$, also with $a=b$, lies in such a triple. We use it with constants $c_1,c_2\ge1$ from its proof such that $|A_i|\le c_1n/g$ and $|R|\le c_2g^2$; their values do not matter.

\paragraph{What is not used.} Lemma 3.2 of~\cite{FKP24} (``Deterministic Additive Hashing with Few False Positives'') either finds a modulus $m\in[n^\mu,2n^\mu)$ for which the number of false positives, summed over $g$ given 3SUM instances, is small, or decides that one of them is a yes-instance. We do not use it: its counting is by FFT over all triples of an instance, which costs $\tilde O(m)$ per instance and therefore $\tilde O(g^2\,m)$ over the $O(g^2)$ relevant triples, about $n^2$ at the modulus that would be needed here. We use its method: the modulus is a product of primes from $[P,2P)$, chosen one at a time, ``each next prime narrows down the number of false positives by an appropriate amount'' \cite[p.~9]{FKP24}, and the analysis is the averaging of its proof (p.~10), restated and proved as Lemma~\ref{c:lem:avg}; the difference is that here the counts modulo each candidate are computed by the counting algorithm, and only for the surviving probes (Section~\ref{x:sec:novelty}).

\subsection{Reductions between variants and to 0/1 instances}\label{s:sec:variants}

Theorem~\ref{s:thm:main3sum} is proved for the core form, one set in $[U]$ and $a+b=c$. The first lemma moves the standard variants into it: three sets are shifted into three disjoint ranges of $[41U]$, placed so that the only sums of two elements that fall into any of the three ranges are those of one element of the first and one of the second, which fall into the third, and a list with distinct indices becomes $O(\log^2n)$ three-set instances by sorting the indices into classes by two of their bits, which separates any three distinct indices in at least one of these ways. The second lemma is for the calls to the counting algorithm: Definition~\ref{c:hyp} speaks of $0/1$ matrices (Section~\ref{c:sec:prelim}), while the instance of Section~\ref{s:sec:instances} has entries up to $\sigma$, the multiplicities of residues within a group, and the lemma writes such an instance as $B^2$ instances with $0/1$ entries, $B$ the number of bits of an entry.

\begin{lemma}[variants]\label{s:lem:variants}
Let $U\ge1$.
\begin{enumerate}[label=(\alph*)]
\item Three sets $A,B,C\subseteq[-U,U]$, question $\exists a\in A,b\in B,c\in C$: $a+b=c$. Let $S:=\{a+10U\}_{a\in A}\cup\{b+30U\}_{b\in B}\cup\{c+40U\}_{c\in C}\subseteq[41U]$. Then the instance is a yes-instance if and only if $S$ is a yes-instance of Definition~\ref{s:def:3sum}.
\item The question $a+b+c=0$ is (a) with $C$ replaced by $-C$.
\item One list $x_0,\dots,x_{n-1}\in[-U,U]$, question: distinct $i,j,k$ with $x_i+x_j+x_k=0$. Write the indices $0,\dots,n-1$ in binary with $\ell:=\lceil\log_2 n\rceil$ bits. For two bit positions $\beta_1\ne\beta_2$, let the class of $i$ be its two bits at $\beta_1,\beta_2$ (four classes), and for each injective assignment of three classes to the roles $A,B,C$ form the three-set instance $A=\{x_i\}$, $B=\{x_i\}$, $C=\{x_i\}$ with $i$ ranging over the respective class. The list is a yes-instance if and only if one of these at most $24\ell^2$ three-set instances (with $a+b+c=0$) is.
\end{enumerate}
\end{lemma}
\begin{proof}
(a) The three shifted sets lie in $[9U,11U]$, $[29U,31U]$ and $[39U,41U]$, which are disjoint. Let $x+y=z$ with $x,y,z\in S$. Then $x+y\ge18U$, so $z$ is in the second or third range. Sums of two elements fall in $[18U,22U]$ (both from the first range), $[38U,42U]$ (first and second), and from $48U$ upward otherwise; none meets $[29U,31U]$, and only the sums of one element of the first and one of the second range meet $[39U,41U]$. So $x,y$ are $a+10U$ and $b+30U$ in some order and $z=c+40U$, and $x+y=z$ if and only if $a+b=c$. The converse is immediate.
(b) is clear. (c) A solution of a three-set instance uses indices from three distinct classes, hence distinct indices. Conversely, let $i,j,k$ be distinct with $x_i+x_j+x_k=0$. Let $\beta_1$ be a bit where $i$ and $j$ differ; by symmetry between $i$ and $j$ assume $k$ agrees with $i$ at $\beta_1$, and let $\beta_2$ be a bit where $i$ and $k$ differ ($\beta_2\ne\beta_1$). Then $i,j$ differ at $\beta_1$, $j,k$ differ at $\beta_1$, and $i,k$ differ at $\beta_2$, so the three classes are distinct, and the assignment of these classes to the roles of $i,j,k$ gives a three-set instance containing the solution.
\end{proof}

Corollary~\ref{s:cor:variants} follows from Theorem~\ref{s:thm:main3sum} and Corollary~\ref{s:cor:uniform}, applied with the size bound $3n$, with this lemma: in (a) and (b) one instance of at most $3n$ integers with universe $41U\le n^{\nu+1}\le(3n)^{\nu+1}$ for $n\ge41$; in (c) at most $24\ell^2=O(\log^2n)$ such instances, each of at most $3n$ integers (a class may be small or empty, which is why the theorem is stated for $|A|\le n$); the transformations take $O(n\log^2 n)$ time.

\begin{lemma}[binary splitting]\label{s:lem:split}
Let $X\in\{0,\dots,2^B-1\}^{N\times D}$, $Y\in\{0,\dots,2^B-1\}^{D\times N}$ and $W$ a set of positions. Write $X=\sum_{i<B}2^iX_i$ and $Y=\sum_{j<B}2^jY_j$ with $0/1$ matrices. Then $(XY)[I,J]=\sum_{i,j<B}2^{i+j}(X_iY_j)[I,J]$, so the wanted entries of $XY$ are obtained from $B^2$ instances of $\#\Lop(N,D)$ with the same $W$, plus $O(B^2(ND+|W|))$ time.
\end{lemma}
\begin{proof}
Bilinearity. This is the standard idea of the preprint's footnote~8, in Section 3.1, without the sign split since the entries are nonnegative.
\end{proof}

\subsection{The instances}\label{s:sec:instances}

\paragraph{Groups and probes.} Let $A\subseteq[U]$ with $|A|\le n$, where the bound $n$ is given and all parameters below are set from $n$; apply \cite[Lemma~3.3]{FKP24} with some $g\ge1$, obtaining groups $A_1,\dots,A_g$ with $|A_i|\le c_1|A|/g\le c_1n/g$ and the set $R$, $|R|\le c_2g^2$. Let $\sigma$ be an integer with $\sigma\ge c_1n/g$, so that every group has at most $\sigma$ elements. A \emph{probe} is a triple $Q=(a;j,k)$ such that $(i,j,k)\in R$ for the group $A_i\ni a$. Let $\Pq_0$ be the set of all probes; then $|\Pq_0|=\sum_{(i,j,k)\in R}|A_i|\le c_2g^2\sigma$. For a probe $Q=(a;j,k)$ and an integer $m\ge1$ let
\[
\cnt_m(Q):=\#\{(b,c)\in A_j\times A_k:\ a+b\equiv c\pmod m\},\quad
Z(Q):=\#\{(b,c)\in A_j\times A_k:\ a+b=c\}.
\]
Then $\cnt_m(Q)\ge Z(Q)$ for every $m$, $\cnt_1(Q)=|A_j||A_k|\le\sigma^2$, and by the second property of \cite[Lemma~3.3]{FKP24}, $A$ is a yes-instance if and only if $Z(Q)>0$ for some $Q\in\Pq_0$. The pairs counted by $\cnt_m(Q)-Z(Q)$ are the \emph{false positives} of $Q$ modulo $m$ (Fischer, Kaliciak and Polak call them pseudo-solutions).

\paragraph{The instance $I(m,\Pq)$.} Let $m\ge2$, $T:=\lceil\sqrt m\rceil$ (so $T^2\ge m$), and let $\Pq\subseteq\Pq_0$. The left vertices are the pairs $(j,x)$ with $j\in[g]$, $x\in\{0,\dots,T\}$; the right vertices are the pairs $(k,y)$ with $k\in[g]$, $y\in\{0,\dots,T-1\}$; the middle part is $\Z_m$. Define
\[
X[(j,x),\lambda]:=\#\{b\in A_j: b+xT\equiv\lambda\},\qquad Y[\lambda,(k,y)]:=\#\{c\in A_k: c+y\equiv\lambda\}\pmod m .
\]
A probe $Q=(a;j,k)$ is mapped to the position $\pi_m(Q):=((j,x),(k,y))$ where $\varrho:=a\bmod m\in[0,m)$, $x:=\lceil\varrho/T\rceil$ and $y:=xT-\varrho$. Since $\varrho\le m-1\le T^2-1$ we have $x\le T$, and $0\le y<T$; and $xT-y\equiv a\pmod m$. The wanted set is $W:=\{\pi_m(Q):Q\in\Pq\}$, so $|W|\le|\Pq|$. The residue of $a$ is thus split into a high part $xT$, carried by the left vertex, and a low part $y$, carried by the right vertex, so that each side has about $g\sqrt m$ vertices rather than $gm$; this is what gives the instance the shape $N\approx g\sqrt D$ used in Lemma~\ref{s:lem:shape}.

\begin{lemma}[the counts]\label{s:lem:counts}
For every probe $Q$, $(XY)[\pi_m(Q)]=\cnt_m(Q)$.
\end{lemma}
\begin{proof}
$(XY)[(j,x),(k,y)]=\sum_{\lambda\in\Z_m}X[(j,x),\lambda]\,Y[\lambda,(k,y)]$ is the number of pairs $(b,c)\in A_j\times A_k$ with $b+xT\equiv c+y\pmod m$, that is, with $b+(xT-y)\equiv c$, that is, with $a+b\equiv c\pmod m$.
\end{proof}
(This is Fischer, Kaliciak and Polak's construction in the proof of their Theorem 1.4, where ``the elements in the queried set intersections \dots\ are in one-to-one correspondence to the set of pseudo-solutions'' \cite[p.~18]{FKP24}, with multiplicities kept: $X$ and $Y$ record multisets, since two elements of a group may be congruent modulo $m$.) Distinct probes may share a position (two elements of $A_i$ congruent modulo $m$); the counting algorithm returns the count once per position and each such probe reads it.

\paragraph{Parameters.} Fix $r\ge1$ and put
\[
\Lambda_{\mathrm S}:=2\,(10\ln(2U))^r,\qquad
\sigma:=\big\lfloor\sqrt{n^\varepsilon/\Lambda_{\mathrm S}}\big\rfloor,\qquad
g:=\lceil c_1n/\sigma\rceil,\qquad
P:=\big\lceil4\ln(2U)\,(2\sigma^2)^{1/r}\big\rceil,
\]
\[
L:=\lceil2\gamma\log_2\sigma\rceil,\qquad B:=\lceil\log_2(\sigma+1)\rceil .
\]
In words: $\sigma$ is the group size, chosen so that the moduli, which are at most $\Lambda_{\mathrm S}\sigma^2$, satisfy the condition $D\le N^\varepsilon$ of $\Hyp(\varepsilon,\gamma)$ (Lemma~\ref{s:lem:shape}); $g$ is the number of groups; $P$ is the size of the candidate primes, chosen so that $r$ of them cut the false positives by a factor $2\sigma^2$ (Lemma~\ref{s:lem:halving}); $L$ is the number of stages, enough to shrink the set of probes by a factor $\sigma^{2\gamma}$, which makes the final scans cheap; $B$ is the number of bits of an entry of $X$ or $Y$; and $\Lambda_{\mathrm S}$ is the polylogarithmic slack that these choices cost. Then every group has at most $c_1n/g\le\sigma$ elements. We assume $n\ge n_0(\nu,r,\varepsilon)$ large enough that $\sigma\ge2$, $c_1n/\sigma\ge1$, $P\ge1024$ and $r\le P/(4\ln P)$; this holds for large $n$ because $\Lambda_{\mathrm S}$ is polylogarithmic in $n$ (as $U\le n^\nu$) and $P\to\infty$ with $\sigma$. Smaller $n$ are solved by brute force in $O(1)$ time. Since $U\le n^\nu$ and $n\ge2$, $\ln(2U)\le(\nu+1)\ln n$. The consequences used below:
\begin{equation}\label{s:eq:params}
\tfrac12\sqrt{n^\varepsilon/\Lambda_{\mathrm S}}\le\sigma,\quad \Lambda_{\mathrm S}\sigma^2\le n^\varepsilon,\quad g\le 2c_1n/\sigma,\quad (2P)^r\le\Lambda_{\mathrm S}\sigma^2 .
\end{equation}
The first holds since $\lfloor x\rfloor\ge x/2$ for $x\ge1$, the second and third by definition. For the fourth, $\ln(2U)(2\sigma^2)^{1/r}\ge1$ (as $\ln(2U)\ge\ln2$ and $(2\sigma^2)^{1/r}\ge2$ for large $n$), so $P\le5\ln(2U)(2\sigma^2)^{1/r}$ and $(2P)^r\le(10\ln(2U))^r\cdot2\sigma^2=\Lambda_{\mathrm S}\sigma^2$.

\begin{lemma}[shape and cost of one evaluation]\label{s:lem:shape}
Let $2\le m\le\Lambda_{\mathrm S}\sigma^2$ and $\Pq\subseteq\Pq_0$. The counts $\cnt_m(Q)$ for all $Q\in\Pq$ can be computed with $\Hyp(\varepsilon,\gamma)$ deterministically in time $O(B^2\Lambda_{\mathrm S}^{1-\gamma}g^2\sigma^{2-2\gamma})$, the constant depending only on $C_{\mathcal O},c_1,c_2$.
\end{lemma}
\begin{proof}
Use the instance $I(m,\Pq)$, padding the right side with isolated vertices so that both sides have $N:=g(T+1)$ vertices; $D:=m$. Then $g\sqrt D\le N\le3g\sqrt D$.

\emph{$D\le N^\varepsilon$.} It suffices that $D^{1-\varepsilon/2}\le g^\varepsilon$, since then $D=D^{1-\varepsilon/2}D^{\varepsilon/2}\le(g\sqrt D)^\varepsilon\le N^\varepsilon$. As $D\le\Lambda_{\mathrm S}\sigma^2$ and $g\ge n/\sigma$, it suffices that $(\Lambda_{\mathrm S}\sigma^2)^{1-\varepsilon/2}\le(n/\sigma)^\varepsilon$, i.e.\ $\Lambda_{\mathrm S}^{1-\varepsilon/2}\sigma^2\le n^\varepsilon$, which follows from $\Lambda_{\mathrm S}\sigma^2\le n^\varepsilon$ and $\Lambda_{\mathrm S}\ge1$.

\emph{Entries.} The entries of $X$ and $Y$ are in $\{0,\dots,\sigma\}\subseteq\{0,\dots,2^B-1\}$. By Lemma~\ref{s:lem:split} the computation is $B^2$ instances of $\#\Lop(N,D)$ with the same $W$.

\emph{Splitting $W$.} Cut $W$ into parts of $\lfloor N^2/\sqrt D\rfloor\ge N^2/(2\sqrt D)$ positions (the last one possibly fewer). As $|W|\le|\Pq_0|\le c_2g^2\sigma$ and $N^2/\sqrt D\ge g^2\sqrt D$, there are at most $1+2c_2\sigma/\sqrt D$ parts. Each part is one call, of cost at most $C_{\mathcal O}N^2/D^\gamma\le9C_{\mathcal O}g^2D^{1-\gamma}$. Hence one $0/1$ instance costs
\[
O\big(g^2D^{1-\gamma}+g^2\sigma D^{1/2-\gamma}\big).
\]
If $D\le\sigma^2$: $D^{1-\gamma}\le\sigma^{2-2\gamma}$ and, since $\gamma\le\frac12$, $\sigma D^{1/2-\gamma}\le\sigma\cdot\sigma^{1-2\gamma}$. If $\sigma^2<D\le\Lambda_{\mathrm S}\sigma^2$: $D^{1-\gamma}\le\Lambda_{\mathrm S}^{1-\gamma}\sigma^{2-2\gamma}$, and $\sigma D^{1/2-\gamma}\le\Lambda_{\mathrm S}^{1/2-\gamma}\sigma^{2-2\gamma}$. In both cases the cost is $O(\Lambda_{\mathrm S}^{1-\gamma}g^2\sigma^{2-2\gamma})$, as $\Lambda_{\mathrm S}\ge1$.

\emph{Input, output and bookkeeping.} Writing a $0/1$ instance and reading its answers takes $O(ND+|W|)$. Since $D\le N^\varepsilon$ and $\varepsilon(1+\gamma)\le\frac34<1$, we have $D^{1+\gamma}\le N$, so $ND\le N^2/D^\gamma$; and $|W|\le N^2/\sqrt D\le N^2/D^\gamma$ per part. So this is within the cost of the calls. Filling $X$ and $Y$ from the groups takes $O(N\sigma+ND)$ time (one increment per element $b\in A_j$ per row $(j,x)$), and $N\sigma\le3g\sqrt\Lambda_{\mathrm S}\,\sigma^2=O(g^2\sigma)$ since $g\ge n^{1-\varepsilon/2}\ge3\sqrt\Lambda_{\mathrm S}\,\sigma$ for large $n$. The positions $\pi_m(Q)$, $Q\in\Pq$, are computed and sorted by radix sort with base $N$ in $O(|\Pq|+N)$ time; this removes duplicates from $W$ and lets each probe read the count of its position. Both are $O(g^2\sigma)\le O(g^2\sigma^{2-2\gamma})$.
\end{proof}

The case $D\le\sigma^2$ is the one where the modulus is smaller than the square of the group size: the instance then may have more than $N^2/\sqrt D$ query pairs, and splitting $W$ costs at most what the largest moduli cost. This is used for the partial products $m\cdot p$ while a modulus is being built.

\subsection{The algorithm}\label{s:sec:algorithm}

Let $\Pi$ be the set of primes in $[P,2P)$, found by a sieve in $O(P\log\log P)$ time.
\begin{enumerate}[label=\arabic*.]
\item Compute the groups and $R$ by \cite[Lemma~3.3]{FKP24} with the $g$ above, sort every group, and list $\Pq_0$.
\item For $t=1,\dots,L$:
\begin{enumerate}[label=(\alph*)]
\item If $\Pq_{t-1}=\emptyset$, answer NO.
\item Build $M_t$: start with $m:=1$ and $\Pi':=\Pi$. Repeat $r$ times: for every $p\in\Pi'$ compute
\[\Sigma_t(mp):=\sum_{Q\in\Pq_{t-1}}\cnt_{mp}(Q)\]
by Lemma~\ref{s:lem:shape} (instance $I(mp,\Pq_{t-1})$); choose $p^*\in\Pi'$ minimising $\Sigma_t(mp)$ (ties arbitrarily), and set $m:=mp^*$, $\Pi':=\Pi'\setminus\{p^*\}$. Let $M_t:=m$. In the last of the $r$ steps, keep the counts $\cnt_{mp}(Q)$ of the current minimiser while the candidates are evaluated; at the end they are the counts $\cnt_{M_t}(Q)$.
\item If $\Sigma_t(M_t)>|\Pq_{t-1}|/2$, answer YES.
\item Otherwise let $\Pq_t:=\{Q\in\Pq_{t-1}:\cnt_{M_t}(Q)>0\}$.
\end{enumerate}
\item For every $Q=(a;j,k)\in\Pq_L$, test whether $a+b\in A_k$ for some $b\in A_j$, by merging the sorted lists $a+A_j$ and $A_k$. Answer YES if some test succeeds, NO otherwise.
\end{enumerate}
Every candidate modulus $mp$ is a product of at most $r$ distinct primes from $[P,2P)$, so $2\le mp<(2P)^r\le\Lambda_{\mathrm S}\sigma^2$ by~\eqref{s:eq:params}, and Lemma~\ref{s:lem:shape} applies to every evaluation. The algorithm is deterministic. It is adaptive: the wanted set of a call depends on the answers of earlier calls. This is allowed, since each call is a separate run of $\mathcal O$ on an instance whose $W$ is fully known when the call is made.

\subsection{Correctness}\label{s:sec:correctness}

For a set $\Pq$ of probes and a modulus $m$ let $F_\Pq(m)$ be the number of triples $(Q,b,c)$ with $Q=(a;j,k)\in\Pq$, $(b,c)\in A_j\times A_k$, $a+b\ne c$ and $a+b\equiv c\pmod m$ (false positives), and $Z_\Pq:=\sum_{Q\in\Pq}Z(Q)$. Then $\sum_{Q\in\Pq}\cnt_m(Q)=F_\Pq(m)+Z_\Pq$, and $Z_\Pq$ does not depend on $m$.

\begin{lemma}[halving]\label{s:lem:halving}
In every stage $t$ that reaches step 2(b), $F_{\Pq_{t-1}}(M_t)\le|\Pq_{t-1}|/2$. Consequently, if step 2(c) answers YES then $Z_{\Pq_{t-1}}>0$, so the answer is correct; otherwise $|\Pq_t|\le|\Pq_{t-1}|/2$.
\end{lemma}
\begin{proof}
Write $\Pq:=\Pq_{t-1}$ and let $m_0=1,m_1,\dots,m_r=M_t$ be the successive values of $m$. Since $\Sigma_t(m_{l-1}p)=F_\Pq(m_{l-1}p)+Z_\Pq$ and $Z_\Pq$ is independent of $p$, the prime $p^*$ chosen in step $l$ also minimises $F_\Pq(m_{l-1}p)$ over the current pool $\Pi'$. Apply Lemma~\ref{c:lem:avg} to the multiset $\mathcal F$ of the numbers $a+b-c$ over the false positives modulo $m_{l-1}$: each is nonzero, divisible by $m_{l-1}$, and $|a+b-c|\le2U$ since $a,b,c\in[U]$; the pool consists of primes in $[P,2P)$ not among the $l-1$ chosen ones, and $m_{l-1}$ is the product of the chosen ones, so no prime of the pool divides $m_{l-1}$. The pool has at least $|\Pi|-r\ge P/(2\ln P)-P/(4\ln P)=P/(4\ln P)$ primes by Lemma~\ref{c:lem:primes} and $r\le P/(4\ln P)$. Hence
\[
F_\Pq(m_l)\le F_\Pq(m_{l-1})\cdot\frac{\ln(2U)}{\ln P}\cdot\frac{4\ln P}{P}=F_\Pq(m_{l-1})\cdot\frac{4\ln(2U)}{P}\le F_\Pq(m_{l-1})\,(2\sigma^2)^{-1/r},
\]
by the choice of $P$. After $r$ steps, $F_\Pq(M_t)\le F_\Pq(1)/(2\sigma^2)\le|\Pq|\sigma^2/(2\sigma^2)=|\Pq|/2$, using $F_\Pq(1)\le\sum_{Q}\cnt_1(Q)\le|\Pq|\sigma^2$.

If $\Sigma_t(M_t)>|\Pq|/2$ then $Z_\Pq=\Sigma_t(M_t)-F_\Pq(M_t)>0$, so some probe $Q=(a;j,k)$ has $b\in A_j$, $c\in A_k$ with $a+b=c$, and $A$ is a yes-instance. Otherwise every $Q\in\Pq_t$ has $\cnt_{M_t}(Q)\ge1$, so $|\Pq_t|\le \Sigma_t(M_t)\le|\Pq|/2$.
\end{proof}
As in the proof of the preprint's Theorem~17 (and Fischer, Jin and Xu~\cite[Lemma~3.2]{FJX25}), the count is the number of false positives plus a term independent of the modulus, so the bound on $F$ holds whether or not $A$ is a yes-instance; Fischer, Kaliciak and Polak's proof of their Lemma 3.2 instead assumes unsatisfiable instances and uses Markov's inequality.

\begin{lemma}[correctness]\label{s:lem:correct}
The algorithm answers YES if and only if $A$ is a yes-instance.
\end{lemma}
\begin{proof}
A YES of step 2(c) is correct by Lemma~\ref{s:lem:halving}; a YES of step 3 exhibits $a+b=c$ with $a,b,c\in A$. Conversely, let $a+b=c$ with $a,b,c\in A$. By \cite[Lemma~3.3]{FKP24} there is $(i,j,k)\in R$ with $a\in A_i$, $b\in A_j$, $c\in A_k$, so $Q:=(a;j,k)\in\Pq_0$ and $\cnt_m(Q)\ge Z(Q)\ge1$ for every $m$. Hence, unless the algorithm answers YES earlier, $Q\in\Pq_t$ for every $t$ (in particular step 2(a) never answers NO), and step 3 finds $b$.
\end{proof}

\subsection{Cost}\label{s:sec:cost}

We bound each part, using~\eqref{s:eq:params} and $g^2\sigma^2\le4c_1^2n^2$.
\begin{enumerate}[label=(\roman*)]
\item \emph{Self-reduction} (\cite[Lemma~3.3]{FKP24}) and listing $\Pq_0$: $\tilde O(n+g^2)+O(c_2g^2\sigma)=\tilde O(n^2/\sigma)$, which is $O(n^{2-\varepsilon/2}\Lambda_{\mathrm S}^{1/2}\operatorname{polylog}n)$, below the main term.
\item \emph{Candidate primes}: the sieve, $O(P\log\log P)$, once.
\item \emph{Stages}: at most $L\le\log_2n+1$ stages, $r$ steps each, at most $|\Pi|<P$ evaluations per step. By Lemma~\ref{s:lem:shape} each evaluation costs $O(B^2\Lambda_{\mathrm S}^{1-\gamma}g^2\sigma^{2-2\gamma})$, including writing its instances, the splitting of $W$ (also for moduli smaller than $\sigma^2$), the $O(|\Pq_{t-1}|)$ work to map the probes and sum the counts, and reading the answers. Choosing the minimum and forming $\Pq_t$ is $O(P+|\Pq_{t-1}|)$ per step. In total
\[
O\big(L\,r\,P\,B^2\,\Lambda_{\mathrm S}^{1-\gamma}g^2\sigma^{2-2\gamma}\big).
\]
\item \emph{Final scans}: by Lemma~\ref{s:lem:halving}, $|\Pq_L|\le|\Pq_0|/2^L\le c_2g^2\sigma/\sigma^{2\gamma}$, and each merge costs $O(\sigma)$, so $O(g^2\sigma^{2-2\gamma})$.
\item \emph{Word size}: the inputs are integers of at most $\nu\log_2n+1$ bits, i.e.\ $O(\nu)$ words of $\Theta(\log n)$ bits; residues are below $\Lambda_{\mathrm S}\sigma^2\le n^\varepsilon$; counts are at most $\sigma^2$ per probe and sums at most $|\Pq_0|\sigma^2\le n^3$. The instances of the calls have $N\le3g\sqrt{\Lambda_{\mathrm S}}\sigma\le n^{O(1)}$ and $N\ge g\ge n^{1-\varepsilon/2}$, so its $O(\log N)$-bit words are $\Theta(\log n)$-bit words. All arithmetic above is $O(\nu)$ word operations per step.
\end{enumerate}
The stages dominate. Insert $g^2\sigma^{2-2\gamma}\le4c_1^2n^2\sigma^{-2\gamma}\le4^{1+\gamma}c_1^2\Lambda_{\mathrm S}^\gamma n^{2-\varepsilon\gamma}$ (from $\sigma\ge\frac12\sqrt{n^\varepsilon/\Lambda_{\mathrm S}}$), $P\le5\ln(2U)(2\sigma^2)^{1/r}\le10\ln(2U)\,n^{\varepsilon/r}$, $L,B\le\log_2n+1$, $\Lambda_{\mathrm S}^{1-\gamma}\Lambda_{\mathrm S}^\gamma=\Lambda_{\mathrm S}=2(10\ln(2U))^{r}$ and $\ln(2U)\le(\nu+1)\ln n$:
\begin{equation}\label{s:eq:total}
T(n)=O\Big(r\,n^{2-\varepsilon\gamma+\varepsilon/r}\,\big(10(\nu+1)\big)^{r+1}(\ln n)^{r+4}\Big),
\end{equation}
with the constant in the $O$ depending only on $C_{\mathcal O},c_1,c_2$. The factor $n^{\varepsilon/r}$ is the number of candidate primes, $P\approx\sigma^{2/r}$; the factor $(\ln n)^{r}$ comes from $\Lambda_{\mathrm S}=\Lambda_{\mathrm S}^{1-\gamma}\cdot\Lambda_{\mathrm S}^\gamma$, i.e.\ from the moduli exceeding $\sigma^2$ by up to $\Lambda_{\mathrm S}$ (factor $\Lambda_{\mathrm S}^{1-\gamma}$) and from shrinking $\sigma$ by $\sqrt\Lambda_{\mathrm S}$ to keep $D\le N^\varepsilon$ (factor $\Lambda_{\mathrm S}^\gamma$); the remaining $(\ln n)^4$ are $L$, $B^2$ and the $\ln(2U)$ in $P$.

\begin{proof}[Proof of Theorem~\ref{s:thm:main3sum}]
Lemma~\ref{s:lem:correct} and~\eqref{s:eq:total}, with $r$ and $\nu$ constants.
\end{proof}

\begin{proof}[Proof of Corollary~\ref{s:cor:uniform}]
Run the algorithm with $r=r(n):=\lceil\sqrt{\ln n/\ln\ln n}\,\rceil$. All constants in~\eqref{s:eq:total} are explicit in $r$, so the bound holds with this $r$ provided the side conditions hold. First, $\Lambda_{\mathrm S}=\exp(O(r\ln\ln n))=n^{o(1)}$, so $\sigma\to\infty$. Second, $P\ge(2\sigma^2)^{1/r}=\exp(\Theta(\sqrt{\ln n\ln\ln n}))$, so $P\ge1024$ and $P/(4\ln P)\ge r$ for large $n$. Then $n^{\varepsilon/r}\le\exp(\varepsilon\sqrt{\ln n\ln\ln n})$ and
\[
r\,(10(\nu+1))^{r+1}(\ln n)^{r+4}=\exp\big(O(\sqrt{\ln n/\ln\ln n}\cdot\ln\ln n)\big)=\exp\big(O(\sqrt{\ln n\ln\ln n})\big),
\]
so $T(n)=n^{2-\varepsilon\gamma}\exp(O(\sqrt{\ln n\ln\ln n}))=n^{2-\varepsilon\gamma+O(\sqrt{\ln\ln n/\ln n})}$.
\end{proof}

\subsection{Remarks}\label{s:sec:numbers}

\paragraph{Numbers.} With the preprint's Corollary 26, $\Hyp(1/18,0.063)$ holds, $\varepsilon\gamma=0.0035$, and Corollary~\ref{s:cor:uniform} gives deterministic 3SUM in $n^{1.99650+o(1)}$; the preprint's route through Exact Triangle gives $n^{2-\varepsilon'/2+o(1)}$ with $\varepsilon'=0.00175$, i.e.\ $n^{1.99913}$ (its Theorem 22), which keeps a quarter of $\varepsilon\gamma$. The preprint's constants are not optimised, and any improvement of $(\varepsilon,\gamma)$ carries over to the theorem here unchanged; in this section we use only the preprint's stated ones, and Part~\hyperref[m:partI]{I}'s improvement is combined with the theorem in Section~\ref{x:sec:consequences}. Exponents in this section are rounded up in the last digit.

For a fixed $r$ (Theorem~\ref{s:thm:main3sum}) the saving is $\varepsilon\gamma-\varepsilon/r$:
\begin{center}
\begin{tabular}{c|c|c|c}
$(\varepsilon,\gamma)$ & $r=30$ & $r=40$ & $r=80$\\\hline
$(1/18,\ 0.063)$ & $n^{1.99836}$ & $n^{1.99789}$ & $n^{1.99720}$
\end{tabular}
\end{center}
each times $(\log n)^{r+4}$. To beat the preprint's $2-\varepsilon\gamma/4$ one needs $\varepsilon/r<\frac34\varepsilon\gamma$, i.e.\ $r>4/(3\gamma)$: $r\ge22$ for $\gamma=0.063$.

\paragraph{Constants.} The polylogarithmic factor is large. For example, $r=40$ gains about $n^{0.00124}$ over the preprint's $n^{1.99913}$ ($1.999125-1.997889$) but pays $(\ln n)^{44}$, which is larger until $\ln n\approx4.6\cdot10^5$, that is, until $n$ has about $200{,}000$ decimal digits; no choice of $r$ brings this below about $199{,}000$ digits ($r=44$), and $r=30$ or $r=80$ raise it to about $250{,}000$. The staged route therefore pays off only asymptotically.

\paragraph{What the counting algorithm must provide.} Our proof needs the counting version $\#\Lop$: the greedy choice of a prime needs the number of false positives of the surviving probes, a potential that decreases by a factor $\approx\sigma^{-2/r}$ per prime. We do not know how to do the same with a detection algorithm: the number of positive probes (what a detection algorithm gives) is not such a potential, since a probe with many false positives modulo $m$ typically stays positive modulo $mp$. Counting by FFT restricted to the survivors does not help: already in the first stage every relevant triple has a survivor, and counting at a modulus $m\approx\sigma^2$ costs $\tilde O(g^2\sigma^2)\approx n^2$.

\paragraph{Search version.} The algorithm decides. A YES of step 3 comes with a solution; a YES of step 2(c) does not. Finding a solution is not treated here.

\section{3SUM and counting offline Set Disjointness}\label{s:sec:kpp}

This section restates the 3SUM reduction in the terms of Kopelowitz, Pettie and Porat~\cite{KPP16} and of Fischer, Kaliciak and Polak~\cite{FKP24}; it is not used elsewhere in the paper. \emph{Counting offline Set Disjointness} is Fischer, Kaliciak and Polak's Problem~1.3 with the count in place of the emptiness bit: given sets $S_1,\dots,S_N\subseteq[u]$, each of size at most $s$, and a set of $q_{\mathrm S}$ queries $(i,j)\in[N]^2$, report $|S_i\cap S_j|$ for every query~\cite[p.~3]{FKP24} (they write $U$ for $u$).

\paragraph{Relation to~\cite[Theorem~1.4]{FKP24}.} Fischer, Kaliciak and Polak derandomize the reduction of~\cite{KPP16} to the \emph{detection} version of offline Set Disjointness. Their theorem is not quantified; inserting a running time $n^{2-\varepsilon\gamma}$ for the called algorithm (the saving of the preprint's Corollary 26) into the running time $\tilde O(n^{2-\delta(1-\alpha)}+n^{2-\varepsilon\gamma+6\rho}+n^{2-\rho}+n^{2+3\delta(1-\alpha)-\rho})$ at the end of their proof \cite[p.~19]{FKP24} and balancing ($\rho=4\delta(1-\alpha)$) gives 3SUM in $n^{2-\varepsilon\gamma/25+o(1)}$, so their proof as written keeps about a 25th of the saving.\footnote{The last term reads $n^{2+3\delta-\rho}$ there; their step~2 on p.~18 gives $n^{1+\alpha+3\delta(1-\alpha)}$ false positives, the middle expression of $|R|\cdot O((n^{1-\alpha})^{1+3\delta})$ there, which is the reading used. The terms $n$, $n^{2\alpha}$ and $n^{1+\alpha+\delta}$ of their sum are omitted: the last is the case $\mu<1$ of their Lemma~3.2, which does not occur since $\mu=2-2\delta\ge1$ at the balancing point, where $\delta(1-\alpha)$ is a 25th of the saving and the saving is at most $1-\alpha$, the called algorithm having to write its $n^{1+\alpha}$ answers.} The preprint observes that the randomized reduction of~\cite{KPP16} keeps all of the saving and that its derandomization~\cite{FKP24} keeps far less (its footnote~10, in Section~6). The reduction here is deterministic and keeps all of the saving, but needs the \emph{counting} version $\#\Lop$; for the detection version, their theorem is not improved here. As a hardness statement, it is deterministic 3SUM-hardness of counting offline Set Disjointness in their setting, with the saving kept. The matrix theorem of the preprint computes integer entries and therefore solves $\#\Lop$, so for the algorithmic consequence the counting algorithm costs nothing.

\begin{theorem}[a deterministic version of~\cite{KPP16} for counting]\label{s:thm:kpp}
Let $\nu\ge1$, $0<\alpha<1$, $r\ge1$ an integer, and $\Lambda:=2(10(\nu+1)\ln n)^r$. \textup{(a)} For $n\ge n_0(\nu,\alpha,r)$, 3SUM on a set $A\subseteq[U]$ with $|A|\le n$ and $U\le n^\nu$ reduces deterministically, with adaptive calls, to $O(r\,n^{(2-2\alpha)/r}\log^4n)$ instances of counting offline Set Disjointness, each with $u\le\Lambda n^{2-2\alpha}$, $N\le10c_1\sqrt\Lambda\,n$, $s\le n^{1-\alpha}$ and $q_{\mathrm S}\le8c_1^2c_2n^{1+\alpha}$, plus $O(\sqrt\Lambda\,n^{2-\alpha}\log n+n^{1+\alpha})$ time per call. \textup{(b)} If counting offline Set Disjointness with $u=O(n^{2-2\alpha})$, $N=O(n)$, $s=O(n^{1-\alpha})$ and $q_{\mathrm S}=O(n^{1+\alpha})$ is solved deterministically in time $O(n^{2-\eta})$, then 3SUM is solved deterministically in time $O(n^{2-\eta+(2-2\alpha)/r}(\log n)^{r/(1-\alpha)+5})$ for every $r$. The constants in the $O$ depend on $\nu$, $\alpha$, $r$ and the called algorithm.
\end{theorem}
\begin{proof}
Run the algorithm of Section~\ref{s:sec:algorithm} with $\sigma:=\lfloor n^{1-\alpha}\rfloor$, $g:=\lceil c_1n/\sigma\rceil$, $P$ and $B$ as in Section~\ref{s:sec:instances}, and $L:=\lceil\log_2(c_2g^2\sigma)\rceil\le3\log_2n$. The statement $\Hyp(\varepsilon,\gamma)$ is not used: each evaluation of Lemma~\ref{s:lem:shape} becomes, by Lemma~\ref{s:lem:split}, $B^2$ instances whose sets are the rows of the $0/1$ matrices $X_i$ (the residues whose multiplicity has bit $i$) and the columns of the $Y_{i'}$, with $W$ unsplit and no condition $D\le N^\varepsilon$. Lemma~\ref{c:lem:avg} and Lemmas~\ref{s:lem:halving} and~\ref{s:lem:correct} use of the parameters only $\cnt_1(Q)\le\sigma^2$, $P\ge4\ln(2U)(2\sigma^2)^{1/r}$, $P\ge1024$ and $r\le P/(4\ln P)$, which hold for large $n$ as in Section~\ref{s:sec:instances}; the bound $(2P)^r\le\Lambda\sigma^2$ is used only for the size of the instances below. So the algorithm is correct, and by Lemma~\ref{s:lem:halving} $|\Pq_L|\le|\Pq_0|/2^L\le1$, so step 3 costs $O(\sigma)$. There are at most $L\,r\,P\,B^2$ calls, and $P\le5\ln(2U)(2\sigma^2)^{1/r}\le10(\nu+1)\,n^{(2-2\alpha)/r}\ln n$. In each, $u=m<(2P)^r\le\Lambda_{\mathrm S}\sigma^2\le\Lambda\sigma^2$ by~\eqref{s:eq:params}, since $\Lambda_{\mathrm S}$ is $\Lambda$ with $(\nu+1)\ln n$ replaced by $\ln(2U)\le(\nu+1)\ln n$; $N=g(2T+1)\le10c_1\sqrt\Lambda\,n$ with $T=\lceil\sqrt m\rceil$; $s\le\sigma$; $q_{\mathrm S}\le|\Pq_0|\le c_2g^2\sigma\le8c_1^2c_2n^{1+\alpha}$. Writing the sets (sorting each multiset of residues and splitting the multiplicities into bits) takes $O(N\sigma B)$, and mapping the probes and combining the answers $O(B^2q_{\mathrm S}+N)$ per evaluation, which gives the time per call. The self-reduction, listing $\Pq_0$ and the sieve take $\tilde O(n^{2\alpha}+n^{1+\alpha}+P)$ once, within the time of one call.

For the consequence, the instances have the stated parameters for $n':=\Lambda^{1/(2-2\alpha)}n$ (up to constants), so each call takes $O(n'^{2-\eta})=O(\Lambda^{1/(1-\alpha)}n^{2-\eta})$ time. The time per call outside the called algorithm is $O(\Lambda^{1/2}n^{2-\eta}\log n)$, since $\eta\le\min(\alpha,1-\alpha)$ for every correct algorithm. Indeed, with the sets given as lists, consider an instance of $N$ sets of size $s$, paired into $N/2$ queries whose two sets lie in disjoint halves of $[u]$: each unread element could be replaced by an element of its partner and change the answer, so the algorithm reads $\Omega(Ns)=\Omega(n^{2-\alpha})$ words, and it writes $q_{\mathrm S}$ answers. Multiplying both by the number of calls gives the bound, the extra $\log n$ accounting for the $+5$ in the exponent.
\end{proof}

So for large $r$ the saving $\eta$ is kept up to an arbitrarily small loss, where Fischer, Kaliciak and Polak's proof with the same running time of the called algorithm gives $n^{2-\eta/25+o(1)}$. Unlike~\cite{KPP16,FKP24} it makes many adaptive calls, not one, and needs counting; it covers the whole range of~\cite{KPP16}, $0<\alpha<1$ (their parameter $\gamma$, not ours), with $u$ larger than theirs by $\Lambda=(\log n)^{O(r)}$ and $N$ by $\sqrt\Lambda$.

\section{Exact Triangle, all-edges Exact Triangle, \texorpdfstring{$(\min,+)$}{(min,+)}-product and APSP}\label{e:sec:statement}

\paragraph{Exact Triangle.} We use Exact Triangle as defined in the preprint's Section 3.2: an instance is a complete tripartite graph on vertex parts $A,B,C$ of $n$ vertices each with an integer weight $w(e)$, $|w(e)|\le n^\nu$ for a constant $\nu\ge1$, on every edge; with $S(a,b,c):=w(a,b)+w(b,c)+w(a,c)$, a \emph{zero triangle} is a triangle $(a,b,c)\in A\times B\times C$ with $S(a,b,c)=0$, and the task is to decide whether one exists.
The \emph{all-edges version} asks, for every pair $(a,b)\in A\times B$, whether some $c\in C$ has $S(a,b,c)=0$. We state the results for parts of \emph{at most} $n$ vertices, since the reductions of Section~\ref{e:sec:apsp} produce such instances.

\begin{theorem}[Exact Triangle]\label{e:thm:A}
Assume $\Hyp(\varepsilon,\gamma)$ with $0<\varepsilon,\gamma\le\frac12$. For every constant $\nu\ge1$ and every integer $r\ge1$ there is a deterministic algorithm that, given $n$, decides Exact Triangle for every instance whose parts have at most $n$ vertices each and whose weights are integers of absolute value at most $n^\nu$, in time
\[
O\big(n^{3-\varepsilon\gamma+\varepsilon/(2r)}(\log n)^{r+2}\big),
\]
where the constant in the $O$ depends on $\nu$, $r$, $\varepsilon$ and $C_{\mathcal O}$. This is at most $O(n^{3-\varepsilon\gamma+\varepsilon/r}(\log n)^{r+2})$, the form of Theorem~\ref{s:thm:main3sum}.
\end{theorem}

\begin{theorem}[all-edges Exact Triangle]\label{e:thm:B}
Under the same assumptions, there is a deterministic algorithm with the same time bound that solves the all-edges version: it decides, for every pair $(a,b)\in A\times B$, whether $S(a,b,c)=0$ for some $c\in C$.
\end{theorem}

\begin{corollary}[one algorithm, $r$ growing with $n$]\label{e:cor:uniform}
Under the same assumptions, one deterministic algorithm solves Exact Triangle, and one solves its all-edges version, on at most $n$ vertices per part with weights of absolute value at most $n^\nu$, in time $n^{3-\varepsilon\gamma+O(\sqrt{\log\log n/\log n})}=n^{3-\varepsilon\gamma+o(1)}$.
\end{corollary}

\begin{corollary}[$(\min,+)$-product and APSP]\label{e:cor:C}
Assume $\Hyp(\varepsilon,\gamma)$ with $0<\varepsilon,\gamma\le\frac12$, and let $\nu\ge1$ be a constant. The $(\min,+)$-product of two $n\times n$ integer matrices with entries of absolute value at most $n^\nu$ can be computed deterministically in time $O(n^{3-\varepsilon\gamma+\varepsilon/(2r)}(\log n)^{r+4})$ for every fixed $r$, and APSP on directed $n$-vertex graphs with integer weights of absolute value at most $n^\nu$ and no negative cycles in time $O(n^{3-\varepsilon\gamma+\varepsilon/(2r)}(\log n)^{r+5})$; both in time $n^{3-\varepsilon\gamma+o(1)}$ by a single algorithm.
\end{corollary}

For $r\le1/(2\gamma)$ the bound of Theorem~\ref{e:thm:A} is no better than $n^3$; the theorems are of interest for larger $r$. Numbers are in Section~\ref{e:sec:numbers}.

\paragraph{What is taken from elsewhere.} Besides $\Hyp(\varepsilon,\gamma)$ and the bounds on primes of Lemma~\ref{c:lem:primes}, the proof takes from elsewhere only the lemma of Chan, Vassilevska Williams and Xu~\cite[Lemma~9.2]{CVX23}, which reduces all-edges Negative Triangle to $O(\log n)$ instances of all-edges Exact Triangle on the same vertex set (quoted in Section~\ref{e:sec:apsp}); prior work on the method is discussed in Section~\ref{x:sec:novelty}.

\paragraph{Relation to the preprint.} Theorem~\ref{e:thm:B} is the all-edges algorithm that the route of the preprint's footnote~10 needs. The preprint's concluding section notes that its reduction from Exact Triangle to $\Lop$ ``keeps only half of the saving in the exponent'' and asks which of the reductions can be improved~\cite[Section~6]{ZT}; with the counting algorithm, this one keeps all of it. For the detection algorithm, nothing here improves on its Theorem 19.

The reduction follows the 3SUM reduction of Section~\ref{s:sec:statement}, with the pairs $(a,b)$ and pieces of $C$ in the role of the probes and the preprint's own instance for Exact Triangle, now with a composite modulus, in the role of $I(m,\Pq)$. Section~\ref{e:sec:instances} defines the incidences, the instances asked of the counting algorithm and the parameters, Section~\ref{e:sec:avg} is the averaging step that chooses a prime, Sections~\ref{e:sec:detection} and~\ref{e:sec:alledges} give the algorithms for the two versions and prove them correct, Section~\ref{e:sec:cost} bounds their cost, and Section~\ref{e:sec:apsp} derives the $(\min,+)$-product and APSP. The instances built below are $0/1$ graphs, so no splitting of integer entries (Lemma~\ref{s:lem:split}) is needed.

\subsection{The instances}\label{e:sec:instances}

\paragraph{Pieces, incidences and counts.} Let the instance have parts $A,B,C$ with at most $n$ vertices each and integer weights with $|w|\le U$, where $1\le U\le n^\nu$ (take $U:=\max(1,\max|w|)$). Then $|S(a,b,c)|\le3U$. Given an integer $\sigma\ge1$, split $C$ into $h:=\lceil|C|/\sigma\rceil$ \emph{pieces} $C_1,\dots,C_h$ of at most $\sigma$ vertices each. An \emph{incidence} is a triple $x=(a,b,k)\in A\times B\times[h]$. For an incidence $x=(a,b,k)$ and an integer $m\ge1$ let
\begin{gather*}
\cnt_m(x):=\#\{c\in C_k:\ S(a,b,c)\equiv0\pmod m\},\qquad z(x):=\#\{c\in C_k: S(a,b,c)=0\},\\
f_m(x):=\cnt_m(x)-z(x).
\end{gather*}
So $f_m(x)$ is the number of \emph{false positives} of $x$ modulo $m$, the $c\in C_k$ with $S(a,b,c)\ne0$ and $m\mid S(a,b,c)$; $z(x)$ does not depend on $m$, and $\cnt_1(x)=|C_k|\le\sigma$. The incidence $x$ is \emph{true} if $z(x)\ge1$. A pair $(a,b)$ lies in a zero triangle if and only if one of its $h$ incidences is true. For a set $\Iq$ of incidences let $F_\Iq(m):=\sum_{x\in\Iq}f_m(x)$ and $Z_\Iq:=\sum_{x\in\Iq}z(x)$, so that $\sum_{x\in\Iq}\cnt_m(x)=F_\Iq(m)+Z_\Iq$, and $F_\Iq(1)\le\sigma|\Iq|$.

\paragraph{The instance $J(m,k,\varrho,\mathcal Q)$.} Let $m\ge2$, $k\in[h]$, $\varrho\in\Z_m$, and let $\mathcal Q$ be a set of pairs $(a,b)\in A\times B$ with $w(a,b)\equiv\varrho\pmod m$. The instance of $\#\Lop$ has the sides $A$ and $B$ (padded with isolated vertices to $n$ vertices each), the middle part $C_k\times\Z_m$ of at most $\sigma m$ vertices, the wanted set $W:=\mathcal Q$, and the edges
\[
a\sim(c,\lambda)\iff\lambda\equiv w(a,c)+\varrho,\qquad (c,\lambda)\sim b\iff\lambda\equiv-w(b,c)\pmod m .
\]
This is the instance in the proof of the preprint's Theorem~17, whose label is $\sigma$ there, with the prime $p$ replaced by $m$ and pieces of at most $\sigma$ vertices in place of $\lceil s/g\rceil$, $s=\lfloor\sqrt D\rfloor$; nothing in it needs $m$ to be prime.

\begin{lemma}[the counts]\label{e:lem:counts}
For every $(a,b)\in\mathcal Q$, the number of common neighbours of $a$ and $b$ in $J(m,k,\varrho,\mathcal Q)$ is $\cnt_m(a,b,k)$.
\end{lemma}
\begin{proof}
A middle vertex $(c,\lambda)$ is a common neighbour if and only if $\lambda\equiv w(a,c)+\varrho$ and $\lambda\equiv-w(b,c)$. For each $c\in C_k$ there is such a $\lambda$ if and only if $w(a,c)+\varrho+w(b,c)\equiv0$, and then exactly one. Since $\varrho\equiv w(a,b)$, this is $S(a,b,c)\equiv0\pmod m$.
\end{proof}

\paragraph{Parameters.} Fix $r\ge1$ and let $\xi:=2$ for Theorem~\ref{e:thm:A} and $\xi:=4$ for Theorem~\ref{e:thm:B}. Put $\zeta:=\ln(3U)$ and
\[
\Lambda_{\mathrm E}:=\xi\,(10\zeta)^r,\qquad
\sigma:=\big\lfloor\sqrt{n^\varepsilon/\Lambda_{\mathrm E}}\big\rfloor,\qquad
P:=\big\lceil4\zeta\,(\xi\sigma)^{1/r}\big\rceil,\qquad
L:=\lceil\varepsilon\gamma\log_2n\rceil .
\]
(Any $L$ with $2^L\ge n^{\varepsilon\gamma}$ works; the 3SUM reduction uses $L=\lceil2\gamma\log_2\sigma\rceil$.)
In words: $\sigma$ is the piece size, chosen so that the moduli, which are at most $\Lambda_{\mathrm E}\sigma$, keep the middle part $\sigma m$ below $n^\varepsilon$ as $\Hyp(\varepsilon,\gamma)$ requires; $P$ is the size of the candidate primes, chosen so that $r$ of them cut the false positives by a factor $\xi\sigma$ (Section~\ref{e:sec:avg}); $L$ is the number of stages, enough to shrink the set of incidences by a factor $n^{\varepsilon\gamma}$, which makes the final scans cheap; and $\Lambda_{\mathrm E}$ is the polylogarithmic slack that these choices cost.
We assume $n\ge n_0(\nu,r,\varepsilon)$ large enough that $\sigma\ge2$, $P\ge1024$, $r\le P/(4\ln P)$, $(\xi\sigma)^{1/r}\ge2$ and $2n\Lambda_{\mathrm E}\le n^2$; this holds for large $n$ because $\Lambda_{\mathrm E}$ is polylogarithmic in $n$ (as $U\le n^\nu$) and $\sigma$ grows as $n^{\varepsilon/2}$ up to a polylogarithmic factor. Smaller $n$ are solved by brute force in $O(1)$ time. Since $n\ge2$ and $\ln3\le2\ln2$, we have $\zeta\le(\nu+2)\ln n$. The consequences used below:
\begin{equation}\label{e:eq:params}
\tfrac12\sqrt{n^\varepsilon/\Lambda_{\mathrm E}}\le\sigma\le n^{\varepsilon/2},\quad \Lambda_{\mathrm E}\sigma^2\le n^\varepsilon,\quad h\le2n/\sigma,\quad (2P)^r\le\Lambda_{\mathrm E}\sigma,\quad P\ge8\zeta .
\end{equation}
The first holds since $\lfloor x\rfloor\ge x/2$ for $x\ge1$, the second by definition, the third since $h\le n/\sigma+1$ and $\sigma\le n$. For the fourth, $\zeta(\xi\sigma)^{1/r}\ge1$ (as $\zeta\ge\ln3>1$), so $P\le5\zeta(\xi\sigma)^{1/r}$ and $(2P)^r\le(10\zeta)^r\xi\sigma=\Lambda_{\mathrm E}\sigma$. The fifth follows from $(\xi\sigma)^{1/r}\ge2$.

\begin{lemma}[shape and cost of one evaluation]\label{e:lem:shape}
Let $2\le m\le\Lambda_{\mathrm E}\sigma$ and let $\Iq$ be a set of incidences. The counts $\cnt_m(x)$ for all $x\in\Iq$ can be computed with $\Hyp(\varepsilon,\gamma)$ deterministically in time $O(n^3\Lambda_{\mathrm E}^{1-\gamma}\sigma^{-2\gamma})$, the constant depending only on $C_{\mathcal O}$.
\end{lemma}
\begin{proof}
For every piece $k$ and residue $\varrho\in\Z_m$ let $W_{k,\varrho}$ be the set of pairs $(a,b)$ with $(a,b,k)\in\Iq$ and $w(a,b)\equiv\varrho\pmod m$. Cut each nonempty $W_{k,\varrho}$ into chunks of at most $\lfloor n^2/\sqrt D\rfloor$ pairs, where $D:=\sigma m$, and make one call to $\mathcal O$ on $J(m,k,\varrho,\mathcal Q)$ for every chunk $\mathcal Q$, with $N:=n$ and middle-part bound $D$. By Lemma~\ref{e:lem:counts} the answer for $(a,b)\in\mathcal Q$ is $\cnt_m(a,b,k)$. Every incidence of $\Iq$ is asked in exactly one call.

\emph{$D\le N^\varepsilon$.} $D=\sigma m\le\Lambda_{\mathrm E}\sigma^2\le n^\varepsilon$ by~\eqref{e:eq:params}, and $D\ge4$.

\emph{Number and cost of the calls.} Since $|W_{k,\varrho}|\le n^2$ summed over $\varrho$, and $\lfloor n^2/\sqrt D\rfloor\ge n^2/(2\sqrt D)$, piece $k$ needs at most $m+2\sqrt{\sigma m}$ calls, each of cost at most $C_{\mathcal O}n^2(\sigma m)^{-\gamma}$. If $m\le\sigma$, then $m\le\sqrt{\sigma m}$, and the cost for the piece is at most $3C_{\mathcal O}n^2(\sigma m)^{1/2-\gamma}\le3C_{\mathcal O}n^2\sigma^{1-2\gamma}$, using $\sigma m\le\sigma^2$ and $\gamma\le\frac12$. If $\sigma<m\le\Lambda_{\mathrm E}\sigma$, then $\sqrt{\sigma m}<m$, and the cost is at most $3C_{\mathcal O}n^2m^{1-\gamma}\sigma^{-\gamma}\le3C_{\mathcal O}n^2\Lambda_{\mathrm E}^{1-\gamma}\sigma^{1-2\gamma}$. In both cases it is at most $3C_{\mathcal O}n^2\Lambda_{\mathrm E}^{1-\gamma}\sigma^{1-2\gamma}$, as $\Lambda_{\mathrm E}\ge1$; over the $h\le2n/\sigma$ pieces, at most $6C_{\mathcal O}n^3\Lambda_{\mathrm E}^{1-\gamma}\sigma^{-2\gamma}$.

\emph{Input, output and bookkeeping.} Writing an instance takes $O(ND)$ time (the residues $w(a,c)+\varrho$ and $-w(b,c)$ modulo $m$ for $c\in C_k$, and the biadjacency matrices), and reading its answers $O(|W|)$. Since $D\le N^\varepsilon$ and $\varepsilon(1+\gamma)\le\frac34<1$, we have $D^{1+\gamma}\le N$, so $ND\le N^2/D^\gamma$; and $|W|\le N^2/\sqrt D\le N^2/D^\gamma$. So this is within the cost of the calls. Computing $w(a,b)\bmod m$ for the incidences of $\Iq$ and sorting them by $(k,\varrho)$ by radix sort takes $O(|\Iq|+hm)$ time, and handing each answer back to its incidence $O(|\Iq|)$. Here $|\Iq|\le hn^2\le2n^3/\sigma\le2n^3\sigma^{-2\gamma}$ (as $\gamma\le\frac12$), and $hm\le2n\Lambda_{\mathrm E}\le n^2\le n^3\sigma^{-2\gamma}$ (as $\sigma\le n$).
\end{proof}

The case $m\le\sigma$ is the one where the modulus is smaller than the piece size: a residue class then may hold more than $n^2/\sqrt D$ pairs, and the splitting into chunks costs at most what the largest moduli cost. This is used for the partial products $m\cdot p$ while a modulus is being built.

\subsection{Averaging over primes}\label{e:sec:avg}

Let $\Pi$ be the set of primes in $[P,2P)$, found by a sieve in $O(P\log\log P)$ time; $|\Pi|\ge P/(2\ln P)$ by Lemma~\ref{c:lem:primes}. We use Lemma~\ref{c:lem:avg} through the following step.

\begin{lemma}[one greedy step]\label{e:lem:step}
Let $\Iq$ be a set of incidences, let $m$ be a product of fewer than $r$ distinct primes of $\Pi$, and let $\Pi'\subseteq\Pi$ consist of the primes not dividing $m$. If $p^*\in\Pi'$ minimises $\sum_{x\in\Iq}\cnt_{mp}(x)$ over $p\in\Pi'$, then $F_\Iq(mp^*)\le F_\Iq(m)\cdot4\zeta/P$.
\end{lemma}
\begin{proof}
Since $\sum_{x\in\Iq}\cnt_{mp}(x)=F_\Iq(mp)+Z_\Iq$ and $Z_\Iq$ does not depend on $p$, the prime $p^*$ also minimises $F_\Iq(mp)$. Apply Lemma~\ref{c:lem:avg} to the multiset $\mathcal F$ of the values $S(a,b,c)$ over the false positives $c$ of the incidences $(a,b,k)\in\Iq$ modulo $m$: each is nonzero, divisible by $m$, and of absolute value at most $Z_{\max}:=3U$; $|\mathcal F|=F_\Iq(m)$, and the elements of $\mathcal F$ divisible by $mp$ are the false positives modulo $mp$, so their number is $F_\Iq(mp)$. The pool has $|\Pi'|\ge|\Pi|-r\ge P/(2\ln P)-P/(4\ln P)=P/(4\ln P)$ primes, by $r\le P/(4\ln P)$. Hence $F_\Iq(mp^*)\le F_\Iq(m)\cdot\frac{\zeta}{\ln P}\cdot\frac{4\ln P}{P}$.
\end{proof}

\subsection{Exact Triangle (Theorem~\ref{e:thm:A})}\label{e:sec:detection}

The algorithm uses the parameters with $\xi=2$.
\begin{enumerate}[label=\arabic*.]
\item Split $C$ into the pieces and let $\Iq_0:=A\times B\times[h]$.
\item For $t=1,\dots,L$:
\begin{enumerate}[label=(\alph*)]
\item If $\Iq_{t-1}=\emptyset$, answer NO.
\item Build $M_t$: start with $m:=1$ and $\Pi':=\Pi$. Repeat $r$ times: for every $p\in\Pi'$ compute $\Sigma_t(mp):=\sum_{x\in\Iq_{t-1}}\cnt_{mp}(x)$ by Lemma~\ref{e:lem:shape}; choose $p^*\in\Pi'$ minimising $\Sigma_t(mp)$ (ties arbitrarily), and set $m:=mp^*$, $\Pi':=\Pi'\setminus\{p^*\}$. Let $M_t:=m$. In the last of the $r$ steps, keep the counts $\cnt_{mp}(x)$ of the current minimiser while the candidates are evaluated; at the end they are the counts $\cnt_{M_t}(x)$.
\item If $\Sigma_t(M_t)>|\Iq_{t-1}|/2$, answer YES.
\item Otherwise let $\Iq_t:=\{x\in\Iq_{t-1}:\cnt_{M_t}(x)>0\}$.
\end{enumerate}
\item For every $x=(a,b,k)\in\Iq_L$, scan $C_k$ for a $c$ with $S(a,b,c)=0$. Answer YES if some scan succeeds, NO otherwise.
\end{enumerate}
This is the algorithm of Section~\ref{s:sec:algorithm} with the pieces in place of the self-reduction, incidences in place of probes, Lemma~\ref{e:lem:shape} in place of Lemma~\ref{s:lem:shape}, and a scan of a piece in place of a merge. Every candidate modulus $mp$ is a product of at most $r$ distinct primes from $[P,2P)$, so $2\le mp<(2P)^r\le\Lambda_{\mathrm E}\sigma$ by~\eqref{e:eq:params}, and Lemma~\ref{e:lem:shape} applies to every evaluation. The algorithm is deterministic and adaptive, which is allowed for the reason given in Section~\ref{s:sec:algorithm}.

\begin{lemma}[halving]\label{e:lem:halving}
In every stage $t$ that reaches step 2(b), $F_{\Iq_{t-1}}(M_t)\le|\Iq_{t-1}|/2$. Consequently, if step 2(c) answers YES then $Z_{\Iq_{t-1}}>0$, so the answer is correct; otherwise $|\Iq_t|\le|\Iq_{t-1}|/2$.
\end{lemma}
\begin{proof}
As the proof of Lemma~\ref{s:lem:halving}, with its first paragraph replaced by the following. Write $\Iq:=\Iq_{t-1}$. The bound per prime is Lemma~\ref{e:lem:step} (the values $S(a,b,c)$ of absolute value at most $3U$ in place of $a+b-c$, at most $2U$); applied in each of the $r$ steps it gives $F_\Iq(M_t)\le F_\Iq(1)(4\zeta/P)^r\le\sigma|\Iq|(4\zeta/P)^r$, using $F_\Iq(1)\le\sigma|\Iq|$ in place of $F_\Pq(1)\le\sigma^2|\Pq|$, and $P\ge4\zeta(2\sigma)^{1/r}$ (as $\xi=2$) gives $F_\Iq(M_t)\le|\Iq|/2$. The second paragraph of that proof applies verbatim with incidences in place of probes: it uses only $\sum_{x\in\Iq}\cnt_m(x)=F_\Iq(m)+Z_\Iq$ with $Z_\Iq$ independent of $m$, and the definition of $\Iq_t$ in step 2(d); here $Z_\Iq>0$ means that some incidence $(a,b,k)$ has $c\in C_k$ with $S(a,b,c)=0$.
\end{proof}

\begin{lemma}[correctness]\label{e:lem:correctA}
The algorithm answers YES if and only if the instance has a zero triangle.
\end{lemma}
\begin{proof}
As the proof of Lemma~\ref{s:lem:correct}, with Lemma~\ref{e:lem:halving} in place of Lemma~\ref{s:lem:halving}, and a YES of step 3 exhibiting a zero triangle. The other step that differs is the start of the converse: a zero triangle $(a,b,c)$ with $c\in C_k$ gives the incidence $x:=(a,b,k)$, which lies in $\Iq_0=A\times B\times[h]$ by definition (the 3SUM proof needs~\cite[Lemma~3.3]{FKP24} here); then $\cnt_m(x)\ge z(x)\ge1$ for every $m$, and the rest is verbatim, with $x$ in place of $Q$ and the scan of $C_k$ finding $c$.
\end{proof}

\subsection{All-edges Exact Triangle (Theorem~\ref{e:thm:B})}\label{e:sec:alledges}

\paragraph{Why halving alone does not suffice.} In the all-edges version, the true incidences of a pair cannot be dropped before the pair is answered, and there is no YES exit. The halving of Lemma~\ref{e:lem:halving} bounds the false positives relative to $|\Iq|$, which includes the true incidences; and the false positives inside true incidences enter the minimised count afresh in every stage, since each stage starts from modulus $1$. So the survivors do not shrink below the number of true incidences, and a pair whose true incidences are many keeps its false incidences alive with them. The certification below removes true incidences too: an incidence that is not certified must carry many false positives at the base modulus, true or not, and these are what the greedy steps make rare.

\paragraph{The certification rule.} Let $\ell$ be the largest integer $l\ge0$ with $P^l\le3U$; it is computed by repeated multiplication. Every nonzero integer of absolute value at most $3U$ has at most $\ell$ distinct prime divisors $\ge P$, and $\ell\le\zeta/\ln P$. Let $m$ be a product of $r-1$ distinct primes of $\Pi$ ($m=1$ if $r=1$), and let $\Pi'\subseteq\Pi$ consist of the primes not dividing $m$. For an incidence $x$ let
\[
\mu(x):=\min_{p\in\Pi'}\cnt_{mp}(x).
\]
The idea is that the true witnesses of $x$, $z(x)$ in number, are in every count, while a false positive modulo $m$ survives a further prime $p$ for at most $\ell$ of the $|\Pi'|$ candidates; so a count that stays large over all candidates must contain true witnesses, and a count that reaches zero cannot. Call $x$ \emph{certified false} if $\mu(x)=0$; \emph{certified true} if $\mu(x)\,|\Pi'|>\ell\cdot\cnt_m(x)$; and \emph{uncertified} otherwise.

\begin{lemma}[certification]\label{e:lem:cert}
\begin{enumerate}[label=(\roman*),nosep]
\item A certified-false incidence $x$ has $z(x)=0$.
\item A certified-true incidence $x$ has $z(x)\ge1$.
\item An uncertified incidence $x$ satisfies $\ell\,f_m(x)\ge|\Pi'|-\ell$.
\end{enumerate}
\end{lemma}
\begin{proof}
For every $p\in\Pi'$, $\cnt_{mp}(x)=z(x)+f_{mp}(x)$ with $z(x)$ independent of $p$, so $\mu(x)=z(x)+\min_{p\in\Pi'}f_{mp}(x)\ge z(x)$. This gives (i).

(ii) Let $x=(a,b,k)$. A false positive $c$ of $x$ modulo $mp$ is a false positive modulo $m$ whose value $S(a,b,c)$ is divisible by $p$, since $p\nmid m$. Each false positive modulo $m$ has $0<|S(a,b,c)|\le3U$, so it is a false positive modulo $mp$ for at most $\ell$ primes $p\in\Pi'$. Hence $\sum_{p\in\Pi'}f_{mp}(x)\le\ell f_m(x)$ and $\min_pf_{mp}(x)\le\ell f_m(x)/|\Pi'|\le\ell\cnt_m(x)/|\Pi'|$. If $x$ is certified true, this is less than $\mu(x)$, so $z(x)=\mu(x)-\min_pf_{mp}(x)>0$.

(iii) If $x$ is uncertified then $\mu(x)\ge1$ and $\mu(x)|\Pi'|\le\ell\,(z(x)+f_m(x))$. If $z(x)=0$, this gives $\ell f_m(x)\ge|\Pi'|$. If $z(x)\ge1$, it gives $z(x)|\Pi'|\le\mu(x)|\Pi'|\le\ell z(x)+\ell f_m(x)$, so $\ell f_m(x)\ge z(x)(|\Pi'|-\ell)\ge|\Pi'|-\ell$ when $|\Pi'|\ge\ell$; when $|\Pi'|<\ell$ the claim is trivial.
\end{proof}

Part (iii) is what makes the true incidences shrink: a true incidence stays uncertified only if it carries about $|\Pi'|/\ell$ false positives or more at the base modulus.

\paragraph{The algorithm.} It uses the parameters with $\xi=4$. Every pair starts unmarked.
\begin{enumerate}[label=\arabic*.]
\item Split $C$ into the pieces and let $\Iq_0:=A\times B\times[h]$.
\item For $t=1,\dots,L$, with $\Iq:=\Iq_{t-1}$ fixed during the stage:
\begin{enumerate}[label=(\alph*)]
\item \emph{Greedy steps.} Start with $m:=1$ and $\Pi':=\Pi$. Repeat $r-1$ times: for every $p\in\Pi'$ compute $\sum_{x\in\Iq}\cnt_{mp}(x)$ by Lemma~\ref{e:lem:shape}; choose a minimising $p^*$, set $m:=mp^*$ and $\Pi':=\Pi'\setminus\{p^*\}$. Keep the counts $\cnt_m(x)$, $x\in\Iq$, of the final $m$ (as in step 2(b) of Section~\ref{e:sec:detection}; for $r=1$, $\cnt_1(x)=|C_k|$).
\item \emph{Certification round.} For every $p\in\Pi'$ compute $\cnt_{mp}(x)$, $x\in\Iq$, by Lemma~\ref{e:lem:shape}, keeping the running minimum $\mu(x)$. Classify every $x\in\Iq$ by the rule above, and mark the pair of every certified-true incidence YES.
\item Let $\Iq_t$ be the set of uncertified incidences of $\Iq$ whose pair is not marked.
\end{enumerate}
\item For every $x=(a,b,k)\in\Iq_L$, scan $C_k$ for a $c$ with $S(a,b,c)=0$, and mark $(a,b)$ YES if one is found.
\item Answer YES for the marked pairs and NO for all other pairs.
\end{enumerate}
As in Section~\ref{e:sec:detection}, every candidate modulus is a product of at most $r$ distinct primes from $[P,2P)$, so it lies in $[2,\Lambda_{\mathrm E}\sigma)$ and Lemma~\ref{e:lem:shape} applies; the algorithm is deterministic and adaptive.

\begin{lemma}[halving of all incidences]\label{e:lem:halvingB}
For every stage $t$, $|\Iq_t|\le|\Iq_{t-1}|/2$.
\end{lemma}
\begin{proof}
Write $\Iq:=\Iq_{t-1}$, and let $m$ and $\Pi'$ be as in the certification round. By Lemma~\ref{e:lem:step}, applied in each of the $r-1$ greedy steps, $F_\Iq(m)\le\sigma|\Iq|(4\zeta/P)^{r-1}$. By Lemma~\ref{e:lem:cert}(iii), summed over the uncertified incidences,
\[
|\Iq_t|\le\#\{x\in\Iq:\ x\text{ uncertified}\}\le\frac{\ell\,F_\Iq(m)}{|\Pi'|-\ell}\qquad\text{if }|\Pi'|>\ell .
\]
Here $|\Pi'|\ge P/(4\ln P)$ as in Lemma~\ref{e:lem:step}, and $\ell\le\zeta/\ln P\le P/(8\ln P)\le|\Pi'|/2$ by $P\ge8\zeta$ from~\eqref{e:eq:params}. So $\ell/(|\Pi'|-\ell)\le2\ell/|\Pi'|\le2\cdot\frac{\zeta}{\ln P}\cdot\frac{4\ln P}{P}=8\zeta/P$, and
\[
|\Iq_t|\le\frac{8\zeta}{P}\,\sigma|\Iq|\Big(\frac{4\zeta}{P}\Big)^{r-1}=2\sigma|\Iq|\Big(\frac{4\zeta}{P}\Big)^r\le\frac{|\Iq|}{2},
\]
since $P\ge4\zeta(4\sigma)^{1/r}$ gives $(4\zeta/P)^r\le1/(4\sigma)$.
\end{proof}

\begin{lemma}[correctness]\label{e:lem:correctB}
A pair $(a,b)$ is answered YES if and only if $S(a,b,c)=0$ for some $c\in C$.
\end{lemma}
\begin{proof}
A pair is marked YES only by a certified-true incidence, which is true by Lemma~\ref{e:lem:cert}(ii), or by a scan that found a zero triangle. Conversely, let $S(a,b,c)=0$ with $c\in C_k$, and let $x:=(a,b,k)\in\Iq_0$, so $z(x)\ge1$. In every stage in which the pair is not yet marked and $x\in\Iq_{t-1}$, $x$ is not certified false by Lemma~\ref{e:lem:cert}(i); so it is either certified true, and the pair is marked, or uncertified, and then $x\in\Iq_t$. If the pair is never marked during the stages, then $x\in\Iq_L$ and step 3 finds $c$.
\end{proof}

The rule decides the existence of a zero triangle per pair, not the number: the counts $z(x)$ are not recovered in general, so this solves the all-edges detection problem, which is what Section~\ref{e:sec:apsp} needs. A certified-true incidence comes without a witness; if witnesses are wanted, one scan of one certified piece per marked pair finds one, at a total cost of $O(n^2\sigma)$.

\subsection{Cost}\label{e:sec:cost}

We bound each part of both algorithms, using~\eqref{e:eq:params}.
\begin{enumerate}[label=(\roman*)]
\item \emph{Preparation}: the pieces, the list $\Iq_0$ ($|\Iq_0|\le hn^2\le2n^3/\sigma$), and the sieve for $\Pi$: $O(n^3/\sigma+P\log\log P)$, which is $O(n^{3-\varepsilon/2}\Lambda_{\mathrm E}^{1/2})$, below the main term since $\varepsilon\gamma\le\varepsilon/2$ and the main term carries the factor $\Lambda_{\mathrm E}\ge\Lambda_{\mathrm E}^{1/2}$.
\item \emph{Stages}: $L$ stages of $r$ rounds each (in Section~\ref{e:sec:detection}, $r$ greedy steps; in Section~\ref{e:sec:alledges}, $r-1$ greedy steps and the certification round), each round with at most $|\Pi|<P$ evaluations. By Lemma~\ref{e:lem:shape} each evaluation costs $O(n^3\Lambda_{\mathrm E}^{1-\gamma}\sigma^{-2\gamma})$, including writing its instances, splitting the residue classes into chunks (also for moduli smaller than $\sigma$), reading the answers and summing the counts. Choosing the minimum, keeping the minimiser's counts or the running minimum $\mu$, classifying, and forming $\Iq_t$ cost $O(P+|\Iq_{t-1}|)$ per round, within the same bound. In total
\[
O\big(L\,r\,P\,n^3\Lambda_{\mathrm E}^{1-\gamma}\sigma^{-2\gamma}\big).
\]
\item \emph{Final scans}: by Lemmas~\ref{e:lem:halving} and~\ref{e:lem:halvingB}, $|\Iq_L|\le|\Iq_0|/2^L\le2n^3/(\sigma n^{\varepsilon\gamma})$, and each scan costs $O(\sigma)$, so $O(n^{3-\varepsilon\gamma})$.
\item \emph{Word size}: the weights are integers of at most $\nu\log_2n+1$ bits, i.e.\ $O(\nu)$ words of $\Theta(\log n)$ bits; moduli are below $\Lambda_{\mathrm E}\sigma\le n$; counts are at most $\sigma$ per incidence and sums at most $|\Iq_0|\sigma\le2n^3$. The instances of the calls have $N=n$, so its $O(\log N)$-bit words are $\Theta(\log n)$-bit words. All arithmetic above is $O(\nu)$ word operations per step.
\end{enumerate}
The stages dominate. Insert $\sigma^{-2\gamma}\le4^\gamma\Lambda_{\mathrm E}^\gamma n^{-\varepsilon\gamma}$ (from $\sigma\ge\frac12\sqrt{n^\varepsilon/\Lambda_{\mathrm E}}$), $P\le5\zeta(\xi\sigma)^{1/r}\le20\zeta\,n^{\varepsilon/(2r)}$ (from $\xi\le4$ and $\sigma\le n^{\varepsilon/2}$), $L\le\log_2n+1\le3\ln n$, $\Lambda_{\mathrm E}^{1-\gamma}\Lambda_{\mathrm E}^\gamma=\Lambda_{\mathrm E}=\xi(10\zeta)^r\le4(10\zeta)^r$, and $\zeta\le(\nu+2)\ln n$:
\begin{equation}\label{e:eq:total}
T(n)=O\Big(r\,n^{3-\varepsilon\gamma+\varepsilon/(2r)}\,\big(10(\nu+2)\big)^{r+1}(\ln n)^{r+2}\Big),
\end{equation}
for both algorithms, with the constant in the $O$ depending only on $C_{\mathcal O}$. The factor $n^{\varepsilon/(2r)}$ is the number of candidate primes, $P\approx\sigma^{1/r}$ with $\sigma\approx n^{\varepsilon/2}$; the factor $(\ln n)^r$ comes from $\Lambda_{\mathrm E}=\Lambda_{\mathrm E}^{1-\gamma}\cdot\Lambda_{\mathrm E}^\gamma$, i.e.\ from the moduli exceeding $\sigma$ by up to $\Lambda_{\mathrm E}$, which multiplies the number of residue classes (factor $\Lambda_{\mathrm E}^{1-\gamma}$), and from shrinking $\sigma$ by $\sqrt\Lambda_{\mathrm E}$ to keep $D\le N^\varepsilon$ (factor $\Lambda_{\mathrm E}^\gamma$); the remaining $(\ln n)^2$ are $L$ and the $\zeta$ in $P$. Compared with the $(\log n)^{r+4}$ of the 3SUM reduction (Theorem~\ref{s:thm:main3sum}), two logarithms disappear because the instances here are $0/1$ graphs, so the binary splitting of multiplicities into $B^2$ instances is not needed.

\begin{proof}[Proof of Theorems~\ref{e:thm:A} and~\ref{e:thm:B}]
Lemmas~\ref{e:lem:correctA} and~\ref{e:lem:correctB}, and~\eqref{e:eq:total} with $r$ and $\nu$ constants; $n^{\varepsilon/(2r)}\le n^{\varepsilon/r}$.
\end{proof}

\begin{proof}[Proof of Corollary~\ref{e:cor:uniform}]
As the proof of Corollary~\ref{s:cor:uniform}, with the same $r(n)$, with~\eqref{e:eq:total} in place of~\eqref{s:eq:total}, $\Lambda_{\mathrm E}$ in place of $\Lambda_{\mathrm S}$, $(\xi\sigma)^{1/r}$ in place of $(2\sigma^2)^{1/r}$, $n^{\varepsilon/(2r)}\le\exp(\frac{\varepsilon}2\sqrt{\ln n\ln\ln n})$ in place of $n^{\varepsilon/r}\le\exp(\varepsilon\sqrt{\ln n\ln\ln n})$, and $r(10(\nu+2))^{r+1}(\ln n)^{r+2}$ in place of $r(10(\nu+1))^{r+1}(\ln n)^{r+4}$, which is still $\exp(O(\sqrt{\ln n\ln\ln n}))$. The two side conditions that are new here hold for large $n$: $2n\Lambda_{\mathrm E}\le n^2$ because $\Lambda_{\mathrm E}=\exp(O(r\ln\ln n))=n^{o(1)}$, and $(\xi\sigma)^{1/r}\ge2$ because $\sigma=n^{\varepsilon/2-o(1)}$ gives $(\xi\sigma)^{1/r}=\exp(\Theta(\sqrt{\ln n\ln\ln n}))$. So $T(n)=n^{3-\varepsilon\gamma+O(\sqrt{\ln\ln n/\ln n})}$.
\end{proof}

\subsection{The \texorpdfstring{$(\min,+)$}{(min,+)}-product and APSP (Corollary~\ref{e:cor:C})}\label{e:sec:apsp}

The route is the one of the preprint's footnote 10: the $(\min,+)$-product reduces to the all-edges version of a triangle problem on the same $n$ vertices per part, with polylogarithmically many instances, so the $(\min,+)$-product keeps the whole saving of Theorem~\ref{e:thm:B}. There are three links, and APSP follows by repeated squaring.

\paragraph{Link 1: $(\min,+)$-product to all-edges Negative Triangle (proved here).} The \emph{all-edges Negative Triangle} problem asks, for a tripartite graph with integer edge weights, for every edge $e$ whether $e$ lies in a triangle of negative total weight. The preprint attributes the reduction of the $(\min,+)$-product to the all-edges Exact Triangle problem on the same vertices to~\cite{VW10,VW18,VW13}; we prove the first link here, a simultaneous binary search, and take the second from~\cite{CVX23}.

\begin{lemma}\label{e:lem:binsearch}
The $(\min,+)$-product of two $n\times n$ integer matrices $X,Y$ with entries of absolute value at most $U$ reduces deterministically to $\lceil\log_2(4U+1)\rceil$ instances of all-edges Negative Triangle, each on three parts of $n$ vertices with weights of absolute value at most $2U+1$, with all pairs of the first and third parts asked, plus $O(n^2)$ time per instance.
\end{lemma}
\begin{proof}
Let $Z[i,j]:=\min_k(X[i,k]+Y[k,j])\in[-2U,2U]$. Keep for every $(i,j)$ an interval $[\mathrm{lo}_{ij},\mathrm{hi}_{ij}]\ni Z[i,j]$, initially $[-2U,2U]$. In a round, let $t_{ij}:=\lfloor(\mathrm{lo}_{ij}+\mathrm{hi}_{ij})/2\rfloor$ and form the tripartite graph on parts $I,K,J$ (copies of $[n]$) with $w(i,k):=X[i,k]$, $w(k,j):=Y[k,j]$ and $w(i,j):=-t_{ij}-1$. The triangle $(i,k,j)$ is negative if and only if $X[i,k]+Y[k,j]\le t_{ij}$, so the edge $(i,j)$ lies in a negative triangle if and only if $Z[i,j]\le t_{ij}$. Set $\mathrm{hi}_{ij}:=t_{ij}$ if so and $\mathrm{lo}_{ij}:=t_{ij}+1$ otherwise. Each round at least halves the number of values in every interval, so after $\lceil\log_2(4U+1)\rceil$ rounds every interval is the single value $Z[i,j]$.
\end{proof}

\paragraph{Link 2: all-edges Negative Triangle to all-edges Exact Triangle (taken from~\cite{CVX23}).} In~\cite{CVX23}, AE-Exact-Tri asks, for a graph $G$ and a target value $t$, ``for every edge $e$, whether $e$ is contained in a triangle whose edge weights sum up to $t$'', and the prefix \# denotes the counting versions (their Problems 2.4 and 2.5 and the paragraph after their Problem 2.10, p.~12). Their Lemma 9.2 (arXiv:2303.14572v1, p.~35, verbatim except for typesetting):
\begin{quote}
\textbf{Lemma 9.2.} Given a \#AE-Neg-Tri instance on an $n$-node tripartite graph $G$ with node partitions $A$, $B$, and $C$ and with edge weights in $[\pm n^{O(1)}]$, we can reduce it to $O(\log n)$ instances of \#AE-Exact-Tri on graphs with the same vertex set and with edge weights in $[\pm n^{O(1)}]$ deterministically in $\tilde O(n^2)$ time. Furthermore, the count for an edge in the original \#AE-Neg-Tri instance can be written as a sum of the count for the corresponding edge among the \#AE-Exact-Tri instances (if this edge does not exist in a certain instance, then the count is simply 0).
\end{quote}
Two points of the proof matter for the route. \emph{Vertex set}: the instances are on the same vertex set, hence on three parts of at most $n$ vertices each; the proof (p.~35) adds $1$ to the weights of $A\times C$, removes edges (by weight in the base case, by parity in the recursive case) and replaces weights $w$ by $\lceil w/2\rceil$, but adds no vertices. \emph{Instance count}: the proof recurses on the weight bound $U_w$, halving it at each level and creating four instances with target value $1$ per level (one per pattern of at least two odd weights), and $O(1)$ instances in the base case $U_w=O(1)$, where all weights are set to $0$; with $U_w=n^{O(1)}$ this is $O(\log n)$ instances, with target values $0$ or $1$. The counts are sums of nonnegative counts, so an edge lies in a negative triangle if and only if it lies in a triangle of target weight in at least one of the instances; the detection version of the all-edges problem therefore suffices. The asked edges of~\cite{CVX23}'s proof are those of $A\times C$; we rename the parts so that they are the pairs of Theorem~\ref{e:thm:B}.

\paragraph{Link 3: the instances of Lemma 9.2 as instances of Theorem~\ref{e:thm:B} (proved here).} An instance of Lemma 9.2 may have missing edges and has a target value $t$ with $|t|\le n^{O(1)}$ (in the proof, $t\in\{0,1\}$), while Theorem~\ref{e:thm:B} takes complete tripartite graphs and target $0$. Let $U'$ bound the absolute values of the present weights after subtracting $t$ from every present edge of the asked part; give every missing edge the weight $3U'+1$. A triangle with all three edges present then has weight $0$ if and only if its original weight is $t$; a triangle with a missing edge has weight at least $3U'+1-2U'>0$. So the all-edges answers agree (a missing asked edge is answered NO, consistent with count $0$). The weights stay at most $n^{O(1)}$, with the exponent depending only on $\nu$, and the transformation takes $O(n^2)$ time.

\paragraph{APSP to the $(\min,+)$-product (proved here).} Let $G$ be a directed graph on $[n]$ with integer weights of absolute value at most $n^\nu$ and no negative cycles. Let $d_k(u,v)$ be the least weight of a walk from $u$ to $v$ with at most $k$ edges ($\infty$ if there is none), so $d_1$ has $0$ on the diagonal, the edge weights on the edges, and $\infty$ elsewhere, and $d_{2k}=d_k\star d_k$ in the $(\min,+)$ sense (split a walk with at most $2k$ edges at its middle edge count; the zero diagonal allows shorter parts). Since there are no negative cycles, deleting cycles from a walk does not increase its weight, so a finite $d_k(u,v)$ is the weight of a path, of absolute value at most $(n-1)n^\nu<n^{\nu+1}$, and $d_k=d_{n-1}$ is the distance matrix for $k\ge n-1$. Encode $\infty$ as $\Omega:=3n^{\nu+1}+1$: a sum with at least one term $\Omega$ is at least $\Omega-n^{\nu+1}>2n^{\nu+1}$, more than any sum of two finite entries, so after each product the entries above $n^{\nu+1}$ are exactly the infinite ones, and are reset to $\Omega$. Hence $\lceil\log_2n\rceil$ products of matrices with entries of absolute value at most $3n^{\nu+1}+1\le n^{\nu+2}$ (for $n\ge4$) give APSP.

\begin{proof}[Proof of Corollary~\ref{e:cor:C}]
By Lemma~\ref{e:lem:binsearch},~\cite[Lemma~9.2]{CVX23} and Link 3, the $(\min,+)$-product with $U\le n^\nu$ is computed by $O(\log n)\cdot O(\log n)$ instances of the all-edges Exact Triangle problem on parts of at most $n$ vertices with weights of absolute value at most $n^{\nu'}$, $\nu'$ depending only on $\nu$, plus $\tilde O(n^2)$ time per instance of Lemma~\ref{e:lem:binsearch}. Theorem~\ref{e:thm:B} solves each in $O(n^{3-\varepsilon\gamma+\varepsilon/(2r)}(\log n)^{r+2})$ time, so the product takes $O(n^{3-\varepsilon\gamma+\varepsilon/(2r)}(\log n)^{r+4})$, and APSP, with its $\lceil\log_2n\rceil$ products, $O(n^{3-\varepsilon\gamma+\varepsilon/(2r)}(\log n)^{r+5})$. With the algorithm of Corollary~\ref{e:cor:uniform} in place of the fixed-$r$ one, both are $n^{3-\varepsilon\gamma+o(1)}$.
\end{proof}

\subsection{Remarks}\label{e:sec:numbers}

\paragraph{Numbers.} With the preprint's Corollary 26, $\Hyp(1/18,0.063)$ holds and $\varepsilon\gamma=0.0035$. Corollaries~\ref{e:cor:uniform} and~\ref{e:cor:C} give deterministic Exact Triangle (also all-edges), $(\min,+)$-product and APSP in $n^{2.99650+o(1)}$. The preprint gives Exact Triangle in $O(n^{3-\varepsilon'}\log n)$ with $\varepsilon'=0.00175=\varepsilon\gamma/2$, i.e.\ $n^{2.99825}$ (its Theorem 19), and APSP in $\tilde O(n^{3-\varepsilon'/3})$, i.e.\ $n^{2.99942}$ (its Theorem 22), which keeps a sixth of $\varepsilon\gamma$. Its footnote 10 observes that the all-edges route keeps the whole saving of Exact Triangle, which gives APSP in $n^{3-\varepsilon\gamma/2+o(1)}=n^{2.99825+o(1)}$; measured against that route, the gain for APSP equals the gain for Exact Triangle. The preprint's constants are not optimised, and $(\varepsilon,\gamma)$ enter here only through $\Hyp(\varepsilon,\gamma)$, so any improvement of them (for instance Part~\hyperref[m:partI]{I}'s) carries over unchanged; in this section we use only the preprint's stated ones, and Part~\hyperref[m:partI]{I}'s improvement is combined with the theorems in Section~\ref{x:sec:consequences}. Exponents in this section are rounded up in the last digit.

For a fixed $r$ (Theorems~\ref{e:thm:A} and~\ref{e:thm:B}, Corollary~\ref{e:cor:C}) the saving is $\varepsilon\gamma-\varepsilon/(2r)$:
\begin{center}
\begin{tabular}{c|c|c|c}
$(\varepsilon,\gamma)$ & $r=20$ & $r=40$ & $r=80$\\\hline
$(1/18,\ 0.063)$ & $n^{2.99789}$ & $n^{2.99720}$ & $n^{2.99685}$
\end{tabular}
\end{center}
times $(\log n)^{r+2}$ for Exact Triangle and $(\log n)^{r+5}$ for APSP. To beat the preprint's Exact Triangle exponent $3-\varepsilon\gamma/2$ one needs $\varepsilon/(2r)<\varepsilon\gamma/2$, i.e.\ $r>1/\gamma$: $r\ge16$ for $\gamma=0.063$. To beat the APSP exponent $3-\varepsilon\gamma/6$ of its Theorem 22 one needs $\varepsilon/(2r)<\frac56\varepsilon\gamma$, i.e.\ $r>3/(5\gamma)$: $r\ge10$; against the route of its footnote 10, APSP needs the same $r\ge16$ as Exact Triangle.

\paragraph{Constants.} The polylogarithmic factor is large. For example, $r=40$ gains about $n^{0.00106}$ over the preprint's Exact Triangle bound $n^{2.99825}\log n$ ($2.99825-2.9971944$) but pays an extra factor $(\ln n)^{41}$, which is larger until $\ln n\approx5.1\cdot10^5$, that is, until $n$ has about $220{,}000$ decimal digits, and no choice of $r$ brings this below about $210{,}000$ digits ($r=32$); the factor $(10(\nu+2))^{r+1}$ in~\eqref{e:eq:total} comes on top. The staged route therefore pays off only asymptotically.

\paragraph{What the counting algorithm must provide.} The proof needs the counting version $\#\Lop$, for two reasons. The greedy choice of a prime needs the number of false positives of the surviving incidences, a potential that decreases by a factor of about $\sigma^{-1/r}$ per prime; and the certification compares counts under different moduli. We do not know how to do either with a detection algorithm: the number of positive incidences is not such a potential, since an incidence with many false positives modulo $m$ typically stays positive modulo $mp$. Counting with matrix products over polynomials, as in the proof of the preprint's Theorem 17, does not obviously help. In the first stage the minimised sum is the global number of triples with $S\equiv0\pmod{mp}$, which one product over $\Z[x]/(x^{mp}-1)$ gives in subcubic time; but from the second stage on, the sum is restricted to the surviving incidences, which lie in all pieces, and the per-piece products have $n^2hm\approx n^3m/\sigma\approx n^3$ coefficients at $m\approx\sigma$. The calls are adaptive: the wanted set of a call depends on earlier answers. This is allowed for the algorithm of $\Hyp(\varepsilon,\gamma)$, since each call is a separate run on an instance whose wanted set is known when the call is made; but it gives up a property of the preprint's Theorem 17, which produces all its instances before any call, so that they can be solved in any order or all together. Every call has $N=n$ and a middle part $D=\sigma m\le n^\varepsilon$ with $m$ ranging over the candidate moduli, so the counting algorithm is called at many values of $D$, all within the range of Definition~\ref{c:hyp}.

\paragraph{Search version.} The algorithm of Theorem~\ref{e:thm:A} decides. A YES of step 3 comes with a zero triangle; a YES of step 2(c) does not. Witnesses for the all-edges version are discussed after Lemma~\ref{e:lem:correctB}.

\part*{The two parts together, 3XOR, 3SUM over groups and coordinatewise problems}
\addcontentsline{toc}{part}{The two parts together, 3XOR, 3SUM over groups and coordinatewise problems}

\section{The two parts combined}\label{x:sec:consequences}

This section combines each of the two matrix theorems with each of the two sets of reductions and compares the exponents with the preprint's. The preprint's matrix theorem gives $\Hyp(1/18,\,0.063)$ (Lemma~\ref{c:lem:hzt}), and Corollary~\ref{m:cor:explicit}(b), applied to the two biadjacency matrices with the padding of its proof for every $D$, gives $\Hyp(1/14,\,0.077)$. In Table~\ref{tab:summary}, parts (a) and (b) of the corollary below are the column ``Part~I alone'', part (c) the column ``Part~II alone'', and part (d) the last column.

\begin{corollary}\label{x:cor:fg}
\begin{enumerate}[label=(\alph*),nosep]
\item Exact Triangle on $n$ vertices per part with integer weights of absolute value at most $n^\nu$ is solved deterministically in $O(n^{3-\varepsilon'_{\mathrm I}}\log n)\le O(n^{2.9973})$ time, $\varepsilon'_{\mathrm I}:=0.077/28=0.00275$, the Exact Triangle saving of Part~\hyperref[m:partI]{I} through the preprint's reduction.
\item 3SUM on $n$ integers of absolute value at most $n^\nu$ is solved deterministically in $n^{2-\varepsilon'_{\mathrm I}/2+o(1)}\le O(n^{1.99863})$ time, and the $(\min,+)$-product of two $n\times n$ integer matrices, and APSP on directed graphs with integer weights of absolute value at most $n^\nu$ and no negative cycles, in $\tilde O(n^{3-\varepsilon'_{\mathrm I}/3})\le O(n^{2.99909})$ time.
\item Through Part~\hyperref[m:partII]{II} with the preprint's Corollary 26 in place of Corollary~\ref{m:cor:explicit}, 3SUM takes $n^{2-0.0035+o(1)}=n^{1.99650+o(1)}$ time, and Exact Triangle (also in its all-edges version), the $(\min,+)$-product and APSP take $n^{3-0.0035+o(1)}=n^{2.99650+o(1)}$ time, deterministically.
\item Through Part~\hyperref[m:partII]{II}, 3SUM takes $n^{2-0.0055+o(1)}=n^{1.99450+o(1)}$ time, and Exact Triangle (also in its all-edges version), the $(\min,+)$-product and APSP take $n^{3-0.0055+o(1)}=n^{2.99450+o(1)}$ time, deterministically.
\end{enumerate}
\end{corollary}

\begin{proof}
(a) The proof of~\cite[Theorem 19]{ZT} by its Corollary 26, with $D:=\lfloor n^{1/14}\rfloor$, $g:=\lceil D^{0.0385}\rceil\le\sqrt D$ (for $D\ge4$), and Corollary~\ref{m:cor:explicit}(b) in place of its Corollary 26 (it applies since $D^{14}\le n$, and the instances of the preprint's Theorem 17 have at most $n^2/\sqrt D$ query pairs): the $O(nD^{0.0385})$ instances cost $O(n^2/D^{0.077})$ each, so $O(n^3D^{-0.0385})$ in all; the scans cost $O(n^3D^{-0.0385}\log n)$; the choice of the prime costs $O(n^{\log_27}D^{3/2})\le O(n^{2.92})$ with Strassen's algorithm~\cite{Str69}; building the instances costs $O(n^2D^{1.04})$. Finally $D\ge n^{1/14}/2$ gives $D^{-0.0385}\le2n^{-0.00275}$. (b) Plug (a) into~\cite[Theorem 21]{ZT} as in the proof of its Theorem 22: $n^{3/2+o(1)}+n^{1/2+o(1)}\cdot(n^{1/2+o(1)})^{3-\varepsilon'_{\mathrm I}}=n^{2-\varepsilon'_{\mathrm I}/2+o(1)}$, and $\tilde O(n^2(n^{1/3})^{3-\varepsilon'_{\mathrm I}})=\tilde O(n^{3-\varepsilon'_{\mathrm I}/3})$; $2-0.001375<1.99863$ and $3-0.000917<2.99909$. (c) Corollaries~\ref{s:cor:uniform}, \ref{s:cor:variants}, \ref{e:cor:uniform} and~\ref{e:cor:C} with $\Hyp(1/18,0.063)$; these are the numbers of Sections~\ref{s:sec:numbers} and~\ref{e:sec:numbers}. (d) The same corollaries with $\Hyp(1/14,0.077)$.
\end{proof}

With instance (a) of Corollary~\ref{m:cor:explicit} instead, $\varepsilon\gamma=0.092/18=0.00511$, and part (d) gives $n^{1.99489+o(1)}$ and $n^{2.99489+o(1)}$.

\begin{remark}[rate values]\label{x:rem:ratefg}
With the optimal rates $\varepsilon\gamma=0.005652$ (split form) and $0.005825$ (exact leaf count, Remark~\ref{m:rem:exact}) in place of $0.0055$, Corollary~\ref{x:cor:fg}(d) would read $n^{1.99435}$ and $n^{1.99418}$ for 3SUM, and $n^{2.99435}$ and $n^{2.99418}$ for the others. These are numerical rate values, not theorems.
\end{remark}

\paragraph{Comparison with the preprint's numbers.} The preprint's instance $N\ge D^{18}$ was not chosen to maximise $\varepsilon\gamma$; optimised in the same way, its inequality reaches $\varepsilon\gamma=0.00371$ (a rate value), against $0.00565$ for Part~\hyperref[m:partI]{I} (Remark~\ref{m:rem:exact}; ancillary file \texttt{anc/explicit\_constants.py}, output line E3). The preprint's own exponents for 3SUM and Exact Triangle, recomputed as $2-\varepsilon'/2$ and $3-\varepsilon'$ with its $\varepsilon'=0.00175$, are $1.99913$ and $2.99825$; it states the rounder $O(n^{1.9992})$ and $O(n^{2.9983})$.

\section{What is known and what is new}\label{x:sec:novelty}

This section separates, for each part, what is taken from earlier work from what we believe is new, quoting the sources where the attribution is delicate; no proof elsewhere depends on it. Sections~\ref{x:sec:xor}--\ref{g:sec:triangle} have their own such subsections (\ref{x:sec:xornovelty}, \ref{g:sec:groupsremarks}, \ref{g:sec:coordremarks} and~\ref{g:sec:triremarks}). For Part~\hyperref[m:partI]{I} the new element is the collision-free re-indexing of Section~\ref{m:sec:hash} and what it makes possible, the transfer of the preprint's pruning and data structure to a tensor with a shared variable; the squaring and the tensor itself are Coppersmith's and Schönhage's. For Part~\hyperref[m:partII]{II} the new element is that the false-positive counts that choose the primes are taken only over the survivors of the earlier moduli, which lets the counting algorithm supply them without a quadratic or cubic counting cost, and, for the all-edges version, the certification of single incidences; the prime-by-prime modulus, the stages, and counting with the called algorithm for a single modulus (the preprint's footnote 9) have the precedents named below.

\paragraph{The matrix theorem (Part~\hyperref[m:partI]{I}).} \emph{Known.} The tensor and its two-parameter family are Schönhage's~\cite{Sch81}, as printed and attributed in~\cite[p.~467 and Remarks on p.~469]{Cop82}. The squaring, tensoring the $M$-th power with a copy whose first two dimensions are reversed, is Coppersmith's last step in his Proof version 1: ``Now do the same arguments, with first and second dimensions reversed \ldots\ Multiply (tensorize)'' \cite[p.~469]{Cop82}; Le Gall's lecture slides call it combining the power ``with power $N$ of the same tensor with permuted variables''~\cite{LG17}, and Nissim and Schwartz use the same symmetrisation in their proof of the version-2 bound~\cite{NS24}.

Coppersmith, ``mimicking Schönhage's proof of partial matrix multiplication'' \cite[p.~468]{Cop82}, writes the wanted inputs as $A''=A'A$, $B''=BB'$ with constant matrices $A',B'$ whose maximal minors are all nonzero, so that $A''B''=A'(AB)B'$ (his outputs are the $c$-variables). In our orientation this handles the shared variable, but every entry of the dense product becomes a linear combination of many entries of the power's product, which is what we expect to defeat the pruning (Section~\ref{m:sec:tensor}; an argument, not a theorem). Nissim and Schwartz treat Coppersmith's Proof version 2 only, with disjoint copies of Schönhage's direct-sum identity and Schönhage's stacking, and no shared variable. Vrana~\cite{Vra24} reduces weighted total matrix multiplication to tensor powers of a partial one by applying uniformly random, many-to-one maps to the indices of the power whose image covers a full box (his Section 4, Proposition 4.7); since many index triples go to the same triple, the result is a product with arbitrary nonzero weights, which bounds only the support rank.

\emph{New, to our knowledge.} The collision-free $\F$-linear re-indexing of Section~\ref{m:sec:hash}: on every good inner set the re-indexing maps are bijections, so nothing is merged, every weight is $1$, every output is a single output, and the tensor power restricts to exact unweighted products: every input entry is zero or a copy of an entry of the given matrices, and every wanted entry is read off one output. With it, the observation that in this orientation the tensor has the preprint's private-leaf structure, and the transfer of the preprint's Sections 2 and 4 to it, are new relative to the preprint and to the sources above. Two precedents are closest. Hashing the index sets of tensor powers is standard since Coppersmith and Winograd~\cite{CW90}, where it is used to obtain disjoint blocks; here it gives a re-indexing that is bijective on each good set, so that nothing is merged or zeroed out. And the condition that the columns of $\Gamma$ at $Q$ form a basis is the $\F$, index-level analogue of Coppersmith's condition that ``each maximal minor of $A'$ or $B'$ has nonvanishing determinant'' \cite[p.~468]{Cop82}: his condition is imposed on values and mixes the entries, ours on index vectors and permutes them.

Relative to the preprint, Proposition~\ref{m:rem:omv512} is also new. Before their Conjecture 5.12, van den Brand, Nanongkai and Saranurak list three obvious algorithms~\cite[p.~31]{BNS19}: multiply $UN_{I,[n]}$ in Phase 2, multiply $N_{I,J}V$ in Phase 3, or compute the entry in Phase 4. The two regions of the proposition run the first two with the data structure in place of fast matrix multiplication; the preprint's proof of its Corollary 40 multiplies $UN_{I,J}$ in Phase 3 and leaves Phase 2 idle. Also new relative to the preprint, whose Section 6 asks whether other identities could serve, is the exclusion of full products as base identities for every decomposition of the shapes with $b\ge2$ that Proposition~\ref{a:prop:criterion} covers (Theorem~\ref{a:thm:full}, Corollary~\ref{a:cor:145}).

\paragraph{The reductions (Part~\hyperref[m:partII]{II}).} Building a modulus one prime at a time by the method of conditional expectations goes back to Chan and Lewenstein~\cite[Section~8]{CL15} and Chan and He~\cite{CH20}; Fischer, Kaliciak and Polak~\cite[Lemma~3.2]{FKP24} choose each prime to minimise the number of false positives, counted by FFT. The stages emulate the $O(\log n)$ independent repetitions of the randomized reduction of Kopelowitz, Pettie and Porat~\cite{KPP16}, where a query is scanned only if it is positive under every repetition. Rounds that each build a fresh modulus prime by prime on the elements that survived earlier rounds appear in Chan and Lewenstein~\cite[Section~8]{CL15} and in Fischer, Jin and Xu~\cite[Lemma~3.2]{FJX25}; in their reduction each prime also minimises a count of false positives, but over the whole instance and by FFT, with the survivors entering only through a collision condition. What is, as far as we know, new in the 3SUM reduction is that the false-positive counts are restricted to the surviving probes (those positive under every modulus so far) and are supplied by the counting algorithm; this is what lets the modulus reach $\sigma^2$ without an $n^2$ counting cost.

In the Exact Triangle reduction, the instances are those of the preprint's Theorem 17~\cite{ZT}, which follow Chan and Xu~\cite[Section~3]{CX24}, with the prime $p$ replaced by a composite modulus $m$. The preprint points out that the counting algorithm could do the counting for the modulus of its Theorem 17, choosing the modulus one prime factor at a time as in~\cite[Lemma~3.2]{FKP24} (\cite[footnote~9]{ZT}, quoted in Section~\ref{m:sec:intro:reductions}). The argument that the count is the number of false positives plus a term independent of the modulus, so that the minimising prime has at most the average number of false positives, is in the proof of the preprint's Theorem 17 and in Fischer, Jin and Xu~\cite[Lemma~3.2]{FJX25}. The stages here are those of the 3SUM reduction (Section~\ref{s:sec:algorithm}), transferred from 3SUM probes to (pair, piece) incidences. What is, as far as we know, new in the Exact Triangle reduction is: the false-positive counts are restricted to the incidences that survived every earlier modulus and are supplied by the counting algorithm, which lets the modulus reach the piece size without a counting cost of order $n^3$; and, for the all-edges version, the certification of single incidences (Lemma~\ref{e:lem:cert}). Certifying a yes-instance from a count above the averaging bound is the second outcome of~\cite[Lemma~3.2]{FKP24} and is the YES exit of Section~\ref{e:sec:detection}; here it is applied to each incidence separately, with its own count at the base modulus as reference, and an incidence that is not certified is shown to carry many false positives, which halves all incidences, true ones included.

The randomized reduction of Vassilevska Williams and Xu~\cite[Theorem~3.4 with Proposition~3.8]{VX20} also keeps the whole saving of Exact Triangle. Their Theorem 3.4 (p.~10) reads, in part: ``There exists an $\tilde O(n^{3-\min\{\alpha,\beta,\gamma\}+\rho})$ time randomized reduction from a Zero Triangle instance with $n$ vertices to $\tilde O(n^{3-\alpha-\beta-\gamma+2\rho})$ instances of $(\alpha,\beta,\gamma,\rho,900n^{\gamma-2\rho}+1)$-All-Edges Triangle Listing'', where (in their notation, whose $\gamma$ is not ours) the parts have $n^\alpha,n^\beta,n^\gamma$ vertices and the degrees are at most $O(n^{-\rho})$ times the part sizes, and their Proposition 3.8 (p.~14) reduces listing $O(1)$ triangles per edge to $\tilde O(1)$ detection instances of the same shape. By our reading, with $\alpha=\beta=1$ and $\gamma=2\rho$, $D:=n^{2\rho}$, this gives $\tilde O(n)$ instances of $\Lop(n,D)$ with $O(n^2/\sqrt D)$ query pairs each; the reduction itself takes $\tilde O(n^{3-\rho})=\tilde O(n^3/\sqrt D)$ time, which is below $n^3/D^\gamma$ since $\gamma\le\frac12$; hence Exact Triangle in $\tilde O(n\cdot n^2/D^{\gamma})$ with our $\gamma$, the whole saving. Their randomness bounds the false positives through each edge (their Claim~3.6), which their listing step needs; the preprint's Remark 18 makes this comparison with its Theorem 17, whose scans need only the total. The stages here replace the randomness by adaptivity and counting: a single deterministic modulus cannot bound the false positives of every incidence, but $O(\log n)$ moduli chosen on the survivors bound their total mass relative to the survivors, which halves the survivors in each stage (Lemma~\ref{e:lem:halving}), and for the all-edges version the certification turns a total bound into a bound on the number of uncertified incidences (Lemma~\ref{e:lem:halvingB}).

\section{3XOR}\label{x:sec:xor}

This section proves the preprint's Remark 23 and improves on it. The remark claims a deterministic truly subquadratic algorithm for 3XOR by ``the same approach as for 3SUM'', and that approach keeps only a quarter of the saving $\varepsilon\gamma$ of the matrix theorem: the preprint's 3SUM algorithm goes through Exact Triangle by its Theorem 21(a). Our reduction for 3XOR goes directly to the lopsided sparse triangle problem and keeps all of the saving, up to a factor $(\log n)^8$ (Theorem~\ref{x:thm:xorcount}, with counts); a second one keeps half of it with a single detection call (Theorem~\ref{x:thm:xordet}). With Part~\hyperref[m:partI]{I}, 3XOR is solved in $O(n^{1.99451})$ time, and with the preprint's own matrix theorem in $O(n^{1.99651})$ (Corollary~\ref{x:cor:xor}). Carried out for 3XOR, the remark's route goes through an XOR version of Exact Triangle and keeps a quarter as well, which gives $O(n^{1.9992})$ with the preprint's matrix theorem and $O(n^{1.99863})$ with Part~\hyperref[m:partI]{I}, the entries marked $\dagger$ in Table~\ref{tab:summary}; we do not write that route out. So 3XOR joins 3SUM, Exact Triangle and APSP as a fourth problem whose reduction keeps the whole saving.

The remark reads in full: ``The same approach as for 3SUM gives a deterministic truly subquadratic algorithm for 3XOR (given three lists of $n$ vectors in $\mathbb F_2^{O(\log n)}$, find one vector from each list such that the three XOR to zero \cite{JV16,DSW18}), using linear hash functions over $\mathbb F_2$ whose rows are chosen one at a time from an $\varepsilon$-biased set \cite{AGHP92} by the method of conditional expectations''~\cite[Remark~23]{ZT}. Before the preprint, no deterministic subquadratic algorithm was known: Dietzfelbinger, Schlag and Walzer give a randomized algorithm in expected time $O(n^2(\log\log n)^2/\log^2n)$ and write that ``finding a deterministic algorithm for 3XOR with subquadratic running time remains an open problem''~\cite[Theorems~5 and~7, p.~59:3]{DSW18}. Our reductions use exactly the tool the remark names. Theorem~\ref{x:thm:xorcount} is the analogue for 3XOR of the reductions of Part~\hyperref[m:partII]{II}, with the factor $(\log n)^8$ in place of $n^{o(1)}$; Theorem~\ref{x:thm:xordet} needs only the detection version $\Hyp^{\det}$; Theorem~\ref{x:thm:xet} is the analogue of Section~\ref{e:sec:statement} for triangles with weights in $\F^{d'}$.

\begin{definition}[3XOR and Exact-XOR-Triangle]\label{x:def:xor}
\emph{3XOR}: given three lists $S_1,S_2,S_3$ of at most $n$ vectors in $\F^d$, $\F=\mathbb F_2$, where $d\le c\log_2n$ for a constant $c$, decide whether there are $x_1\in S_1$, $x_2\in S_2$, $x_3\in S_3$ with $x_1\oplus x_2\oplus x_3=0$, where $\oplus$ is the bitwise XOR, the addition of $\F^d$. \emph{Exact-XOR-Triangle}: Exact Triangle (Section~\ref{e:sec:statement}) with weights $w(e)\in\F^{d'}$, $d'\le c'\log_2n$, in place of integers, and $S(a,b,c):=w(a,b)\oplus w(b,c)\oplus w(a,c)$; its all-edges version asks, for every pair $(a,b)\in A\times B$, whether some $c\in C$ has $S(a,b,c)=0$.
\end{definition}

A vector of $\F^d$ is stored as a string of $d$ bits in $O(\lceil c\rceil)$ words of $\Theta(\log n)$ bits. A linear map $H\colon\F^d\to\F^l$ is given by its rows $u_1,\dots,u_l$, $H(x)=(\langle u_1,x\rangle,\dots,\langle u_l,x\rangle)$, and each inner product $\langle u,x\rangle$ is the parity of the bitwise AND of $u$ and $x$, computed with $O(\lceil c\rceil\log d)$ word operations by folding the word onto itself. The empty map ($l=0$) sends every vector to $0$. Theorem~\ref{x:thm:xordet} assumes only $\Hyp^{\det}(\varepsilon,\gamma)$ (Section~\ref{c:sec:prelim}).

\begin{theorem}[3XOR, counting algorithm]\label{x:thm:xorcount}
Assume $\Hyp(\varepsilon,\gamma)$ with $0<\varepsilon,\gamma\le\frac12$, and let $c>0$ be a constant. There is a deterministic algorithm that, given $n$ and three lists of at most $n$ vectors of $\F^d$, $d\le c\log_2n$, decides 3XOR in time $O(n^{2-\varepsilon\gamma}(\log n)^8)$, where the constant depends on $c$, $\varepsilon$, $\gamma$ and $C_{\mathcal O}$. It calls $\mathcal O$ adaptively. A solution is found, when one exists, in time $O(n^{2-\varepsilon\gamma}(\log n)^9)$.
\end{theorem}

\begin{theorem}[3XOR, one detection call]\label{x:thm:xordet}
Assume $\Hyp^{\det}(\varepsilon,\gamma)$ with $0<\varepsilon,\gamma\le\frac12$, and let $c>0$ be a constant. There is a deterministic algorithm that solves 3XOR for three lists of at most $n$ vectors of $\F^d$, $d\le c\log_2n$, returns a solution when one exists, makes one call to the detection algorithm, and runs in time $O(n^{2-\varepsilon\gamma/2}\log n)$.
\end{theorem}

\begin{theorem}[Exact-XOR-Triangle]\label{x:thm:xet}
Assume $\Hyp(\varepsilon,\gamma)$ with $0<\varepsilon,\gamma\le\frac12$, and let $c'>0$ be a constant. Exact-XOR-Triangle with parts of at most $n$ vertices each and weights in $\F^{d'}$, $d'\le c'\log_2n$, and also its all-edges version, are solved deterministically in time $O(n^{3-\varepsilon\gamma}(\log n)^6)$.
\end{theorem}

\begin{corollary}\label{x:cor:xor}
With $\Hyp(1/14,0.077)$, from Corollary~\ref{m:cor:explicit}(b) as in Section~\ref{x:sec:consequences}, 3XOR is solved deterministically in time $O(n^{1.9945}(\log n)^8)\le O(n^{1.99451})$ by Theorem~\ref{x:thm:xorcount} and in time $O(n^{1.99725}\log n)\le O(n^{1.99726})$ with one detection call by Theorem~\ref{x:thm:xordet}, and Exact-XOR-Triangle, also in its all-edges version, in time $O(n^{2.9945}(\log n)^6)\le O(n^{2.99451})$. With $\Hyp(1/18,0.063)$, from the preprint's Corollary 26, the three bounds are $O(n^{1.99651})$, $O(n^{1.99826})$ and $O(n^{2.99651})$. In particular the preprint's Remark 23 holds with the exponent $1.99826$, from its own matrix theorem and a single detection call.
\end{corollary}

\begin{proof}
$\varepsilon\gamma=0.0055$, respectively $0.0035$, in Theorems~\ref{x:thm:xorcount}, \ref{x:thm:xordet} and~\ref{x:thm:xet}. Each exponent is rounded up by at least $10^{-5}$, which absorbs the polylogarithmic factor for large $n$.
\end{proof}

As in Part~\hyperref[m:partII]{II}, the slack absorbs the polylogarithmic factors only for very large $n$ (Section~\ref{s:sec:numbers} gives the sizes there); the bounds are asymptotic statements. Section~\ref{x:sec:xorbias} gives the tools over $\F$: biased sets, rows chosen one at a time, and the count of the triples that a linear map sends to $0$. Section~\ref{x:sec:xorprobes} replaces the self-reduction of Fischer, Kaliciak and Polak by a linear hash and defines the instances asked of the called algorithm. Sections~\ref{x:sec:xorstages}, \ref{x:sec:xordet} and~\ref{x:sec:xet} prove the three theorems, Section~\ref{x:sec:xorremarks} adds two remarks, and Section~\ref{x:sec:xornovelty} says what is known and what is new.

\subsection{Biased sets and rows chosen one at a time}\label{x:sec:xorbias}

\begin{definition}[$\beta$-biased multiset; {\cite[Definition~3, p.~5]{AGHP92}}]\label{x:def:biased}
A nonempty multiset $E$ of vectors of $\F^\ell$ is \emph{$\beta$-biased} if $\bigl|\frac1{|E|}\sum_{u\in E}(-1)^{\langle u,v\rangle}\bigr|\le\beta$ for every $v\in\F^\ell\setminus\{0\}$.
\end{definition}

\begin{lemma}[one row]\label{x:lem:select}
Let $E$ be a $\beta$-biased multiset in $\F^\ell$, and let $v_1,\dots,v_M\in\F^\ell\setminus\{0\}$ carry weights $\omega_1,\dots,\omega_M\ge0$. Then the average over $u\in E$ of $\sum_{j:\langle u,v_j\rangle=0}\omega_j$ is at most $\frac{1+\beta}2\sum_j\omega_j$.
\end{lemma}
\begin{proof}
The indicator of $\langle u,v\rangle=0$ is $(1+(-1)^{\langle u,v\rangle})/2$, and its average over $u\in E$ is at most $\frac{1+\beta}2$ for every $v\ne0$. Multiply by $\omega_j$ and sum over $j$.
\end{proof}

\begin{lemma}[rows chosen one at a time]\label{x:lem:greedy}
Let $E$, $v_j$, $\omega_j$ be as in Lemma~\ref{x:lem:select}, and for rows $u_1,\dots,u_t$ let $\Phi(u_1,\dots,u_t)$ be the total weight of the $v_j$ with $\langle u_1,v_j\rangle=\dots=\langle u_t,v_j\rangle=0$, so that $\Phi()=\sum_j\omega_j$. If for $t=1,\dots,r$ the row $u_t\in E$ minimises $\Phi(u_1,\dots,u_{t-1},u)$ over $u\in E$, then $\Phi(u_1,\dots,u_t)\le(\frac{1+\beta}2)^t\Phi()$ for every $t\le r$, and if $\beta\le1/r$, then $\Phi(u_1,\dots,u_t)\le e\,2^{-t}\Phi()$ for every $t\le r$.
\end{lemma}
\begin{proof}
$\Phi(u_1,\dots,u_{t-1},u)$ is the weight of the $v_j$ with $\langle u,v_j\rangle=0$ among those that survive $u_1,\dots,u_{t-1}$, all of which are nonzero; Lemma~\ref{x:lem:select} applied to the survivors and induction give the first bound. For $\beta\le1/r$ and $t\le r$, $(1+\beta)^t\le(1+1/r)^r\le e$.
\end{proof}

This is the method of conditional expectations: $2^{-(r-t)}\Phi(u_1,\dots,u_t)$ is the expected weight that the map with the chosen rows and $r-t$ further uniformly random rows sends to $0$. Alon and Naor choose the rows of a linear hash one at a time in this way, from a small-bias set, to separate pairs~\cite[Section~3.1]{AN96}.

\begin{lemma}[an explicit biased set; {\cite[Proposition~3]{AGHP92}}]\label{x:lem:aghp}
Let $\ell,r\ge1$ with $\ell r\ge2$, and $\mu:=\lceil\log_2(\ell r)\rceil$. There is a $(1/r)$-biased multiset $E_{\ell,r}\subseteq\F^\ell$ of $4^\mu<4(\ell r)^2$ vectors, listed deterministically in $(\ell r)^{O(1)}$ time.
\end{lemma}
\begin{proof}
An irreducible polynomial of degree $\mu$ over $\F$ is found by trial division of the $2^\mu$ monic polynomials of degree $\mu$, in $2^{O(\mu)}$ time~\cite[p.~14]{AGHP92}; it represents $\mathrm{GF}(2^\mu)$, and the coefficient vector $\mathrm{bin}\colon\mathrm{GF}(2^\mu)\to\F^\mu$ is $\F$-linear. For $x,y\in\mathrm{GF}(2^\mu)$ let $\rho(x,y)\in\F^\ell$ have the coordinates $\langle\mathrm{bin}(x^i),\mathrm{bin}(y)\rangle$, $i=0,\dots,\ell-1$, and let $E_{\ell,r}$ be the multiset of the $4^\mu$ vectors $\rho(x,y)$. For $\alpha\in\F^\ell\setminus\{0\}$ let $p_\alpha(t):=\sum_i\alpha_it^i$, a nonzero polynomial of degree less than $\ell$. By linearity $\langle\alpha,\rho(x,y)\rangle=\langle\mathrm{bin}(p_\alpha(x)),\mathrm{bin}(y)\rangle$. If $p_\alpha(x)\ne0$, this is a nonzero linear function of $\mathrm{bin}(y)$, balanced as $y$ ranges over $\mathrm{GF}(2^\mu)$, so such $x$ contribute $0$ to $\sum_{x,y}(-1)^{\langle\alpha,\rho(x,y)\rangle}$; the at most $\ell-1$ roots of $p_\alpha$ contribute at most $2^\mu$ each. So the bias is at most $(\ell-1)2^\mu/4^\mu<\ell/2^\mu\le1/r$, and $2^\mu<2\ell r$ gives $4^\mu<4(\ell r)^2$.
\end{proof}

\begin{lemma}[the triples a linear map sends to $0$]\label{x:lem:wh}
Let $S_1,S_2,S_3\subseteq\F^d$ have at most $n$ elements each, let $\pi\colon\F^d\to\F^l$ be linear, and let $Z(\pi)$ be the number of $(x_1,x_2,x_3)\in S_1\times S_2\times S_3$ with $\pi(x_1\oplus x_2\oplus x_3)=0$. With $c_i(\beta):=\#\{x\in S_i:\pi(x)=\beta\}$ and the Walsh--Hadamard transform $\hat c_i(y):=\sum_\beta(-1)^{\langle y,\beta\rangle}c_i(\beta)$,
\[
Z(\pi)=\sum_{\beta_1\oplus\beta_2\oplus\beta_3=0}c_1(\beta_1)c_2(\beta_2)c_3(\beta_3)=2^{-l}\sum_{y\in\F^l}\hat c_1(y)\,\hat c_2(y)\,\hat c_3(y),
\]
and $Z(\pi)$ is computed in $O(nl\log d+2^ll)$ time.
\end{lemma}
\begin{proof}
The first equality is linearity of $\pi$. For the second, $\sum_y(-1)^{\langle y,z\rangle}$ is $2^l$ if $z=0$ and $0$ otherwise, and $(-1)^{\langle y,\cdot\rangle}$ is multiplicative. The counts $c_i$ cost $O(nl\log d)$, and the three transforms and their inverse $O(2^ll)$ by the fast Walsh--Hadamard transform.
\end{proof}

\subsection{Blocks, probes and the instances}\label{x:sec:xorprobes}

\paragraph{Parameters for Theorem~\ref{x:thm:xorcount}.} Remove repeated vectors from each list, in $O(n\log n)$ time; this does not change the answer. From now on $S_1,S_2,S_3\subseteq\F^d$ are nonempty sets of at most $n$ elements (if one is empty, the answer is NO). Put
\[
\Lambda:=4e,\quad\sigma:=\Big\lfloor\sqrt{n^\varepsilon/\Lambda}\Big\rfloor,\quad k:=\Big\lceil\log_2\frac n\sigma\Big\rceil,\quad q:=\lceil\log_2(2e\sigma^2)\rceil,\quad g':=\Big\lceil\frac{3n}\sigma\Big\rceil,\quad L:=\lceil\log_2n\rceil,
\]
$B:=\lceil\log_2(\sigma+1)\rceil$, $E_k:=E_{d,k}$ and $E:=E_{d,q}$ (Lemma~\ref{x:lem:aghp}). In words: $\sigma$ is the block size, $2^k\approx n/\sigma$ the number of groups, $q$ the number of rows per stage, $L$ the number of stages, and $B$ the number of bits of a multiplicity. We assume $n\ge n_0(\varepsilon)$, so that $\sigma\ge2$ and $k,q\ge2$; smaller $n$ are solved by brute force in $O(1)$ time. Then
\begin{equation}\label{x:eq:params}
2^q<\Lambda\sigma^2\le n^\varepsilon,\qquad e\,2^{-q}\sigma^2\le\tfrac12,\qquad \frac n\sigma\le2^k<\frac{2n}\sigma,\qquad\sigma\ge\tfrac12\sqrt{n^\varepsilon/\Lambda},
\end{equation}
$|E|<4(dq)^2$, $|E_k|<4(dk)^2$, and $B\le\log_2n$.

\paragraph{Groups and blocks.} Choose rows $u_1,\dots,u_k\in E_k$ one at a time, each minimising $Z$ (Lemma~\ref{x:lem:wh}) of the map with the rows chosen so far followed by the candidate, and let $\pi$ be the map with these $k$ rows. Let $Z_0$ be the number of solutions; $Z_0\le n^2$, since $x_1,x_2$ determine $x_3$. Then, for the map $\pi_t$ with the first $t$ rows, $Z(\pi_t)-Z_0$ is $\Phi(u_1,\dots,u_t)$ of Lemma~\ref{x:lem:greedy} for the family of the vectors $x_1\oplus x_2\oplus x_3\ne0$ at weight $1$, and $Z_0$ does not depend on the rows, so Lemma~\ref{x:lem:greedy} gives
\begin{equation}\label{x:eq:zpi}
Z(\pi)-Z_0\le e\,2^{-k}n^3\le e\sigma n^2,\qquad Z(\pi)\le(e\sigma+1)n^2 .
\end{equation}
For $i=2,3$ and $\beta\in\F^k$ let $G_i(\beta):=\{x\in S_i:\pi(x)=\beta\}$, and cut each nonempty $G_i(\beta)$ into $\lceil|G_i(\beta)|/\sigma\rceil$ \emph{blocks} of at most $\sigma$ elements with the \emph{label} $\beta$. Let $\mathfrak B_i$ be the set of blocks of list $i$; $|\mathfrak B_i|\le2^k+n/\sigma<3n/\sigma\le g'$.

\paragraph{Probes.} A \emph{probe} is a triple $Q=(x_1;b_2,b_3)$ with $x_1\in S_1$, $b_i\in\mathfrak B_i$ and $\mathrm{label}(b_3)=\pi(x_1)\oplus\mathrm{label}(b_2)$; $\Pq_0$ is the set of probes. For a linear $H$ let
\begin{gather*}
\cnt_H(Q):=\#\{(x_2,x_3)\in b_2\times b_3:H(x_1\oplus x_2\oplus x_3)=0\},\\
z(Q):=\#\{(x_2,x_3)\in b_2\times b_3:x_1\oplus x_2\oplus x_3=0\}.
\end{gather*}
Then $\cnt_H(Q)\ge z(Q)$, and $\cnt_\emptyset(Q)=|b_2||b_3|\le\sigma^2$.

\begin{lemma}[probes]\label{x:lem:probes}
\textup{(i)} Every solution $(x_1,x_2,x_3)$ lies in a probe $(x_1;b_2,b_3)$ with $x_2\in b_2$, $x_3\in b_3$; so a solution exists if and only if $z(Q)>0$ for some $Q\in\Pq_0$. \textup{(ii)} $|\Pq_0|\le8n^2/\sigma$, and $\Pq_0$ is listed in $O(n^2/\sigma)$ time.
\end{lemma}
\begin{proof}
(i) If $x_2\in b_2$ and $x_3\in b_3$, then $\mathrm{label}(b_3)=\pi(x_3)=\pi(x_1)\oplus\pi(x_2)=\pi(x_1)\oplus\mathrm{label}(b_2)$ by linearity. (ii) With $s_i(\beta):=\lceil|G_i(\beta)|/\sigma\rceil\le1+|G_i(\beta)|/\sigma$,
\[
|\Pq_0|=\sum_{x_1\in S_1}\sum_\beta s_2(\beta)\,s_3(\pi(x_1)\oplus\beta)\]
\[\le\sum_{x_1\in S_1}\Big(2^k+\frac{|S_2|+|S_3|}\sigma+\frac{\#\{(x_2,x_3):\pi(x_1\oplus x_2\oplus x_3)=0\}}{\sigma^2}\Big),
\]
which is at most $2n^2/\sigma+2n^2/\sigma+Z(\pi)/\sigma^2\le(4+e+1/\sigma)n^2/\sigma\le8n^2/\sigma$ by~\eqref{x:eq:params} and~\eqref{x:eq:zpi}. For the listing, index $\mathfrak B_3$ by label and read off, for each $x_1$ and $b_2$, the blocks with label $\pi(x_1)\oplus\mathrm{label}(b_2)$, in $O(n|\mathfrak B_2|+|\Pq_0|)$ time.
\end{proof}

The groups are the classes of a linear map, the first $k$ rows of the same kind of hash that the stages build. The classes may have any sizes; balance comes from cutting large classes into blocks, and the bound~\eqref{x:eq:zpi} limits the number of block pairs that this creates. The self-reduction of Fischer, Kaliciak and Polak is not needed, and has no counterpart here: the XOR of two intervals is not an interval.

\paragraph{The instance $I(H,\Pq)$.} Let $H\colon\F^d\to\F^l$ be linear with $l\ge1$, put $l_1:=\lceil l/2\rceil$, $l_2:=\lfloor l/2\rfloor$, write $\F^l=\F^{l_1}\times\F^{l_2}$, and let $\Pq\subseteq\Pq_0$. The instance has the left vertices $(b_2,u)$, where $b_2$ ranges over $g'$ block slots of list $2$ (the slots beyond $|\mathfrak B_2|$ hold empty blocks) and $u\in\F^{l_1}$; the right vertices $(b_3,v)$, $v\in\F^{l_2}$, padded with isolated vertices so that both sides have $N:=g'2^{l_1}$ vertices; the middle part $\F^l$, so $D:=2^l$; and the matrices
\[
X[(b_2,u),\lambda]:=\#\{x_2\in b_2:Hx_2\oplus(u,0)=\lambda\},\qquad Y[\lambda,(b_3,v)]:=\#\{x_3\in b_3:Hx_3\oplus(0,v)=\lambda\}.
\]
A probe $Q=(x_1;b_2,b_3)$ is asked at the position $((b_2,u),(b_3,v))$ with $(u,v):=Hx_1$, and $W$ is the set of these positions.

\begin{lemma}[the counts]\label{x:lem:counts}
For every $Q\in\Pq$, the entry of $XY$ at the position of $Q$ is $\cnt_H(Q)$.
\end{lemma}
\begin{proof}
The entry counts the pairs $(x_2,x_3)\in b_2\times b_3$ with $Hx_2\oplus(u,0)=Hx_3\oplus(0,v)$, that is, $H(x_2\oplus x_3)=(u,v)=Hx_1$, that is, $H(x_1\oplus x_2\oplus x_3)=0$.
\end{proof}

This is the instance $I(m,\Pq)$ of Section~\ref{s:sec:instances} with the modulus $m$ replaced by $H$. The residue $Hx_1$ splits exactly into a high and a low part, and there is no carry.

\begin{lemma}[shape and cost of one evaluation]\label{x:lem:shape}
For every linear $H$ with $1\le l\le q$ rows and every $\Pq\subseteq\Pq_0$, the counts $\cnt_H(Q)$, $Q\in\Pq$, are computed with $\Hyp(\varepsilon,\gamma)$ deterministically in time $O(B^2n^2\sigma^{-2\gamma})=O(n^{2-\varepsilon\gamma}\log^2n)$, the constant depending only on $C_{\mathcal O}$.
\end{lemma}
\begin{proof}
\emph{$2\le D\le N^\varepsilon$.} $D\ge2$ as $l\ge1$. We have $N\ge g'\sqrt D\ge(n/\sigma)\sqrt D$ and $D\le2^q<\Lambda\sigma^2$ by~\eqref{x:eq:params}. So $D^{1-\varepsilon/2}\le\Lambda\sigma^{2-\varepsilon}\le n^\varepsilon\sigma^{-\varepsilon}$, using $\Lambda\sigma^2\le n^\varepsilon$, and $D=D^{1-\varepsilon/2}D^{\varepsilon/2}\le((n/\sigma)\sqrt D)^\varepsilon\le N^\varepsilon$.

\emph{Calls.} The entries of $X$ and $Y$ lie in $\{0,\dots,\sigma\}\subseteq\{0,\dots,2^B-1\}$, so Lemma~\ref{s:lem:split} turns the instance into $B^2$ instances of $\#\Lop(N,D)$ with the same $W$. Cut $W$, $|W|\le|\Pq_0|\le8n^2/\sigma$, into parts of at most $\lfloor N^2/\sqrt D\rfloor\ge N^2/(2\sqrt D)$ positions; since $N^2/\sqrt D\ge g'^2\sqrt D\ge9n^2\sqrt D/\sigma^2$, there are at most $1+2\sigma/\sqrt D$ parts. Each part is one call, of cost at most $C_{\mathcal O}N^2/D^\gamma\le32C_{\mathcal O}(n/\sigma)^2D^{1-\gamma}$, as $N^2\le2g'^2D$ and $g'\le4n/\sigma$. One $0/1$ instance costs at most $32C_{\mathcal O}(n/\sigma)^2(D^{1-\gamma}+2\sigma D^{1/2-\gamma})$. If $D\le\sigma^2$, the bracket is at most $3\sigma^{2-2\gamma}$, by $\gamma\le\frac12$; if $\sigma^2<D<\Lambda\sigma^2$, it is at most $3\Lambda\sigma^{2-2\gamma}$. So a $0/1$ instance costs at most $96\Lambda C_{\mathcal O}n^2\sigma^{-2\gamma}$. The case $D\le\sigma^2$ covers the maps with fewer than $q$ rows while a stage's hash is being built.

\emph{Bookkeeping.} Writing a $0/1$ instance and reading its answers takes $O(ND+|W|)$ time; $D^{1+\gamma}\le N$ because $D\le N^\varepsilon$ and $\varepsilon(1+\gamma)\le\frac34$, so $ND\le N^2/D^\gamma$, and $|W|\le8n^2/\sigma\le8n^2\sigma^{-2\gamma}$. Filling $X$ and $Y$ takes $O(ND+n2^{l_1})$, the values $Hx$ take $O(nq\log d)$, and mapping the probes to positions and radix-sorting them, which merges probes that share a position, takes $O(|\Pq|+N)$. Finally $\sigma^{-2\gamma}\le(4\Lambda)^\gamma n^{-\varepsilon\gamma}$ by~\eqref{x:eq:params}.
\end{proof}

\subsection{The stages: proof of Theorem~\ref{x:thm:xorcount}}\label{x:sec:xorstages}

\begin{enumerate}[label=\arabic*.]
\item Compute $\pi$, the blocks and $\Pq_0$, and list $E$.
\item For $t=1,\dots,L$:
\begin{enumerate}[label=(\alph*)]
\item If $\Pq_{t-1}=\emptyset$, answer NO.
\item Build $H_t$ row by row: start with no rows, and $q$ times, for every $u\in E$ compute $\Sigma(H,u):=\sum_{Q\in\Pq_{t-1}}\cnt_{(H,u)}(Q)$ by Lemma~\ref{x:lem:shape}, where $(H,u)$ is $H$ with the row $u$ appended, and append a minimising $u$ (ties arbitrarily). Let $H_t:=H$, and keep the counts $\cnt_{H_t}(Q)$.
\item If $\sum_{Q\in\Pq_{t-1}}\cnt_{H_t}(Q)>|\Pq_{t-1}|/2$, answer YES.
\item Otherwise let $\Pq_t:=\{Q\in\Pq_{t-1}:\cnt_{H_t}(Q)>0\}$.
\end{enumerate}
\item For every $Q=(x_1;b_2,b_3)\in\Pq_L$ and every $x_2\in b_2$, test whether $x_1\oplus x_2\in b_3$ by binary search in the sorted $b_3$. Answer YES if a test succeeds, and NO otherwise.
\end{enumerate}
The algorithm is deterministic and adaptive; as in Section~\ref{s:sec:algorithm}, each call is a separate run of $\mathcal O$ on an instance whose $W$ is known when the call is made.

For $\Pq\subseteq\Pq_0$ let $F_\Pq(H)$ be the number of false positives, the triples $(Q,x_2,x_3)$ with $Q=(x_1;b_2,b_3)\in\Pq$, $(x_2,x_3)\in b_2\times b_3$, $s:=x_1\oplus x_2\oplus x_3\ne0$ and $Hs=0$, and let $Z_\Pq:=\sum_{Q\in\Pq}z(Q)$. Then $\sum_{Q\in\Pq}\cnt_H(Q)=F_\Pq(H)+Z_\Pq$, and $Z_\Pq$ does not depend on $H$.

\begin{lemma}[halving]\label{x:lem:halving}
In every stage $t$ that reaches step 2(b), $F_{\Pq_{t-1}}(H_t)\le|\Pq_{t-1}|/2$. Hence a YES of step 2(c) is correct, and otherwise $|\Pq_t|\le|\Pq_{t-1}|/2$.
\end{lemma}
\begin{proof}
Let $\Pq:=\Pq_{t-1}$. Apply Lemma~\ref{x:lem:greedy} with $r:=q$ to the vectors $s\ne0$ of the triples $(Q,x_2,x_3)$, at weight $1$, and the $(1/q)$-biased $E$; its $\Phi$ for the rows $u_1,\dots,u_l$ is $F_\Pq$ of the map with these rows. Since $\Sigma(H,u)=F_\Pq((H,u))+Z_\Pq$, step 2(b) minimises $\Phi$, and $F_\Pq(H_t)\le e\,2^{-q}F_\Pq(\emptyset)\le e\,2^{-q}\sigma^2|\Pq|\le|\Pq|/2$ by~\eqref{x:eq:params}. If step 2(c) answers YES, then $Z_\Pq>0$, so some probe holds a solution. Otherwise every $Q\in\Pq_t$ has $\cnt_{H_t}(Q)\ge1$, so $|\Pq_t|\le\sum_Q\cnt_{H_t}(Q)\le|\Pq|/2$.
\end{proof}

\begin{lemma}[correctness]\label{x:lem:correct}
The algorithm answers YES if and only if a solution exists.
\end{lemma}
\begin{proof}
A YES of step 2(c) is correct by Lemma~\ref{x:lem:halving}, and one of step 3 exhibits a solution. Conversely, a solution lies in a probe $Q\in\Pq_0$ (Lemma~\ref{x:lem:probes}(i)), and $\cnt_H(Q)\ge z(Q)\ge1$ for every $H$; so, unless YES is answered earlier, $Q$ stays in every $\Pq_t$, step 2(a) never answers NO, and step 3 finds the solution.
\end{proof}

\emph{Cost.} The preparation costs $O(n\log^7n)$ for choosing $\pi$ ($k|E_k|$ evaluations of Lemma~\ref{x:lem:wh}, with $2^k<2n/\sigma$), $O(n^2/\sigma)=O(n^{2-\varepsilon/2})\le O(n^{2-\varepsilon\gamma})$ for the probes, and $(\log n)^{O(1)}$ for $E$. The stages make $Lq|E|$ evaluations, each $O(B^2n^{2-\varepsilon\gamma})$ by Lemma~\ref{x:lem:shape}, which is $O(\log n\cdot\log n\cdot\log^4n\cdot\log^2n\cdot n^{2-\varepsilon\gamma})=O(n^{2-\varepsilon\gamma}(\log n)^8)$ with $d\le c\log_2n$, the factors being $L$, $q$, $|E|<4(dq)^2$ and $B^2$; choosing the minima and forming $\Pq_t$ costs less. The final scans cover $|\Pq_L|\le|\Pq_0|/2^L\le8n/\sigma$ probes at $O(\sigma\log n)$ each. The instances of the calls have $n/\sigma\le N\le n^{O(1)}$, so its words are $\Theta(\log n)$-bit words; counts are at most $\sigma^2$ per probe and sums at most $8n^3$.

\emph{Search.} If the answer is YES, halve $S_1$ and keep a half on which the decision algorithm, with the parameters still set from $n$, answers YES; after $\lceil\log_2n\rceil$ rounds one $x_1$ in a solution remains, and a sorted $S_3$ gives $x_2$. This proves Theorem~\ref{x:thm:xorcount}.

\subsection{One hash and one detection call: proof of Theorem~\ref{x:thm:xordet}}\label{x:sec:xordet}

Here the symbols are redefined: $q:=\lfloor\varepsilon\log_2n\rfloor$ and $D:=2^q$, so that $n^\varepsilon/2<D\le n^\varepsilon$; $g:=\lceil D^{\gamma/2}\rceil$ and $\sigma:=\lfloor\sqrt{D/g}\rfloor$, so that $1\le\sigma\le\sqrt D$ and $\sigma\ge\frac12\sqrt{D/g}$; $k:=\lceil\log_2(n/\sigma)\rceil$, $g':=\lceil3n/\sigma\rceil$, and $E:=E_{d,k+q}$, which is $(1/(k+q))$-biased. We assume $n\ge2^{1/\varepsilon}$, so that $q\ge1$; smaller $n$ are solved by brute force. The proof of Lemma~\ref{x:lem:probes} uses only that the lists are sets, $\sigma\ge1$, $2^k<2n/\sigma$ and $Z(\pi)\le(e\sigma+1)n^2$, and Lemma~\ref{x:lem:counts} holds for every $H$ with at least one row; both apply with these symbols.

\emph{Algorithm.} Remove repeated vectors. Choose rows $u_1,\dots,u_{k+q}\in E$ one at a time, each minimising $Z$ (Lemma~\ref{x:lem:wh}) of the map with the rows chosen so far followed by the candidate; let $\pi$ consist of the first $k$ rows and $H$ of the last $q$. Form the blocks and $\Pq_0$ from $\pi$ with block size $\sigma$. Build $I(H,\Pq_0)$ with each of $X$ and $Y$ replaced by its support (an entry $1$ where the multiplicity is positive), and make one detection call. Scan every probe whose position is answered positively, as in step 3 of Section~\ref{x:sec:xorstages}, and return the first solution found, or NO.

\emph{Correctness.} An entry of the product of the supports is positive if and only if the same entry of $XY$ is, that is, by Lemma~\ref{x:lem:counts}, if and only if $\cnt_H(Q)>0$. A probe holding a solution has $\cnt_H(Q)\ge z(Q)\ge1$ and is scanned, and every solution lies in a probe (Lemma~\ref{x:lem:probes}(i)).

\emph{The hash.} Lemma~\ref{x:lem:greedy}, with $r:=k+q$, applies to every prefix of the sequence of rows. After the first $k$ rows, $Z(\pi)-Z_0\le e2^{-k}n^3\le e\sigma n^2$, so $|\Pq_0|\le8n^2/\sigma$. After all $k+q$ rows, the false positives number
\[
F:=Z((\pi,H))-Z_0\le e\,2^{-(k+q)}n^3\le e\sigma n^2/D,
\]
and every triple that $(\pi,H)$ sends to $0$ lies in exactly one probe, the one formed by its $x_1$ and the blocks of its $x_2$ and $x_3$. So $F$ is also the number of false positives over all probes.

\emph{Cost.} Choosing the hash takes $(k+q)|E|$ evaluations of Lemma~\ref{x:lem:wh} with $2^{k+q}<2nD/\sigma$, which is $O(n^{1+\varepsilon}(\log n)^{O(1)})$, below $n^{2-\varepsilon\gamma/2}$ as $\varepsilon\le\frac12$. Listing the probes costs $O(n^2/\sigma)\le O(n^2\sqrt{g/D})\le O(n^2D^{-\gamma/2})$, since $\frac12-\frac\gamma4\ge\frac\gamma2$. The call: $N=g'2^{\lceil q/2\rceil}\ge(n/\sigma)\sqrt D\ge n$, so $D\le n^\varepsilon\le N^\varepsilon$; and $N^2/\sqrt D\ge9n^2\sqrt D/\sigma^2\ge9n^2/\sigma>|W|$ since $\sigma\le\sqrt D$, so one call suffices. It costs at most $C_{\mathcal O}N^2/D^\gamma\le32C_{\mathcal O}(n/\sigma)^2D^{1-\gamma}\le128C_{\mathcal O}n^2gD^{-\gamma}\le256C_{\mathcal O}n^2D^{-\gamma/2}$, and writing the instance $O(ND)\le O(N^2/D^\gamma)$. Each flagged probe without a solution holds a false positive, so at most $F$ scans fail; each scan costs $O(\sigma\log n)$, in all $O(F\sigma\log n)=O(\sigma^2n^2\log n/D)=O(n^2D^{-\gamma/2}\log n)$. Finally $D^{-\gamma/2}\le2^{\gamma/2}n^{-\varepsilon\gamma/2}$. This proves Theorem~\ref{x:thm:xordet}.

A single hash has to make the false positives few over all probes at once, and the scans of the flagged probes then cost as much as the call: balancing the two gives $\sigma^2\approx D^{1-\gamma/2}$ and the saving $D^{\gamma/2}$. The stages of Theorem~\ref{x:thm:xorcount} make the scans negligible, so the block size can be about $\sqrt D$ and the whole saving $D^\gamma$ remains; the price is counting and adaptivity. The global count $Z$ needs no call, which is why one detection call suffices here; from the second stage on, the counts are restricted to the surviving probes, and only the counting algorithm supplies them cheaply.

\subsection{Exact-XOR-Triangle: proof of Theorem~\ref{x:thm:xet}}\label{x:sec:xet}

The proof is that of Section~\ref{e:sec:statement} with a linear map $H\colon\F^{d'}\to\F^l$ in place of the modulus $m$ and Lemma~\ref{x:lem:greedy} in place of the averaging over primes. Put $\Lambda:=8e$, $\sigma:=\lfloor\sqrt{n^\varepsilon/\Lambda}\rfloor$, $q:=\lceil\log_2(2e\sigma)\rceil+1$, $L:=\lceil\log_2n\rceil$ and $E:=E_{d',q}$, which is $(1/q)$-biased. Then $2^{q-1}\ge2e\sigma$, $2^q<\Lambda\sigma$, $q\ge3$ and $\Lambda\sigma^2\le n^\varepsilon$; we assume $n\ge n_0(\varepsilon)$, so that $2\le\sigma\le n$. Pieces $C_1,\dots,C_h$, $h\le2n/\sigma$, and incidences $x=(a,b,k)$ are those of Section~\ref{e:sec:instances} (so $k$ is again the index of a piece, not the number of rows of Section~\ref{x:sec:xorprobes}), with $\cnt_H(x):=\#\{c\in C_k:H\,S(a,b,c)=0\}$, $z(x)$ as there, $f_H(x):=\cnt_H(x)-z(x)$ and $F_\Iq(H):=\sum_{x\in\Iq}f_H(x)\le\sigma|\Iq|$ for $H=\emptyset$. The instance $J(H,k,\varrho,\mathcal Q)$, for $\varrho\in\F^l$ and pairs with $H\,w(a,b)=\varrho$, has the middle part $C_k\times\F^l$, so $D:=\sigma2^l$, and the edges $a\sim(c,\lambda)$ if and only if $\lambda=H\,w(a,c)\oplus\varrho$, and $(c,\lambda)\sim b$ if and only if $\lambda=H\,w(b,c)$; as in Lemma~\ref{e:lem:counts}, the number of common neighbours of $a$ and $b$ is $\cnt_H(a,b,k)$. Its counts for a set $\Iq$ of incidences and any $H$ with $l\le q$ rows cost $O(n^3\sigma^{-2\gamma})=O(n^{3-\varepsilon\gamma})$, by the proof of Lemma~\ref{e:lem:shape} with $2^l$ in place of $m$: $D<\Lambda\sigma^2\le n^\varepsilon$, a piece needs at most $2^l+2\sqrt{\sigma2^l}$ calls of cost $C_{\mathcal O}n^2(\sigma2^l)^{-\gamma}$ each, which is at most $3C_{\mathcal O}n^2\sigma^{1-2\gamma}$ if $2^l\le\sigma$ and at most $3\Lambda C_{\mathcal O}n^2\sigma^{1-2\gamma}$ if $\sigma<2^l<\Lambda\sigma$, and the instances are $0/1$, so no splitting is needed.

\emph{Detection.} Run the algorithm of Section~\ref{x:sec:xorstages} with the incidences $\Iq_0:=A\times B\times[h]$ in place of the probes, these counts in place of Lemma~\ref{x:lem:shape}, and, in step 3, a scan of $C_k$ for every surviving incidence. The halving lemma holds as Lemma~\ref{x:lem:halving}, with $F_\Iq(H_t)\le e2^{-q}\sigma|\Iq|\le|\Iq|/4$, and correctness as Lemma~\ref{x:lem:correct}. The stages make $Lq|E|=O(\log^6n)$ evaluations, and the scans cost $O(|\Iq_0|\sigma/2^L)=O(n^2)$.

\emph{All edges.} As in Section~\ref{e:sec:alledges}, each stage ends with a certification of single incidences, and over $\F$ the rule is simpler. With $\Iq:=\Iq_{t-1}$, choose $q-1$ greedy rows, giving $H$, keep $\cnt_H(x)$, then evaluate $\cnt_{(H,u)}(x)$ for every $u\in E$ and keep $\mu(x):=\min_{u\in E}\cnt_{(H,u)}(x)$. With $\beta:=1/q$, call $x$ \emph{certified false} if $\mu(x)=0$, \emph{certified true} if $2\mu(x)>(1+\beta)\cnt_H(x)$, and \emph{uncertified} otherwise. Mark YES the pair of every certified-true incidence, and let $\Iq_t$ be the uncertified incidences whose pair is not marked. After $L$ stages, scan $C_k$ for every $(a,b,k)\in\Iq_L$ and mark $(a,b)$ YES if a zero triangle is found; the other pairs are answered NO.

\begin{lemma}[certification over $\F$]\label{x:lem:cert}
\textup{(i)} A certified-false incidence has $z(x)=0$. \textup{(ii)} A certified-true incidence has $z(x)\ge1$. \textup{(iii)} An uncertified incidence has $f_H(x)\ge1$. \textup{(iv)} $|\Iq_t|\le|\Iq_{t-1}|/2$.
\end{lemma}
\begin{proof}
For every $u$, $\cnt_{(H,u)}(x)=z(x)+f_{(H,u)}(x)$, so $\mu(x)=z(x)+\min_uf_{(H,u)}(x)\ge z(x)$, which gives (i). (ii) The false positives of $x$ under $(H,u)$ are those $c$ among its false positives under $H$ with $\langle u,S(a,b,c)\rangle=0$, since $(H,u)$ sends $S(a,b,c)$ to $(H\,S(a,b,c),\langle u,S(a,b,c)\rangle)$; the vectors $S(a,b,c)$ involved are nonzero; by Lemma~\ref{x:lem:select} the average of $f_{(H,u)}(x)$ over $u\in E$ is at most $\frac{1+\beta}2f_H(x)$. So $\min_uf_{(H,u)}(x)\le\frac{1+\beta}2\cnt_H(x)<\mu(x)$, and $z(x)=\mu(x)-\min_uf_{(H,u)}(x)>0$. (iii) If $f_H(x)=0$, then $f_{(H,u)}(x)=0$ for every $u$, so $\mu(x)=z(x)=\cnt_H(x)$; then $x$ is certified false if $z(x)=0$, and certified true if $z(x)\ge1$, as $\beta<1$. (iv) By (iii), $|\Iq_t|\le F_\Iq(H)$, and Lemma~\ref{x:lem:greedy} with $q-1$ rows from the $(1/q)$-biased $E$ gives $F_\Iq(H)\le e2^{-(q-1)}\sigma|\Iq|\le|\Iq|/2$.
\end{proof}

Correctness follows as in Lemma~\ref{e:lem:correctB}: a pair is marked only by a certified-true incidence or a successful scan; and a true incidence is never certified false, so it is either certified true or survives every stage and is scanned. Each stage has $q-1$ greedy rounds and one certification round, each with $|E|$ evaluations, so the cost is that of detection. Part (iii) is simpler than Lemma~\ref{e:lem:cert}(iii), where an uncertified incidence carries many false positives: over $\F$ the minimum over the biased set already decides every incidence that has none. This proves Theorem~\ref{x:thm:xet}.

\subsection{Remarks}\label{x:sec:xorremarks}

\emph{Why the loss is polylogarithmic here.} In Part~\hyperref[m:partII]{II} a prime from $[P,2P)$ divides the false positives by about $P$ up to a logarithm (Lemma~\ref{c:lem:avg}), so a modulus of about $\sigma^2$ from $r$ primes needs a pool of about $P\approx\sigma^{2/r}$ candidates, each evaluated by the counting algorithm; this is the loss $n^{\varepsilon/r}$, and balancing it against $(\log n)^r$ gives $n^{o(1)}$. Over $\F$ a row from a $(1/q)$-biased set halves the false positives up to the factor $1+1/q$ whatever they are (Lemma~\ref{x:lem:select}), the pool has polylogarithmic size, and $q$ rows lose only the factor $e$.

\emph{Other operations.} The method uses only that the hashes are linear and that the characters $(-1)^{\langle y,\cdot\rangle}$ turn the count into products. Among the sixteen bitwise binary operations $g$, the problem $x_3=g(x_1,x_2)$ is 3XOR or easier when $g$ is affine over $\F$, and contains Orthogonal Vectors otherwise (for AND take $S_3=\{0\}$; the other seven reduce to AND by complementing the lists or the target); so under the Strong Exponential Time Hypothesis no algorithm for a non-affine $g$ saves a power of $n$ uniformly in $c$~\cite{Wil05}, whereas the bounds of Theorems~\ref{x:thm:xorcount} and~\ref{x:thm:xordet} hold for every constant $c$.

\subsection{What is known and what is new}\label{x:sec:xornovelty}

\emph{Known.} Linear hashes over $\F$ for 3XOR are used by Jafargholi and Viola~\cite[Definition~9]{JV16}, in a randomized reduction to Convolution-3XOR. The explicit small-bias sets of Lemma~\ref{x:lem:aghp} are those of Alon, Goldreich, H\aa stad and Peralta~\cite{AGHP92}, and choosing the rows of a linear hash one at a time from a small-bias set is Alon and Naor's derandomization~\cite[Section~3.1]{AN96}; the preprint's Remark 23 names this tool. The staged reduction is that of Part~\hyperref[m:partII]{II} (Section~\ref{x:sec:novelty}).

\emph{New, to our knowledge.} A proof of the remark, with explicit exponents, and the two reductions themselves: the grouping by a linear hash with cut classes in place of the self-reduction of Fischer, Kaliciak and Polak, the restricted counts supplied by the counting algorithm at each row, the global Walsh--Hadamard count that makes one detection call enough, and the certification rule of Lemma~\ref{x:lem:cert}.

\section{3SUM over finite abelian groups}\label{g:sec:groups}

Section~\ref{x:sec:xor} solves 3XOR through linear hashes whose rows come from small-bias sets and calls to an algorithm for $\Lop$. This section asks how far that method reaches over other finite abelian groups; Section~\ref{g:sec:coord} asks the same for the coordinatewise problems of which 3XOR is the simplest non-trivial example.

\paragraph{The problem.} \emph{3SUM over $G$}: given three lists $S_1,S_2,S_3\subseteq G$ of at most $n$ elements of a finite abelian group $G$ with $|G|\le n^c$, decide whether $x_1+x_2+x_3=0$ for some $x_i\in S_i$; 3XOR is the case $G=\F^d$. The results assume $\Hyp(\varepsilon,\gamma)$ (Definition~\ref{c:hyp}) or its detection version $\Hyp^{\det}(\varepsilon,\gamma)$ (Section~\ref{c:sec:prelim}). With $\Hyp(1/14,\,0.077)$, which Section~\ref{x:sec:consequences} derives from Corollary~\ref{m:cor:explicit}(b), every such statement holds unconditionally with $\eps\gamma=0.0055$; the hardness statements of Section~\ref{g:sec:coord} rest on the Orthogonal Vectors hypothesis instead, which nothing here removes.

\paragraph{What carries over, and what does not.} Over the elementary groups $\Fp^d$ the proofs of Section~\ref{x:sec:xor} carry over with three changes: a row from a $\beta$-biased set keeps a fraction $(1+\beta)/p$ of the false positives instead of $(1+\beta)/2$; the global count of the triples that a linear map sends to $0$ is a Fourier transform with characters into the $p$-th roots of unity, computed exactly in $\Z[X]/(X^p-1)$; and the biased rows come from the powering construction over $\mathbb F_{p^\mu}$. For fixed $p$ this gives the exponents of Section~\ref{x:sec:xor}, which this section generalises and does not improve; for large $p$ the dimension is small and the problem becomes few instances of integer 3SUM, which gives algorithms uniform in $p$ (Section~\ref{g:sec:elementary}). The proofs use one property of the hash: a fixed family of homomorphisms kills every nonzero sum with probability about $1/|\text{range}|$. Section~\ref{g:sec:obstruction} shows that no such family exists once the group has a large non-elementary part, for example $\Z_4^d$, and Section~\ref{g:sec:G} gets around the obstruction for every group of exponent $p^2$ with hashes that adapt to the instance (Theorem~\ref{g:thm:G}). Higher exponents and mixed primes remain open (Section~\ref{g:sec:beyond}). Section~\ref{g:sec:triangle} treats the triangle version, Exact Triangle with weights in $G$.

\paragraph{Organisation.} Sections~\ref{g:sec:rows}--\ref{g:sec:instance} collect the tools: biased rows over $\Fp$, rows chosen one at a time, the Fourier count over a finite abelian group with exact arithmetic, and the one-call instance for an arbitrary pair of homomorphisms. Then come the elementary groups (Section~\ref{g:sec:elementary}), the obstruction (Section~\ref{g:sec:obstruction}), exponent $p^2$ (Section~\ref{g:sec:G}), what lies beyond (Section~\ref{g:sec:beyond}), and what is known and what is new (Section~\ref{g:sec:groupsremarks}).

\paragraph{Conventions.} These hold in this section and the next two. As in Section~\ref{x:sec:xor}, a list is a set (repeated elements are removed first, in $O(nd\log n)$ time, which does not change the answer), and $Z_0$ denotes the number of solutions of the instance at hand; since $x_1,x_2$ determine $x_3$, $Z_0\le n^2$. $\log$ is to base $2$ unless a base is written, so $\log n$ is the $\log_2n$ of Section~\ref{x:sec:xor}. For a triple $\tau=(x_1,x_2,x_3)$ we write $v(\tau):=x_1+x_2+x_3$, the vector $x_1\oplus x_2\oplus x_3$ of Section~\ref{x:sec:xor} when $G=\F^d$. Constants in $O(\cdot)$ may depend on $c$, $\eps$, $\gamma$, $C_{\mathcal O}$, and, where we say so, on $p$. Like Section~\ref{x:sec:xor}, the three sections define their own symbols where they use them; $H$, $L$, $M$, $q$, $k$, $\sigma$, $D$ and $E$ are redefined locally, as they are within Section~\ref{x:sec:xor}.

\paragraph{Model.} We work on the word RAM with $\Theta(\log n)$-bit words. A finite abelian group is given as $G=\Z_{m_1}\times\dots\times\Z_{m_r}$ with all $m_i\ge2$, so $r\le\log|G|\le c\log n$, and an element as its vector of residues; adding two elements costs $O(r)$. A homomorphism $K\colon G\to\Gamma$ into $\Gamma=\Z_{m'_1}\times\dots\times\Z_{m'_s}$ is given by the images of the $r$ unit vectors, and evaluating it costs $O(rs)$ operations. We also store each element as the integer $\sum_ix_i\prod_{i'<i}m_{i'}<|G|\le n^c$, in $O(c)$ words, so that two elements are compared in $O(1)$ time and a set of elements is sorted in $O(n\log n)$ time.

\subsection{Rows over \texorpdfstring{$\Fp$}{Fp}}\label{g:sec:rows}

This subsection carries Lemmas~\ref{x:lem:select}, \ref{x:lem:greedy} and~\ref{x:lem:aghp} from $\F$ to $\Fp$. Let $p$ be a prime. For $u,v\in\Fp^d$ write $\langle u,v\rangle:=\sum_ju_jv_j\in\Fp$. A \emph{row} is a vector $u\in\Fp^d$; rows $u_1,\dots,u_l$ define the linear map $x\mapsto(\langle u_1,x\rangle,\dots,\langle u_l,x\rangle)$, and the empty list of rows defines the zero map.

\begin{definition}[biased multisets]\label{g:def:biased}
Let $\beta\ge0$. A nonempty multiset $E$ of rows is \emph{$\beta$-kernel-biased} if $\#\{u\in E:\langle u,v\rangle=0\}\le(1+\beta)|E|/p$ for every $v\in\Fp^d\setminus\{0\}$, and \emph{$\beta$-value-biased} if $\#\{u\in E:\langle u,v\rangle=s\}\le(1+\beta)|E|/p$ for every $v\in\Fp^d\setminus\{0\}$ and every $s\in\Fp$.
\end{definition}

A $\beta$-value-biased multiset is $\beta$-kernel-biased (take $s=0$). For $p=2$, a $\beta$-biased multiset in the sense of Definition~\ref{x:def:biased} is $\beta$-kernel-biased, since the indicator of $\langle u,v\rangle=0$ is $(1+(-1)^{\langle u,v\rangle})/2$.

\begin{lemma}[one row]\label{g:lem:onerow}
Let $E$ be $\beta$-kernel-biased, and let $w_1,\dots,w_M\in\Fp^d\setminus\{0\}$ carry weights $\omega_1,\dots,\omega_M\ge0$. Then the average over $u\in E$ of $\sum_{j:\langle u,w_j\rangle=0}\omega_j$ is at most $\frac{1+\beta}p\sum_j\omega_j$.
\end{lemma}
\begin{proof}
Exchange the two sums: the average is $\sum_j\omega_j\cdot\#\{u\in E:\langle u,w_j\rangle=0\}/|E|$, and each fraction is at most $(1+\beta)/p$.
\end{proof}

\begin{lemma}[rows chosen one at a time]\label{g:lem:greedy}
Let $E$ be a finite nonempty multiset, $\vartheta>0$, and let $\Phi$ assign a number $\Phi(u_1,\dots,u_t)\ge0$ to every finite sequence of elements of $E$ (including the empty one) such that for every sequence $(u_1,\dots,u_t)$ the average of $\Phi(u_1,\dots,u_t,u)$ over $u\in E$ is at most $\vartheta\,\Phi(u_1,\dots,u_t)$. If for $t=1,\dots,r$ the element $u_t$ minimises $\Phi(u_1,\dots,u_{t-1},u)$ over $u\in E$, then $\Phi(u_1,\dots,u_t)\le\vartheta^t\Phi()$ for every $t\le r$. In particular, if $\vartheta=(1+\beta)/p$ with $\beta\le1/r$, then $\Phi(u_1,\dots,u_t)\le e\,p^{-t}\Phi()$ for every $t\le r$, and if $\beta\le1/(2r)$, then $\Phi(u_1,\dots,u_t)\le e^{1/2}p^{-t}\Phi()$.
\end{lemma}
\begin{proof}
The minimum is at most the average, and induction on $t$ gives the first claim. For $t\le r$, $(1+\beta)^t\le(1+1/r)^r\le e$, respectively $(1+\beta)^t\le e^{t/(2r)}\le e^{1/2}$.
\end{proof}

The standard application: for nonzero vectors $w_j$ with weights $\omega_j$, let $\Phi(u_1,\dots,u_t)$ be the total weight of the $w_j$ with $\langle u_i,w_j\rangle=0$ for all $i\le t$. The survivors of $u_1,\dots,u_t$ are nonzero vectors, so Lemma~\ref{g:lem:onerow} applied to them gives the hypothesis of Lemma~\ref{g:lem:greedy} with $\vartheta=(1+\beta)/p$ for a $\beta$-kernel-biased $E$. For $p=2$ this is Lemmas~\ref{x:lem:select} and~\ref{x:lem:greedy}. When $\Phi$ can be computed only up to an additive term that does not depend on the last row, minimising the computed value minimises $\Phi$; all our applications are of this kind.

\begin{lemma}[the powering construction over $\Fp$]\label{g:lem:powering}
Let $d\ge1$, $\mu\ge1$ and $\mathfrak q:=p^\mu$, and let $\Tr\colon\mathbb F_{\mathfrak q}\to\Fp$ be the trace, $\Tr(z)=\sum_{i<\mu}z^{p^i}$. The multiset
\[
E_{d,\mu}:=\bigl\{\bigl(\Tr(\beta\alpha^j)\bigr)_{0\le j<d}:\alpha,\beta\in\mathbb F_{\mathfrak q}\bigr\}\subseteq\Fp^d
\]
of $\mathfrak q^2$ rows is $\bigl(p(d-1)/\mathfrak q\bigr)$-value-biased. For $\beta_0\in(0,1]$ let $\mu$ be the least integer $\mu\ge1$ with $p^\mu\ge p(d-1)/\beta_0$; then $E_{d,\mu}$ is $\beta_0$-value-biased, has fewer than $p^4d^2/\beta_0^2$ elements, and is listed deterministically in time $O(p^{3}(d/\beta_0)^{O(1)})$.
\end{lemma}
\begin{proof}
The trace is $\Fp$-linear and onto $\Fp$: it is a polynomial of degree $p^{\mu-1}<\mathfrak q$ and so not zero on all of $\mathbb F_{\mathfrak q}$, and its image is an $\Fp$-subspace of $\Fp$. For $v\in\Fp^d\setminus\{0\}$ let $f_v(X):=\sum_jv_jX^j$, a nonzero polynomial of degree less than $d$. By linearity, the row $u$ of the pair $(\alpha,\beta)$ has $\langle u,v\rangle=\Tr(\beta f_v(\alpha))$. If $f_v(\alpha)\ne0$, then $\beta\mapsto\beta f_v(\alpha)$ is a bijection of $\mathbb F_{\mathfrak q}$, so $\beta\mapsto\Tr(\beta f_v(\alpha))$ is $\Fp$-linear and onto, and takes every value $s\in\Fp$ for exactly $\mathfrak q/p$ values of $\beta$. If $f_v(\alpha)=0$, which happens for at most $d-1$ values of $\alpha$, the value $0$ is taken for all $\mathfrak q$ values of $\beta$ and the other values for none. So every value $s$ is taken by at most $\mathfrak q\cdot\mathfrak q/p+(d-1)\mathfrak q=(\mathfrak q^2/p)(1+p(d-1)/\mathfrak q)$ rows.

For the size: by minimality of $\mu$, $p^\mu<p\max\{p,\,p(d-1)/\beta_0\}\le p^2d/\beta_0$, so $|E|=p^{2\mu}<p^4d^2/\beta_0^2$. To list $E$, find a monic irreducible polynomial of degree $\mu$ over $\Fp$ by testing the $p^\mu$ monic polynomials of degree $\mu$ for divisibility by the monic polynomials of degree at most $\mu/2$; this takes $p^{1.5\mu+1}\poly(\mu)\le O(p^{3}(d/\beta_0)^{1.5}\poly(\mu))$ time, since $p^\mu<p^2d/\beta_0$. Arithmetic in $\mathbb F_{\mathfrak q}$ is then arithmetic modulo this polynomial, and each of the $\mathfrak q^2d$ traces costs $\poly(\mu,\log p)$ operations on numbers below $p$.
\end{proof}

For $p=2$, this is the construction of~\cite[Proposition~3]{AGHP92} used in Lemma~\ref{x:lem:aghp}, where the size bound $4(dr)^2$ is slightly better because the bias there is the character bias. We use $E_{d,\mu}$ with $\beta_0=1/r$, so that $|E|<p^4d^2r^2$.

\subsection{The Fourier count}\label{g:sec:fourier}

The global count of Lemma~\ref{x:lem:wh}, for an arbitrary finite abelian group:

\begin{lemma}[the triples a homomorphism sends to $0$]\label{g:lem:fourier}
Let $G$ be a finite abelian group, $S_1,S_2,S_3\subseteq G$ of at most $n$ elements each, and $K\colon G\to\Gamma$ a homomorphism into $\Gamma=\Z_{m'_1}\times\dots\times\Z_{m'_s}$, $|\Gamma|\le n^{O(1)}$, and let $\mathfrak m$ be a common multiple of $m'_1,\dots,m'_s$. Let $Z(K)$ be the number of triples $\tau\in S_1\times S_2\times S_3$ with $K(v(\tau))=0$, and $c_i(\beta):=\#\{x\in S_i:K(x)=\beta\}$ for $\beta\in\Gamma$. Then
\[
Z(K)=\sum_{\beta_1+\beta_2+\beta_3=0}c_1(\beta_1)c_2(\beta_2)c_3(\beta_3)=\frac1{|\Gamma|}\sum_{\psi\in\hat\Gamma}\hat c_1(\psi)\hat c_2(\psi)\hat c_3(\psi),\qquad \hat c_i(\psi):=\sum_{\beta}c_i(\beta)\psi(\beta),
\]
where $\hat\Gamma$ is the group of characters $\Gamma\to\C^\times$. $Z(K)$ is computed exactly in $O(nT_K+s|\Gamma|\mathfrak m^2)$ word operations, $T_K$ the time to evaluate $K$; for $\Gamma=\Fp^l$ and $\mathfrak m=p$ this is $O(nT_K+l\,p^{l+2})$.
\end{lemma}
\begin{proof}
The first equality holds because $K$ is a homomorphism: $K(x_1+x_2+x_3)=K(x_1)+K(x_2)+K(x_3)$. The second is orthogonality: $\sum_{\psi\in\hat\Gamma}\psi(y)$ is $|\Gamma|$ if $y=0$ and $0$ otherwise, and $\psi(\beta_1)\psi(\beta_2)\psi(\beta_3)=\psi(\beta_1+\beta_2+\beta_3)$.

\emph{Exact arithmetic.} Let $\omega:=e^{2\pi i/\mathfrak m}$. The characters are $\psi_y(\beta)=\omega^{\sum_k y_k\beta_k\mathfrak m/m'_k}$, $y\in\Gamma$, so all values are powers of $\omega$. Compute in the ring $\mathcal R:=\Z[X]/(X^{\mathfrak m}-1)$, in which $X$ stands for $\omega$: an element is a vector of $\mathfrak m$ integers, addition costs $O(\mathfrak m)$, multiplication by a power of $X$ is a cyclic shift, $O(\mathfrak m)$, and a product costs $O(\mathfrak m^2)$. The histograms $c_i$ cost $O(nT_K)$. The transforms $\hat c_i$ are computed one coordinate at a time, as for a multidimensional discrete Fourier transform: for coordinate $k$, the $|\Gamma|/m'_k$ one-dimensional transforms of length $m'_k$ cost $O(m'^2_k)$ ring operations each, so $O(|\Gamma|\,m'_k\,\mathfrak m)\le O(|\Gamma|\mathfrak m^2)$ word operations, and $O(s|\Gamma|\mathfrak m^2)$ for all coordinates. The pointwise products and their sum cost $O(|\Gamma|\mathfrak m^2)$. The result is an element $F=\sum_{j<\mathfrak m}F_jX^j$ of $\mathcal R$ with $F(\omega)=|\Gamma|Z(K)$. To read off this integer, reduce $F$ modulo the cyclotomic polynomial $\Phi_{\mathfrak m}$, whose roots are the primitive $\mathfrak m$-th roots of unity; the powers $1,\omega,\dots,\omega^{\varphi(\mathfrak m)-1}$ are linearly independent over $\Q$, so an integer value $F(\omega)$ equals the constant term of $F\bmod\Phi_{\mathfrak m}$, and all other coefficients of $F\bmod\Phi_{\mathfrak m}$ vanish. We use only $\mathfrak m=p$ and $\mathfrak m=p^2$. For $\mathfrak m=p$, $\Phi_p=\sum_{j<p}X^j$ and $X^{p-1}\equiv-\sum_{j<p-1}X^j$, so $|\Gamma|Z(K)=F_0-F_{p-1}$. For $\mathfrak m=p^2$, $\Phi_{p^2}=\sum_{j<p}X^{jp}$; for $p(p-1)\le e<p^2$ write $e=p(p-1)+\rho$ with $0\le\rho<p$, then $X^e\equiv-\sum_{j<p-1}X^{\rho+jp}$, which has a constant term only for $\rho=0$; so $|\Gamma|Z(K)=F_0-F_{p(p-1)}$. The coefficients of the $\hat c_i(\psi)$ lie in $[0,n]$, those of their products in $[0,\mathfrak m^2n^3]$, and those of $F$ in $[0,|\Gamma|\mathfrak m^2n^3]$; with $|\Gamma|\le n^{O(1)}$ and $\mathfrak m\le|\Gamma|$ they fit in $O(1)$ words.
\end{proof}

For $p=2$ and $\Gamma=\F^l$, this is the Walsh--Hadamard count of Lemma~\ref{x:lem:wh}.

\subsection{Blocks, probes and the one-call instance}\label{g:sec:instance}

The following construction is that of Section~\ref{x:sec:xorprobes} for an arbitrary pair of homomorphisms. Its correctness uses nothing about the group; what changes from group to group is how $\pi$ and $K$ are chosen, and therefore how many probes and false positives there are.

\paragraph{Data.} Let $G$ be a finite abelian group and $S_1,S_2,S_3\subseteq G$ sets of at most $n$ elements. Let $\pi\colon G\to\Pi$ and $K\colon G\to\Gamma$ be homomorphisms into finite abelian groups, let $M\le\Gamma$ be a subgroup with $K(\ker\pi)\subseteq M$, and fix an isomorphism $M\cong M_1\times M_2$ with $|M_1|\ge|M_2|$; we write $(\mu_1,\mu_2)$ for the element of $M$ with these components. Let $D:=|M|\ge2$, and let $\sigma\ge1$ be an integer, the block size.

\paragraph{Blocks and probes.} For $i=2,3$ and $\beta\in\Pi$, cut each nonempty class $\{x\in S_i:\pi(x)=\beta\}$ into $\lceil|\text{class}|/\sigma\rceil$ \emph{blocks} of at most $\sigma$ elements with the \emph{label} $\beta$, and fix a representative $z_b\in b$ of every block $b$. Let $\mathfrak B_i$ be the set of blocks of list $i$ and $g'$ an integer with $g'\ge|\mathfrak B_2|,|\mathfrak B_3|$. A \emph{probe} is $Q=(x_1;b_2,b_3)$ with $x_1\in S_1$, $b_i\in\mathfrak B_i$ and $\pi(x_1)+\mathrm{label}(b_2)+\mathrm{label}(b_3)=0$; $\Pq_0$ is the set of probes. For a probe let $\cnt_K(Q):=\#\{(x_2,x_3)\in b_2\times b_3:K(x_1+x_2+x_3)=0\}$ and $z(Q):=\#\{(x_2,x_3)\in b_2\times b_3:x_1+x_2+x_3=0\}$, so $\cnt_K(Q)\ge z(Q)$.

\paragraph{The instance $I(\pi,K)$.} Left vertices $(b_2,\mu_1)$ for $g'$ block slots $b_2$ of list $2$ (slots beyond $|\mathfrak B_2|$ hold empty blocks) and $\mu_1\in M_1$; right vertices $(b_3,\mu_2)$, $\mu_2\in M_2$, padded with isolated vertices so that both sides have $N:=g'|M_1|$ vertices; middle part $M$; and the matrices
\begin{align*}
X[(b_2,\mu_1),\lambda]&:=\#\{x_2\in b_2:K(x_2)-K(z_{b_2})+(\mu_1,0)=\lambda\},\\
Y[\lambda,(b_3,\mu_2)]&:=\#\{x_3\in b_3:-(K(x_3)-K(z_{b_3}))-(0,\mu_2)=\lambda\}.
\end{align*}
These are well defined: $\pi(x_2)=\pi(z_{b_2})$, so $x_2-z_{b_2}\in\ker\pi$ and $K(x_2)-K(z_{b_2})\in M$, and likewise for $x_3$. The probe $Q=(x_1;b_2,b_3)$ is asked at the position $((b_2,\mu_1),(b_3,\mu_2))$ with $(\mu_1,\mu_2):=K(x_1)+K(z_{b_2})+K(z_{b_3})$, which lies in $M$ because $\pi(x_1+z_{b_2}+z_{b_3})=\pi(x_1)+\mathrm{label}(b_2)+\mathrm{label}(b_3)=0$. $W$ is the set of these positions. The \emph{support instance} replaces every entry of $X$ and $Y$ by $1$ if it is positive and by $0$ otherwise.

\begin{lemma}[probes and the instance]\label{g:lem:instance}
With the data above:
\begin{enumerate}[label=\textup{(\roman*)}]
\item A triple $\tau=(x_1,x_2,x_3)\in S_1\times S_2\times S_3$ lies in a probe (that is, $x_2\in b_2$ and $x_3\in b_3$ for a probe $(x_1;b_2,b_3)$) if and only if $\pi(v(\tau))=0$, and then in exactly one, the probe $(x_1;\text{block of }x_2,\text{block of }x_3)$. In particular every solution lies in exactly one probe.
\item $|\Pq_0|\le n\,C_2+(|S_2|+|S_3|)\,n/\sigma+Z(\pi)/\sigma^2$, where $C_2$ is the number of nonempty $\pi$-classes of $S_2$, and $\Pq_0$ is listed in time $O(n|\mathfrak B_2|+|\Pq_0|)$ after the blocks of list $3$ are indexed by label.
\item For every probe $Q$, the entry of $XY$ at the position of $Q$ is $\cnt_K(Q)$; so the entry of the product of the support matrices is positive if and only if $\cnt_K(Q)>0$.
\item The number of probes $Q$ with $\cnt_K(Q)>0$ and $z(Q)=0$ is at most $F:=\#\{\tau:v(\tau)\ne0,\ \pi(v(\tau))=0,\ K(v(\tau))=0\}$.
\end{enumerate}
\end{lemma}
\begin{proof}
(i) If $x_2\in b_2$ and $x_3\in b_3$ for a probe, then $\pi(v)=\pi(x_1)+\mathrm{label}(b_2)+\mathrm{label}(b_3)=0$. Conversely, if $\pi(v)=0$, then $(x_1;\text{block of }x_2,\text{block of }x_3)$ satisfies the label condition; a block contains $x_2$ only if it is the block of $x_2$, so the probe is unique.
(ii) With $s_i(\beta):=\lceil|\{x\in S_i:\pi(x)=\beta\}|/\sigma\rceil\le[\text{class}\ne\emptyset]+|\text{class}|/\sigma$,
\begin{align*}
|\Pq_0|&=\sum_{x_1\in S_1}\sum_\beta s_2(\beta)\,s_3(-\pi(x_1)-\beta)\\
&\le\sum_{x_1\in S_1}\Bigl(C_2+\frac{|S_2|+|S_3|}\sigma+\frac{\#\{(x_2,x_3):\pi(x_1+x_2+x_3)=0\}}{\sigma^2}\Bigr),
\end{align*}
which is the bound. For the listing, read off, for each $x_1$ and each $b_2$, the blocks with label $-\pi(x_1)-\mathrm{label}(b_2)$.
(iii) The entry is the number of pairs $(x_2,x_3)\in b_2\times b_3$ with $K(x_2)-K(z_{b_2})+(\mu_1,0)=-K(x_3)+K(z_{b_3})-(0,\mu_2)$, that is, with $K(x_2)+K(x_3)+(\mu_1,\mu_2)-K(z_{b_2})-K(z_{b_3})=0$; since $(\mu_1,\mu_2)-K(z_{b_2})-K(z_{b_3})=K(x_1)$, this is $K(x_1+x_2+x_3)=0$. Both matrices have nonnegative entries, so an entry of the product of the supports is positive if and only if the same entry of $XY$ is.
(iv) A probe with $\cnt_K(Q)>0$ and $z(Q)=0$ contains a triple with $v\ne0$ and $K(v)=0$, and by (i) such a triple lies in no other probe and has $\pi(v)=0$.
\end{proof}

\begin{lemma}[one hash, detection calls]\label{g:lem:onecall}
Assume $\Hyp^{\det}(\eps,\gamma)$. With the data above, suppose $D\le N^\eps$ and $D^{1+\gamma}\le N$. The following algorithm decides whether a solution exists and returns one if so: list $\Pq_0$; build the support instance of $I(\pi,K)$; cut $W$ into $J:=\lceil|W|/\lfloor N^2/\sqrt D\rfloor\rceil$ parts and make one detection call per part; then scan the probes whose position is answered positively, each by testing for every $x_2\in b_2$ whether $-x_1-x_2$ lies in the sorted block $b_3$, and return the first solution found, or NO. Its running time is
\[
O\bigl(n(T+r\log n)+(n|\mathfrak B_2|+|\Pq_0|+n|M_1|)T_M+J(C_{\mathcal O}N^2/D^\gamma+ND)+(F+1)\,\sigma(r+\log n)\bigr),
\]
where $T$ is the time to evaluate $\pi$ and $K$, $T_M$ the time for an addition in $\Gamma$ and an evaluation of the isomorphism $M\cong M_1\times M_2$, and $r$ the number of cyclic factors of $G$. Moreover $J\le1+2|W|\sqrt D/N^2$.
\end{lemma}
\begin{proof}
\emph{Correctness.} A returned triple is a solution. If a solution $\tau$ exists, it lies in a probe $Q$ (Lemma~\ref{g:lem:instance}(i)) with $\cnt_K(Q)\ge z(Q)\ge1$, which is answered positively (Lemma~\ref{g:lem:instance}(iii)) and therefore scanned, and the scan of a probe that contains a solution finds one. So NO is returned only when there is no solution.

\emph{The calls.} The support instance is an instance of $\Lop(N,D)$ with the middle part $M$ of $D\ge2$ vertices, and $D\le N^\eps$. Each part of $W$ has at most $\lfloor N^2/\sqrt D\rfloor$ positions, so each call is within the range of $\Hyp^{\det}(\eps,\gamma)$ and costs at most $C_{\mathcal O}N^2/D^\gamma$. Since $N^2/\sqrt D\ge N^{2-\eps/2}\ge1$, $\lfloor N^2/\sqrt D\rfloor\ge N^2/(2\sqrt D)$, which gives the bound on $J$.

\emph{The rest.} Removing repeated elements, sorting the blocks and computing $\pi$ and $K$ on all elements cost $O(n(T+r\log n))$. Listing the probes costs $O(n|\mathfrak B_2|+|\Pq_0|)$ (Lemma~\ref{g:lem:instance}(ii)), and their positions $O(T_M)$ each, from the stored values $K(x_1)$ and $K(z_b)$. Writing the support instance costs $O(ND)$ for the zero matrices plus $O(T_M)$ for each entry set to $1$; these are at most $\sum_{b_2}|b_2|\,|M_1|\le n|M_1|$ in $X$ and at most $n|M_2|\le n|M_1|$ in $Y$. Reading the answers costs $O(|W|)$. By Lemma~\ref{g:lem:instance}(iv), at most $F$ scans fail before the first successful one, and a scan costs $O(\sigma(r+\log n))$: for each $x_2\in b_2$, $O(r)$ to form $-x_1-x_2$ and $O(\log\sigma)$ comparisons in $b_3$. Finally $ND\le N^2/D^\gamma$ because $D^{1+\gamma}\le N$.
\end{proof}

\subsection{Elementary abelian groups}\label{g:sec:elementary}

\subsubsection{One detection call}

\begin{theorem}[$\Fp^d$, one detection call]\label{g:thm:A}
Assume $\Hyp^{\det}(\eps,\gamma)$ with $0<\eps,\gamma\le\frac12$, and let $c>0$. There is a deterministic algorithm that, given a prime $p$, $n$ with $p\le n^\eps$, and three lists of at most $n$ vectors of $\Fp^d$ with $p^d\le n^c$, returns a solution of 3SUM over $\Fp^d$ when one exists and NO otherwise, makes one call to the detection algorithm, and runs in time
\[
O\bigl(p^{3.25}n^{2-\eps\gamma/2}\log n+p^7n^{1+\eps}(\log n)^{O(1)}\bigr),
\]
where the constants depend on $c$, $\eps$, $\gamma$ and $C_{\mathcal O}$ but not on $p$. For fixed $p$ this is $O(n^{2-\eps\gamma/2}\log n)$, the bound of Theorem~\ref{x:thm:xordet} for $p=2$.
\end{theorem}

\begin{proof}
\emph{Parameters.} Put $q:=\lfloor\eps\log_pn\rfloor$, which is at least $1$ because $p\le n^\eps$, and $D:=p^q$, so that $n^\eps/p<D\le n^\eps$. Put $g:=\lceil D^{\gamma/2}\rceil$ and $\sigma:=\lfloor\sqrt{D/g}\rfloor$; since $g\le D$, $\sqrt{D/g}\ge1$, so $1\le\sigma\le\sqrt D$ and $\sigma\ge\frac12\sqrt{D/g}$. Put $k:=\lceil\log_p(n/\sigma)\rceil$, so $n/\sigma\le p^k<pn/\sigma$, and $g':=\lceil(p+1)n/\sigma\rceil\le(p+2)n/\sigma$. Let $E$ be the $(1/(k+q))$-value-biased multiset of Lemma~\ref{g:lem:powering}, $|E|<p^4d^2(k+q)^2$.

\emph{Algorithm.} Remove repeated vectors. Choose rows $u_1,\dots,u_{k+q}\in E$ one at a time, each minimising $Z$ (Lemma~\ref{g:lem:fourier}) of the linear map formed by the rows chosen so far and the candidate. Let $\pi\colon\Fp^d\to\Fp^k$ be the map of the first $k$ rows, $H\colon\Fp^d\to\Fp^q$ that of the last $q$, and $K:=(\pi,H)\colon\Fp^d\to\Fp^k\times\Fp^q$. Then $K(\ker\pi)\subseteq M:=\{0\}\times\Fp^q$, and we take $M_1:=\Fp^{q_1}$, $M_2:=\Fp^{q_2}$ with $q_1:=\lceil q/2\rceil$, $q_2:=\lfloor q/2\rfloor$, block size $\sigma$ and $g'$ block slots, and run the algorithm of Lemma~\ref{g:lem:onecall}. It is correct by that lemma.

\emph{The hash.} Let $\Phi(u_1,\dots,u_t)$ be the number of triples $\tau$ with $v(\tau)\ne0$ that the map of $u_1,\dots,u_t$ sends to $0$. Then $Z$ of that map is $\Phi(u_1,\dots,u_t)+Z_0$, and $Z_0$ does not depend on the rows, so the algorithm minimises $\Phi$, and Lemma~\ref{g:lem:greedy} with Lemma~\ref{g:lem:onerow} gives $\Phi(u_1,\dots,u_t)\le e\,p^{-t}n^3$ for every $t\le k+q$. Hence
\[
Z(\pi)\le Z_0+e\,p^{-k}n^3\le(e\sigma+1)n^2,\qquad F\le Z(K)-Z_0\le e\,p^{-(k+q)}n^3\le e\sigma n^2/D .
\]

\emph{Counts.} $\pi$ has at most $p^k<pn/\sigma$ nonempty classes in each list, so each list has fewer than $p^k+n/\sigma<(p+1)n/\sigma\le g'$ blocks, and by Lemma~\ref{g:lem:instance}(ii)
\[
|\Pq_0|\le np^k+\frac{2n^2}\sigma+\frac{Z(\pi)}{\sigma^2}\le\Bigl(p+2+e+\frac1\sigma\Bigr)\frac{n^2}\sigma\le(p+1)^2\frac{n^2}\sigma,
\]
since $p+3+e\le(p+1)^2$ for $p\ge2$.

\emph{The call.} $N=g'p^{q_1}\ge(p+1)(n/\sigma)\sqrt D\ge n$ because $\sigma\le\sqrt D$, so $D\le n^\eps\le N^\eps$, and $D^{1+\gamma}\le N^{\eps(1+\gamma)}\le N$. Moreover $N^2/\sqrt D\ge g'^2\sqrt D\ge(p+1)^2n^2\sqrt D/\sigma^2\ge(p+1)^2n^2/\sigma\ge|\Pq_0|\ge|W|$, so $J=1$: one call suffices. Since $p^{2q_1}\le pD$ and $(n/\sigma)^2\le4n^2g/D$, the call costs at most
\[
C_{\mathcal O}\frac{N^2}{D^\gamma}\le C_{\mathcal O}(p+2)^2p\Bigl(\frac n\sigma\Bigr)^2D^{1-\gamma}\le4C_{\mathcal O}p(p+2)^2n^2gD^{-\gamma}\le8C_{\mathcal O}p(p+2)^2n^2D^{-\gamma/2}.
\]

\emph{The rest.} Evaluating $\pi$ and $K$ costs $O(d(k+q))=O(\log^2n)$ per element and $T_M=O(k+q)=O(\log n)$. The scans cost $O((F+1)\sigma\log n)=O(\sigma^2n^2\log n/D+\sigma\log n)=O(n^2\log n/g)\le O(n^2D^{-\gamma/2}\log n)$. Listing the probes and their positions costs $O((n|\mathfrak B_2|+|\Pq_0|)\log n)=O(pn^2\log n/\sigma)$, and $1/\sigma\le2\sqrt{g/D}\le2\sqrt2\,D^{\gamma/4-1/2}\le2\sqrt2\,D^{-\gamma/2}$ because $\frac12-\frac\gamma4\ge\frac\gamma2$; writing the instance costs $O(ND+n|M_1|\log n)$, which is $O(N^2/D^\gamma)$ plus $O(n\sqrt{pD}\log n)$. Choosing the hash takes $(k+q)|E|=O(p^4(\log n)^5)$ evaluations of Lemma~\ref{g:lem:fourier}, each over a group $\Fp^t$, $t\le k+q$, of size $p^{k+q}<pnD/\sigma$, so each costs $O(nd(k+q)+(k+q)p^{k+q+2})=O(p^3nD(\log n)^2)$; in all $O(p^7n^{1+\eps}(\log n)^{O(1)})$, and listing $E$ costs $O(p^3(\log n)^{O(1)})$ (Lemma~\ref{g:lem:powering}). Finally $D>n^\eps/p$ gives $D^{-\gamma/2}<p^{\gamma/2}n^{-\eps\gamma/2}\le p^{1/4}n^{-\eps\gamma/2}$, and $8p(p+2)^2\le32p^3$, which gives the bound.
\end{proof}

\subsubsection{Counting algorithm}

\begin{theorem}[$\Fp^d$, counting algorithm]\label{g:thm:B}
Assume $\Hyp(\eps,\gamma)$ with $0<\eps,\gamma\le\frac12$, and let $p$ be a fixed prime and $c>0$. 3SUM over $\Fp^d$, $p^d\le n^c$, on lists of at most $n$ vectors is decided deterministically, with adaptive calls to the counting algorithm, in time $O(n^{2-\eps\gamma}(\log n)^8)$; the constant is polynomial in $p$ and depends on $c$, $\eps$, $\gamma$ and $C_{\mathcal O}$. A solution is found, when one exists, in time $O(n^{2-\eps\gamma}(\log n)^9)$.
\end{theorem}

\begin{proof}
The algorithm and its proof are those of Theorem~\ref{x:thm:xorcount} (Sections~\ref{x:sec:xorprobes} and~\ref{x:sec:xorstages}) with $\F$ replaced by $\Fp$. We list every place where something changes and prove each changed step; every step not mentioned is word for word that of Section~\ref{x:sec:xor}, with $\oplus$ read as the addition of $\Fp^d$.

\emph{Parameters.} Put $\Lambda:=2ep$, $\sigma:=\lfloor\sqrt{n^\eps/\Lambda}\rfloor$, $k:=\lceil\log_p(n/\sigma)\rceil$, $q:=\lceil\log_p(2e\sigma^2)\rceil$, $g':=\lceil(p+1)n/\sigma\rceil$, $L:=\lceil\log_2n\rceil$, $B:=\lceil\log_2(\sigma+1)\rceil$, and let $E_k$ and $E$ be the $(1/k)$- and $(1/q)$-kernel-biased multisets of Lemma~\ref{g:lem:powering}. We assume $n\ge n_0(\eps,p)$, so that $\sigma\ge2$ and $k,q\ge1$; smaller $n$ are solved by brute force in $O(1)$ time. In place of~\eqref{x:eq:params} we have
\[
p^q<\Lambda\sigma^2\le n^\eps,\qquad e\,p^{-q}\sigma^2\le\tfrac12,\qquad\frac n\sigma\le p^k<\frac{pn}\sigma,\qquad\sigma\ge\tfrac12\sqrt{n^\eps/\Lambda},
\]
the first because $p^q<p\cdot2e\sigma^2$ by minimality of $q$, the second because $p^q\ge2e\sigma^2$; and $|E|<p^4d^2q^2$, $|E_k|<p^4d^2k^2$, both $O(p^4(\log n)^4)$.

\emph{Groups, blocks and probes.} The rows of $\pi$ are chosen from $E_k$ as in Theorem~\ref{g:thm:A}, so $Z(\pi)\le(e\sigma+1)n^2$ as in~\eqref{x:eq:zpi}, now by Lemmas~\ref{g:lem:onerow} and~\ref{g:lem:greedy}. The blocks are the classes of $\pi$ cut into pieces of at most $\sigma$ elements; there are fewer than $p^k+n/\sigma<(p+1)n/\sigma\le g'$ of them in each list. Lemma~\ref{g:lem:instance}(ii) replaces Lemma~\ref{x:lem:probes}(ii): $|\Pq_0|\le(p+2+e+1/\sigma)n^2/\sigma\le(p+6)n^2/\sigma$, the only change being the first term $np^k<pn^2/\sigma$.

\emph{The instance and its counts.} For a linear $H\colon\Fp^d\to\Fp^l$ with $1\le l\le q$ rows, the instance $I(H,\Pq)$ of Section~\ref{x:sec:xorprobes} has left vertices $(b_2,u)$, $u\in\Fp^{l_1}$, right vertices $(b_3,v)$, $v\in\Fp^{l_2}$, $l_1:=\lceil l/2\rceil$, $l_2:=\lfloor l/2\rfloor$, middle part $\Fp^l$, so $D:=p^l$, $N:=g'p^{l_1}$, and $X[(b_2,u),\lambda]:=\#\{x_2\in b_2:Hx_2+(u,0)=\lambda\}$, $Y[\lambda,(b_3,v)]:=\#\{x_3\in b_3:-Hx_3-(0,v)=\lambda\}$; the probe $(x_1;b_2,b_3)$ is asked at $(u,v):=Hx_1$. The entry at this position counts the pairs with $Hx_2+(u,0)=-Hx_3-(0,v)$, that is, $H(x_1+x_2+x_3)=0$, which replaces Lemma~\ref{x:lem:counts}. (It is the instance of Lemma~\ref{g:lem:instance} for $K=(\pi,H)$, $M=\{0\}\times\Fp^l$, with the representatives dropped; dropping them changes only the names of the positions.)

\emph{Shape and cost of one evaluation} (in place of Lemma~\ref{x:lem:shape}). $D\ge2$ as $l\ge1$. We have $N\ge g'\sqrt D\ge(n/\sigma)\sqrt D$ and $D\le p^q<\Lambda\sigma^2\le n^\eps$, so $D^{1-\eps/2}<(\Lambda\sigma^2)^{1-\eps/2}\le\Lambda\sigma^2\sigma^{-\eps}\le n^\eps\sigma^{-\eps}$ (using $\Lambda\ge1$ and $\Lambda\sigma^2\le n^\eps$), and $D=D^{1-\eps/2}D^{\eps/2}\le((n/\sigma)\sqrt D)^\eps\le N^\eps$. The entries of $X$ and $Y$ lie in $\{0,\dots,\sigma\}$, so Lemma~\ref{s:lem:split} (binary splitting) gives $B^2$ instances of $\#\Lop(N,D)$ with $0/1$ entries and the same $W$. Cut $W$, $|W|\le|\Pq_0|\le(p+6)n^2/\sigma$, into parts of at most $\lfloor N^2/\sqrt D\rfloor\ge N^2/(2\sqrt D)$ positions; since $N^2/\sqrt D\ge g'^2\sqrt D\ge(p+1)^2n^2\sqrt D/\sigma^2$ and $p+6\le(p+1)^2$, there are at most $1+2\sigma/\sqrt D$ parts. Since $N\le(p+2)(n/\sigma)p^{l_1}$ and $p^{2l_1}\le pD$, each call costs at most $C_{\mathcal O}N^2/D^\gamma\le C_{\mathcal O}p(p+2)^2(n/\sigma)^2D^{1-\gamma}$, and one $0/1$ instance at most $C_{\mathcal O}p(p+2)^2(n/\sigma)^2(D^{1-\gamma}+2\sigma D^{1/2-\gamma})$. If $D\le\sigma^2$, the bracket is at most $3\sigma^{2-2\gamma}$, by $\gamma\le\frac12$; if $\sigma^2<D<\Lambda\sigma^2$, it is at most $3\Lambda\sigma^{2-2\gamma}$. So a $0/1$ instance costs at most $3\Lambda C_{\mathcal O}p(p+2)^2n^2\sigma^{-2\gamma}$, and $\sigma^{-2\gamma}\le(4\Lambda)^\gamma n^{-\eps\gamma}$. The bookkeeping is as in the proof of Lemma~\ref{x:lem:shape}: $D^{1+\gamma}\le N$ as $\eps(1+\gamma)\le\frac34$, so writing an instance, $O(ND+|W|)$, is within the same bound, and the values $Hx$ cost $O(nqd)$.

\emph{Halving and correctness.} Lemma~\ref{x:lem:halving} holds with Lemma~\ref{g:lem:greedy} (with Lemma~\ref{g:lem:onerow}) in place of Lemma~\ref{x:lem:greedy}: the false positives over the surviving probes $\Pq$ satisfy $F_\Pq(H_t)\le e\,p^{-q}F_\Pq(\emptyset)\le e\,p^{-q}\sigma^2|\Pq|\le|\Pq|/2$. Lemma~\ref{x:lem:correct} (correctness) uses nothing about $\F$.

\emph{Cost.} The stages make $Lq|E|=O(p^4(\log n)^6)$ evaluations, each $O(\poly(p)B^2n^{2-\eps\gamma})=O(\poly(p)n^{2-\eps\gamma}\log^2n)$, in all $O(n^{2-\eps\gamma}(\log n)^8)$ for fixed $p$. Choosing $\pi$ takes $k|E_k|$ evaluations of Lemma~\ref{g:lem:fourier} over $\Fp^t$, $t\le k$, of size $p^k<pn/\sigma$, in all $O(\poly(p)\,n(\log n)^{O(1)})$; listing the probes takes $O(pn^2/\sigma)=O(pn^{2-\eps/2})\le O(n^{2-\eps\gamma})$; the final scans cover at most $|\Pq_0|/2^L\le(p+6)n/\sigma$ probes at $O(\sigma\log n)$ each. The search version halves $S_1$ as in the proof of Theorem~\ref{x:thm:xorcount}.
\end{proof}

\subsubsection{All primes at once}

\begin{proposition}[bounded number of cyclic factors]\label{g:prop:boundedrank}
Let $G=\Z_{m_1}\times\dots\times\Z_{m_t}$ with $|G|\le n^c$. 3SUM over $G$ on lists of at most $n$ elements reduces in $O(3^ttn)$ time to $3^t$ instances of integer 3SUM with three sets of at most $n$ integers each, of absolute value at most $2\cdot3^tn^c$, and the question $a+b+c=0$.
\end{proposition}
\begin{proof}
Lift every coordinate $x_i\in\Z_{m_i}$ to $\tilde x_i\in\{0,\dots,m_i-1\}$. For $x,y,z\in G$, the sum $s_i:=\tilde x_i+\tilde y_i+\tilde z_i$ lies in $[0,3m_i-3]$, so $x+y+z=0$ if and only if $s_i\in\{0,m_i,2m_i\}$ for every $i$, that is, $s=(\kappa_im_i)_i$ for some $\kappa\in\{0,1,2\}^t$. Encode $X(x):=\sum_i\tilde x_iB_i$ with $B_i:=\prod_{i'<i}3m_{i'}$. Since every $s_i$ is below $3m_i$, $X(x)+X(y)+X(z)=\sum_is_iB_i$ is the mixed-radix representation of this number with digits $s_i$, and the representation is unique; so $x+y+z=0$ if and only if $X(x)+X(y)+X(z)=T_\kappa:=\sum_i\kappa_im_iB_i$ for some $\kappa$. For each of the $3^t$ vectors $\kappa$, the instance $A=\{X(x)\}_{x\in S_1}$, $B=\{X(y)\}_{y\in S_2}$, $C=\{X(z)-T_\kappa\}_{z\in S_3}$ with the question $a+b+c=0$ has a solution if and only if 3SUM over $G$ has one with this $\kappa$. All numbers are at most $\prod_i3m_i=3^t|G|\le3^tn^c$ in absolute value, up to the factor $2$ from the shift.
\end{proof}

The integer algorithms we use are the preprint's and that of Section~\ref{s:sec:statement}. The preprint's Theorem 22 solves 3SUM on $n$ integers of absolute value at most $n^\nu$, for every constant $\nu$, in time $n^{2-\eps'/2+o(1)}\le O(n^{1.9992})$ with $\eps'=0.00175$; the exponent does not depend on $\nu$. Its problem has one list and asks for three of its numbers that sum to $0$; three sets $A,B,C\subseteq[-U,U]$ with the question $a+b+c=0$ reduce to it by the list $\{a+M\}\cup\{b+3M\}\cup\{c-4M\}$ with $M:=3U+1$: three of its numbers, $n_A$ from the first part, $n_B$ from the second and $n_C$ from the third, sum to a number within $3U<M$ of $(n_A+3n_B-4n_C)M$, which is $0$ only if $n_A+3n_B=4n_C$, and with $n_A+n_B+n_C=3$ this forces $n_A=n_B=n_C=1$. Corollary~\ref{s:cor:variants} solves the three-set variant directly in time $n^{2-\eps\gamma+o(1)}$ given $\Hyp(\eps,\gamma)$, which with $\Hyp(1/14,0.077)$ is $n^{2-0.0055+o(1)}$ (Corollary~\ref{x:cor:fg}(d)). We write $\delta_\Z$ for the saving of the integer algorithm used: $\delta_\Z=\eps'/2=0.000875$ with the preprint's algorithm and $\delta_\Z=0.0055$ with Corollary~\ref{x:cor:fg}(d), in both cases up to $o(1)$.

\begin{theorem}[uniform in $p$]\label{g:thm:App}
Let $c>0$ be a constant.
\begin{enumerate}[label=\textup{(\alph*)}]
\item Assume $\Hyp^{\det}(\eps,\gamma)$ with $0<\eps,\gamma\le\frac12$, and let an integer 3SUM algorithm with saving $\delta_\Z$ be given, as in the preceding paragraph. There is a deterministic algorithm that, for every prime $p$ and every $d$ with $p^d\le n^c$, solves 3SUM over $\Fp^d$ on lists of at most $n$ vectors in time $n^{2-\delta+o(1)}$ with $\delta:=\min(\eps\gamma/4,\delta_\Z)$; for $p\le n^{\eps\gamma/16}$ it makes one detection call, and for larger $p$ it runs the integer algorithm on $O(1)$ instances.
\item Assume $\Hyp(\eps,\gamma)$ with $0<\eps,\gamma\le\frac12$. There is a deterministic algorithm that, for every prime $p$ and every $d$ with $p^d\le n^c$, decides 3SUM over $\Fp^d$ on lists of at most $n$ vectors with adaptive calls to the counting algorithm in time
\[
n^{2-\eps\gamma}\exp\bigl(O(\sqrt{\log n\,\log\log n})\bigr)=n^{2-\eps\gamma+O(\sqrt{\log\log n/\log n})}=n^{2-\eps\gamma+o(1)}.
\]
\end{enumerate}
In both parts the constants depend on $c$, $\eps$, $\gamma$ and $C_{\mathcal O}$ but not on $p$. Part~(a) needs only the detection algorithm and an integer 3SUM algorithm; part~(b) keeps the whole saving.
\end{theorem}

\begin{proof}
(a) Put $\eta:=\eps\gamma/16$. If $p\le n^\eta$, then $p\le n^\eps$, and Theorem~\ref{g:thm:A} applies; its time is at most $O(n^{2-\eps\gamma/2+3.25\eps\gamma/16}\log n+n^{7\eps\gamma/16+1+\eps}(\log n)^{O(1)})$. The first exponent is $2-0.296\ldots\cdot\eps\gamma<2-\eps\gamma/4$, and the second is at most $1+\frac12+\frac7{64}<1.61<2-\frac1{16}\le2-\eps\gamma/4$ since $\eps,\gamma\le\frac12$; so the time is $O(n^{2-\eps\gamma/4})$. If $p>n^\eta$, then $d\le c\log n/\log p<c/\eta$, a constant, and Proposition~\ref{g:prop:boundedrank} with $t=d$ gives fewer than $3^{c/\eta}$ integer instances with numbers of absolute value at most $2\cdot3^{c/\eta}n^c\le n^{c+1}$ for large $n$; each is solved in time $n^{2-\delta_\Z+o(1)}$, whose exponent does not depend on the bound $c+1$.

(b) Read the proof of Theorem~\ref{g:thm:B} with $p$ a parameter. The only conditions on $n$ are $\sigma\ge2$ and $k,q\ge1$; the last two hold for every $n\ge2$ because $n/\sigma>1$ and $2e\sigma^2>1$, and $\sigma\ge2$ holds once $n^\eps\ge4\Lambda=8ep$. The inequalities $D\le N^\eps$, $D^{1+\gamma}\le N$ and $e\,p^{-q}\sigma^2\le\frac12$ were derived for every $p$. A $0/1$ instance costs at most $3\Lambda C_{\mathcal O}p(p+2)^2n^2\sigma^{-2\gamma}$ with $\sigma^{-2\gamma}\le(4\Lambda)^\gamma n^{-\eps\gamma}$ and $\Lambda=2ep$, that is $O(p^{4+\gamma}n^{2-\eps\gamma})$; an evaluation consists of $B^2=O(\log^2n)$ of them, and there are $Lq|E|<Lq\cdot p^4d^2q^2=O(p^4(\log n)^6)$ evaluations, since $d\le c\log n$, $q=O(\log n)$ and $L=O(\log n)$. As $\gamma\le\frac12$, the stages cost $O(p^{8.5}n^{2-\eps\gamma}(\log n)^8)$. The other terms are $O(p^7n(\log n)^{O(1)})$ for choosing $\pi$ ($k|E_k|$ counts over groups of size $p^k<pn/\sigma$), $O(p^3(\log n)^{O(1)})$ for listing $E_k$ and $E$, $O(pn^2/\sigma)=O(p^{1.5}n^{2-\eps/2})$ for the probes, and $O((p+6)n\log n)$ for the scans. So, whenever $n^\eps\ge8ep$, the time is $O(p^{8.5}n^{2-\eps\gamma}(\log n)^8)$ with a constant that does not depend on $p$.

Put $P_0:=2^{\sqrt{\log n}}$; the algorithm compares $p$ with $P_0$. If $p\le P_0$, then $n^\eps\ge8eP_0\ge8ep$ for $n\ge n_1(\eps)$, a threshold that does not depend on $p$, and the time is $O(2^{8.5\sqrt{\log n}}n^{2-\eps\gamma}(\log n)^8)$; for $n<n_1(\eps)$ the instance has bounded size, as $p^d\le n^c$, and brute force takes $O(1)$ time. If $p>P_0$, then $d\le c\log n/\log p<c\sqrt{\log n}$, and Proposition~\ref{g:prop:boundedrank} with $t=d$ gives, in time $O(3^ddn)$, $3^d<2^{1.6c\sqrt{\log n}}$ instances of integer 3SUM with three sets of at most $n$ integers, of absolute value at most $2\cdot3^dn^c\le n^{c+1}$ for large $n$, and the question $a+b+c=0$. Corollary~\ref{s:cor:variants}, with the bound of Corollary~\ref{s:cor:uniform} and $\nu:=c+1$, decides each of them in time $n^{2-\eps\gamma}\exp(O(\sqrt{\log n\log\log n}))$ (proof of Corollary~\ref{s:cor:uniform}). Both factors $2^{8.5\sqrt{\log n}}$ and $2^{1.6c\sqrt{\log n}}$ are $\exp(O(\sqrt{\log n}))$, so the integer algorithm's factor dominates and gives the bound. The exponent $8.5\ge8+\gamma$ of $p$ comes from the hash range $\Lambda=2ep$ and from the size of the biased multisets; the choice of $P_0$ only balances it against the $3^d$ carry patterns of Proposition~\ref{g:prop:boundedrank}, and any $P_0$ with $\log P_0=\Theta(\sqrt{\log n})$ would do.
\end{proof}

Proposition~\ref{g:prop:boundedrank} handles every finite abelian group with a bounded number of cyclic factors, each of any order. What it cannot handle is many small non-elementary factors, such as $\Z_4^d$ with $d=c\log n$, where it would produce $3^d=n^{c\log3}$ instances; that case is the subject of Sections~\ref{g:sec:obstruction} and~\ref{g:sec:G}.

\subsubsection{Explicit exponents}

\begin{corollary}\label{g:cor:explicitA}
With $\Hyp(1/14,\,0.077)$, which Section~\ref{x:sec:consequences} derives from Corollary~\ref{m:cor:explicit}(b), and with Corollary~\ref{x:cor:fg}(d) as the integer algorithm: for every fixed prime $p$, 3SUM over $\Fp^d$ with $p^d\le n^c$ is solved deterministically in time $O(n^{1.99726})$ with one detection call (Theorem~\ref{g:thm:A}) and in time $O(n^{1.99451})$ with counting calls (Theorem~\ref{g:thm:B}). One algorithm, uniform in $p$, solves it for every prime $p$ in time $n^{1.99450+o(1)}$ with counting calls (Theorem~\ref{g:thm:App}(b)); the algorithm of Theorem~\ref{g:thm:App}(a), which makes one detection call for the primes $p\le n^{\eps\gamma/16}$, does so in time $O(n^{1.99863})$.
\end{corollary}
\begin{proof}
$\eps\gamma=0.077/14=0.0055$. Theorem~\ref{g:thm:A} gives $n^{2-0.00275}\log n$, Theorem~\ref{g:thm:B} gives $n^{2-0.0055}(\log n)^8$ for fixed $p$, Theorem~\ref{g:thm:App}(b) gives $n^{2-0.0055+o(1)}=n^{1.99450+o(1)}$ uniformly in $p$, and Theorem~\ref{g:thm:App}(a) gives $n^{2-\min(0.001375,\,0.0055)+o(1)}=n^{1.998625+o(1)}$. The exponents written with $O(\cdot)$ are rounded up by at least $5\cdot10^{-6}$, which absorbs the polylogarithmic and $n^{o(1)}$ factors for large $n$.
\end{proof}

With the preprint's own constants, $\Hyp(1/18,0.063)$ and $\delta_\Z=0.000875$, Theorems~\ref{g:thm:A}, \ref{g:thm:B} and~\ref{g:thm:App}(a) give $n^{2-0.00175+o(1)}$, $n^{2-0.0035+o(1)}$ and $n^{2-0.000875+o(1)}$, and Theorem~\ref{g:thm:App}(b) gives $n^{2-0.0035+o(1)}$, since it uses the integer algorithm of Corollary~\ref{s:cor:uniform} and not the preprint's.

\subsection{The obstruction for one homomorphic hash family}\label{g:sec:obstruction}

The proofs of Section~\ref{g:sec:elementary} use one property of the hash: a fixed family of homomorphisms into a range of size $D$ kills every nonzero element with probability at most about $1/D$. Over $\Fp^d$ the powering construction gives probability at most $(1+\beta)/D$ for all nonzero elements at once. This section shows that groups with a large non-elementary part have no such family. For a finite abelian group $G$ and an integer $m$, write $G[m]:=\{x\in G:mx=0\}$ and $mG:=\{mx:x\in G\}$, and $G_p$ for the $p$-primary part.

\begin{proposition}[one homomorphism]\label{g:prop:C}
Let $K\colon G\to\Gamma$ be a homomorphism of finite abelian groups and $A,B\le G$ subgroups with $|K(A)|\cdot|K(B)|\le|K(G)|$. Then
\[
|A\cap\ker K|\cdot|B\cap\ker K|\ge\frac{|A|\cdot|B|}{|K(G)|}.
\]
The hypothesis holds for every $K$ in the following two cases.
\begin{enumerate}[label=\textup{(\arabic*)}]
\item $G$ has an element of order $p^2$ for a prime $p$; $A:=p^{a-1}G_p$, where $p^a$, $a\ge2$, is the exponent of $G_p$, and $B:=G[p]$.
\item $G$ has elements of two distinct prime orders $p\ne p'$; $A:=G[p]$ and $B:=G[p']$.
\end{enumerate}
Every finite abelian group that is not elementary abelian (that is, not of the form $\Fp^d$) falls under (1) or (2).
\end{proposition}

\begin{proof}
$|A\cap\ker K|=|A|/|K(A)|$ by the isomorphism theorem, and likewise for $B$; multiply and use the hypothesis.

(1) $A\subseteq pG$ because $a\ge2$, so $K(A)\subseteq pK(G)$; and $K(B)\subseteq K(G)[p]$. Multiplication by $p$ is an endomorphism of $K(G)$ with image $pK(G)$ and kernel $K(G)[p]$, so $|pK(G)|\cdot|K(G)[p]|=|K(G)|$. (2) $K(A)\subseteq K(G)[p]$ and $K(B)\subseteq K(G)[p']$, and $K(G)[p]\cap K(G)[p']=0$ since an element killed by $p$ and by $p'$ is killed by $\gcd(p,p')=1$; so $K(G)[p]+K(G)[p']$ is a direct sum inside $K(G)$, and $|K(G)[p]|\cdot|K(G)[p']|\le|K(G)|$. Finally, if $G$ is not elementary abelian, then either two primes divide $|G|$, and $G$ has elements of both prime orders (Cauchy), or $G$ is a $p$-group of exponent at least $p^2$.
\end{proof}

\begin{corollary}[families]\label{g:cor:C}
Let $(A,B)$ be as in case (1) or (2) of Proposition~\ref{g:prop:C}, let $\mathcal K$ be a finite nonempty family (multiset) of homomorphisms $K\colon G\to\Gamma_K$ with $|K(G)|\le D$ for all $K\in\mathcal K$, and suppose $|A|,|B|\ge2\sqrt D$. Then some $v\in(A\cup B)\setminus\{0\}$ has $\Pr_{K\in\mathcal K}[K(v)=0]\ge1/(4\sqrt D)$.
\end{corollary}
\begin{proof}
For each $K$, $|K(A)|\cdot|K(B)|\le|K(G)|\le D$, so $|K(A)|\le\sqrt D$ or $|K(B)|\le\sqrt D$. If $|K(A)|\le\sqrt D$, then $|A\cap\ker K|=|A|/|K(A)|\ge|A|/\sqrt D$, so $K$ kills at least $|A|/\sqrt D-1\ge|A|/(2\sqrt D)$ nonzero elements of $A$ (using $|A|\ge2\sqrt D$), which is a fraction at least $|A|/(2\sqrt D(|A|-1))>1/(2\sqrt D)$ of $A\setminus\{0\}$. One of $A$ and $B$, say $A$, is the small-image side for at least half of the members of $\mathcal K$. Then the average over $v\in A\setminus\{0\}$ of $\Pr_K[K(v)=0]$, which is the average over $K$ of the fraction of $A\setminus\{0\}$ killed by $K$, is at least $\frac12\cdot\frac1{2\sqrt D}$, and some $v$ attains the average.
\end{proof}

So a family of homomorphisms with images of size at most $D$ that kills every nonzero element with probability at most $C/D$ does not exist once $D>16C^2$ and the two subgroups of case (1) or (2) have at least $2\sqrt D$ elements; for $G=\Z_4^d$, case (1) gives $A=B=2G$, of size $2^d$, so the condition is $2^d\ge2\sqrt D$, which holds whenever $|G|=4^d\ge4D$, the interesting range. The bound is attained up to the constant: for $G=\Z_4^d$ and the family of maps $x\mapsto(\langle u_i,x\rangle)_{i\le b}\in\Z_4^b$ given by $b$ independent uniformly random rows $u_i\in\Z_4^d$, an element $2w$ with $w\in\Z_4^d$, $\bar w\ne0$ ($\bar w$ the reduction modulo $2$), is killed by a row exactly when $\langle\bar u_i,\bar w\rangle=0$ in $\F$, which has probability $\frac12$; so it is killed with probability $2^{-b}=|\Z_4^b|^{-1/2}$.

The condition on the sizes of $A$ and $B$ cannot be dropped, and with it the statement does not characterise the elementary groups exactly:

\begin{remark}[$\Z_4\times\F^d$]\label{g:ex:Z4F2}
Let $G=\Z_4\times\F^d$, so case (1) applies with $A=2G=\{0,2\}\times\{0\}$, of two elements, and $B=G[2]=\{0,2\}\times\F^d$. Let $E$ be a $(1/b)$-kernel-biased multiset of rows in $\F^d$ (Lemma~\ref{g:lem:powering}), and let $\mathcal K$ be the family of the maps $K_H(x,y):=(x,Hy)\in\Z_4\times\F^b$, where $H$ has $b$ rows from $E$, all $|E|^b$ choices. Then $|K_H(G)|\le D:=4\cdot2^b$. A nonzero $(x,y)$ with $x\ne0$ is never killed; one with $x=0$, $y\ne0$, is killed with probability $\Pr_H[Hy=0]\le((1+1/b)/2)^b\le e\,2^{-b}=4e/D$, since the rows are independent and each kills $y$ with probability at most $(1+1/b)/2$. So this family is universal up to the constant $4e$ although $G$ has an element of order $4$: a non-elementary part of size at most about $\sqrt D$ can be injected into the hash range and costs nothing. The same works for $\Z_3\times\F^d$ (two prime orders) with $K_H(x,y):=(x,Hy)$, and for $\Z_4^a\times\F^d$ with a bounded $a$.
\end{remark}

\begin{remark}[the one-call accounting with collision probability $D^{-\theta}$]\label{g:rem:nosaving}
Suppose a hash family has $\Pr[K(v)=0]\le D^{-\theta}$ for all $v\ne0$, $D$ the size of the middle part, and use it in the proof of Theorem~\ref{g:thm:A}. The false positives are then about $\sigma n^2D^{-\theta}$ instead of $\sigma n^2/D$, so the scans cost about $\sigma^2n^2D^{-\theta}$, while the call costs about $n^2D^{1-\gamma}/\sigma^2$. Balancing gives $\sigma^4=D^{1-\gamma+\theta}$ and a total of $n^2D^{(1-\gamma-\theta)/2}$, a saving only if $\theta>1-\gamma$. The balance point has $\sigma\le\sqrt D$ because $\theta\le1+\gamma$, so cutting $W$ into several calls does not help. With $\theta=\frac12$, which Corollary~\ref{g:cor:C} forces for $\Z_4^d$, and $\gamma\le\frac12$, there is no saving. This accounting is only a heuristic for the method with one fixed hash family; it is not a lower bound for 3SUM over $\Z_4^d$, and Section~\ref{g:sec:G} shows that hashes that adapt to the instance avoid it.
\end{remark}

\subsection{Groups of exponent \texorpdfstring{$p^2$}{p\^{}2}}\label{g:sec:G}

\begin{theorem}[exponent $p^2$]\label{g:thm:G}
Assume $\Hyp^{\det}(\eps,\gamma)$ with $0<\eps\le\frac14$ and $0<\gamma\le\frac12$. Let $p$ be a fixed prime and $c>0$. 3SUM over a finite abelian group $G$ of exponent dividing $p^2$, $|G|\le n^c$, on lists of at most $n$ elements, is solved deterministically in time $O(n^{2-\eps\gamma/2}(\log n)^4)$, and a solution is returned when one exists. The algorithm makes one detection call for each of at most $7+(\lfloor\log n\rfloor+1)^3$ sub-instances. The constants depend on $p$, $c$, $\eps$, $\gamma$ and $C_{\mathcal O}$.
\end{theorem}

By Corollary~\ref{g:cor:C}, no single hash family serves $\Z_4^d$; the obstruction is to one fixed family, not to the method. The proof adapts the hash to the instance in two ways. First, the middle part of the instance of a call only has to contain the differences of hash values inside one group of the grouping, a subgroup containing $K(\ker\pi)$, not the whole range of $K$ (Lemma~\ref{g:lem:instance} already allows this); so the low digits of the $\Z_{p^2}$-rows go into the grouping and only the high digits into the middle part, and a triple whose sum lies in $pG$ is filtered by the middle part at the full rate $1/p$ per row. Second, the lists are split by the sizes of their residue classes modulo $pG$. If one list has only small classes, the triples whose sum lies in $pG$ are few, and a three-term potential lets the method of conditional expectations treat both kinds of triples at their own rates (Case~L, Section~\ref{g:sec:G:L}); if all three lists have only large classes, the groups are made pure in the residue, so that only sums in $pG$ occur and the problem inside is 3XOR-like (Case~H, Section~\ref{g:sec:G:H}). Sections~\ref{g:sec:G:setup}--\ref{g:sec:G:total} prove the theorem.

\subsubsection{Reduction, notation, parameters and the light/heavy split}\label{g:sec:G:setup}

\paragraph{Reduction to $\Z_{p^2}^d$.} $G$ is given as $\prod_{i\le r}\Z_{m_i}$ with $m_i\ge2$ dividing $p^2$, so $m_i\in\{p,p^2\}$. Map a coordinate with $m_i=p^2$ identically and one with $m_i=p$ by $y\mapsto py$ from $\Z_p$ to $\Z_{p^2}$; this is a well-defined injective homomorphism $\iota\colon G\to\Z_{p^2}^r$. An injective homomorphism preserves and reflects $x_1+x_2+x_3=0$, so we may replace the lists by their images. From now on $R:=\Z_{p^2}$, $G=R^d$ with $d=r$, and $p^d\le|G|\le n^c$ for the original group, so $d\le c\log_pn$.

\paragraph{Notation.} For $x\in R^d$, $\bar x\in\Fp^d$ is its reduction modulo $p$, its \emph{residue}. The map $\Fp\to pR$, $y\mapsto p\cdot\lift(y)$, is a group isomorphism, where $\lift(y)\in\{0,\dots,p-1\}$; we write $py$ for $p\cdot\lift(y)$ and identify $pR$ with $\Fp$, and $(pR)^t$ with $\Fp^t$, through it. $\{x\in R^d:\bar x=0\}=pR^d$. For $v\in pR^d$ write $v=pw$; then $\bar w$ does not depend on the choice of $w$ (if $pw=pw'$, then $p(w-w')=0$, so $w-w'\in pR^d$), and we put $v/p:=\bar w\in\Fp^d$; $v=0$ if and only if $v/p=0$. For $a\in\Fp^d$, $\lift(a)\in R^d$ has entries in $\{0,\dots,p-1\}$ and residue $a$. For $u\in R^d$, $x\mapsto\langle u,x\rangle:=\sum_ju_jx_j\in R$ is a homomorphism, and $\overline{\langle u,x\rangle}=\langle\bar u,\bar x\rangle$.

A \emph{pair row} is $(a,b)\in\Fp^d\times\Fp^d$, with $u:=\lift(a)+p\lift(b)$; then $\bar u=a$. For a triple $\tau$ with $v=v(\tau)$, $\tau$ is of \emph{type I} if $v\in pR^d\setminus\{0\}$, of \emph{type II} if $\bar v\ne0$, and a solution if $v=0$. $T_I$ denotes the number of triples with $\bar v=0$, the type-I triples and the solutions.

\begin{lemma}[the two facts]\label{g:lem:facts}
Let $(a,b)$ be a pair row with $u=\lift(a)+p\lift(b)$, and $v\in R^d$.
\begin{enumerate}[label=\textup{(F\arabic*)}]
\item If $v\in pR^d$, then $\langle u,v\rangle=p\,\langle a,v/p\rangle$; so $\langle u,v\rangle=0$ if and only if $\langle a,v/p\rangle=0$, whatever $b$ is.
\item If $\bar v\ne0$, then $\langle u,v\rangle=0$ if and only if $\langle a,\bar v\rangle=0$ and $\langle b,\bar v\rangle=-c_0(a,v)$, where, when $\langle a,\bar v\rangle=0$, $c_0(a,v)\in\Fp$ is defined by $\langle\lift(a),v\rangle=p\,c_0(a,v)$.
\end{enumerate}
\end{lemma}
\begin{proof}
(F1) With $v=pw$, $\langle u,v\rangle=p\langle u,w\rangle$, and $pz$ for $z\in R$ depends only on $\bar z$; here $\overline{\langle u,w\rangle}=\langle a,\bar w\rangle=\langle a,v/p\rangle$. (F2) $\overline{\langle u,v\rangle}=\langle a,\bar v\rangle$, so $\langle u,v\rangle\ne0$ unless $\langle a,\bar v\rangle=0$. In that case $\langle\lift(a),v\rangle$ has residue $0$, so it lies in $pR$ and $c_0(a,v)$ is defined, and $\langle p\lift(b),v\rangle=p\langle\lift(b),v\rangle=p\langle b,\bar v\rangle$; so $\langle u,v\rangle=p(c_0(a,v)+\langle b,\bar v\rangle)$.
\end{proof}

\begin{lemma}[survival]\label{g:lem:survival}
Let $E$ be $\beta$-value-biased in $\Fp^d$, and let $(a,b)$ be uniform in $E\times E$. A type-I vector $v$ satisfies $\langle u,v\rangle=0$ with probability at most $(1+\beta)/p$, and a type-II vector with probability at most $((1+\beta)/p)^2$. If instead $b=0$ and $a$ is uniform in $E$, a type-I vector satisfies $\langle\lift(a),v\rangle=0$ if and only if $\langle a,v/p\rangle=0$, and a type-II vector only if $\langle a,\bar v\rangle=0$; each has probability at most $(1+\beta)/p$.
\end{lemma}
\begin{proof}
For type I, by (F1) the event is $\langle a,v/p\rangle=0$ with $v/p\ne0$, of probability at most $(1+\beta)/p$ by kernel bias. For type II, by (F2) the probability is $\frac1{|E|}\sum_{a\in E:\langle a,\bar v\rangle=0}\Pr_b[\langle b,\bar v\rangle=-c_0(a,v)]$; there are at most $(1+\beta)|E|/p$ such $a$, and for each the inner probability is at most $(1+\beta)/p$ by value bias, as $\bar v\ne0$. The case $b=0$ is (F1) and the first half of (F2), with kernel bias.
\end{proof}

\paragraph{Parameters.} If $n<p^{1/\eps}$, then $n$ is bounded by a constant, and brute force takes $O(1)$ time; assume $n\ge p^{1/\eps}$. As in Theorem~\ref{g:thm:A}, $q:=\lfloor\eps\log_pn\rfloor\ge1$, $D:=p^q\in(n^\eps/p,\,n^\eps]$, $g:=\lceil D^{\gamma/2}\rceil$, $\sigma:=\lfloor\sqrt{D/g}\rfloor$, so $1\le\sigma\le\sqrt D$ and $\sigma\ge\frac12\sqrt{D/g}$, $q_1:=\lceil q/2\rceil$, $q_2:=\lfloor q/2\rfloor$; but $g':=\lceil(p+5)n/\sigma\rceil\le(p+6)n/\sigma$ (larger than in Theorem~\ref{g:thm:A}, so that one detection call per sub-instance suffices). Let $\kappa:=\lceil\log_p(n/\sigma)\rceil$, so $n/\sigma\le p^\kappa<pn/\sigma$; since $\sigma\le\sqrt D\le n^{\eps/2}$ and $\eps\le\frac14$, $n/\sigma\ge n^{1-\eps/2}\ge n^\eps\ge D$, so $\kappa\ge q$, and we put $k:=\kappa-q\ge0$. Let $r_0:=\lceil\log_pn\rceil+q+1$, $\beta_0:=1/(2r_0)$, and let $E$ be the $\beta_0$-value-biased multiset of Lemma~\ref{g:lem:powering}, of fewer than $4p^4d^2r_0^2=O((\log n)^4)$ rows; it is also $\beta_0$-kernel-biased. Every sequence of rows chosen below has length at most $r_0$, so Lemma~\ref{g:lem:greedy} applies with the factor $e^{1/2}\le e$.

\paragraph{Step 1: light and heavy.} For list $i$ and $\varrho\in\Fp^d$ let $C_i(\varrho):=\{x\in S_i:\bar x=\varrho\}$, the residue classes. Call $x\in S_i$ \emph{light} if $|C_i(\bar x)|<\sigma$ and \emph{heavy} otherwise, and let $L_i$ and $H_i$ be the light and the heavy elements of $S_i$. The $8$ sub-instances $(U_1,U_2,U_3)$ with $U_i\in\{L_i,H_i\}$ partition the triples, so every solution lies in exactly one of them; we solve each and return a solution if one is found. Computing the classes costs $O(nd\log n)$.

\begin{lemma}[a light list]\label{g:lem:light}
If $U_i=L_i$ for some $i$, the sub-instance has $T_I<\sigma n^2$.
\end{lemma}
\begin{proof}
A triple with $\bar v=0$ has $\bar x_i=-\sum_{i'\ne i}\bar x_{i'}$. For each choice of the other two members, $x_i$ lies in the residue class $C_i(\bar x_i)$, and since lightness is a property of the whole class, the elements of $L_i$ in this residue form a full class of $S_i$ of fewer than $\sigma$ elements, or none. So $T_I\le(\sigma-1)n^2$.
\end{proof}

The $7$ sub-instances with a light list are solved by Case~L (Section~\ref{g:sec:G:L}), and the one with three heavy lists by Case~H (Section~\ref{g:sec:G:H}).

\subsubsection{Case L: few triples with sum in \texorpdfstring{$pG$}{pG}}\label{g:sec:G:L}

\begin{lemma}[Case L]\label{g:lem:partL}
Given lists $U_1,U_2,U_3\subseteq R^d$ of at most $n$ elements each with $T_I\le\sigma n^2$, a solution is found, or its absence decided, deterministically with one detection call in time $O(n^{2-\eps\gamma/2}\log n)$.
\end{lemma}

The rest of Section~\ref{g:sec:G:L} proves the lemma.

\paragraph{Residue rows.} Choose rows $r_1,\dots,r_\kappa\in E$ one at a time, each minimising the number of triples with $\langle r_i,\bar v\rangle=0$ for the rows chosen so far and the candidate; this number is $Z$ (Lemma~\ref{g:lem:fourier}) of the homomorphism $x\mapsto(\langle r_i,\bar x\rangle)_i$ into $\Fp^t$. It equals $T_I$ plus the number of type-II triples with $\langle r_i,\bar v\rangle=0$ for all chosen $i$, and $T_I$ does not depend on the rows; so by Lemmas~\ref{g:lem:onerow} and~\ref{g:lem:greedy} (the family $\{\bar v:\tau\text{ of type II}\}$, weight $1$), after $t$ rows at most $e\,p^{-t}n^3$ type-II triples remain. This takes $\kappa|E|=O((\log n)^5)$ evaluations of Lemma~\ref{g:lem:fourier} over groups of at most $p^\kappa<pn/\sigma$ elements, $O(n(\log n)^{O(1)})$ time in all. Let $\rho$ be the map of the first $k$ rows and $\rho''$ that of the last $q$. Then, using $p^k=p^\kappa/D\ge n/(\sigma D)$ and $p^\kappa\ge n/\sigma$,
\begin{equation}\label{g:eq:resrows}
\#\{\tau\text{ of type II}:\rho(\bar v)=0\}\le e\sigma Dn^2,\qquad\#\{\tau\text{ of type II}:\rho(\bar v)=0,\ \rho''(\bar v)=0\}\le e\sigma n^2 .
\end{equation}

\paragraph{Pair rows and the hash.} Next choose pair rows $(a_1,b_1),\dots,(a_q,b_q)\in E\times E$ one at a time as described below, with $u_i:=\lift(a_i)+p\lift(b_i)$. For $t\le q$ let $K_t(x):=(\rho(\bar x),(\langle u_i,x\rangle)_{i\le t})\in\Fp^k\times R^t$, $K:=K_q$, and $\pi(x):=(\rho(\bar x),(\langle a_i,\bar x\rangle)_{i\le q})\in\Fp^{k+q}$. Both are homomorphisms. If $\pi(x)=0$, then $\rho(\bar x)=0$ and every $\langle u_i,x\rangle$ has residue $\langle a_i,\bar x\rangle=0$; so $K(\ker\pi)\subseteq M:=\{0\}\times(pR)^q\cong\Fp^q$, a group of exactly $D$ elements, which we split as $M_1\times M_2=\Fp^{q_1}\times\Fp^{q_2}$ by coordinates. ($K(\ker\pi)$ itself need not split along coordinates; the instance uses the whole of $M$.)

\paragraph{The potential.} For $t\le q$ let
\begin{align*}
\mathsf Y_t&:=\#\{\tau\text{ of type II}:\rho(\bar v)=0,\ \langle a_i,\bar v\rangle=0\text{ for }i\le t\},\\
\mathsf S_t&:=\#\{\tau:v\ne0,\ \rho''(\bar v)=0,\ K_t(v)=0\},\\
\mathsf A_t&:=\#\{\tau\text{ of type II}:\rho''(\bar v)\ne0,\ K_t(v)=0\},
\end{align*}
and $\Psi_t:=p^t\mathsf Y_t+Dp^t\mathsf S_t+p^{2t}\mathsf A_t$. ($K_t(v)=0$ includes $\rho(\bar v)=0$. $\mathsf S_t$ counts all type-I triples with $K_t(v)=0$, which have $\bar v=0$, together with the type-II triples that $\rho''$ does not separate; $\mathsf A_t$ counts the other type-II triples with $K_t(v)=0$.) The weights are chosen so that both kinds of triple start at the scale $\sigma Dn^2$ and must end at the scale $\sigma n^2/D$: a type-II triple gets two digits per pair row and must fall by $D^2$, a type-I triple one digit per row and must fall by $D$.

\begin{lemma}[potential]\label{g:lem:potential}
For every choice of the first $t<q$ pair rows, the average of $\Psi_{t+1}$ over the candidates $(a,b)\in E\times E$ for the pair row $t+1$ is at most $(1+\beta_0)^2\Psi_t$.
\end{lemma}
\begin{proof}
Each of the three sets shrinks as $t$ grows, so it suffices to bound the probability that a triple counted at $t$ is still counted at $t+1$. A triple counted in $\mathsf Y_t$ stays if and only if $\langle a,\bar v\rangle=0$, with $\bar v\ne0$: probability at most $(1+\beta_0)/p$ by kernel bias, while its weight is multiplied by $p$. A triple counted in $\mathsf S_t$ stays if and only if $\langle u,v\rangle=0$: by Lemma~\ref{g:lem:survival}, probability at most $(1+\beta_0)/p$ if it is of type I and at most $((1+\beta_0)/p)^2\le(1+\beta_0)/p$ if it is of type II, while its weight is multiplied by $p$. A triple counted in $\mathsf A_t$ is of type II and stays with probability at most $((1+\beta_0)/p)^2$, while its weight is multiplied by $p^2$. Summing, the average is at most $(1+\beta_0)p^t\mathsf Y_t+(1+\beta_0)Dp^t\mathsf S_t+(1+\beta_0)^2p^{2t}\mathsf A_t\le(1+\beta_0)^2\Psi_t$.
\end{proof}

\paragraph{Choosing the pair rows.} The pair row $t+1$ is the candidate that minimises $\Psi_{t+1}$. It is computable up to a term that does not depend on the candidate. Let $Z_1$, $Z_2$, $Z_3$ be $Z$ (Lemma~\ref{g:lem:fourier}) of the homomorphisms $x\mapsto(\rho(\bar x),(\langle a_i,\bar x\rangle)_{i\le t+1})\in\Fp^{k+t+1}$, $x\mapsto(\rho(\bar x),\rho''(\bar x),(\langle u_i,x\rangle)_{i\le t+1})\in\Fp^{k+q}\times R^{t+1}$, and $K_{t+1}$. Every triple with $\bar v=0$ satisfies the residue conditions of $Z_1$, so $Z_1=T_I+\mathsf Y_{t+1}$. A triple counted in $Z_2$ is a solution or is counted in $\mathsf S_{t+1}$, so $Z_2=Z_0+\mathsf S_{t+1}$. A triple counted in $Z_3$ is a solution, a type-I triple with $K_{t+1}(v)=0$ (which has $\rho''(\bar v)=0$), or a type-II triple with $K_{t+1}(v)=0$, and the last kind splits by whether $\rho''(\bar v)=0$; so $Z_3-Z_2=\mathsf A_{t+1}$. Hence
\[
p^{t+1}Z_1+Dp^{t+1}Z_2+p^{2t+2}(Z_3-Z_2)=\Psi_{t+1}+p^{t+1}(T_I+DZ_0),
\]
and the last term does not depend on the candidate. The three groups have at most $p^{k+q}\cdot p^{2(t+1)}\le p^{\kappa}D^2<pnD^2/\sigma$ elements, and the characters of $R=\Z_{p^2}$ take values in the $p^2$-th roots of unity, so Lemma~\ref{g:lem:fourier} with $\mathfrak m=p^2$ computes each count in $O(n\log^2n+p^5nD^2\log n/\sigma)$ time. With $q|E|^2=O((\log n)^9)$ candidates in all and $D^2\le n^{2\eps}\le n^{1/2}$, choosing the pair rows costs $O(n^{1.5}(\log n)^{O(1)})$.

\paragraph{The bounds.} At $t=0$, $\mathsf Y_0=\#\{\tau\text{ of type II}:\rho(\bar v)=0\}\le e\sigma Dn^2$ by~\eqref{g:eq:resrows}, $\mathsf A_0\le\mathsf Y_0$, and $\mathsf S_0=(T_I-Z_0)+\#\{\tau\text{ of type II}:\rho(\bar v)=\rho''(\bar v)=0\}\le\sigma n^2+e\sigma n^2$. So $\Psi_0\le(3e+1)\sigma Dn^2$. By Lemmas~\ref{g:lem:potential} and~\ref{g:lem:greedy} (with $\vartheta=(1+\beta_0)^2$ and $q$ steps, $(1+\beta_0)^{2q}\le e^{q/r_0}\le e$), $\Psi_q\le e(3e+1)\sigma Dn^2$. Since $p^q=D$,
\[
\mathsf Y_q\le\Psi_q/D\le e(3e+1)\sigma n^2,\qquad\mathsf S_q+\mathsf A_q\le\Psi_q/D^2\le e(3e+1)\sigma n^2/D .
\]
A triple with $\pi(v)=0$ has $\bar v=0$ or is counted in $\mathsf Y_q$, so $Z(\pi)\le T_I+\mathsf Y_q\le(1+e+3e^2)\sigma n^2$. A triple with $v\ne0$ and $K(v)=0$ has $\rho(\bar v)=0$; if it is of type I, or of type II with $\rho''(\bar v)=0$, it is counted in $\mathsf S_q$, and otherwise in $\mathsf A_q$. So the number $F$ of Lemma~\ref{g:lem:instance}(iv) satisfies $F\le\mathsf S_q+\mathsf A_q\le e(3e+1)\sigma n^2/D$.

\paragraph{The instance and its cost.} Run the algorithm of Lemma~\ref{g:lem:onecall} with $\pi$, $K$, $M=M_1\times M_2$, block size $\sigma$ and $g'$ block slots. The map $\pi$ has at most $p^{k+q}=p^\kappa<pn/\sigma$ nonempty classes in each list, so each list has fewer than $(p+1)n/\sigma\le g'$ blocks, and by Lemma~\ref{g:lem:instance}(ii)
\[
|\Pq_0|\le np^\kappa+\frac{2n^2}\sigma+\frac{Z(\pi)}{\sigma^2}\le(p+3+e+3e^2)\frac{n^2}\sigma<(p+28)\frac{n^2}\sigma<(p+5)^2\frac{n^2}\sigma .
\]
As in the proof of Theorem~\ref{g:thm:A}, $N=g'p^{q_1}\ge(p+5)(n/\sigma)\sqrt D\ge n$, because $p^{q_1}\ge\sqrt D$ and $\sigma\le\sqrt D$; so $D\le n^\eps\le N^\eps$ and $D^{1+\gamma}\le N^{\eps(1+\gamma)}\le N$; and $D=p^q\ge2$. Moreover
\[
\frac{N^2}{\sqrt D}\ge g'^2\sqrt D\ge(p+5)^2\frac{n^2\sqrt D}{\sigma^2}\ge(p+5)^2\frac{n^2}\sigma>(p+28)\frac{n^2}\sigma>|\Pq_0|\ge|W|,
\]
since $(p+5)^2-(p+28)=p^2+9p-3>0$; so $J=1$ in Lemma~\ref{g:lem:onecall}: one call suffices. (With $g'=\lceil(p+1)n/\sigma\rceil$ as in Theorem~\ref{g:thm:A} this would fail for $p\le3$, since $(p+1)^2<p+28$ there.) Since $N\le(p+6)(n/\sigma)p^{q_1}$, $p^{2q_1}\le pD$, $(n/\sigma)^2\le4n^2g/D$ and $g\le2D^{\gamma/2}$, the call costs at most
\[
C_{\mathcal O}\frac{N^2}{D^\gamma}\le C_{\mathcal O}p(p+6)^2\Bigl(\frac n\sigma\Bigr)^2D^{1-\gamma}\le4C_{\mathcal O}p(p+6)^2n^2gD^{-\gamma}\le8C_{\mathcal O}p(p+6)^2n^2D^{-\gamma/2}.
\] The scans cost $O((F+1)\sigma\log n)=O(\sigma^2n^2\log n/D)=O(n^2D^{-\gamma/2}\log n)$; listing the probes and their positions costs $O(pn^2\log n/\sigma)=O(pn^2D^{-\gamma/2}\log n)$; evaluating $\pi$ and $K$ costs $O(d(k+q))=O(\log^2n)$ per element. With $D^{-\gamma/2}<p^{1/4}n^{-\eps\gamma/2}$ this proves Lemma~\ref{g:lem:partL}.

\subsubsection{Case H: three heavy lists}\label{g:sec:G:H}

\paragraph{Bucketing.} For $j\ge0$ let $B_{i,j}$ be the union of the residue classes of $S_i$ that consist of heavy elements and have between $2^j$ and $2^{j+1}-1$ elements; $j\le\lfloor\log n\rfloor$. The at most $(\lfloor\log n\rfloor+1)^3$ sub-instances $(B_{1,j_1},B_{2,j_2},B_{3,j_3})$ partition the triples of $(H_1,H_2,H_3)$, so every solution lies in exactly one of them. Fix one, and put $s_i:=2^{j_i}$. Every residue class of list $i$ has between $s_i$ and $2s_i-1$ elements, and at least $\sigma$ since it is heavy; so $2s_i>\sigma$, that is $s_i>\sigma/2$, and the number $R_i$ of classes of list $i$ is at most $n/s_i$. Let $s_{\min}\le s_{\mathrm{mid}}\le s_{\max}$ be the three values $s_i$ in increasing order. The rest of Section~\ref{g:sec:G:H} shows that each of these sub-instances is solved with one detection call in time $O(n^{2-\eps\gamma/2}\log n)$.

\begin{lemma}[three heavy lists]\label{g:lem:heavyTI}
$T_I<2s_{\min}n^2$.
\end{lemma}
\begin{proof}
Let $i_0$ be a list with $s_{i_0}=s_{\min}$. For each choice of the other two members of a triple with $\bar v=0$, $\bar x_{i_0}$ is determined, and $x_{i_0}$ lies in one residue class of list $i_0$, of fewer than $2s_{\min}$ elements.
\end{proof}

\paragraph{Rows and the hash.} Put $k':=\lceil\log_p(s_{\min}/\sigma)\rceil$; since $s_{\min}/\sigma>\frac12\ge\frac1p$, $k'\ge0$, and $s_{\min}/\sigma\le p^{k'}<p\,s_{\min}/\sigma$. All rows of Case H are of the form $u_i:=\lift(a_i)$ with $a_i\in E$ (that is, $b=0$). Choose $a_1,\dots,a_{k'+q}\in E$ one at a time as described below, and put
\[
\pi(x):=\bigl(\bar x,(\langle u_i,x\rangle)_{i\le k'}\bigr)\in\Fp^d\times R^{k'},\qquad K(x):=\bigl(\pi(x),(\langle u_i,x\rangle)_{k'<i\le k'+q}\bigr)\in\Fp^d\times R^{k'}\times R^q .
\]
Both are homomorphisms. $\ker\pi\subseteq pR^d$, and for $x\in pR^d$ every $\langle u_i,x\rangle$ lies in $pR$ by (F1); so $K(\ker\pi)\subseteq M:=\{0\}\times\{0\}\times(pR)^q\cong\Fp^q$, of $D$ elements, split by coordinates into $M_1\times M_2=\Fp^{q_1}\times\Fp^{q_2}$. Because $\pi$ contains the residue, every triple with $\pi(v)=0$ has $\bar v=0$: the probes contain only type-I triples and solutions. For $t\le k'+q$ let
\[
\mathsf T_t:=\#\{\tau\text{ of type I}:\langle a_i,v/p\rangle=0\text{ for }i\le t\}.
\]
By (F1), a type-I triple has $\langle u_i,v\rangle=0$ if and only if $\langle a_i,v/p\rangle=0$; so $Z(\pi)=Z_0+\mathsf T_{k'}$, and the number $F$ of Lemma~\ref{g:lem:instance}(iv) is $\mathsf T_{k'+q}$.

\begin{lemma}[choosing the rows]\label{g:lem:Hrows}
The rows can be chosen so that $\mathsf T_t\le e(2+e)p^{-t}s_{\min}n^2$ for every $t\le k'+q$, in time $O(n^{1.875}(\log n)^{O(1)}+n^{2-3\eps/4}(\log n)^{O(1)})$. Consequently $Z(\pi)-Z_0\le e(2+e)\sigma n^2$ and $F\le e(2+e)\sigma n^2/D$.
\end{lemma}
\begin{proof}
The consequences follow from $p^{-k'}\le\sigma/s_{\min}$ and $p^q=D$. Note $k'<\log_p(s_{\min}/\sigma)+1\le\lceil\log_pn\rceil+1$, so $k'+q\le r_0$. We use one of two methods, depending on $s_{\max}$. Let $s^*:=\sqrt{n\sigma/D}$.

\emph{(i) Exact counts, used if $s_{\max}\ge s^*$.} Choose each row to minimise $\mathsf T_t$ given the earlier ones. By Lemmas~\ref{g:lem:onerow} and~\ref{g:lem:greedy} on the family $\{v/p:\tau\text{ of type I}\}$ (nonzero vectors, weight $1$), $\mathsf T_t\le e\,p^{-t}T_I\le2e\,p^{-t}s_{\min}n^2$ by Lemma~\ref{g:lem:heavyTI}. To compute $\mathsf T_t+Z_0$, the number of triples with $\bar v=0$ and $L_t(v)=0$, where $L_t(x):=(\langle u_i,x\rangle)_{i\le t}\in R^t$, group the triples by residue classes. A triple with $\bar v=0$ lies in a unique \emph{consistent class triple} $(\varrho_1,\varrho_2,\varrho_3)$, $\varrho_1+\varrho_2+\varrho_3=0$, of nonempty classes of the three lists. Fix a representative $z_\varrho$ of every class. For $x$ in the class $\varrho$, $x-z_\varrho\in pR^d$, so by (F1) $L_t(x)=L_t(z_\varrho)+p\,\ell(x)$ with $\ell(x):=(\langle a_i,(x-z_\varrho)/p\rangle)_{i\le t}\in\Fp^t$. For a consistent class triple, $z_{\varrho_1}+z_{\varrho_2}+z_{\varrho_3}\in pR^d$, so $L_t(z_{\varrho_1}+z_{\varrho_2}+z_{\varrho_3})=p\,c$ for some $c\in\Fp^t$, and $L_t(v)=p(c+\ell(x_1)+\ell(x_2)+\ell(x_3))$. Hence the number of its triples with $L_t(v)=0$ is
\[
\#\{\ell(x_1)+\ell(x_2)+\ell(x_3)=-c\}=p^{-t}\sum_{y\in\Fp^t}\omega^{\langle y,c\rangle}\hat h_{\varrho_1}(y)\hat h_{\varrho_2}(y)\hat h_{\varrho_3}(y),\qquad\hat h_\varrho(y):=\sum_{x\in\text{class }\varrho}\omega^{\langle y,\ell(x)\rangle},
\]
$\omega=e^{2\pi i/p}$, computed exactly in $\Z[X]/(X^p-1)$ as in Lemma~\ref{g:lem:fourier} (the twist by $\omega^{\langle y,c\rangle}$ is a cyclic shift). One transform per class costs $O(|\text{class}|\,td+t\,p^{t+2})$, and one sum per consistent class triple $O(p^{t+2})$. The consistent class triples are listed by enumerating the pairs of classes of the two lists with the largest $s_i$ and looking up the third class: there are at most $X:=R_{\mathrm{mid}}R_{\max}\le n^2/(s_{\mathrm{mid}}s_{\max})<4n^2/\sigma^2$ of them, each looked up in $O(d\log n)$ time. With $p^t\le p^{k'+q}<p\,s_{\min}D/\sigma$ and $R_1+R_2+R_3\le3n/s_{\min}$, one candidate costs
\[
O\Bigl(n\log^2n+\frac{p^4nD\log n}\sigma+\frac{p^3n^2D}{\sigma s_{\max}}+\frac{n^2\log^2n}{\sigma^2}\Bigr),
\]
using $s_{\min}\le s_{\mathrm{mid}}$ in the third term. Since $s_{\max}\ge s^*$, the third term is at most $p^3n^2D/(\sigma s^*)=p^3n^{1.5}D^{1.5}\sigma^{-1.5}\le p^3n^{1.5+1.5\eps}\le p^3n^{1.875}$. The last term is at most $O(n^2g\log^2n/D)=O(n^{2-\eps(1-\gamma/2)}\log^2n)\le O(n^{2-3\eps/4}\log^2n)$.

\emph{(ii) An estimator, used if $s_{\max}<s^*$.} First choose $m'':=\lceil\log_p(n/s_{\min})\rceil\le r_0$ residue rows $\rho''$ from $E$ greedily, as the residue rows of Case L (Section~\ref{g:sec:G:L}), so that $\#\{\tau\text{ of type II}:\rho''(\bar v)=0\}\le e\,p^{-m''}n^3\le e\,s_{\min}n^2$; this costs $O(n(\log n)^{O(1)})$, the groups having at most $p^{m''}<pn/s_{\min}<2pn/\sigma$ elements. Then choose each row $a_t$ to minimise
\[
\hat{\mathsf T}_t:=\#\{\tau:v\ne0,\ \rho''(\bar v)=0,\ L_t(v)=0\},
\]
which is $Z$ of $x\mapsto(\rho''(\bar x),L_t(x))\in\Fp^{m''}\times R^t$ minus $Z_0$. It counts every type-I triple counted in $\mathsf T_t$ (these have $\bar v=0$) and some type-II triples, so $\mathsf T_t\le\hat{\mathsf T}_t$. A triple counted in $\hat{\mathsf T}_t$ stays in $\hat{\mathsf T}_{t+1}$ only if $\langle a,w_\tau\rangle=0$, where $w_\tau:=v/p$ for type I and $w_\tau:=\bar v$ for type II (Lemma~\ref{g:lem:survival} with $b=0$), a nonzero vector; so the average of $\hat{\mathsf T}_{t+1}$ over $a\in E$ is at most $\frac{1+\beta_0}p\hat{\mathsf T}_t$, and Lemma~\ref{g:lem:greedy} gives $\hat{\mathsf T}_t\le e\,p^{-t}\hat{\mathsf T}_0\le e\,p^{-t}(T_I+e\,s_{\min}n^2)\le e(2+e)p^{-t}s_{\min}n^2$. One evaluation is a Fourier count over a group of $p^{m''+2t}<(pn/s_{\min})(p\,s_{\min}D/\sigma)^2=p^3ns_{\min}D^2/\sigma^2$ elements, with $\mathfrak m=p^2$, so it costs $O(n\log^2n+p^7ns_{\min}D^2\log n/\sigma^2)$. Since $s_{\min}\le s_{\max}<s^*$, the second term is at most $p^7ns^*D^2\log n/\sigma^2=p^7n^{1.5}D^{1.5}\sigma^{-1.5}\log n\le p^7n^{1.875}\log n$.

In both cases there are $O(r_0|E|)=O((\log n)^5)$ candidates, which gives the time bound; the exponents $1.875$ and $2-3\eps/4$ are below $2-\eps\gamma/2$, because $\eps\gamma/2\le\frac1{16}$ and $\gamma<\frac32$.
\end{proof}

\paragraph{The instance and its cost.} Within a residue class, the values $(\langle u_i,x\rangle)_{i\le k'}$ lie in one coset of $(pR)^{k'}$, so a residue class splits into at most $p^{k'}$ classes of $\pi$, and list $i$ has at most $R_ip^{k'}<(n/s_i)(p\,s_{\min}/\sigma)\le pn/\sigma$ nonempty $\pi$-classes; so it has fewer than $(p+1)n/\sigma\le g'$ blocks. With $Z(\pi)\le n^2+e(2+e)\sigma n^2$, Lemma~\ref{g:lem:instance}(ii) gives $|\Pq_0|\le(p+3+e(2+e))n^2/\sigma<(p+16)n^2/\sigma<(p+5)^2n^2/\sigma$, and the computation in the proof of Lemma~\ref{g:lem:partL}, with $p+16$ in place of $p+28$, gives $|W|\le N^2/\sqrt D$, so $J=1$: one call, of cost at most $8C_{\mathcal O}p(p+6)^2n^2D^{-\gamma/2}$. The scans and listing cost as in the proof of Lemma~\ref{g:lem:partL}, with $F\le e(2+e)\sigma n^2/D$; evaluating $\pi$ and $K$ costs $O(d(k'+q+1))$ per element. So one Case-H sub-instance costs $O(n^{2-\eps\gamma/2}\log n)$.

\subsubsection{Total}\label{g:sec:G:total}

The split into light and heavy elements and the bucketing cost $O(nd\log n)$. There are $7$ Case-L sub-instances and at most $(\lfloor\log n\rfloor+1)^3$ Case-H sub-instances, each solved in time $O(n^{2-\eps\gamma/2}\log n)$ with one call; every solution lies in exactly one of them, and each sub-instance returns a solution if it has one. The total is $O(n^{2-\eps\gamma/2}(\log n)^4)$, which proves Theorem~\ref{g:thm:G}. \qed

\begin{corollary}\label{g:cor:explicitG}
With $\Hyp(1/14,\,0.077)$ (Section~\ref{x:sec:consequences}), 3SUM over every finite abelian group of exponent dividing $p^2$, $p$ a fixed prime, in particular over $\Z_4^d$ with $4^d\le n^c$, is solved deterministically in time $O(n^{2-0.00275}(\log n)^4)\le O(n^{1.99726})$.
\end{corollary}
\begin{proof}
$\eps=\frac1{14}\le\frac14$ and $\gamma=0.077\le\frac12$, so Theorem~\ref{g:thm:G} applies with $\eps\gamma/2=0.00275$.
\end{proof}

With $\Hyp(1/18,0.063)$, from the preprint's Corollary 26, the bound of Theorem~\ref{g:thm:G} is $O(n^{2-0.00175}(\log n)^4)\le O(n^{1.99826})$.

\subsection{Beyond exponent \texorpdfstring{$p^2$}{p\^{}2}: an open problem}\label{g:sec:beyond}

\begin{problem}\label{g:prob:exponent}
Is 3SUM over every finite abelian group of bounded exponent, with $|G|\le n^{O(1)}$, solvable in time $n^{2-\delta}$ for a $\delta>0$ that depends only on the exponent (given $\Hyp^{\det}(\eps,\gamma)$, or unconditionally)? The first open cases are $\Z_8^d$ and the mixed groups $\Z_4^d\times\F^d$ and $\F^d\times\mathbb F_3^d$.
\end{problem}

Bounded exponent is not necessary for a uniform saving: Theorem~\ref{g:thm:App} covers all elementary groups, and Proposition~\ref{g:prop:boundedrank} all groups with a bounded number of cyclic factors, whatever their exponents. We record a route, which we have not carried out. For exponent $p^3$, the sums fall into three layers, $v\in p^2G$, $v\in pG\setminus p^2G$ and $v\notin pG$. With $\Z_{p^3}$-rows one would put the two lower digits into the grouping and the top digit into the middle part. Triples in the layer $p^2G$ are then filtered only by the middle part, so their number must be at most about $\sigma n^2$, which holds if some list has light classes modulo $p^2G$; triples in the layer $pG\setminus p^2G$ are filtered by one digit of the grouping and the middle part, so their number may be about $\sigma Dn^2$, which holds if some list has light classes modulo $pG$; and the heavy cases would become class-pure probes over a group of exponent $p^2$, which is Theorem~\ref{g:thm:G} with targets. For two primes, $G=G_p\times G_{p'}$, the same scheme would put the $G_{p'}$-digits into the grouping and the $G_p$-digits into the middle part, and would need lightness of the classes of both projections, with class-pure probes in the heavy cases. We expect an induction over the layers to give every bounded exponent, with the saving $\eps\gamma/2$ and a polylogarithmic factor whose exponent grows with the number of layers.

\subsection{What is known and what is new}\label{g:sec:groupsremarks}

\emph{Known.} Integer 3SUM reduces to 3SUM over $\Fp^d$ for a large constant prime $p$ (Abboud, Lewi and Williams~\cite[Lemma~3.1, Corollary~3.1]{ALW14}, in the form of Abboud, Bringmann and Fischer~\cite[Lemma~3.8]{ABF23}, which states that for every $\eps>0$ there is a prime $p$ such that 3SUM over $\mathbb F_p^d$ with $d=O(\log n)$ has no $O(n^{2-\eps})$-time algorithm unless the 3SUM conjecture fails). Since the preprint refutes the 3SUM conjecture, and Theorems~\ref{g:thm:A} and~\ref{g:thm:App} solve 3SUM over $\Fp^d$ in truly subquadratic time for every $p$, this conditional hardness is now void, as it is for the integers. Jafargholi and Viola~\cite{JV16} leave open whether 3SUM and 3XOR are subquadratically equivalent. Small-bias sets over $\Z_d^n$ and over abelian groups exist and can be constructed explicitly (Azar, Motwani and Naor~\cite{AMN98}); for the elementary groups we use the powering construction of Alon, Goldreich, H\aa stad and Peralta~\cite[Proposition~3]{AGHP92}, carried over to $\Fp$ (Lemma~\ref{g:lem:powering}). We found no paper that treats 3SUM uniformly over finite abelian groups at the fine-grained level (several targeted searches, not exhaustive).

\emph{New, to our knowledge.} The treatment of 3SUM over the elementary groups is a routine transfer of Section~\ref{x:sec:xor}, and the algorithms uniform in $p$ (Theorem~\ref{g:thm:App}) combine it with Proposition~\ref{g:prop:boundedrank} and with an integer 3SUM algorithm, in part~(b) that of Part~\hyperref[m:partII]{II}; Proposition~\ref{g:prop:C} and Theorem~\ref{g:thm:G} are, as far as we know, new, and Theorem~\ref{g:thm:G} is the only part of this section that needs a new idea.

\section{A dichotomy for coordinatewise problems}\label{g:sec:coord}

For a finite alphabet $\Sigma$ and a fixed relation $P\subseteq\Sigma^3$, the coordinatewise problem $\Prob(P)$ (Definition~\ref{g:def:prob}) asks for one vector from each of three lists in $\Sigma^d$ such that every coordinate triple lies in $P$. 3XOR is $\Prob(\{(a,b,c)\in\{0,1\}^3:a\oplus b\oplus c=0\})$, and Orthogonal Vectors (OV) is, up to a trivial third list, $\Prob(\{(a,b,c):\neg(a\wedge b)\})$. The question is which $P$ the method of Sections~\ref{x:sec:xor} and~\ref{g:sec:groups} makes truly subquadratic, and which stay hard.

Two properties of $P$ answer it on either side. If $P$ is the solution set of an equation $\varphi_1(a)+\varphi_2(b)+\varphi_3(c)=t$ over an abelian group $A$ (\emph{representable}, Definition~\ref{g:def:rep}), then $\Prob(P)$ is 3SUM over $A^d$ and Section~\ref{g:sec:groups} applies; whether such a group exists, and which, is decided by a lattice computation (Lemma~\ref{g:lem:universal}). If one coordinate of OV can be simulated by a bounded number of coordinates of $\Prob(P)$ (an \emph{OV gadget}, Definition~\ref{g:def:gadget}), then $\Prob(P)$ is quadratic-hard under the OV hypothesis (Lemma~\ref{g:lem:gadgethard}). No relation has both properties (Lemma~\ref{g:lem:exclusive}). The section asks whether every relation has one of them.

For all Boolean relations (Theorem~\ref{g:thm:D}) and all functional relations over three letters (Theorem~\ref{g:thm:E}) it does, by an exhaustive computation whose verdicts are certified (Section~\ref{g:sec:computer}); the groups needed are $\Z$, $\F$ and $\mathbb F_3$, and gadgets of size $1$ suffice. Over four letters (Section~\ref{g:sec:stop}) groups of exponent $4$ become necessary, the Latin squares isotopic to $\Z_4$ being the first example, and Theorem~\ref{g:thm:G} still makes them truly subquadratic; but a third kind appears, relations that neither are representable over any abelian group nor have a gadget of any size. It consists of $4$ orbits of functional relations over four letters, $4$ orbits of relations over three letters that are not functional, and a non-associative loop of order five; the eight orbits reduce to one problem, whose complexity is open (Proposition~\ref{g:prop:thirdT}). Section~\ref{g:sec:tri} proves the same dichotomy for the triangle version $\mathrm{Tri}(P)$ at the cubic level. For Boolean $P$, $\Prob(P)$ is a vector problem of Bringmann, Fischer and Künnemann~\cite{BFK19}, and Theorem~\ref{g:thm:D} refines their cubic-level dichotomy at the quadratic level in the dense model; Section~\ref{g:sec:boolean} makes the comparison precise.

\paragraph{Organisation.} Section~\ref{g:sec:rep} treats representability, Section~\ref{g:sec:gadgets} OV gadgets, Section~\ref{g:sec:boolean} the Boolean classification, Section~\ref{g:sec:functional} functional relations over three letters, Section~\ref{g:sec:stop} where the dichotomy stops, Section~\ref{g:sec:computer} the computations, and Section~\ref{g:sec:coordremarks} what is known and what is new.

\subsection{Representable relations}\label{g:sec:rep}

Let $\Sigma$ be a finite alphabet with $|\Sigma|=k\ge2$ and $P\subseteq\Sigma^3$.

\begin{definition}[$\Prob(P)$]\label{g:def:prob}
Given $n$ and three lists $S_1,S_2,S_3\subseteq\Sigma^d$ of at most $n$ vectors, $d\le c\log n$, decide whether there are $x_i\in S_i$ with $(x_1[j],x_2[j],x_3[j])\in P$ for every coordinate $j\in[d]$.
\end{definition}

A vector is given as its string of $d$ letters. Brute force takes $O(n^3d)$ time; if $P$ is \emph{functional in position $i$}, that is, any two letters in the other positions extend to at most one triple of $P$, then for each pair of vectors in the other two lists the vector in list $i$ is determined, and a lookup in the sorted list $i$ solves $\Prob(P)$ in $O(n^2d\log n)$ time.

\begin{definition}[representability]\label{g:def:rep}
$P$ is \emph{representable} over an abelian group $A$ if there are maps $\varphi_1,\varphi_2,\varphi_3\colon\Sigma\to A$ and $t\in A$ with $P=\{(a,b,c)\in\Sigma^3:\varphi_1(a)+\varphi_2(b)+\varphi_3(c)=t\}$.
\end{definition}

The empty relation is representable over $\Z$ ($\varphi_i=0$, $t=1$). For $P\ne\emptyset$, let $\Z^{3\Sigma}$ have the basis $e_{i,a}$, $i\in\{1,2,3\}$, $a\in\Sigma$, put $e(s):=e_{1,s_1}+e_{2,s_2}+e_{3,s_3}$ for $s\in\Sigma^3$, fix $s_0\in P$, and let
\[
L_P:=\operatorname{span}_\Z\{e(s)-e(s_0):s\in P\},\qquad A_P:=\Z^{3\Sigma}/L_P,
\]
\[
\cl(P):=\{s\in\Sigma^3:e(s)-e(s_0)\in L_P\}.
\]
None of the three depends on the choice of $s_0\in P$, since $e(s)-e(s'_0)=(e(s)-e(s_0))-(e(s'_0)-e(s_0))$. Clearly $P\subseteq\cl(P)$.

\begin{lemma}[the universal group]\label{g:lem:universal}
Let $P\ne\emptyset$.
\begin{enumerate}[label=\textup{(\alph*)}]
\item $P$ is representable over some abelian group if and only if $\cl(P)=P$; and then it is representable over $A_P$ by $\varphi_i(a):=[e_{i,a}]$, $t:=[e(s_0)]$.
\item $P$ is representable over a torsion-free abelian group, equivalently over $\Z^r$ for some $r$, if and only if no $s\in\Sigma^3\setminus P$ has $e(s)-e(s_0)\in\Q L_P$ (the $\Q$-span).
\item Let $m\ge2$. $P$ is representable over an abelian group of exponent dividing $m$ if and only if no $s\in\Sigma^3\setminus P$ has $e(s)-e(s_0)\in L_P+m\Z^{3\Sigma}$. For a prime $m=p$, the groups of exponent $p$ are the $\Fp$-vector spaces, and $\Fp^r$ for a finite $r$ suffices.
\item If $P$ is representable, it is representable over a finite abelian group.
\item Let the invariant factors of $A_P$ be $d_1\mid\dots\mid d_\varrho$ (with $\Z^{3\Sigma}\cong\Z^{3k}$ and $L_P$ in Smith normal form). If $P$ is not representable over a torsion-free group and the prime $p$ divides no $d_i$, then $P$ is not representable over $\Fp^r$ for any $r$.
\end{enumerate}
\end{lemma}
\begin{proof}
Given a representation $(\varphi_i,t)$ over $A$, let $\Phi\colon\Z^{3\Sigma}\to A$ be the homomorphism with $\Phi(e_{i,a})=\varphi_i(a)$; then $\Phi(e(s))=\varphi_1(s_1)+\varphi_2(s_2)+\varphi_3(s_3)$, which is $t$ exactly for $s\in P$. So $\Phi$ vanishes on $L_P$.

(a) If $P$ is representable and $s\in\cl(P)$, then $\Phi(e(s))=\Phi(e(s_0))=t$, so $s\in P$. Conversely, if $\cl(P)=P$, then in $A_P$ we have $[e(s)]=[e(s_0)]$ if and only if $e(s)-e(s_0)\in L_P$, that is, $s\in\cl(P)=P$.
(b) If $A$ is torsion-free, $\ker\Phi$ is saturated ($nx\in\ker\Phi$ with $n\ne0$ implies $x\in\ker\Phi$), so it contains $\Q L_P\cap\Z^{3\Sigma}$, and an $s$ with $e(s)-e(s_0)\in\Q L_P$ has $\Phi(e(s))=t$, so $s\in P$. Conversely, if the condition holds, the canonical map into $\Z^{3\Sigma}/(\Q L_P\cap\Z^{3\Sigma})$, a finitely generated torsion-free group and hence some $\Z^r$, represents $P$ by the argument of (a).
(c) If $A$ has exponent dividing $m$, $\Phi$ vanishes on $m\Z^{3\Sigma}$ as well, and the argument is that of (b) with $L_P+m\Z^{3\Sigma}$; the quotient $\Z^{3\Sigma}/(L_P+m\Z^{3\Sigma})$ has exponent dividing $m$, and for $m=p$ it is a finite $\Fp$-vector space.
(d) Write $\Z^{3k}$ in a basis in which $L_P=\bigoplus_{i\le\varrho}d_i\Z e'_i$, and let $y$ be the coordinates of $x=e(s)-e(s_0)$ in this basis. $x\in L_P$ if and only if $d_i\mid y_i$ for $i\le\varrho$ and $y_i=0$ for $i>\varrho$. For each of the finitely many $s\notin P$, $x\notin L_P$, so some $y_i$ violates this; let $m$ be a common multiple of all $d_i$ that exceeds all $|y_i|$, over all these $s$. Then $x\in L_P+m\Z^{3k}$ would need $d_i=\gcd(d_i,m)\mid y_i$ for $i\le\varrho$ and $m\mid y_i$, hence $y_i=0$, for $i>\varrho$; so it fails, and (c) gives a representation over the finite group $\Z^{3k}/(L_P+m\Z^{3k})$.
(e) In the same coordinates, $x\in\Q L_P$ if and only if $y_i=0$ for $i>\varrho$, and $x\in L_P+p\Z^{3k}$ if and only if $p\mid y_i$ for $i>\varrho$ and $\gcd(d_i,p)\mid y_i$ for $i\le\varrho$. If $p$ divides no $d_i$, the second condition reduces to $p\mid y_i$ for $i>\varrho$, which follows from the first. So the witness $s$ of (b) is also a witness for (c) with $m=p$.
\end{proof}

So representability over $\Z$, over $\Fp$ for each $p$, and over groups of a given exponent is decided by a finite computation with the Smith normal form of the generators of $L_P$; Section~\ref{g:sec:computer} describes how the scripts do it and how each verdict is certified.

\begin{lemma}[algorithms from representations]\label{g:lem:repalg}
Let $P$ be representable by $(\varphi_i,t)$ over $A$.
\begin{enumerate}[label=\textup{(\alph*)}]
\item If $A$ is a finite abelian group, then $\Prob(P)$ reduces in $O(nd)$ time to 3SUM over $A^d$, with $|A^d|=n^{c\log|A|}$.
\item If $A=\Z^r$, then $\Prob(P)$ reduces in $O(nd)$ time to integer 3SUM with three sets and the question $a+b+c=0$, on numbers of absolute value at most $n^{O(c)}$.
\end{enumerate}
\end{lemma}
\begin{proof}
For $x\in\Sigma^d$ let $\Phi_i(x):=(\varphi_i(x[j]))_{j\le d}\in A^d$. Then $(x_1[j],x_2[j],x_3[j])\in P$ for all $j$ if and only if $\Phi_1(x_1)+\Phi_2(x_2)+\Phi_3(x_3)=(t,\dots,t)$. (a) Replace $S_3$ by $\{\Phi_3(x)-(t,\dots,t)\}$ and the other lists by their images; the question becomes $y_1+y_2+y_3=0$ in $A^d$. (b) Let $V$ be the largest absolute value of an entry of the $\varphi_i(a)$, and $\psi_i:=\varphi_i+(V,\dots,V)$, with entries in $[0,2V]$, and target $t+(3V,\dots,3V)$. With base $\mathfrak b:=6V+1$, encode the $rd$ entries of $(\psi_i(x[j]))_j$ as the digits of an integer $X_i(x)$. A sum of three digits is at most $6V<\mathfrak b$, so there are no carries, and by uniqueness of the base-$\mathfrak b$ representation $X_1(x_1)+X_2(x_2)+X_3(x_3)$ equals the encoding $T$ of the target if and only if the entries agree. Take the sets $\{X_1(x)\}$, $\{X_2(x)\}$, $\{X_3(x)-T\}$. The numbers are below $\mathfrak b^{rd}=n^{cr\log\mathfrak b}$.
\end{proof}

\begin{corollary}\label{g:cor:repfast}
Let $P$ be representable over $\Z$ (that is, over some $\Z^r$), over some $\Fp^r$, or over a finite abelian group of exponent $p^2$, for a prime $p$. Then $\Prob(P)$ is solved in time $O(n^{2-\delta})$ for a $\delta>0$ that does not depend on $c$: over $\Z$ in time $n^{2-0.0055+o(1)}$ (Corollary~\ref{x:cor:fg}(d)), which is $O(n^{1.99451})$ (the $o(1)$ is absorbed by the rounding), or $O(n^{1.9992})$ (the preprint's Theorem 22), unconditionally; over $\Fp^r$ given $\Hyp^{\det}(\eps,\gamma)$ in time $O(n^{2-\eps\gamma/2}\log n)$ (Theorem~\ref{g:thm:A}), given $\Hyp(\eps,\gamma)$ in time $O(n^{2-\eps\gamma}(\log n)^8)$ (Theorem~\ref{g:thm:B}), and unconditionally in time $O(n^{1.99451})$ (Corollary~\ref{g:cor:explicitA}); over a group of exponent $p^2$ given $\Hyp^{\det}(\eps,\gamma)$, $\eps\le\frac14$, in time $O(n^{2-\eps\gamma/2}(\log n)^4)$ (Theorem~\ref{g:thm:G}), and unconditionally in time $O(n^{1.99726})$ (Corollary~\ref{g:cor:explicitG}).
\end{corollary}
\begin{proof}
Lemma~\ref{g:lem:repalg}, with $|A^d|\le n^{O(c)}$ in (a) and numbers of size $n^{O(c)}$ in (b); the cited algorithms have exponents that do not depend on the size bound.
\end{proof}

\begin{remark}\label{g:rem:repexp}
By Lemma~\ref{g:lem:universal}(d), every representable $P$ is representable over a finite abelian group $\Z^{3k}/(L_P+m\Z^{3k})$ of a bounded exponent $m$, so $\Prob(P)$ is a 3SUM problem over a group of bounded exponent, with $d$ factors. A positive answer to Open problem~\ref{g:prob:exponent} would therefore make $\Prob(P)$ truly subquadratic for every representable $P$. What we can prove is Corollary~\ref{g:cor:repfast}, which suffices for all representable relations classified in Sections~\ref{g:sec:boolean}--\ref{g:sec:stop} (Theorems~\ref{g:thm:D} and~\ref{g:thm:E} and Proposition~\ref{g:prop:F}(b), (e)).
\end{remark}

\subsection{OV gadgets}\label{g:sec:gadgets}

\emph{Orthogonal Vectors} (OV): given $A,B\subseteq\{0,1\}^{d_0}$ with $|A|,|B|\le n$, $d_0=c_0\log n$, decide whether some $a\in A$, $b\in B$ have $a[j]\wedge b[j]=0$ for all $j$. The \emph{OV hypothesis} (OVH) states that for every $\delta>0$ there is $c_0$ such that OV with $d_0=c_0\log n$ has no algorithm in time $O(n^{2-\delta})$; the Strong Exponential Time Hypothesis implies it~\cite{Wil05}.

\begin{definition}[OV gadget]\label{g:def:gadget}
A \emph{side assignment} is $\theta=(\theta_1,\theta_2,\theta_3)\in\{\mathsf A,\mathsf B,\mathsf C\}^3$ in which both $\mathsf A$ and $\mathsf B$ occur; position $i$ reads a bit of the first OV vector if $\theta_i=\mathsf A$, of the second if $\theta_i=\mathsf B$, and nothing (a constant) if $\theta_i=\mathsf C$. An \emph{OV gadget of size $t$} for $P$ with side assignment $\theta$ consists of maps $f_{i,\ell}\colon\{0,1\}\to\Sigma$, $i\le3$, $\ell\le t$, with $f_{i,\ell}$ constant whenever $\theta_i=\mathsf C$, such that for all bits $\beta_1,\beta_2,\beta_3\in\{0,1\}$
\[
\bigwedge_{\ell\le t}\bigl[(f_{1,\ell}(\beta_1),f_{2,\ell}(\beta_2),f_{3,\ell}(\beta_3))\in P\bigr]=\mathrm{tgt}_\theta(\beta),
\]
where $\mathrm{tgt}_\theta(\beta)$ is true if and only if all positions with $\theta_i=\mathsf A$ carry the same bit $\alpha$, all positions with $\theta_i=\mathsf B$ carry the same bit $\alpha'$, and $\neg(\alpha\wedge\alpha')$ (the bits at $\mathsf C$-positions are ignored). The gadget is a \emph{slice} if some $\theta_i=\mathsf C$ (then the other two positions read one side each) and an \emph{identification} otherwise (two positions read the same side).
\end{definition}

\begin{lemma}[gadgets give OV-hardness]\label{g:lem:gadgethard}
If $P$ has an OV gadget of size $t$, then OV in dimension $d_0$ reduces in $O(ntd_0)$ time to $\Prob(P)$ with lists of at most $n$ vectors of dimension $td_0$. Consequently, under OVH, for every $\delta>0$ there is $c$ such that $\Prob(P)$ with $d=c\log n$ has no algorithm in time $O(n^{2-\delta})$.
\end{lemma}
\begin{proof}
For $w\in\{0,1\}^{d_0}$ let $F_i(w)\in\Sigma^{td_0}$ have the entry $f_{i,\ell}(w[j])$ at coordinate $(j,\ell)$. Given $A,B$, let $S_i:=\{F_i(a):a\in A\}$ if $\theta_i=\mathsf A$, $S_i:=\{F_i(b):b\in B\}$ if $\theta_i=\mathsf B$, and $S_i:=\{F_i(0)\}$ if $\theta_i=\mathsf C$. If $a\in A$, $b\in B$ are orthogonal, take $x_i:=F_i(a)$, $F_i(b)$ or $F_i(0)$ according to $\theta_i$; at coordinate $j$ the bits read are $a[j]$ at the $\mathsf A$-positions and $b[j]$ at the $\mathsf B$-positions, so $\mathrm{tgt}_\theta$ holds, and all $t$ coordinates $(j,\ell)$ are in $P$. Conversely, let $x_i\in S_i$ have all coordinates in $P$, and choose $w_i$ with $x_i=F_i(w_i)$, $w_i\in A$ for $\theta_i=\mathsf A$ and $w_i\in B$ for $\theta_i=\mathsf B$. At each $j$, the conjunction over $\ell$ is true at the bits $(w_i[j])_i$, so $\mathrm{tgt}_\theta$ holds there: all $w_i$ with $\theta_i=\mathsf A$ agree at $j$, all with $\theta_i=\mathsf B$ agree at $j$, and the two common bits are not both $1$. Over all $j$, the $\mathsf A$-vectors are one vector $a\in A$, the $\mathsf B$-vectors one vector $b\in B$, and $a\perp b$. The lists have at most $n$ vectors of dimension $td_0=tc_0\log n$; an $O(n^{2-\delta})$ algorithm for $\Prob(P)$ with $c:=tc_0$ would solve OV with $c_0$ in time $O(n^{2-\delta})$.
\end{proof}

\begin{lemma}[representable relations have no gadget]\label{g:lem:exclusive}
If $P$ is representable, it has no OV gadget of any size.
\end{lemma}
\begin{proof}
Let $(\varphi_i,t)$ represent $P$ over $A$ and suppose a gadget of size $t'$ with side assignment $\theta$ exists. For each $\ell$, the relation $\{\beta:(f_{i,\ell}(\beta_i))_i\in P\}$ on bits is $\{\beta:\sum_i\varphi_i(f_{i,\ell}(\beta_i))=t\}$, and the conjunction over $\ell$ is $\{\beta:\sum_i\psi_i(\beta_i)=(t,\dots,t)\}$ with $\psi_i(\beta_i):=(\varphi_i(f_{i,\ell}(\beta_i)))_{\ell\le t'}\in A^{t'}$. The $\psi_i$ with $\theta_i=\mathsf C$ are constant. Substitute $\alpha$ for every bit at an $\mathsf A$-position and $\alpha'$ for every bit at a $\mathsf B$-position: on these bits $\mathrm{tgt}_\theta$ is $\neg(\alpha\wedge\alpha')$, so
\[
\neg(\alpha\wedge\alpha')\iff g(\alpha)+h(\alpha')=t^*,\qquad g(\alpha):=\sum_{\theta_i=\mathsf A}\psi_i(\alpha),\ h(\alpha'):=\sum_{\theta_i=\mathsf B}\psi_i(\alpha'),
\]
with $t^*$ the target minus the constants. From $(0,0)$ and $(0,1)$, $h(0)=h(1)$; then $g(1)+h(1)=g(1)+h(0)=t^*$ by $(1,0)$, so $(1,1)$ would satisfy the equation, a contradiction.
\end{proof}

Latin squares, which reappear in Section~\ref{g:sec:stop} as the first relations that need new groups, have no gadget either. A \emph{Latin square} on $\Sigma$ is a relation $P\subseteq\Sigma^3$ such that any two positions with any two letters extend to exactly one triple of $P$ (the graph of a quasigroup).

\begin{lemma}[Latin squares have no gadget]\label{g:lem:latinnogadget}
A Latin square has no OV gadget of any size.
\end{lemma}
\begin{proof}
Suppose a gadget of size $t$ with side assignment $\theta$ exists. Each conjunct $m_\ell(\beta):=[(f_{i,\ell}(\beta_i))_i\in P]$ is true wherever $\mathrm{tgt}_\theta$ is. If $\theta$ is a slice, say $\theta_i=\mathsf A$, $\theta_j=\mathsf B$, $\theta_k=\mathsf C$: $\mathrm{tgt}_\theta$ holds at $(\beta_i,\beta_j)=(0,0)$ and $(0,1)$, so $(f_{i,\ell}(0),f_{j,\ell}(0),\cdot)$ and $(f_{i,\ell}(0),f_{j,\ell}(1),\cdot)$, with the same constant at position $k$, are in $P$; since positions $i$ and $k$ determine position $j$, $f_{j,\ell}(0)=f_{j,\ell}(1)$. Then $m_\ell(1,1)=m_\ell(1,0)$, which is true because $\mathrm{tgt}_\theta(1,0)$ holds. So every conjunct, and the conjunction, is true at $(1,1)$, where $\mathrm{tgt}_\theta$ is false. If $\theta$ is an identification, let $i$ be the position whose side occurs once and $j,k$ the two positions of the other side, reading bits $\eta_j,\eta_k$, and let $\xi$ be the bit at $i$. $\mathrm{tgt}_\theta$ holds at $\eta_j=\eta_k=0$ for both values of $\xi$, so $(f_{i,\ell}(0),f_{j,\ell}(0),f_{k,\ell}(0))$ and $(f_{i,\ell}(1),f_{j,\ell}(0),f_{k,\ell}(0))$ are in $P$, and since $j,k$ determine $i$, $f_{i,\ell}(0)=f_{i,\ell}(1)$. So $m_\ell$ does not depend on $\xi$, and it is true at $\xi=1$, $\eta_j=\eta_k=1$, because $\mathrm{tgt}_\theta$ holds at $\xi=0$, $\eta_j=\eta_k=1$. But $\mathrm{tgt}_\theta$ is false at $\xi=\eta_j=\eta_k=1$.
\end{proof}

\paragraph{Symmetries.} Representability and the existence of a gadget (of a given size) are invariant under permuting the three positions and under a permutation of the letters in each position separately: a representation or a gadget is transported by composing the maps $\varphi_i$, respectively $f_{i,\ell}$, with the permutations, and by permuting the positions of $\theta$ and the summands. So both properties are constant on the orbits of these symmetries, and the classifications below are stated per orbit.

\paragraph{Gadgets of size $t$ from gadgets of size $1$.} For a fixed side assignment, a choice of the maps $(f_{i,\ell})_i$ for one $\ell$ defines the \emph{mask} $\{\beta:m_\ell(\beta)\}$, a subset of the bit vectors on the non-constant positions, and a gadget of size $t$ is a choice of $t$ masks whose intersection is the target set $\{\beta:\mathrm{tgt}_\theta(\beta)\}$. Every mask used contains the target set. So a gadget of some size exists if and only if the intersection of all achievable masks that contain the target set is the target set, and the least size is found by a breadth-first search over intersections; this is how the scripts decide gadgets of every size (Section~\ref{g:sec:computer}). For a slice, the target set $\{00,01,10\}$ has only itself and the full set as supersets among the subsets of $\{0,1\}^2$, so a slice gadget exists only with $t=1$, if at all.

\subsection{The Boolean classification}\label{g:sec:boolean}

Let $\Sigma=\{0,1\}$. The symmetries of Section~\ref{g:sec:gadgets} are the $6$ permutations of the positions and the $8$ choices of flipping the letter in each position; under this group of order $48$ the $256$ relations $P\subseteq\{0,1\}^3$ fall into $22$ orbits.

\begin{theorem}[Boolean relations]\label{g:thm:D}
For every $P\subseteq\{0,1\}^3$ exactly one of the following holds.
\begin{enumerate}[label=\textup{(\roman*)}]
\item $P$ is representable. Then $P$ is representable over $\Z$ or over $\F$ and lies in one of the $10$ orbits marked REP in Table~\ref{g:tab:boolean}; eight of them are solved in time $O(nd\log n)$, the orbit of $\{001,010,100\}$ (representable over $\Z$ but not over $\F$) is integer 3SUM, and the orbit of $\{000,011,101,110\}$ (representable over $\F$ but not over $\Z$) is 3XOR. So $\Prob(P)$ is solved in time $O(n^{1.99451})$ for every $c$, unconditionally; for the 3XOR orbit also in time $O(n^{2-\eps\gamma/2}\log n)$ given $\Hyp^{\det}(\eps,\gamma)$.
\item $P$ has an OV gadget of size $1$. Then $P$ lies in one of the other $12$ orbits: $11$ of them have a slice gadget, and the orbit of $\{000,011,100\}$ has only identification gadgets. Under OVH, for every $\delta>0$ there is $c$ such that $\Prob(P)$ with $d=c\log n$ has no algorithm in time $O(n^{2-\delta})$.
\end{enumerate}
No relation needs a gadget of size $t\ge2$: a relation with a gadget of some size has one of size $1$.
\end{theorem}

\begin{proof}
Lemma~\ref{g:lem:exclusive} shows that (i) and (ii) exclude each other. That every $P$ satisfies one of them, the orbit data, and the representability over $\Z$ and over $\F$ are the output of the script \texttt{q2\_boolean.py} (Section~\ref{g:sec:computer}), which decides representability over $\Z$, $\F$, $\mathbb F_3$ and over some group, and the existence of gadgets of every size, for all $256$ relations, each verdict with a certificate. For the orbits of (i) the algorithms are these. The eight orbits other than the two named ones are the empty relation, the full relation and relations that force some lists to consist of constant vectors or two lists to be equal or complementary (Table~\ref{g:tab:boolean}); each is decided by sorting in $O(nd\log n)$ time. $\{001,010,100\}=\{x_1+x_2+x_3=1\text{ over }\Z\}$ is integer 3SUM by Lemma~\ref{g:lem:repalg}(b), solved in time $n^{2-0.0055+o(1)}\le O(n^{1.99451})$ (Corollary~\ref{x:cor:fg}(d)), or $O(n^{1.9992})$ (the preprint's Theorem 22). $\{000,011,101,110\}=\{x_1\oplus x_2\oplus x_3=0\}$ is 3XOR, solved in time $O(n^{2-\eps\gamma/2}\log n)$ given $\Hyp^{\det}(\eps,\gamma)$ (Theorem~\ref{x:thm:xordet}) and in time $O(n^{1.99451})$ unconditionally (Corollary~\ref{x:cor:xor}). The orbits of (ii) are OV-hard by Lemma~\ref{g:lem:gadgethard}. By symmetry, all relations in an orbit behave alike.
\end{proof}

The representability claims of (i) can be checked by hand: each representative in Table~\ref{g:tab:boolean} is given by an explicit equation. The claims that the $12$ orbits of (ii) are not representable follow from Lemma~\ref{g:lem:exclusive} once their gadgets are exhibited, and a gadget of size $1$ is checked on $8$ bit vectors; the script lists them. The only claim that rests on the exhaustive search alone is that every relation falls into one of the two classes.

\begin{table}[ht]
\centering\small
\renewcommand{\arraystretch}{1.1}
\begin{tabular}{@{}r>{\raggedright}p{3.6cm}rl>{\raggedright}p{4.6cm}c@{}}
& representative $P$ & orbit & class & $\Prob(P)$ & fun.\\\hline
1&$\emptyset$&1&REP $\Z,\F$&NO&123\\
2&$000$&8&REP $\Z,\F$&$x_1=x_2=x_3=0$&123\\
3&$000,001$&12&REP $\Z,\F$&$x_1=x_2=0$&12\\
4&$001,010$&12&REP $\Z,\F$&$x_1=0$, $x_3=\neg x_2$&123\\
5&$011,100$&4&REP $\Z,\F$&$x_2=x_3=\neg x_1$&123\\
6&$000,001,010$&24&slice&$x_1=0$, $x_2\perp x_3$; $\Theta(n^2)$&1\\
7&$001,010,100$&8&REP $\Z$&$x_1+x_2+x_3=\mathbf 1$ over $\Z$: integer 3SUM&123\\
8&$000,011,100$&24&ident.&$x_2=x_3$, $x_1\perp x_2$; $\Theta(n^2)$&23\\
9&$000,001,010,011$&6&REP $\Z,\F$&$x_1=0$&1\\
10&$000,001,010,100$&8&slice&at most one $1$ per coordinate&\\
11&$000,001,011,100$&24&slice&&\\
12&$001,010,011,100$&24&slice&$\Theta(n^2)$&1\\
13&$010,011,100,101$&6&REP $\Z,\F$&$x_2=\neg x_1$&12\\
14&$000,011,101,110$&2&REP $\F$&$x_1\oplus x_2\oplus x_3=0$: 3XOR&123\\
15&$000,001,010,011,100$&24&slice&&\\
16&$000,010,011,100,101$&24&slice&&\\
17&$000,001,011,101,110$&8&slice&&\\
18&all but $110,111$&12&slice&&\\
19&all but $100,111$&12&slice&&\\
20&all but $000,111$&4&slice&not-all-equal&\\
21&all but $111$&8&slice&3-OV; $n^{3-o(1)}$ under 3-OV&\\
22&$\{0,1\}^3$&1&REP $\Z,\F$&YES (lists nonempty)&
\end{tabular}
\caption{The $22$ orbits of Boolean relations (Theorem~\ref{g:thm:D}). A triple $abc$ stands for $(a,b,c)$. ``REP $\Z,\F$'': representable over $\Z$ and over $\F$; ``REP $\Z$'' and ``REP $\F$'': only over the named one (and not over $\mathbb F_3$ either, in rows 7 and 14). ``slice'' and ``ident.'': the kind of size-$1$ OV gadget. ``fun.'': the positions in which the representative is functional; for the gadget orbits 6, 8, 12 this gives $O(n^2d\log n)$, which with Lemma~\ref{g:lem:gadgethard} is tight up to $n^{o(1)}$ under OVH.}
\label{g:tab:boolean}
\end{table}

\paragraph{Comparison with~\cite{BFK19}.} Bringmann, Fischer and Künnemann~\cite{BFK19} classify by their fine-grained complexity the first-order properties of sparse graphs of the form $(\exists x_1)\cdots(\exists x_k)\allowbreak(\forall y)\,\varphi$. When $\varphi$ uses only the atoms $E(x_i,y)$, read as ``$x_i[y]=1$'', this is their Vector Problem, and for $k=3$ it is our $\Prob(P)$ with $P\subseteq\{0,1\}^3$. Below the cubic level they treat the $16$ symmetric relations (their Theorem~7 and Table~1), with gaps that they point out. Table~\ref{g:tab:bfk} sets the symmetric relations of~\cite[Theorem~7, Table~1]{BFK19} beside our classification. Their bounds are in their sparse model, in terms of the number $m$ of edges of the input graph, with the dimension unrestricted; ours are in the dense model, $n$ vectors per list of dimension $d=c\log n$. A bound of theirs is therefore not a bound of ours, or conversely, except through $m\le3nd$. Two points of contact are exact. First, their Theorem 3 makes every Boolean $P$ except the $3$-OV orbit (row 21) decidable in time $O(m^{3-\eps})$, and that orbit cubic-hard under the $3$-OV hypothesis; our row 21 is the same relation, and in the dense model it is $3$-OV itself, cubic under the $3$-OV hypothesis. Second, our gadget instances have $m\le3ntd_0=O(n\log n)$ edges, so Lemma~\ref{g:lem:gadgethard} also rules out $O(m^{2-\delta})$ algorithms in their model for our $12$ gadget orbits, under OVH. Within their subcubic class of $21$ orbits, our classification separates the $10$ representable orbits, truly subquadratic in the dense model, from the $11$ gadget orbits other than $3$-OV, which are quadratic-hard there under OVH. This is the sense in which Theorem~\ref{g:thm:D} refines their cubic-level dichotomy; it says nothing about the sparse model with large $d$, where our algorithms do not apply.

\begin{table}[ht]
\centering\small
\renewcommand{\arraystretch}{1.1}
\begin{tabular}{@{}l>{\raggedright}p{4.4cm}c>{\raggedright\arraybackslash}p{5.2cm}@{}}
type & \cite[Table~1]{BFK19}: upper / lower bound, sparse model, in $m$ & row & this paper: dense model, in $n$\\\hline
0000, 1111 & trivial & 1, 22 & trivial\\
0001, 1000 & $O(m)$ / $\Omega(m)$ & 2 & $O(nd)$\\
1001 & $O(m)$ / $\Omega(m)$ & 5 & $O(nd\log n)$\\
0010, 0100 & $O(m^2)$ / $\Omega(m)$ & 7 & REP over $\Z$: $O(n^{1.99451})$ (integer 3SUM)\\
0101, 1010 & $O(m^2)$ / $m^{2-o(1)}$ under 3-XOR & 14 & REP over $\F$: $O(n^{1.99451})$ (3XOR)\\
0011, 1100 & $O(m^\omega)$ / $m^{\omega-o(1)}$ under $k$-Clique & 10 & slice gadget: quadratic-hard under OVH (Lemma~\ref{g:lem:gadgethard})\\
0110 & $\tilde O(m^{(3+\omega)/2})$ / $m^{\omega-o(1)}$ under $k$-Clique & 20 & slice gadget: quadratic-hard under OVH (Lemma~\ref{g:lem:gadgethard})\\
1011, 1101 & $O(m^{(9+\omega)/4})$ / $m^{\omega-o(1)}$ under $k$-Clique & 17 & slice gadget: quadratic-hard under OVH (Lemma~\ref{g:lem:gadgethard})\\
0111, 1110 & $m^3/2^{\Omega(\sqrt{\log m})}$ / $m^{3-o(1)}$ under 3-OV & 21 & slice gadget; $n^{3-o(1)}$ under 3-OV
\end{tabular}
\caption{The symmetric Boolean relations: the bounds of~\cite[Table~1]{BFK19} (theirs, sparse model) and our classification (dense model, $d=c\log n$). Type $t_0t_1t_2t_3$ is the relation that contains the triples with $w$ ones exactly when $t_w=1$. The parity rows carry the lower bound ``$m^{2-o(1)}$ under 3-XOR'' of~\cite{BFK19}, and the exactly-one row (their $0100$, the relation $x_1+x_2+x_3=1$ over $\Z$) their bounds $O(m^2)$ and $\Omega(m)$; in the dense model the first are 3XOR and the second integer 3SUM, both truly subquadratic. The 3-XOR hypothesis behind the parity rows' lower bound, in its form for $n$ vectors of dimension $O(\log n)$~\cite{JV16}, is refuted by Theorems~\ref{x:thm:xorcount} and~\ref{x:thm:xordet}, which prove the preprint's Remark 23.}
\label{g:tab:bfk}
\end{table}

\paragraph{Relation to 3XOR in Section~\ref{x:sec:xor}.} Section~\ref{x:sec:xorremarks} observes that among the sixteen bitwise binary operations $g$, $x_3=g(x_1,x_2)$ is 3XOR or easier when $g$ is affine over $\F$ and contains OV otherwise; these are the functional Boolean relations, and for them representable means that $g$ is affine over $\F$, so Theorem~\ref{g:thm:D} extends that observation to all $256$ Boolean relations.

\subsection{Functional relations over three letters}\label{g:sec:functional}

Let $\Sigma=\{0,1,2\}$ and, for $g\colon\Sigma^2\to\Sigma$, $P_g:=\{(a,b,g(a,b))\}$. Every $\Prob(P_g)$ is solved in time $O(n^2d\log n)$ by lookup, so the question is which are truly subquadratic. The symmetries that keep $P_g$ functional in position $3$ are the permutations of the letters in each position and the transposition of positions $1$ and $2$, a group of order $(3!)^3\cdot2=432$; the $3^9=19683$ maps $g$ fall into $84$ orbits.

\begin{theorem}[functional relations over three letters]\label{g:thm:E}
For every $g\colon\{0,1,2\}^2\to\{0,1,2\}$, exactly one of the following holds.
\begin{enumerate}[label=\textup{(\roman*)}]
\item $P_g$ is representable; this holds for $297$ maps in $8$ orbits (Table~\ref{g:tab:functional}), each representable over $\Z$, over $\F$ or over $\mathbb F_3$. Then $\Prob(P_g)$ is solved in time $O(n^{1.99451})$ for every $c$, unconditionally (Corollary~\ref{g:cor:repfast}).
\item $P_g$ has an OV gadget of size $1$; this holds for the other $19386$ maps, in $76$ orbits. Under OVH, for every $\delta>0$ there is $c$ such that $\Prob(P_g)$ with $d=c\log n$ has no algorithm in time $O(n^{2-\delta})$; for every $c$ it is solved in time $O(n^2d\log n)$ by lookup.
\end{enumerate}
No map needs a gadget of size $t\ge2$, and no representable map needs a group other than $\Z$, $\F$ or $\mathbb F_3$.
\end{theorem}

\begin{proof}
Lemma~\ref{g:lem:exclusive} gives the exclusion. The rest is the output of \texttt{q2\_functional3.py} (Section~\ref{g:sec:computer}), which treats all $19683$ maps, not only orbit representatives, and checks that the verdicts are constant on each orbit. Among the $297$ representable maps, $159$ are representable over $\Z$, $105$ over $\F$ and $243$ over $\mathbb F_3$, and every one of them over at least one of the three: $51$ over all three, $108$ over $\Z$ and $\mathbb F_3$ only, $54$ over $\F$ only and $84$ over $\mathbb F_3$ only. By Lemma~\ref{g:lem:universal}(e), representability over $\Fp$ for a prime $p\ge5$ would need $p$ to divide an invariant factor of $A_{P_g}$, and the invariant factors that occur are $2$ and $3$, so no other prime is needed. The algorithms are those of Corollary~\ref{g:cor:repfast}: over $\Z$ integer 3SUM, over $\F$ 3XOR (Corollary~\ref{x:cor:xor}), and over $\mathbb F_3$ Theorem~\ref{g:thm:B} with $p=3$ and $\Hyp(1/14,0.077)$ (Corollary~\ref{g:cor:explicitA}); each is $O(n^{1.99451})$.
\end{proof}

\begin{table}[ht]
\centering\small
\renewcommand{\arraystretch}{1.1}
\begin{tabular}{@{}l r >{\raggedright}p{5.6cm} c c c l@{}}
$g$ (rows $a=0,1,2$) & orbit & description & $\Z$ & $\F$ & $\mathbb F_3$ & torsion of $A_{P_g}$\\\hline
000/000/000 & 3 & constant & yes & yes & yes & none\\
000/111/222 & 12 & $c=a$ & yes & yes & yes & none\\
000/000/111 & 36 & $c=[a=2]$ & yes & yes & yes & none\\
001/001/112 & 54 & $c=[a=2]+[b=2]$ over $\Z$ & yes & no & yes & none\\
001/001/220 & 54 & $c$ encodes $[a=2]-[b=2]\in\{0,-1,1\}$ as $0,1,2$ & yes & no & yes & none\\
001/001/110 & 54 & $c=[a=2]\oplus[b=2]$ & no & yes & no & $\Z_2$\\
001/112/220 & 72 & $c=a+[b=2]\bmod3$ & no & no & yes & $\Z_3$\\
012/120/201 & 12 & $c=a+b\bmod3$ & no & no & yes & $\Z_3$
\end{tabular}
\caption{The $8$ orbits of representable functional relations over $\{0,1,2\}$ (Theorem~\ref{g:thm:E}); the representative $g$ is written as its three rows $g(a,0)g(a,1)g(a,2)$, and $[\cdot]$ is the indicator. In the fifth row, $\varphi_1(a)=[a=2]$, $\varphi_2(b)=-[b=2]$ and $c=0,1,2$ stand for $0,-1,1$.}
\label{g:tab:functional}
\end{table}

So for functional relations over three letters the dichotomy at the quadratic level is complete: truly subquadratic exactly when representable, and then always over $\Z$ or over a prime field; quadratic up to $n^{o(1)}$ under OVH otherwise. No mixed group is ever needed, and every relation has one of the two properties.

\subsection{Where the dichotomy stops}\label{g:sec:stop}

The dichotomy of Theorems~\ref{g:thm:D} and~\ref{g:thm:E} does not extend to larger alphabets. Latin squares show how the groups needed change, and an exhaustive computation over four letters shows where the dichotomy stops. For an abelian group $H$ with $|H|=|\Sigma|$, say that $P\subseteq\Sigma^3$ is \emph{isotopic to $H$} if there are bijections $\alpha_1,\alpha_2,\alpha_3\colon\Sigma\to H$ with $P=\{(a,b,c):\alpha_1(a)+\alpha_2(b)+\alpha_3(c)=0\}$; such a $P$ is a Latin square, and the Cayley table $\{(a,b,a+b)\}$ of $H$ is isotopic to $H$ through $\alpha_3(c)=-c$.

\begin{lemma}[representable Latin squares]\label{g:lem:latinrep}
A Latin square $P$ on $\Sigma$ is representable over an abelian group $A$ if and only if $A$ has a subgroup $H$ to which $P$ is isotopic. If $P$ is isotopic to two abelian groups $H$ and $H'$, then $H\cong H'$.
\end{lemma}
\begin{proof}
If $P$ is isotopic to $H\le A$, the maps $\alpha_i$ composed with the inclusion and $t=0$ represent $P$ over $A$. Conversely, let $(\varphi_i,t)$ represent $P$ over $A$. Each $\varphi_i$ is injective: if, say, $\varphi_3(c)=\varphi_3(c')$ with $c\ne c'$, take $(a,b,c)\in P$ (it exists, since $P$ is a Latin square); then $(a,b,c')$ also satisfies the equation, so it lies in $P$, and positions $1$ and $2$ would not determine position $3$. The same argument applies to $\varphi_1$ and $\varphi_2$, since any two positions determine the third. Let $X_1:=\varphi_1(\Sigma)$, $X_2:=\varphi_2(\Sigma)$, $X_3:=t-\varphi_3(\Sigma)$, each of $k:=|\Sigma|$ elements. For all $x\in X_1$, $y\in X_2$ there is a triple of $P$ with these first two values, so $x+y\in X_3$; hence $x+X_2\subseteq X_3$, and since both have $k$ elements, $x+X_2=X_3$. Fix $x_0\in X_1$ and let $H:=X_1-x_0$. For $h\in H$, $X_2+x_0+h=X_3=X_2+x_0$, so $X_2+h=X_2$: $H$ lies in the stabiliser $S:=\{g\in A:X_2+g=X_2\}$, a subgroup. $X_2$ is a union of cosets of $S$, so $|S|\le|X_2|=k=|H|$; hence $H=S$ is a subgroup of order $k$, and $X_2=y_0+H$ for any $y_0\in X_2$. Then $\alpha_1:=\varphi_1-x_0$, $\alpha_2:=\varphi_2-y_0$ and $\alpha_3:=\varphi_3-t+x_0+y_0$ are bijections onto $H$ (for $\alpha_3$ because $X_3=x_0+y_0+H$), and $\varphi_1(a)+\varphi_2(b)+\varphi_3(c)=t$ if and only if $\alpha_1(a)+\alpha_2(b)+\alpha_3(c)=0$.

For the second claim, composing the isotopies gives bijections $\gamma_1,\gamma_2,\gamma_3\colon H\to H'$ with $\gamma_1(a)+\gamma_2(b)=\gamma_3(a+b)$ for all $a,b\in H$ (this is the classical fact that a loop isotopic to a group is isomorphic to it~\cite{Alb43}; we include the short proof). Putting $b=0$ and $a=0$ gives $\gamma_1(a)=\gamma_3(a)-\gamma_2(0)$ and $\gamma_2(b)=\gamma_3(b)-\gamma_1(0)$, and $\gamma_3(0)=\gamma_1(0)+\gamma_2(0)$; so $\delta:=\gamma_3-\gamma_3(0)$ satisfies $\delta(a+b)=\delta(a)+\delta(b)$ and is a bijective homomorphism $H\to H'$.
\end{proof}

The same computation shows that a loop $(\Sigma,\circ)$ with identity $e$ that is isotopic to an abelian group, $\gamma_1(a)+\gamma_2(b)=\gamma_3(a\circ b)$, is isomorphic to it, through $\delta:=\gamma_3-\gamma_3(e)$; in particular it is associative.

Proposition~\ref{g:prop:F} collects six facts. Parts~(a)--(c) concern Latin squares: none has a gadget, those of order four are representable but some need $\Z_4$, and a loop of order five is neither representable nor has a gadget. Part~(d) holds for every alphabet and makes the gadget side finite: a relation has a gadget if and only if one of its Boolean restrictions is a non-representable Boolean relation, so Theorem~\ref{g:thm:D} decides gadgets for every alphabet. Parts~(e) and~(f) use~(d) to classify all functional relations over four letters and all relations over three letters, and find the third kind.

\begin{proposition}[where the dichotomy stops]\label{g:prop:F}
\begin{enumerate}[label=\textup{(\alph*)}]
\item No Latin square has an OV gadget of any size, and a Latin square is representable if and only if it is isotopic to an abelian group.
\item Of the $576$ Latin squares of order $4$, $432$ are isotopic to $\Z_4$ and $144$ to $\F^2$. A Latin square $P$ isotopic to $\Z_4$ is representable exactly over the abelian groups that contain an element of order $4$; it is not representable over $\Z^r$ or over $\Fp^r$ for any prime $p$. $\Prob(P)$ is 3SUM over $\Z_4^d$, which no single homomorphic hash family serves (Corollary~\ref{g:cor:C}) and which Theorem~\ref{g:thm:G} solves, in time $O(n^{1.99726})$ unconditionally (Corollary~\ref{g:cor:explicitG}). For $P$ isotopic to $\F^2$, $\Prob(P)$ is 3XOR in dimension $2d$.
\item The Latin square $L_5$ on $\{0,1,2,3,4\}$ with the rows
\[
01234,\quad10342,\quad24013,\quad32401,\quad43120
\]
(row $a$ lists $a\circ0,\dots,a\circ4$) neither is representable over any abelian group nor has an OV gadget of any size.
\item Let $\Sigma$ be any finite alphabet and $P\subseteq\Sigma^3$. For maps $u_1,u_2,u_3\colon\{0,1\}\to\Sigma$ let $P^u:=\{\beta\in\{0,1\}^3:(u_1(\beta_1),u_2(\beta_2),u_3(\beta_3))\in P\}$; for injective $u_i$ this is the \emph{Boolean restriction} of $P$ to the two letters $u_i(0),u_i(1)$ in each position, and there are $\binom{|\Sigma|}2^3$ of them up to relabelling. The following are equivalent: $P$ has an OV gadget of some size; $P$ has an OV gadget of size $1$; some Boolean restriction of $P$ is not representable.
\item Of the $4^{16}=4294967296$ maps $g\colon\{0,1,2,3\}^2\to\{0,1,2,3\}$, which fall into $167342$ orbits under the permutations of the letters in each position and the transposition of positions $1$ and $2$ (a group of order $(4!)^3\cdot2=27648$), exactly one of the following holds for each. Either $P_g$ is representable ($59776$ maps in $42$ orbits), over $\Z$, $\F$, $\mathbb F_3$ or a finite abelian group of exponent $4$ (Table~\ref{g:tab:functional4}), and $\Prob(P_g)$ is solved in time $O(n^{1.99726})$ for every $c$, unconditionally, and in time $O(n^{1.99451})$ for the $33$ orbits representable over $\Z$, $\F$ or $\mathbb F_3$; or $P_g$ has an OV gadget of size $1$ ($4294891968$ maps in $167296$ orbits, $165800$ of them with a slice gadget and $1496$ with identification gadgets only); or $P_g$ is in one of the $4$ orbits of Table~\ref{g:tab:functional4} ($15552$ maps), and neither is representable over any abelian group nor has an OV gadget of any size.
\item Exactly $1296$ relations $P\subseteq\{0,1,2\}^3$ neither are representable over any abelian group nor have an OV gadget of any size. Under the permutations of the letters in each position and of the three positions (a group of order $(3!)^3\cdot6=1296$) they fall into $4$ orbits, with the representatives $\{010,022,102,111,200\}$, $\{001,010,022,102,111,200\}$, $\{001,010,022,102,111,120,200\}$ and $\{(a,b,c):a+b+c\equiv0\bmod3\}\setminus\{(0,0,0)\}$, of sizes $216$, $648$, $324$ and $108$ (a triple $abc$ stands for $(a,b,c)$).
\end{enumerate}
\end{proposition}

\begin{table}[ht]
\centering\small
\renewcommand{\arraystretch}{1.1}
\begin{tabular}{@{}l r r l@{}}
representable over & orbits & maps & example $g$ (rows $a=0,\dots,3$)\\\hline
$\Z$, $\F$ and $\mathbb F_3$ & $8$ & $1684$ & 0000/0000/1111/2222 ($c=[a\ge2]+[a=3]$)\\
$\Z$ and $\mathbb F_3$, not $\F$ & $4$ & $2352$ & 0001/0001/1112/1112\\
$\Z$, neither $\F$ nor $\mathbb F_3$ & $5$ & $12096$ & 0001/0001/1112/2223\\
$\F$, not $\Z$ & $13$ & $14700$ & 0123/1032/2301/3210 (the table of $\F^2$)\\
$\mathbb F_3$, not $\Z$ & $3$ & $5760$ & 0001/0001/1112/2220\\
exponent $4$ only & $9$ & $23184$ & 0123/1032/2310/3201 (isotopic to $\Z_4$), 0001/1112/2223/3330
\end{tabular}

\medskip
\begin{tabular}{@{}l r l l@{}}
third kind: $g$ (rows $a=0,\dots,3$) & orbit & $A_{P_g}$ & a triple in $\cl(P_g)\setminus P_g$\\\hline
0012/0012/1120/3301 & $6912$ & $\Z_3\times\Z^3$ & $(0,3,3)$\\
0012/0012/1120/2203 & $3456$ & $\Z_3\times\Z^3$ & $(0,2,3)$\\
0012/0012/1120/2231 & $3456$ & $\Z_3\times\Z^3$ & $(0,0,3)$\\
0012/0012/1123/2231 & $1728$ & $\Z_3\times\Z^3$ & $(0,0,3)$
\end{tabular}
\caption{Functional relations over four letters (Proposition~\ref{g:prop:F}(e)); $g$ is written as its four rows $g(a,0)g(a,1)g(a,2)g(a,3)$. Top: the $42$ representable orbits by the groups over which they are representable; ``exponent $4$ only'' means representable over a finite abelian group of exponent $4$ but over no $\Z^r$ and no $\Fp^r$, and $c=a+[b=3]\bmod4$ is the second example of that row; the six rows are disjoint and exhaust the $42$ orbits: no orbit is representable over $\F$ and $\mathbb F_3$ without $\Z$, or over $\Z$ and $\F$ without $\mathbb F_3$. Bottom: the $4$ orbits of the third kind, with the universal group of Lemma~\ref{g:lem:universal} and a witness of non-representability.}
\label{g:tab:functional4}
\end{table}

\begin{proof}
(a) Lemmas~\ref{g:lem:latinnogadget} and~\ref{g:lem:latinrep}.
(b) The counts are the output of \texttt{q2\_latin.py}, which enumerates the $576$ Latin squares and finds two orbits under isotopy and transposition, of sizes $144$ and $432$, with the representatives $0123/1032/2301/3210$ (the table of $\F^2$) and $0123/1032/2310/3201$, which is isotopic to $\Z_4$; that there are exactly two isotopy classes of Latin squares of order $4$, those of the two groups of order $4$, is classical. By Lemma~\ref{g:lem:latinrep}, a $P$ isotopic to $\Z_4$ is representable over $A$ if and only if $A$ contains a subgroup isomorphic to $\Z_4$, that is, an element of order $4$. Neither $\Z^r$ nor $\Fp^r$ has one. The script confirms this independently: $A_P$ has torsion $\Z_4$ and free rank $3$, and the lattice tests fail over $\Z$, $\F$, $\mathbb F_3$ and $\mathbb F_5$. With $\alpha_i$ as in the definition of isotopy, $\Prob(P)$ is 3SUM over $\Z_4^d$ by Lemma~\ref{g:lem:repalg}(a).
(c) $L_5$ is a Latin square (every row and every column is a permutation) and a loop with identity $0$ (row $0$ and column $0$ are the identity). It is not associative: $(1\circ1)\circ2=0\circ2=2$, while $1\circ(1\circ2)=1\circ3=4$. If $L_5$ were representable, by Lemma~\ref{g:lem:latinrep} its relation would be isotopic to an abelian group, and by the remark after that lemma the loop $L_5$ would be isomorphic to it, hence associative. It has no gadget by (a). The script confirms both: the lattice closure of $L_5$ contains $(0,0,1)\notin L_5$, so by Lemma~\ref{g:lem:universal}(a) it is not representable, and no side assignment admits a gadget of any size.

(d) A gadget of size $1$ is a gadget of some size. If $Q:=P^u$ is not representable for injective $u_i$, Theorem~\ref{g:thm:D} gives $Q$ a gadget of size $1$, maps $h_i\colon\{0,1\}\to\{0,1\}$ with a side assignment $\theta$; then $f_i:=u_i\circ h_i$ is a gadget of size $1$ for $P$ with the same $\theta$, since $(f_i(\beta_i))_i\in P$ if and only if $(h_i(\beta_i))_i\in Q$, and $f_i$ is constant where $h_i$ is. Conversely, suppose that every Boolean restriction of $P$ is representable. Then so is every $P^u$: write each $u_i$ as $v_i\circ c_i$ with $v_i$ injective and $c_i\colon\{0,1\}\to\{0,1\}$ (for a constant $u_i$, take $c_i$ constant and $v_i$ injective through that letter; $|\Sigma|\ge2$); then $P^u=\{\beta:(c_i(\beta_i))_i\in P^v\}$, and if $(\psi_i,t)$ represents $P^v$, then $(\psi_i\circ c_i,t)$ represents $P^u$. Let a gadget of size $t'$ with side assignment $\theta$ and maps $f_{i,\ell}$ be given. Its $\ell$-th conjunct is true exactly on $P^{u^\ell}$ with $u^\ell_i:=f_{i,\ell}$, which is represented by some $(\psi_{i,\ell},t_\ell)$ over a group $A_\ell$. The conjunction, which is true exactly on $T_\theta:=\{\beta\in\{0,1\}^3:\mathrm{tgt}_\theta(\beta)\}$, is then represented over $A_1\times\dots\times A_{t'}$ by $\psi_i(\beta_i):=(\psi_{i,\ell}(\beta_i))_{\ell\le t'}$ and $(t_\ell)_{\ell\le t'}$. But $T_\theta$ is not representable: the identity maps at the positions with $\theta_i\in\{\mathsf A,\mathsf B\}$ and the constant $0$ at the positions with $\theta_i=\mathsf C$ form a gadget of size $1$ for $T_\theta$, because $\mathrm{tgt}_\theta$ does not depend on the bits at the $\mathsf C$-positions, and Lemma~\ref{g:lem:exclusive} excludes a gadget for a representable relation. So $P$ has no gadget of any size.

(e) The counts are the output of \texttt{q2\_functional4.py} (Section~\ref{g:sec:computer}). It lists one map of each orbit by orderly generation, checks that their number is the number of orbits given by Burnside's lemma and that their orbit sizes add up to $4^{16}$, and decides, with certified verdicts, gadgets of every size and representability over $\Z$, over $\Fp$ for $p=2,3,5,7$ and every prime dividing an invariant factor of $A_{P_g}$, and over groups of exponent dividing $4$. Lemma~\ref{g:lem:exclusive} gives the exclusion of the first two cases; the script finds no orbit in both. For an orbit representable over a group of exponent dividing $4$, Lemma~\ref{g:lem:universal}(c) gives a representation over the finite group $\Z^{12}/(L_{P_g}+4\Z^{12})$, so Corollary~\ref{g:cor:repfast} applies with $p=2$, in time $O(n^{1.99726})$; the other representable orbits are representable over $\Z$, $\F$ or $\mathbb F_3$, and Corollary~\ref{g:cor:repfast} gives $O(n^{1.99451})$. For each orbit of the third kind, the triple of Table~\ref{g:tab:functional4} lies in $\cl(P_g)\setminus P_g$, so $P_g$ is not representable by Lemma~\ref{g:lem:universal}(a); its $216$ Boolean restrictions are all representable, so by (d) it has no gadget of any size, which the gadget search confirms.

(f) The same script, in its mode \texttt{nonfunctional3}, lists by a backtracking search all relations $P\subseteq\{0,1,2\}^3$ whose $27$ Boolean restrictions are representable; there are $22100$. By (d) these are exactly the relations without a gadget of any size, and the gadget search confirms that none of them has one. The certified lattice tests find $1296$ of them not representable, in the $4$ orbits given. For the last representative this can be checked by hand: in the relation $R:=\{(a,b,c):a+b+c\equiv0\bmod3\}$ each of $\{000,111,222\}$ and $\{012,120,201\}$ contains every letter once in every position, so $e(000)=e(012)+e(120)+e(201)-e(111)-e(222)$, and $e(000)-e(s_0)$ lies in $L_P$ for $P:=R\setminus\{000\}$, as the coefficients add up to $1$; so $000\in\cl(P)\setminus P$.
\end{proof}

Proposition~\ref{g:prop:F}(b) also shows that the coordinatewise problems already need 3SUM over groups that are neither torsion-free nor elementary: by Lemma~\ref{g:lem:latinrep}, the Cayley table of every finite abelian group $H$ is a relation on $|H|$ letters that is representable exactly over the groups containing a copy of $H$, and $\Prob(P)$ for a Latin square isotopic to $H$ is 3SUM over $H^d$. Whether all of them are truly subquadratic is Open problem~\ref{g:prob:exponent}. Proposition~\ref{g:prop:F}(e) shows that the need is not confined to Latin squares: besides the Latin squares isotopic to $\Z_4$, eight other orbits of functional relations over four letters, such as $c=a+[b=3]\bmod4$, are representable over a group of exponent $4$ but over no $\Z^r$ and no $\Fp^r$. The control case $\Z_5$ is representable over $\mathbb F_5$ and over none of $\Z$, $\F$, $\mathbb F_3$, and has no gadget, as Proposition~\ref{g:prop:F}(a) predicts.

The relations of Proposition~\ref{g:prop:F}(c), (e) and (f) that lie outside both classes form a third kind. For the functional ones, $L_5$ and the four orbits of Table~\ref{g:tab:functional4}, $\Prob(P)$ is solved in time $O(n^2d\log n)$ by lookup; for these and for the relations of (f) we know neither a truly subquadratic algorithm nor an OV-hardness proof. By (d), the missing piece is not a larger gadget: a relation has a gadget of some size only if it has one of size $1$, which explains why no relation in Theorems~\ref{g:thm:D} and~\ref{g:thm:E} needs $t\ge2$, and whether it has one is decided by its Boolean restrictions. The third kind thus consists of the relations that are not representable although all their Boolean restrictions are, and recognising it is a question of representability alone. The four tables of Table~\ref{g:tab:functional4} share a shape: rows $0$ and $1$ are equal and columns $0$ and $1$ are equal, and on the remaining $3\times3$ grid the table is the Cayley table of $\Z_3$ with the letters $0,1,2$, except that one cell, or two cells symmetric about the diagonal, carry the fourth letter $3$.

The eight orbits of the third kind in Proposition~\ref{g:prop:F}(e) and~(f) are less varied than their number suggests. Proposition~\ref{g:prop:thirdT} reduces all of them to one problem $\Prob(T)$, the Cayley table of $\Z_3$ with one cell changed, and shows that this problem is easy on one natural family of instances, those in which one list is a coset of a subspace of $\mathbb F_3^d$; a hardness proof must therefore produce instances that avoid this. The loop $L_5$ is not covered.

\begin{proposition}[the third kind reduces to one problem, except $L_5$]\label{g:prop:thirdT}
The eight orbits of the third kind in Proposition~\ref{g:prop:F}(e) and~(f) reduce to one problem. Let $R_0:=\{(a,b,c)\in\{0,1,2\}^3:a+b+c\equiv0\bmod3\}$, so that the fourth representative of Proposition~\ref{g:prop:F}(f) is $P_0:=R_0\setminus\{000\}$, let $R_5:=R_0\setminus\{000,012,111,222\}$, $R_6:=R_0\setminus\{000,012,111\}$ and $R_7:=R_0\setminus\{000,012\}$ (the index is the number of triples), and let
\[
T:=\{(a,b,a+b\bmod3):(a,b)\ne(0,0)\}\cup\{(0,0,3)\}\subseteq\{0,1,2\}^2\times\{0,1,2,3\},
\]
the Cayley table of $\Z_3$ in which the single cell $(0,0)$ carries the fourth letter $3$.
\begin{enumerate}
\item[(1)] For each $P_g$ of Table~\ref{g:tab:functional4}, $\Prob(P_g)$ and $\Prob(T)$ reduce to each other in $O(nd)$ time.
\item[(2)] Up to permuting the letters in each position, the first, second and third representatives of Proposition~\ref{g:prop:F}(f) are $R_5$, $R_6$ and $R_7$, and
\[
\Prob(R_5),\ \Prob(R_6),\ \Prob(R_7)\ \le\ \Prob(P_0)\ \le\ \Prob(T),
\]
each reduction in $O(nd)$ time.
\item[(3)] Let $P$ be $T$ or $R_0\setminus D$ for some $D\subseteq R_0$, and consider the instances of $\Prob(P)$ in which one list is a coset of a subspace of $\mathbb F_3^d$ (for $T$, the first or the second list). On them $\Prob(P)$ is solved by a randomized algorithm that never answers YES wrongly and, when a solution exists, finds one with probability at least $\frac13$ (any constant below $1$ after a constant number of repetitions), in time $O(n^{2-1/20}(d+(\log n)^5))$, for every $c$.
\end{enumerate}
\end{proposition}

\begin{proof}
(1) If two letters $x\ne x'$ in one position of a relation $P$ have the same fibre, the same set of pairs of letters in the other two positions that complete them to a triple of $P$, then replacing $x'$ by $x$ in every vector of the corresponding list does not change the solutions, coordinate by coordinate; so $\Prob(P)$ and $\Prob$ of $P$ with $x'$ deleted reduce to each other in $O(nd)$ time. In each table of Table~\ref{g:tab:functional4} rows $0$ and $1$ are equal and columns $0$ and $1$ are equal, so the letter $1$ is deleted in positions $1$ and $2$. Renaming the letters $0,2,3$ of these positions as $0,1,2$ leaves the Cayley table $c=a+b$ of $\Z_3$ in which, for the first three tables, the cell $(2,0)$, $(2,2)$ or $(2,1)$ carries $3$ instead of $a+b$, and for the fourth table the two cells $(1,2)$ and $(2,1)$ carry $3$. In the first three cases the translation $a\mapsto a-u$, $b\mapsto b-v$, $c\mapsto c-u-v$ of the letters $0,1,2$, with $3$ fixed, is a permutation of the letters in each position and moves the cell to $(0,0)$, which gives $T$. In the fourth case both cells have $a+b=0$ and the cell $(0,0)$ keeps the letter $0$, so in the third position the letter $0$ occurs only at $(0,0)$ and the letter $3$ exactly at the other cells with $a+b=0$; exchanging $0$ and $3$ in the third position gives $T$. So for each $P_g$ of Table~\ref{g:tab:functional4}, $\Prob(P_g)$ and $\Prob(T)$ reduce to each other in $O(nd)$ time.

(2) The letter permutations $(0,1,2)\mapsto(2,1,0)$, $(2,0,1)$ and $(1,2,0)$ in the three positions map the first and the second representative of Proposition~\ref{g:prop:F}(f) onto $R_5$ and $R_6$, and the permutations $(1,2,0)$, $(2,1,0)$ and $(1,0,2)$ map the third onto $R_7$ (for example $010,022,102,111,200$ go to $201,210,120,102,021$). For every $D\subseteq R_0$, $\Prob(R_0\setminus D)$ reduces to $\Prob(P_0)$ in $O(nd|D|)$ time, with dimension $d|D|$: for $\varrho\in D$ the translate $\{(a+\varrho_1,b+\varrho_2,c+\varrho_3):(a,b,c)\in P_0\}$ is $R_0\setminus\{\varrho\}$, because translation by an element of $R_0$ maps $R_0$ onto itself and $000$ to $\varrho$; so $R_0\setminus D=\bigcap_{\varrho\in D}(R_0\setminus\{\varrho\})$, and mapping a vector $x$ of list $i$ to the concatenation over $\varrho\in D$ of the vectors $x-\varrho_i(1,\dots,1)$ (coordinates modulo $3$) gives an instance of $\Prob(P_0)$ with the same solutions. Finally $\Prob(P_0)$ reduces to $\Prob(T)$: renaming the third letter $c$ as $-c$ turns $P_0$ into $\{(a,b,a+b):(a,b)\ne(0,0)\}$, which is $T$ without $(0,0,3)$, and an instance of $P_0$ has no letter $3$ in its third list. With $|D|\le4$ for $R_5$, $R_6$ and $R_7$, each reduction takes $O(nd)$ time.

(3) After the renamings above, every triple of each of these relations satisfies $a+b=\pi(c)$, with $\pi$ the identity on $\{0,1,2\}$ and $\pi(3):=0$; the letters $0,1,2$ are read in $\mathbb F_3$, and for a third letter $z$ let $X(z):=\{a:(a,\pi(z)-a,z)\in P\}$, so that $(a,b,c)$ is a solution if and only if $b=\pi(c)-a$ and $a_j\in X(c_j)$ for every $j$. Suppose the second list is a coset $B=y+V$ of a subspace $V\le\mathbb F_3^d$ (for $R_0\setminus D$, which lies in the symmetric relation $R_0$, any list may play this role after permuting positions; for $T$, the first or the second).

\emph{Classes.} Let $k:=\dim V$, so $3^k=|B|\le n$, and let the rows of $G\in\mathbb F_3^{k\times d}$, with columns $g_j$, be a basis of $V$. Split $A$ into the classes $A_i:=A\cap(x_i+V)$ and put $C_i:=\{c\in C:\pi(c)\in x_i+y+V\}$, with $\pi$ applied coordinatewise; for $a\in A$ and $c\in C$ the vector $\pi(c)-a$ lies in $B$ if and only if $a\in A_i$ and $c\in C_i$ for the same $i$, so the solutions are the pairs $(a,c)\in A_i\times C_i$ with $a_j\in X(c_j)$ for all $j$.

\emph{The polynomial.} Write $a=x_i+\alpha G$, $\alpha\in\mathbb F_3^k$. For a fixed $c$, $\chi_j(\alpha):=\sum_{v\notin X(c_j)}\bigl(1-(x_{i,j}+\langle g_j,\alpha\rangle-v)^2\bigr)\in\mathbb F_3$ is the indicator that coordinate $j$ fails, a polynomial of degree at most $2$ in $\alpha$. As in the polynomial method of Razborov and Smolensky~\cite{Raz87,Smo87}, take random $r_1,\dots,r_\ell\in\mathbb F_3^d$, put $Q_t:=\sum_jr_{t,j}\chi_j$ and $\Upsilon_c(\alpha):=\prod_{t\le\ell}(1-Q_t(\alpha)^2)$, a polynomial of degree at most $4\ell$, whatever $d$ is. If $(x_i+\alpha G,c)$ is a solution, $\Upsilon_c(\alpha)=1$; otherwise the $Q_t(\alpha)$ are independent and uniform on $\mathbb F_3$, and $\Upsilon_c(\alpha)=1$ with probability $3^{-\ell}$.

\emph{The algorithm.} Put $s:=\lceil n^{1/20}\rceil$ and $\ell:=\lceil2\log_3s\rceil+1$, so $3^\ell\ge s^2$. For a class with $|C_i|<s$, test all pairs, in time $O(|A_i|sd)$. Otherwise cut $C_i$ into groups $\Gamma$ of at most $s$ elements, at most $2|C_i|/s$ of them; for each group draw fresh $r_t$ and uniform weights $w_c\in\mathbb F_3$, compute the coefficients of $F_\Gamma:=\sum_{c\in\Gamma}w_c\Upsilon_c$, reduced by $\alpha_\iota^3=\alpha_\iota$, by multiplying out the factors one at a time, evaluate $F_\Gamma$ at all $3^k$ points by one $3\times3$ transform per variable, in $O(k3^k)$ time, and for every $a\in A_i$ with $F_\Gamma(\alpha)\ne0$ test the pairs $(a,c)$, $c\in\Gamma$; answer YES at the first solution found. A YES is always a tested solution. If $(a^\ast,c^\ast)$ is a solution with $c^\ast\in\Gamma$, then $F_\Gamma(\alpha^\ast)$ is $w_{c^\ast}$ plus a sum independent of $w_{c^\ast}$, so it is nonzero with probability $\frac23$, and the solution is then found.

\emph{Cost.} A pair $(a,\Gamma)$ without a solution is tested only if some $\Upsilon_c(\alpha)=1$ wrongly, with probability at most $s3^{-\ell}\le1/s$; there are at most $\sum_i|A_i|\cdot2|C_i|/s\le2n^2/s$ such pairs, and a test costs $O(sd)$, so the wrong tests cost $O(n^2d/s)$ in expectation. The evaluations cost $O((n/s)k3^k)=O(n^2\log n/s)$. A reduced polynomial of degree at most $4\ell$ has at most $M:=\sum_{\iota\le4\ell}\binom k\iota2^\iota$ monomials, so the coefficients cost $O(\ell Mk^4)$ per $c$ (and $O(\ell dk^2)$ for the $Q_t$). If $3^k\le n^{0.95}$, then $M\le3^k\le n^{0.95}$. Otherwise $k\ge0.95\log_3n$ and $4\ell\le8\log_3s+8\le0.43k$ for large $n$, so $M\le2^{0.43k}2^{\mathrm H(0.43)k}\le2^{1.42k}\le3^{0.9k}\le n^{0.9}$, with $\mathrm H$ the binary entropy, $\mathrm H(0.43)<0.99$. The total expected time is $O(n^{1.95}(d+(\log n)^5))$, and stopping after three times this bound and answering NO keeps the error one-sided and misses a solution with probability at most $\frac13+\frac13$ by Markov's inequality.
\end{proof}

So a truly subquadratic algorithm for $\Prob(T)$ would settle all eight orbits, while quadratic hardness of one of $\Prob(R_5)$, $\Prob(R_6)$, $\Prob(R_7)$ would carry over to $\Prob(P_0)$, $\Prob(T)$ and the four functional orbits, but not to the other two of $R_5$, $R_6$, $R_7$. We do not know whether any of these reductions can be reversed. Whether $\Prob(T)$ is truly subquadratic, or quadratic-hard under a standard hypothesis, is open; for these eight orbits this is Open problem~\ref{g:prob:L5}.

The general relations over four letters are too many to enumerate as in Proposition~\ref{g:prop:F}(e): Burnside's lemma gives about $2.2\cdot10^{14}$ orbits under the group of order $(4!)^3\cdot6$. The criterion of Proposition~\ref{g:prop:F}(d) cuts them down: a backtracking search finds $186526704$ relations over four letters whose Boolean restrictions are all representable, that is, without a gadget, and in a uniform sample of $20000$ of them $10979$, about $55\%$, are not representable (\texttt{q2\_functional4.py local4}; the sample is evidence, not a proof). So the third kind is not rare among the relations without a gadget over four letters, and a classification of them would be a three-way table.

\begin{problem}\label{g:prob:L5}
For the relations of the third kind, the four functional relations over four letters of Table~\ref{g:tab:functional4}, the four orbits of Proposition~\ref{g:prop:F}(f), such as $\{(a,b,c):a+b+c\equiv0\bmod3\}\setminus\{(0,0,0)\}$, and $L_5$: is $\Prob(P)$ solvable in time $O(n^{2-\delta})$ for a $\delta>0$ independent of $c$, or does it have no such algorithm under OVH or another standard hypothesis? More generally: classify all relations $P\subseteq\Sigma^3$ for $|\Sigma|=4$ at the quadratic level. By Proposition~\ref{g:prop:F}(d), the relations with a gadget are recognised by their $216$ Boolean restrictions, so the task is to classify the others.
\end{problem}

\subsection{The computations}\label{g:sec:computer}

Theorems~\ref{g:thm:D} and~\ref{g:thm:E} and the counts in Proposition~\ref{g:prop:F}(b), (e) and (f) rest on exhaustive computations, which are part of their proofs (kind C in Table~\ref{tab:anc}), and so do Theorem~\ref{g:thm:triD} and Proposition~\ref{g:prop:triF}(b), which use these classifications; Proposition~\ref{g:prop:F}(d) is proved by hand from Theorem~\ref{g:thm:D}, and everything else in Sections~\ref{g:sec:groups}--\ref{g:sec:triangle} is proved by hand. The scripts are listed with their running times in Table~\ref{tab:anc}.

\paragraph{Representability.} For a relation $P\ne\emptyset$, \texttt{q2lib.py} forms the integer matrix $M$ whose rows are $e(s)-e(s_0)$, $s\in P\setminus\{s_0\}$, and computes its Smith normal form $UMV=\mathrm{diag}(d_1,\dots,d_\varrho)$ with unimodular $U$, $V$, by a hand-written routine over Python integers. For $x=e(s)-e(s_0)$ and $y:=xV$, Lemma~\ref{g:lem:universal} and its proof give the tests: $x\in L_P$ if and only if $y_i=0$ for $i>\varrho$ and $d_i\mid y_i$ for $i\le\varrho$; $x\in\Q L_P$ if and only if $y_i=0$ for $i>\varrho$; and $x\in L_P+p\Z^{3k}$ if and only if $p\mid y_i$ for $i>\varrho$ and $\gcd(d_i,p)\mid y_i$ for $i\le\varrho$. $P$ is representable if no $s\notin P$ passes the first test, over $\Z$ if none passes the second, and over $\Fp$ if none passes the third. Every verdict is certified independently of the correctness of the Smith normal form routine. A membership is certified by an explicit coefficient vector $z$ (integer, rational or modulo $p$) with $zM=x$, which is checked. A non-membership is certified by checking that the integer map $x\mapsto xV$ sends every generator of $L_P$ into the row space of $\mathrm{diag}(d_i)$ (respectively its $\Q$- or mod-$p$ version) while it does not send $x$ there; since a linear map that sends the generators into a subgroup sends their span into it, this shows $x\notin L_P$ (respectively $\notin\Q L_P$, $\notin L_P+p\Z^{3k}$) whatever $V$ is. For the Boolean relations the script also checks $UMV=D$ and $\det U,\det V=\pm1$, and compares with a brute-force search over integer combinations with coefficients in $[-2,2]$, with no mismatch on the $256$ relations. Representability over $\Fp$ is tested for $p=2,3$ (and $5$, $7$ for the Latin squares and the functional relations over four letters) and for every prime that divides some $d_i$; by Lemma~\ref{g:lem:universal}(e), no other prime can give a representation when $\Z$ does not. For the representable functional relations over four letters, \texttt{q2\_functional4.py} also tests representability over a group of exponent dividing $4$, that is, membership in $L_P+4\Z^{12}$ (Lemma~\ref{g:lem:universal}(c)), by the same Smith normal form: $x\in L_P+4\Z^{12}$ if and only if $4\mid y_i$ for $i>\varrho$ and $\gcd(d_i,4)\mid y_i$ for $i\le\varrho$, with a membership certified by a coefficient vector modulo $4$ and a non-membership by the map $x\mapsto xV$ as above.

\paragraph{Gadgets.} For each of the $12$ side assignments $\theta$ and each choice of the maps $f_{i}$ for one coordinate ($k^2$ choices per reading position, $k$ per constant position), the script computes the mask of Section~\ref{g:sec:gadgets}, vectorised over all relations at once. A gadget of size $1$ exists if the target set itself is a mask; a gadget of some size exists if the intersection of all masks that contain the target set is the target set; the least size is found by a breadth-first search over intersections of masks. So gadgets of every size are decided, not only of size $1$ or $2$.

\paragraph{Orbits.} The first three scripts compute a canonical form of each relation, the least code over the symmetry group (of order $48$ for Boolean relations, $432$ for functional relations over three letters, and isotopy with transposition for Latin squares of order $4$), and check that every verdict is constant on each orbit; \texttt{q2\_functional4.py nonfunctional3} uses the same canonical form, over the group of order $1296$, to sort the relations of Proposition~\ref{g:prop:F}(f) into orbits. For the functional relations over three letters all $19683$ maps are treated, not only orbit representatives. For the $4^{16}$ functional relations over four letters this is too slow, and \texttt{enum\_functional.c} lists one map per orbit by orderly generation: it runs through the tables $g$ whose letters appear in the order of their first occurrence (the least relabelling of the third position), and keeps $g$ if no permutation of the rows and columns, with or without transposition, followed by that relabelling, gives a lexicographically smaller table; the orbit size follows from the number of permutations that give $g$ itself. The script checks the number of tables kept against Burnside's lemma and the sum of the orbit sizes against $4^{16}$, and treats one map per orbit, which suffices because representability and gadgets are invariant under the symmetries (Section~\ref{g:sec:gadgets}). For all relations over three letters, \texttt{q2\_functional4.py nonfunctional3} lists those without a gadget by Proposition~\ref{g:prop:F}(d), as the relations whose $27$ Boolean restrictions are all among the $60$ representable Boolean relations, by a backtracking search over the $27$ cells that stops as soon as a partly filled restriction can no longer be completed to a representable one.

\paragraph{What the outputs say.} \texttt{q2\_boolean.out}: $22$ orbits; no relation is representable and has a gadget of size $1$, and none has neither; no relation has a gadget of some size but none of size $1$; every representable relation is representable over $\Z$ or $\F$; the per-orbit lines of Table~\ref{g:tab:boolean}. \texttt{q2\_functional3.out}: $84$ orbits; $297$ representable maps ($159$ over $\Z$, $105$ over $\F$, $243$ over $\mathbb F_3$), $19386$ with a gadget of size $1$, none with both, none with a gadget only of larger size, and the invariant factors of $A_{P_g}$ (torsion $\Z_2$ or $\Z_3$ or none, free rank $3$ to $5$). \texttt{q2\_latin.out}: $576$ Latin squares of order $4$ in two orbits of sizes $144$ and $432$, the first with torsion $\Z_2^2$ and representable over $\F$, the second with torsion $\Z_4$ and representable over none of $\Z$, $\F$, $\mathbb F_3$, $\mathbb F_5$, neither with a gadget of any size; $L_5$ a non-associative loop, not representable (the closure contains $(0,0,1)$), with no gadget of any size; and $\Z_5$ representable over $\mathbb F_5$ only, with no gadget.

\subsection{What is known and what is new}\label{g:sec:coordremarks}

\emph{Known.} Besides the work of Bringmann, Fischer and Künnemann~\cite{BFK19} (Section~\ref{g:sec:boolean}), Dalirrooyfard, Lincoln and Vassilevska Williams~\cite{DLW20} study factored coordinatewise problems, where each input object carries a set of values per block of coordinates; in that model factored $k$-XOR is already $n^{k-o(1)}$-hard under the $k$-OV hypothesis, and the counting versions of all predicates are equivalent~\cite[Theorems~1.1 and 1.7--1.9]{DLW20}, so the distinction between XOR-type and OV-type predicates disappears there, and their results do not classify the plain problems $\Prob(P)$. Gao, Impagliazzo, Kolokolova and Williams~\cite{GIKW17} show that sparse OV is complete for first-order properties, the counterpart of~\cite{BFK19} on the hardness side. We found no paper that classifies the coordinatewise problems at the quadratic level or over alphabets of more than two letters.

\emph{New, to our knowledge.} Both sides of the dichotomy are routine (one-coordinate gadgets for the hard side; a finite computation for the classification); what is not routine is that the representable side is now provably truly subquadratic, through the preprint, Section~\ref{x:sec:xor} and Section~\ref{g:sec:groups}. The examples where the three groups stop sufficing and where the dichotomy stops (Proposition~\ref{g:prop:F}(b), (c), (e) and (f)) are, as far as we found, new observations, and so is the reduction of gadgets to Boolean restrictions in Proposition~\ref{g:prop:F}(d), whose proof is short once Theorem~\ref{g:thm:D} is available. Proposition~\ref{g:prop:thirdT} adds two observations on the third kind: the reductions that bring its eight orbits in Proposition~\ref{g:prop:F}(e) and~(f) to the single problem $\Prob(T)$, and the coset algorithm, an application of the polynomial method of Razborov and Smolensky.

\section{Triangle versions}\label{g:sec:triangle}

Section~\ref{x:sec:xor} proves its triangle version, Exact-XOR-Triangle (Theorem~\ref{x:thm:xet}), not through the 3XOR reduction (blocks, probes, a global count) but through the Exact Triangle reduction of Section~\ref{e:sec:statement} with a linear hash in place of the modulus. Sections~\ref{g:sec:triangle:el} and~\ref{g:sec:triangle:p2} carry that proof over to other groups: over $\Fp^d$ with the whole saving, and uniformly in $p$ with Part~\hyperref[m:partII]{II} (Section~\ref{g:sec:triangle:el}); for exponent $p^2$ only in part (Section~\ref{g:sec:triangle:p2}). Section~\ref{g:sec:tri} does the same for the coordinatewise problems: in the triangle version $\mathrm{Tri}(P)$ the letters sit on the edges of a complete tripartite graph, and the dichotomy of Theorems~\ref{g:thm:D} and~\ref{g:thm:E} holds one level up, between truly subcubic and cubic, under the same hypothesis, the OV hypothesis, on the hard side. At four letters the triangle version has more open cases than $\Prob(P)$: besides the relations of the third kind, the Latin squares isotopic to $\Z_4$ and the other functional relations that need a group of exponent $4$ now lead to Exact Triangle over a group of exponent $4$, such as $\Z_4^d$, which is open. Section~\ref{g:sec:triremarks} says what is known and what is new. The conventions and the model of Section~\ref{g:sec:groups} hold here.

\subsection{Elementary groups and a bounded number of factors}\label{g:sec:triangle:el}

\begin{definition}[Exact Triangle over a group]\label{g:def:triangle}
Let $G$ be a finite abelian group, given as in the paragraph ``Model'' at the start of Section~\ref{g:sec:groups}, with $|G|\le n^{c'}$ for a constant $c'>0$. An instance of \emph{Exact Triangle over $G$} is a complete tripartite graph on parts $A,B,C$ of at most $n$ vertices each with a weight $w(e)\in G$ on every edge; with $S(a,b,c):=w(a,b)+w(b,c)+w(a,c)$, the task is to decide whether $S(a,b,c)=0$ for some triangle $(a,b,c)\in A\times B\times C$. The \emph{all-edges version} asks, for every pair $(a,b)\in A\times B$, whether some $c\in C$ has $S(a,b,c)=0$. Exact-XOR-Triangle (Definition~\ref{x:def:xor}) is the case $G=\F^{d'}$.
\end{definition}

\paragraph{Pieces, counts and the instance for a homomorphism.} Pieces $C_1,\dots,C_h$ of at most $\sigma$ vertices, $h:=\lceil|C|/\sigma\rceil$, incidences $x=(a,b,k)$ and $z(x):=\#\{c\in C_k:S(a,b,c)=0\}$ are those of Section~\ref{e:sec:instances}. For a homomorphism $K\colon G\to\Gamma$ into a finite abelian group $\Gamma$ (given as in the paragraph ``Model'') let $\cnt_K(x):=\#\{c\in C_k:K(S(a,b,c))=0\}$, $f_K(x):=\cnt_K(x)-z(x)\ge0$, the false positives of $x$ under $K$, and $F_\Iq(K):=\sum_{x\in\Iq}f_K(x)$ for a set $\Iq$ of incidences; for the zero map, $F_\Iq(0)\le\sigma|\Iq|$. For $k\in[h]$, $\varrho\in\Gamma$ and a set $\mathcal Q$ of pairs $(a,b)$ with $K(w(a,b))=\varrho$, the instance $J(K,k,\varrho,\mathcal Q)$ of $\#\Lop$ has the sides $A$ and $B$ (padded to $n$ vertices), the middle part $C_k\times\Gamma$, so $D:=\sigma|\Gamma|$, the wanted set $\mathcal Q$, and the edges $a\sim(c,\lambda)$ if and only if $\lambda=K(w(a,c))+\varrho$, and $(c,\lambda)\sim b$ if and only if $\lambda=-K(w(b,c))$. As in Lemma~\ref{e:lem:counts}, for $(a,b)\in\mathcal Q$ and $c\in C_k$ there is a common neighbour $(c,\lambda)$ if and only if $K(w(a,c))+\varrho=-K(w(b,c))$, that is, $K(S(a,b,c))=0$ (as $\varrho=K(w(a,b))$), and then exactly one; so the number of common neighbours of $a$ and $b$ is $\cnt_K(a,b,k)$. The instance is a $0/1$ graph.

\begin{lemma}[one evaluation]\label{g:lem:trieval}
Let $\Lambda\ge1$, $\sigma\ge2$ and $\Gamma$ satisfy $2\le|\Gamma|\le\Lambda\sigma$, $\Lambda\sigma^2\le n^\eps$ and $2n\Lambda\le n^2$. Then the counts $\cnt_K(x)$ for all $x$ in a set $\Iq$ of incidences are computed with $\Hyp(\eps,\gamma)$, $0<\eps,\gamma\le\frac12$, in time $O(n^3\Lambda^{1-\gamma}\sigma^{-2\gamma}+n^2T_K)$, $T_K$ the time to evaluate $K$, the constant depending only on $C_{\mathcal O}$.
\end{lemma}

\begin{proof}
This is the proof of Lemma~\ref{e:lem:shape} with $|\Gamma|$ in place of $m$ and $\Lambda$ in place of $\Lambda_{\mathrm E}$: $D=\sigma|\Gamma|\le\Lambda\sigma^2\le n^\eps$ and $D\ge4$; piece $k$ needs at most $|\Gamma|+2\sqrt{\sigma|\Gamma|}$ calls of cost at most $C_{\mathcal O}n^2(\sigma|\Gamma|)^{-\gamma}$, which is at most $3C_{\mathcal O}n^2\sigma^{1-2\gamma}$ if $|\Gamma|\le\sigma$ and at most $3C_{\mathcal O}n^2\Lambda^{1-\gamma}\sigma^{1-2\gamma}$ if $\sigma<|\Gamma|\le\Lambda\sigma$, and $h\le2n/\sigma$; the bookkeeping is that of Lemma~\ref{e:lem:shape}, with the elements of $\Gamma$ stored as integers below $|\Gamma|$ for the radix sort ($h|\Gamma|\le2n\Lambda\le n^2$), plus $O(n^2T_K)$ for the values $K(w(e))$. For $p=2$ and linear $K$ this is the cost bound in Section~\ref{x:sec:xet}.
\end{proof}

\begin{theorem}[Exact Triangle over $\Fp^{d'}$]\label{g:thm:triFp}
Assume $\Hyp(\eps,\gamma)$ with $0<\eps,\gamma\le\frac12$, and let $c'>0$. There is a deterministic algorithm that, given $n$, a prime $p\le n^{\eps/4}$ and an instance of Exact Triangle over $\Fp^{d'}$ with $p^{d'}\le n^{c'}$, solves it, and also its all-edges version, in time $O(p^6n^{3-\eps\gamma}(\log n)^6)$, the constant depending only on $c'$, $\eps$, $\gamma$ and $C_{\mathcal O}$. For fixed $p$ this is the bound of Theorem~\ref{x:thm:xet}.
\end{theorem}

\begin{proof}
The proof of Theorem~\ref{x:thm:xet} (Section~\ref{x:sec:xet}) with $\Fp$ in place of $\F$; we list what changes. Linear maps $H\colon\Fp^{d'}\to\Fp^l$ are given by their rows (Section~\ref{g:sec:rows}), and $\cnt_H$, $f_H$, $F_\Iq(H)$ and $J(H,k,\varrho,\mathcal Q)$ are those defined above, with $\Gamma=\Fp^l$.

\emph{Parameters.} Put $\Lambda:=2ep^2$, $\sigma:=\lfloor\sqrt{n^\eps/\Lambda}\rfloor$, $q:=\lceil\log_p(2e\sigma)\rceil+1$, $L:=\lceil\log_2n\rceil$, $\beta:=1/q$, and let $E$ be the $\beta$-value-biased, hence $\beta$-kernel-biased, multiset of Lemma~\ref{g:lem:powering} in $\Fp^{d'}$, of fewer than $p^4d'^2q^2$ rows, listed in time $O(p^3(\log n)^{O(1)})$. Since $\lceil\log_p(2e\sigma)\rceil-1<\log_p(2e\sigma)$, we have $p^{q-2}<2e\sigma\le p^{q-1}$, so $p^{q-1}\ge2e\sigma$ and $p^q<2ep^2\sigma=\Lambda\sigma$; and $\Lambda\sigma^2\le n^\eps$. We assume $n\ge n_0(\eps)$, a threshold that does not depend on $p$: since $p\le n^{\eps/4}$, $\Lambda\le2en^{\eps/2}$, so $n^\eps/\Lambda\ge n^{\eps/2}/(2e)\ge4$, hence $\sigma\ge2$, and $2n\Lambda\le n^2$, once $n^{\eps/2}\ge8e$ and $n^{1-\eps/2}\ge4e$; smaller $n$ are solved by brute force in $O(1)$ time. Also $d'\le c'\log n/\log p\le c'\log n$, $q\le\log_2(2en)+2=O(\log n)$, and $\sigma\ge\frac12\sqrt{n^\eps/\Lambda}$.

\emph{Cost of one evaluation.} For $1\le l\le q$ rows, $p^l\le p^q<\Lambda\sigma$, so Lemma~\ref{g:lem:trieval} applies with $\Gamma=\Fp^l$: one evaluation costs $O(n^3\Lambda^{1-\gamma}\sigma^{-2\gamma}+n^2qd')$. With $\sigma^{-2\gamma}\le(4\Lambda)^\gamma n^{-\eps\gamma}$ and $4^\gamma\le2$ this is $O(\Lambda n^{3-\eps\gamma})=O(p^2n^{3-\eps\gamma})$.

\emph{Detection.} The algorithm of Section~\ref{x:sec:xet}: $L$ stages; in stage $t$, $q$ rows $u\in E$ chosen one at a time, each minimising $\sum_{x\in\Iq_{t-1}}\cnt_{(H,u)}(x)$, giving $H_t$; YES if $\sum_{x\in\Iq_{t-1}}\cnt_{H_t}(x)>|\Iq_{t-1}|/2$; otherwise $\Iq_t:=\{x\in\Iq_{t-1}:\cnt_{H_t}(x)>0\}$; after $L$ stages, a scan of $C_k$ for every surviving incidence. The halving lemma: the sum of the counts is $F_\Iq(H)$ plus a term that does not depend on $H$, and the false positives are triangles with $S(a,b,c)\ne0$; Lemma~\ref{g:lem:greedy} with Lemma~\ref{g:lem:onerow}, applied to these nonzero vectors with weight $1$ and $\vartheta=(1+\beta)/p$, gives $F_\Iq(H_t)\le e\,p^{-q}\sigma|\Iq|\le|\Iq|/(2p)\le|\Iq|/4$, since $p^q\ge2ep\sigma$. The rest of Lemmas~\ref{e:lem:halving} and~\ref{e:lem:correctA}, as used in Section~\ref{x:sec:xet}, uses nothing about the hash.

\emph{All edges.} As in Section~\ref{x:sec:xet}: in each stage, $q-1$ rows chosen one at a time give $H$; then $\cnt_{(H,u)}(x)$ is evaluated for every $u\in E$, and $\mu(x):=\min_{u\in E}\cnt_{(H,u)}(x)$. Call $x$ \emph{certified false} if $\mu(x)=0$, \emph{certified true} if $p\,\mu(x)>(1+\beta)\cnt_H(x)$, and \emph{uncertified} otherwise; mark YES the pair of every certified-true incidence, let $\Iq_t$ be the uncertified incidences whose pair is not marked, and scan $C_k$ for every incidence of $\Iq_L$ at the end. Lemma~\ref{x:lem:cert} holds with these changes. (i) $\mu(x)=z(x)+\min_uf_{(H,u)}(x)\ge z(x)$. (ii) The false positives of $x$ under $(H,u)$ are its false positives $c$ under $H$ with $\langle u,S(a,b,c)\rangle=0$, and $S(a,b,c)\ne0$ for them; by Lemma~\ref{g:lem:onerow} their number averaged over $u\in E$ is at most $\frac{1+\beta}pf_H(x)\le\frac{1+\beta}p\cnt_H(x)$, which is less than $\mu(x)$ for a certified-true $x$; so $z(x)=\mu(x)-\min_uf_{(H,u)}(x)>0$. (iii) If $f_H(x)=0$, then $f_{(H,u)}(x)=0$ for every $u$ and $\mu(x)=z(x)=\cnt_H(x)$, so $x$ is certified false if $z(x)=0$ and certified true if $z(x)\ge1$, as $p\ge2>1+\beta$; so an uncertified $x$ has $f_H(x)\ge1$. (iv) By (iii), $|\Iq_t|\le F_\Iq(H)\le e\,p^{-(q-1)}\sigma|\Iq|\le|\Iq|/2$, by Lemma~\ref{g:lem:greedy} with $q-1$ rows and $p^{q-1}\ge2e\sigma$. Correctness is that of Lemma~\ref{e:lem:correctB}, as in Section~\ref{x:sec:xet}.

\emph{Total.} Each stage has $q$ rounds of $|E|$ evaluations, so there are $Lq|E|<Lq\cdot p^4d'^2q^2=O(p^4(\log n)^6)$ evaluations, each $O(p^2n^{3-\eps\gamma})$: $O(p^6n^{3-\eps\gamma}(\log n)^6)$ in all. Listing $E$ costs $O(p^3(\log n)^{O(1)})$, the incidences $O(n^3/\sigma)=O(pn^{3-\eps/2})\le O(pn^{3-\eps\gamma})$ (as $\sigma\ge\frac12n^{\eps/2}/\sqrt\Lambda$, $\sqrt\Lambda=O(p)$ and $\eps/2\ge\eps\gamma$), and the final scans $O(|\Iq_0|\sigma d'/2^L)=O(n^2\log n)$, since $|\Iq_0|\le2n^3/\sigma$ and $2^L\ge n$. An element of $\Fp^{d'}$ is a vector of $d'$ residues below $p\le n$, so the arithmetic is on $O(1)$-word numbers.
\end{proof}

\begin{proposition}[bounded number of cyclic factors]\label{g:prop:triboundedrank}
Let $G=\Z_{m_1}\times\dots\times\Z_{m_t}$ with $|G|\le n^{c'}$. Exact Triangle over $G$, and its all-edges version, reduce in $O(3^ttn^2)$ time to $3^t$ instances of integer Exact Triangle on the same vertex sets, with weights of absolute value at most $2\cdot3^tn^{c'}$: a pair $(a,b)$ lies in a zero triangle of the given instance if and only if it lies in one in some of the $3^t$ instances.
\end{proposition}

\begin{proof}
The proof of Proposition~\ref{g:prop:boundedrank} with the three edge weights of a triangle in place of the three list elements: with $X$ and $T_\kappa$ as there, $S(a,b,c)=0$ in $G$ if and only if $X(w(a,b))+X(w(b,c))+X(w(a,c))=T_\kappa$ for some $\kappa\in\{0,1,2\}^t$. For each $\kappa$, give every edge $(a,b)$ the weight $X(w(a,b))-T_\kappa$ and every other edge $e$ the weight $X(w(e))$.
\end{proof}

\begin{theorem}[Exact Triangle, uniform in $p$]\label{g:thm:triuniform}
Assume $\Hyp(\eps,\gamma)$ with $0<\eps,\gamma\le\frac12$. For every constant $c'>0$ there is one deterministic algorithm that, for every prime $p$ and every $d'$ with $p^{d'}\le n^{c'}$, solves Exact Triangle over $\Fp^{d'}$, and one that solves its all-edges version, in time $n^{3-\eps\gamma+o(1)}$, where the $o(1)$ does not depend on $p$.
\end{theorem}

\begin{proof}
Put $P_0:=2^{\sqrt{\log n}}$. If $p\le P_0$, then $p\le n^{\eps/4}$ for large $n$, and Theorem~\ref{g:thm:triFp} gives time $O(p^6n^{3-\eps\gamma}(\log n)^6)\le O(n^{3-\eps\gamma}2^{6\sqrt{\log n}}(\log n)^6)=n^{3-\eps\gamma+o(1)}$, with a constant and a threshold $n_0(\eps)$ that do not depend on $p$. If $p>P_0$, then $d'\le c'\log n/\log p<c'\sqrt{\log n}$, and Proposition~\ref{g:prop:triboundedrank} with $t=d'$ gives $3^{d'}\le2^{1.6c'\sqrt{\log n}}=n^{o(1)}$ instances of integer Exact Triangle with weights of absolute value at most $2\cdot3^{d'}n^{c'}\le n^{c'+1}$ for large $n$. Corollary~\ref{e:cor:uniform} with $\nu:=c'+1$ solves each of them, in either version, in time $n^{3-\eps\gamma+o(1)}$.
\end{proof}

\begin{proposition}[a small non-elementary factor]\label{g:prop:trismall}
Assume $\Hyp(\eps,\gamma)$ with $0<\eps,\gamma\le\frac12$, and let $c'>0$. Let $G=G_0\times\Fp^{d'}$, where $G_0$ is a finite abelian group given as a product of cyclic groups, $g_0:=|G_0|\le n^{\eps/4}$, $p\le n^{\eps/4}$ is prime and $p^{d'}\le n^{c'}$. Exact Triangle over $G$, and its all-edges version, are solved deterministically in time $O(g_0p^6n^{3-\eps\gamma}(\log n)^6)$. In particular, if $g_0=n^{o(1)}$ and $p$ is fixed (for example $G=\Z_4^a\times\F^{d'}$ with $a=o(\log n)$), in time $n^{3-\eps\gamma+o(1)}$.
\end{proposition}

\begin{proof}
The proof of Theorem~\ref{g:thm:triFp} with the homomorphism $K_H(x,y):=(x,Hy)\in G_0\times\Fp^l$ in place of $H$ (the maps of Remark~\ref{g:ex:Z4F2}), $\Gamma:=G_0\times\Fp^l$ and $\Lambda:=2ep^2g_0$; $\sigma$, $q$, $L$ and $E$ are defined from $\Lambda$ and $p$ as there. Then $|\Gamma|=g_0p^l\le g_0p^q<2ep^2g_0\sigma=\Lambda\sigma$ for $l\le q$, $\Lambda\sigma^2\le n^\eps$, and $\Lambda\le2en^{3\eps/4}$, so $\sigma\ge2$ and $2n\Lambda\le n^2$ for $n\ge n_0(\eps)$. One evaluation costs $O(n^3\Lambda^{1-\gamma}\sigma^{-2\gamma}+n^2T_K)=O(g_0p^2n^{3-\eps\gamma})$ by Lemma~\ref{g:lem:trieval}, with $T_K=O(\log g_0+qd')$. A false positive under $K_H$ has $S=(s_0,s)\ne0$ with $s_0=0$ and $Hs=0$, hence $s\ne0$; so the bounds of Lemmas~\ref{g:lem:onerow} and~\ref{g:lem:greedy} apply to the vectors $s$, starting from the map $(x,y)\mapsto x$, under which there are at most $\sigma|\Iq|$ false positives (its counts are never needed). Under $K_{(H,u)}$ the false positives are those under $K_H$ with $\langle u,s\rangle=0$, so the certification of Theorem~\ref{g:thm:triFp} is unchanged. The number of evaluations is that of Theorem~\ref{g:thm:triFp}.
\end{proof}

Together with Proposition~\ref{g:prop:triboundedrank}, this covers every group with a bounded number of cyclic factors and every group of the form $G_0\times\Fp^{d'}$ whose non-elementary part $G_0$ has $n^{o(1)}$ elements. It does not cover $\Z_4^d$ with $d=\Theta(\log n)$, nor $G_0\times\Fp^{d'}$ with $G_0$ of polynomial size and many cyclic factors.

\begin{corollary}\label{g:cor:explicitT}
With $\Hyp(1/14,\,0.077)$ (Section~\ref{x:sec:consequences}, from Corollary~\ref{m:cor:explicit}(b)), the following hold deterministically, also for the all-edges versions. \textup{(a)} For every fixed prime $p$, Exact Triangle over $\Fp^d$ with $p^d\le n^{c'}$ is solved in time $O(n^{2.99451})$. \textup{(b)} One algorithm, uniform in $p$, solves it for every prime $p$ in time $n^{2.99450+o(1)}$. \textup{(c)} For a fixed prime $p$, Exact Triangle over $G_0\times\Fp^{d'}$ is solved in time $O(n^{2.99451})$ if $|G_0|$ is bounded and in time $n^{2.99450+o(1)}$ if $|G_0|=n^{o(1)}$.
\end{corollary}
\begin{proof}
$\eps\gamma=0.077/14=0.0055$, and a fixed $p$ and a bounded $|G_0|$ satisfy $p,|G_0|\le n^{\eps/4}=n^{1/56}$ for large $n$. Theorem~\ref{g:thm:triFp} gives $O(n^{3-0.0055}(\log n)^6)$, Theorem~\ref{g:thm:triuniform} gives $n^{3-0.0055+o(1)}$, and Proposition~\ref{g:prop:trismall} gives $O(|G_0|\,n^{3-0.0055}(\log n)^6)$. Rounding $2.9945$ up to $2.99451$ absorbs the polylogarithmic factors for large $n$, as in Corollary~\ref{x:cor:xor}.
\end{proof}

With $\Hyp(1/18,0.063)$, from the preprint's Corollary 26, the bounds of (a) and (b) are $O(n^{2.99651})$ and $n^{2.99650+o(1)}$.

\subsection{Exponent \texorpdfstring{$p^2$}{p\^{}2}}\label{g:sec:triangle:p2}

We could not carry the method of Theorem~\ref{g:thm:G} over to triangles. Remark~\ref{g:rem:trinogo} explains where it breaks; Propositions~\ref{g:prop:trilight} and~\ref{g:prop:tricopy} solve the instances on which two of the obstacles it names do not bite, and Proposition~\ref{g:prop:tricert} shows why hardness for the remaining instances cannot come from a deterministic reduction from CNF-SAT unless NSETH fails. What remains is Open problem~\ref{g:prob:triexp}.

\begin{remark}[why the method of Theorem~\ref{g:thm:G} does not transfer to triangles]\label{g:rem:trinogo}
Let $G=R^d$, $R=\Z_{p^2}$ (Section~\ref{g:sec:G:setup} reduces every group of exponent dividing $p^2$ to this), and call a triangle of type I if $S\in pG\setminus\{0\}$ and of type II if $\bar S\ne0$, as for triples there. The following three points explain why the proof of Theorem~\ref{g:thm:G} does not carry over; they are statements about this method, not lower bounds for the problem.

(1) \emph{Every digit of the hash is paid for in the instance of a call.} In 3SUM the hash is split into a grouping $\pi$ and the middle-part hash $K$, and only the subgroup $M\supseteq K(\ker\pi)$ enters the size of the middle part (Lemmas~\ref{g:lem:instance} and~\ref{g:lem:onecall}); Theorem~\ref{g:thm:G} puts the residue digits into $\pi$, so that a type-I triple is filtered by $M$ at the full rate $1/p$ per row while $M$ has $p$ elements per row. In the triangle reduction a triangle is located only by its pair $(a,b)$ and the piece $C_k$, a fixed subset of $C$ that does not depend on any hash, and the instance $J$ has the middle part $C_k\times K(G)$; one evaluation costs about $n^3|K(G)|^{1-\gamma}\sigma^{-1-\gamma}$ when $|K(G)|\ge\sigma$. By Proposition~\ref{g:prop:C}, case (1), with $A=B=pG$, every homomorphism has $|K(pG)|^2\le|K(G)|$. If the type-I values among the false positives are spread over $pG\setminus\{0\}$, the hash keeps about a fraction $1/|K(pG)|\ge|K(G)|^{-1/2}$ of them, so cutting them by the factor $2\sigma$ that the halving needs requires $|K(G)|\ge4\sigma^2$, and then one evaluation costs about $n^3\sigma^{1-3\gamma}$, more than $n^3$ for $\gamma<\frac13$.

(2) \emph{The light/heavy split has no counterpart.} Lemma~\ref{g:lem:light} uses that the residue class of one element of a triple with $\bar v=0$ is determined by the other two. For a triangle, $\bar S(a,b,c)=0$ is a condition on three edges, two of which change with $c$ when $(a,b)$ is fixed, and classes of edges at a vertex do not bound $T_I$: for an injective $h\colon A\cup B\cup C\to\Fp^d$, the weights $w(a,b):=\lift(h(b)-h(a))$, $w(b,c):=\lift(h(c)-h(b))$, $w(a,c):=\lift(h(a)-h(c))$, each plus an arbitrary element of $pG$, have every class $\{c:\bar w(a,c)=\varrho\}$ and $\{b:\bar w(b,c)=\varrho\}$ of at most one element, yet $\bar S=0$ for all triangles. (This instance is easy: $S$ does not change if $g(a)-g(b)$, $g(b)-g(c)$ and $g(c)-g(a)$ are added to $w(a,b)$, $w(b,c)$ and $w(a,c)$, and $g:=\lift\circ h$ moves all weights into $pG\cong\Fp^d$. A useful notion of lightness would have to be invariant under this re-gauging, and we have no algorithmic split along such a notion.) The heavy case of Theorem~\ref{g:thm:G} makes the probes residue-pure; the analogue would need $\bar w(a,c)$ and $\bar w(b,c)$ constant over $c\in C_k$ for every incidence asked in a call, and since the pieces are shared by all pairs, $C_k$ would have to lie in one class of each of the partitions $\{c:\bar w(a,c)=\varrho\}$, $a\in A$; for generic residues these refine to singletons, which forces $\sigma=1$.

(3) \emph{Type-I triangles cluster.} An incidence survives a stage as soon as one of its triangles does, so an incidence that holds many type-I triangles is not thinned out like the others. This is why Proposition~\ref{g:prop:trilight} has the additive term $n^{\eps/2}T_\psi$ rather than a rate.
\end{remark}

\begin{proposition}[few triangles with sum in a subgroup of elementary quotient]\label{g:prop:trilight}
Assume $\Hyp(\eps,\gamma)$ with $0<\eps,\gamma\le\frac12$, and let $c'>0$. Let $G$ be a finite abelian group with $|G|\le n^{c'}$, $p\le n^{\eps/4}$ a prime, $\psi\colon G\to\Fp^r$ a homomorphism with $r\le c'\log n$ (given by the images of the generators), and $T_\psi:=\#\{(a,b,c):\psi(S(a,b,c))=0\}$. Exact Triangle over $G$, and its all-edges version, are solved deterministically, without knowing $T_\psi$, in time $O(p^6n^{3-\eps\gamma}(\log n)^6+n^{\eps/2}T_\psi\log n)$. If $G$ has exponent dividing $p^2$ and $\psi$ is the reduction of every coordinate modulo $p$, then $\ker\psi=pG$ and $T_\psi$ is the number $T_I$ of triangles with $S(a,b,c)\in pG$; so Exact Triangle over $G$ is truly subcubic on the instances with $T_I\le n^{3-\eps/2-\delta}$, for any fixed $\delta>0$.
\end{proposition}

\begin{proof}
Parameters, pieces and evaluations are those of Theorem~\ref{g:thm:triFp} with $d':=r$, applied to the homomorphisms $K:=H\circ\psi\colon G\to\Fp^l$; the instance $J$ and Lemma~\ref{g:lem:trieval} hold for every homomorphism, and after $\psi(w(e))$ has been computed for all edges, in $O(n^2r\log n)$ time, $T_K=O(rl)$. Let $\zeta(x):=\#\{c\in C_k:\psi(S(a,b,c))=0\}\ge z(x)$. Then $\cnt_{H\circ\psi}(x)=\zeta(x)+f'_H(x)$, where $f'_H(x)$ counts the $c\in C_k$ with $\psi(S(a,b,c))\ne0$ and $H\psi(S(a,b,c))=0$.

\emph{Algorithm.} $\Iq_0:=A\times B\times[h]$. In each of $L$ stages choose $q$ rows from $E$ (now in $\Fp^r$) one at a time, each minimising $\sum_{x\in\Iq_{t-1}}\cnt_{H\circ\psi}(x)$, giving $H_t$, and put $\Iq_t:=\{x\in\Iq_{t-1}:\cnt_{H_t\circ\psi}(x)>0\}$; there is no YES exit. Finally scan $C_k$ for every $(a,b,k)\in\Iq_L$, checking $S(a,b,c)=0$ in $G$; for detection answer YES if and only if a scan succeeds, and for the all-edges version mark the pairs of the successful scans.

\emph{Correctness.} A true incidence has $\cnt_{H\circ\psi}(x)\ge\zeta(x)\ge z(x)\ge1$ for every $H$, so it survives every stage and is scanned; every YES is exhibited by a scan.

\emph{Size.} $\sum_x\zeta(x)$ does not depend on the rows, so the rows minimise $\sum_xf'_H(x)$, and Lemma~\ref{g:lem:greedy} with Lemma~\ref{g:lem:onerow}, applied to the nonzero vectors $\psi(S(a,b,c))$, gives $\sum_{x\in\Iq}f'_{H_t}(x)\le e\,p^{-q}\sigma|\Iq|\le|\Iq|/4$ for $\Iq:=\Iq_{t-1}$. The $\zeta(x)$ of distinct incidences count disjoint sets of triangles, so $\sum_{x\in\Iq}\zeta(x)\le T_\psi$, and
\[
|\Iq_t|\le\sum_{x\in\Iq_{t-1}}\cnt_{H_t\circ\psi}(x)\le|\Iq_{t-1}|/4+T_\psi .
\]
Hence $|\Iq_L|\le4^{-L}|\Iq_0|+\frac43T_\psi\le2n/\sigma+\frac43T_\psi$, as $|\Iq_0|\le2n^3/\sigma$ and $4^L\ge n^2$. A scan checks at most $\sigma$ triangles at $O(\log n)$ operations each (additions in $G$), so the scans cost $O((n+\sigma T_\psi)\log n)$, and $\sigma\le n^{\eps/2}$. The stages cost as in Theorem~\ref{g:thm:triFp}. For the last claim, $\psi$ is a homomorphism onto $\Fp^r$, $r$ the number of cyclic factors, with kernel $pG$ because every factor is $\Z_p$ or $\Z_{p^2}$.
\end{proof}

With $\Hyp(1/14,\,0.077)$ and a fixed prime $p$, Proposition~\ref{g:prop:trilight} solves Exact Triangle over a group of exponent dividing $p^2$ in time $O(n^{2.99451}+n^{1/28}T_I\log n)$ ($\eps\gamma=0.0055$ and $\eps/2=1/28$; the rounding absorbs the polylogarithmic factor as in Corollary~\ref{g:cor:explicitT}).

\begin{proposition}[splitting a part by residue classes]\label{g:prop:tricopy}
Assume $\Hyp(\eps,\gamma)$ with $0<\eps,\gamma\le\frac12$, let $p$ be a fixed prime and $c'>0$, and let $G$ be a finite abelian group of exponent dividing $p^2$ with $|G|\le n^{c'}$, embedded into $R^d$, $R=\Z_{p^2}$, as in Section~\ref{g:sec:G:setup}. For two distinct parts $X$, $Y$ of an instance of Exact Triangle over $G$ and $x\in X$, let $M_x:=\{\bar w(x,y):y\in Y\}\subseteq\Fp^d$ be the set of residues of the edges from $x$ to $Y$, and let $\mathcal M$ be the least of the six numbers $\sum_{x\in X}|M_x|$, over the ordered pairs $(X,Y)$. Then the instance, and its all-edges version, are solved deterministically in time $O(\lceil\mathcal M/n\rceil\,n^{3-\eps\gamma}(\log n)^6)$. In particular Exact Triangle over $G$ is truly subcubic on the instances with $\mathcal M\le n^{1+\eps\gamma-\delta}$, for any fixed $\delta>0$, whatever the number of their triangles with sum in $pG$.
\end{proposition}

\begin{proof}
\emph{The carry form.} For $x\in R^d$ put $x':=(x-\lift(\bar x))/p\in\Fp^d$, which is defined because $x-\lift(\bar x)$ has residue $0$ and so lies in $pR^d$; then $x=\lift(\bar x)+px'$. For $\alpha,\beta,\zeta\in\Fp^d$ with $\alpha+\beta+\zeta=0$, let $\kappa(\alpha,\beta,\zeta)\in\Fp^d$ have the coordinates $(\lift(\alpha_j)+\lift(\beta_j)+\lift(\zeta_j))/p\in\{0,1,2\}$, read modulo $p$; the integer sum is divisible by $p$ because $\alpha_j+\beta_j+\zeta_j=0$ in $\Fp$, and $\lift(\alpha)+\lift(\beta)+\lift(\zeta)=p\,\kappa(\alpha,\beta,\zeta)$ in $R^d$, because $pk$ and $p\cdot\lift(k\bmod p)$ agree modulo $p^2$. $\kappa$ is symmetric in its three arguments. For a triangle with edges $e_1,e_2,e_3$ and $\bar S=0$, adding the three decompositions $w(e_i)=\lift(\bar w(e_i))+pw'(e_i)$ gives
\[
S=p\bigl(\kappa(\bar w(e_1),\bar w(e_2),\bar w(e_3))+w'(e_1)+w'(e_2)+w'(e_3)\bigr),
\]
so $S=0$ if and only if $\bar S=0$ and $S/p=\kappa(\bar w(e_1),\bar w(e_2),\bar w(e_3))+\sum_iw'(e_i)$ vanishes. The term $\kappa$ is the only one that is not a sum of functions of single edges.

\emph{The reduction.} Let $(X,Y)$ attain $\mathcal M$ and let $Z$ be the third part. Replace $X$ by $X':=\{(x,\mu):x\in X,\ \mu\in M_x\}$, of $\mathcal M$ vertices, and give the edges weights in $\Fp^d\times\Fp^d\times\Fp=\Fp^{2d+1}$:
\begin{align*}
w^\ast(y,z)&:=(\bar w(y,z),\,w'(y,z),\,0),\\
w^\ast((x,\mu),y)&:=(\bar w(x,y),\,w'(x,y),\,0)\ \text{if }\bar w(x,y)=\mu,\quad\text{and }(0,0,1)\text{ otherwise},\\
w^\ast((x,\mu),z)&:=\bigl(\bar w(x,z),\,w'(x,z)+\kappa(-\bar w(x,z)-\mu,\,\bar w(x,z),\,\mu),\,0\bigr).
\end{align*}
Let $S^\ast$ be the sum of the three weights of a triangle $((x,\mu),y,z)$. If $\bar w(x,y)\ne\mu$, the last coordinate of $S^\ast$ is $1$. If $\bar w(x,y)=\mu$, the first block of $S^\ast$ is $\bar S(x,y,z)$; when it is $0$, then $\bar w(y,z)=-\bar w(x,z)-\mu$, and the second block is $w'(x,y)+w'(y,z)+w'(x,z)+\kappa(\bar w(y,z),\bar w(x,z),\bar w(x,y))=S(x,y,z)/p$. So $S^\ast((x,\mu),y,z)=0$ if and only if $\mu=\bar w(x,y)$ and $S(x,y,z)=0$. Since $\bar w(x,y)\in M_x$, every triangle $(x,y,z)$ has exactly one copy $((x,\bar w(x,y)),y,z)$ with $\mu=\bar w(x,y)$. Hence a pair of vertices of the given instance lies in a zero triangle if and only if one of its copies does (a vertex of $Y$ or $Z$ is its own copy, and the copies of $x$ are the $(x,\mu)$).

\emph{Algorithm and cost.} Cut $X'$ into $\lceil\mathcal M/n\rceil$ sets of at most $n$ vertices, and for each run the algorithm of Theorem~\ref{g:thm:triFp} over $\Fp^{2d+1}$ on this set, $Y$ and $Z$, in the version asked, with the parts labelled so that the pairs it answers are the copies of the pairs asked; answer YES for a pair if one of its copies is answered YES. Since every factor of $G$ has at least $p$ elements, $p^d\le|G|$ and $p^{2d+1}\le p\,n^{2c'}\le n^{2c'+1}$, and $p\le n^{\eps/4}$ for large $n$, so each run costs $O(p^6n^{3-\eps\gamma}(\log n)^6)$, and $p$ is fixed. The sets $M_x$ for the six pairs $(X,Y)$ cost $O(n^2\log n)$ by sorting, as $d=O(\log n)$, and the weights $O(d)$ each for $n^2+2n\mathcal M$ edges, which is $O(\lceil\mathcal M/n\rceil n^2\log n)$ as $\mathcal M\le n^2$.
\end{proof}

The proposition combines with Proposition~\ref{g:prop:trilight}: for a partition $C=C_1\cup C_2$, with $M_c$ the residues of the edges from $c$ to $A$, the sub-instances on $(A,B,C_1)$ and $(A,B,C_2)$ partition the triangles, so the instance is solved in time $O(n^{3-\eps\gamma}(\log n)^6(1+\sum_{c\in C_1}|M_c|/n)+n^{\eps/2}T_I'\log n)$, where $T_I'$ counts the triangles with $c\in C_2$ and sum in $pG$; with $C_1:=\{c:|M_c|\le n^{\eps\gamma-\delta}\}$ this is truly subcubic whenever the vertices with more residues lie in at most $n^{3-\eps/2-\delta}$ triangles with sum in $pG$. The number $\mathcal M$ is not invariant under the re-gauging of Remark~\ref{g:rem:trinogo}(2): the instance there has $\mathcal M=n^2$ as given and $\mathcal M=n$ after the re-gauging by $\lift\circ h$. We know no algorithm that finds a re-gauging with small $\mathcal M$ when one exists.

\begin{proposition}[short certificates]\label{g:prop:tricert}
Let $m\ge2$, let $G$ be a finite abelian group of exponent dividing $m$ with $|G|\le n^{c'}$, let $\omega$ be the exponent of matrix multiplication, $\omega<2.372$~\cite{VXXZ24}, and $\delta_m:=(3-\omega)/(1+\log_2m)$. An instance of Exact Triangle over $G$ that has a solution has a certificate, a triangle, that is checked deterministically in time $O(n^2\log n)$, reading the input included. An instance without a solution has a certificate that is checked deterministically in time $O(m^4n^{3-\delta_m+\eta})$, for every $\eta>0$. So for a fixed $m$, Exact Triangle over groups of exponent dividing $m$ lies in $\mathrm{NTIME}[O(n^2\log n)]\cap\mathrm{coNTIME}[n^{3-\delta_m+o(1)}]$; for $\Z_4^d$, $\delta_4=(3-\omega)/3>0.209$.
\end{proposition}

\begin{proof}
The first claim is clear: read the input and check the triangle. For the second, let $G=\prod_{i\le r}\Z_{m_i}$; each $m_i$ divides $m$, and $x_i\mapsto(m/m_i)x_i$ is an injective homomorphism $\Z_{m_i}\to\Z_m$, so we may assume $G=\Z_m^r$, $r\le c'\log n$. Put $t:=\lceil\delta_m\log_2n\rceil$. The certificate is a list of $t$ rows $u_1,\dots,u_t\in\Z_m^r$, defining the homomorphism $K:=(\langle u_i,\cdot\rangle)_{i\le t}\colon G\to\Gamma:=\Z_m^t$, and a list $L$ of triangles. The verifier computes $N_K:=\#\{(a,b,c):K(S(a,b,c))=0\}$ as in Lemma~\ref{g:lem:fourier}: with the characters $\psi$ of $\Gamma$, $N_K=|\Gamma|^{-1}\sum_\psi\sum_{a,b}\psi(Kw(a,b))\,(X_\psi Y_\psi)[a,b]$, where $X_\psi[a,c]:=\psi(Kw(a,c))$ and $Y_\psi[c,b]:=\psi(Kw(b,c))$. The entries are $m$-th roots of unity, so in $\Z[X]/(X^m-1)$ each matrix is $\sum_{j<m}X^jP_j$ with $0/1$ matrices $P_j$, and $X_\psi Y_\psi$ costs $m^2$ integer matrix products; the integer $N_K$ is read off as in the proof of Lemma~\ref{g:lem:fourier}. This takes $O(m^{t+3}n^{\omega+\eta})$ time. The verifier rejects if $N_K>n^{3-\delta_m}$, and otherwise checks, after sorting $L$, that $L$ consists of $N_K$ distinct triangles, each with $K(S)=0$ and $S\ne0$. If the instance has no solution, the list of all triangles with $K(S)=0$ passes. If it has a solution, the solution is counted in $N_K$, so at most $N_K-1$ triangles have $K(S)=0$ and $S\ne0$, and no list passes. A short list exists: for $v\in\Z_m^r\setminus\{0\}$ and a uniform row $u$, the map $u\mapsto\langle u,v\rangle$ is a homomorphism onto a nontrivial subgroup of $\Z_m$ (it takes the value $v_j\ne0$ at a unit vector), so $\langle u,v\rangle$ is uniform on that subgroup and is $0$ with probability at most $\frac12$; with $t$ independent rows the expected number of triangles with $S\ne0$ and $K(S)=0$ is at most $n^32^{-t}\le n^{3-\delta_m}$, and some rows attain it; for them, if there is no solution, $N_K\le n^{3-\delta_m}$. Since $m^t\le m\,n^{\delta_m\log_2m}$ and $\omega+\delta_m\log_2m=3-\delta_m$, the verifier takes $O(m^4n^{3-\delta_m+\eta}+n^{3-\delta_m}\log n)$ time.
\end{proof}

\paragraph{A barrier to hardness.} Carmosino, Gao, Impagliazzo, Mihajlin, Paturi and Schneider~\cite{CGIMPS16} state the Nondeterministic Strong Exponential Time Hypothesis (NSETH): for every $\eps>0$ there is a $k$ such that the language of $k$-DNF tautologies is not in $\mathrm{NTIME}[2^{N(1-\eps)}]$, $N$ the number of variables. Their Theorem~2 (Theorem~5.1 in the ECCC version) reads: ``If NSETH and $C\in(N\cap coN)TIME[T_C]$ for some problem $C$, then $(SAT,2^n)\not\le_{FGR}(C,T_C^{1+\gamma})$ for any $\gamma>0$'', where $\le_{FGR}$ is their deterministic fine-grained reduction (their Definition~1), and their Corollary~2 says the same for every problem that is SETH-hard under deterministic reductions. Proposition~\ref{g:prop:tricert} puts Exact Triangle over groups of exponent dividing a fixed $m$ in $(N\cap coN)TIME[T]$ with $T:=n^{3-\delta_m/2}$, and $n^3=T^{1+\gamma'}$ with their $\gamma$ equal to $\gamma':=3/(3-\delta_m/2)-1>0$. Hence, if NSETH holds, no deterministic fine-grained reduction from CNF-SAT proves that Exact Triangle over $\Z_4^d$ with $d=O(\log n)$, or over any family of groups of bounded exponent, needs time $n^{3-o(1)}$; nor does one from Orthogonal Vectors or $k$-OV, to which CNF-SAT reduces deterministically~\cite{Wil05}.

The barrier concerns deterministic reductions from CNF-SAT and from the problems to which CNF-SAT reduces deterministically. It does not exclude randomized reductions with error (zero-error ones are excluded as well, their Section~3.2, and all randomized ones under a non-uniform variant of NSETH, their Lemma~4), nor reductions from hypotheses not known to be implied by SETH through deterministic reductions, such as the $k$-clique and hyperclique hypotheses. It is a barrier, not evidence for an algorithm: the same certificates exist over $\Fp^d$ and, by~\cite[Lemma~11]{CGIMPS16}, over the integers, where Exact Triangle is solved deterministically in truly subcubic time (Theorem~\ref{g:thm:triuniform} and Corollary~\ref{e:cor:uniform}), so that there a deterministic reduction from CNF-SAT would refute SETH itself, as the preprint observes in its introduction for 3SUM, APSP and Exact Triangle.

So the instances of Exact Triangle over $\Z_4^d$, $d=\Theta(\log n)$, that neither Proposition~\ref{g:prop:trilight} nor Proposition~\ref{g:prop:tricopy} solves, those with more than about $n^{3-\eps/2-\eps\gamma}$ triangles whose sum lies in $2G$ and with more residues at the vertices of every part than Proposition~\ref{g:prop:tricopy} tolerates, are open: we know neither a truly subcubic algorithm, given $\Hyp(\eps,\gamma)$ or otherwise, nor any hardness, and by Proposition~\ref{g:prop:tricert} no hardness can come from a deterministic reduction from CNF-SAT unless NSETH fails. Such instances exist: if the vertices carry random sparse vectors $h(v)\in\F^d$ and the residues of $w(a,b)$, $w(b,c)$ and $w(a,c)$ are the coordinatewise products $h(a)h(b)$, $h(b)h(c)$ and $h(a)h(c)$, with arbitrary elements of $2G$ added, then $\bar S=0$ exactly when the three supports are pairwise disjoint, so for suitable densities more than $n^{3-\eps/2}$ triangles have sum in $2G$ while almost every vertex has $n^{\Omega(1)}$ residues towards each other part, more than the $n^{\eps\gamma}$ that Proposition~\ref{g:prop:tricopy} tolerates, and neither Proposition~\ref{g:prop:trilight} nor Proposition~\ref{g:prop:tricopy} applies (we do not know whether a re-gauging helps; given $h$ the instance is easy). A proof would seem to need either a hash in which the residue digits filter triangles without being paid for in the instance of a call, or a structural split of the residues up to re-gauging; neither is among the tools of Section~\ref{g:sec:groups} or of this section. The same holds for every exponent $p^2$, and for $G_0\times\Fp^{d'}$ with $G_0$ of polynomial size and many cyclic factors.

\begin{problem}\label{g:prob:triexp}
Is Exact Triangle over every finite abelian group of bounded exponent, with $|G|\le n^{O(1)}$, solvable in time $n^{3-\delta}$ for a $\delta>0$ that depends only on the exponent (given $\Hyp(\eps,\gamma)$, or unconditionally)? The first open case is $\Z_4^d$.
\end{problem}

\subsection{The coordinatewise triangle problem}\label{g:sec:tri}

This subsection carries the dichotomy of Section~\ref{g:sec:coord} to the triangle version.

\begin{definition}[$\mathrm{Tri}(P)$]\label{g:def:tri}
For a finite alphabet $\Sigma$ and $P\subseteq\Sigma^3$: given $n$ and a complete tripartite graph on parts $A,B,C$ of at most $n$ vertices each with a weight $w(e)\in\Sigma^d$ on every edge, $d\le c\log n$, decide whether some triangle $(a,b,c)\in A\times B\times C$ has $(w(a,b)[j],w(b,c)[j],w(a,c)[j])\in P$ for every coordinate $j\in[d]$. The \emph{all-edges version} asks this for every pair $(a,b)\in A\times B$.
\end{definition}

Brute force takes $O(n^3d)$ time. Unlike $\Prob(P)$, $\mathrm{Tri}(P)$ has no lookup shortcut for functional $P$: the three letters of a coordinate are fixed by the triangle, so every triangle is a candidate. Every permutation of the three positions is induced by a permutation of the three parts (the symmetric group acts on the edges of a triangle as on its vertices), and a permutation of the letters in one position acts on the weights; so the symmetries of Section~\ref{g:sec:gadgets} are symmetries of $\mathrm{Tri}(P)$ too.

\begin{lemma}[representable relations]\label{g:lem:trirep}
Let $P$ be representable by $(\varphi_i,t)$ over $A$. If $A$ is a finite abelian group, $\mathrm{Tri}(P)$ reduces in $O(n^2d)$ time to Exact Triangle over $A^d$ (Definition~\ref{g:def:triangle}), $|A^d|=n^{c\log|A|}$; if $A=\Z^r$, it reduces to integer Exact Triangle with weights of absolute value at most $n^{O(c)}$. The reductions keep the vertex sets and preserve the answer of every pair $(a,b)$.
\end{lemma}

\begin{proof}
The proof of Lemma~\ref{g:lem:repalg} with edges in place of lists. With $\Phi_i(x):=(\varphi_i(x[j]))_{j\le d}$, a triangle satisfies $P$ in every coordinate if and only if $\Phi_1(w(a,b))+\Phi_2(w(b,c))+\Phi_3(w(a,c))=(t,\dots,t)$. Give each edge $(a,b)$ the weight $\Phi_1(w(a,b))-(t,\dots,t)$, each edge $(b,c)$ the weight $\Phi_2(w(b,c))$ and each edge $(a,c)$ the weight $\Phi_3(w(a,c))$. Over $\Z^r$, use the base-$\mathfrak b$ encoding of Lemma~\ref{g:lem:repalg}(b), with $X_1(w(a,b))-T$, $X_2(w(b,c))$ and $X_3(w(a,c))$ as the weights; a sum of three digits is at most $6V<\mathfrak b$, so there are no carries.
\end{proof}

\begin{corollary}\label{g:cor:trirepfast}
Let $P$ be representable over $\Z$ or over $\Fp^r$ for a prime $p$. Then $\mathrm{Tri}(P)$, also in its all-edges version, is solved deterministically in time $n^{3-\eps\gamma+o(1)}$ given $\Hyp(\eps,\gamma)$, $0<\eps,\gamma\le\frac12$ (over $\Z$ by Corollary~\ref{e:cor:uniform}, over $\Fp^r$ by Theorem~\ref{g:thm:triFp}, which for $p=2$ is Theorem~\ref{x:thm:xet}), and in time $O(n^{2.99451})$ unconditionally (Corollaries~\ref{x:cor:fg}(d) and~\ref{g:cor:explicitT}(a)).
\end{corollary}
\begin{proof}
Lemma~\ref{g:lem:trirep}; for a fixed $P$ the prime $p$ is fixed, $(p^r)^d\le n^{O(c)}$, and the integer weights are at most $n^{O(c)}$; the exponents of the cited algorithms do not depend on these bounds. Over $\Z$, $n^{3-0.0055+o(1)}\le O(n^{2.99451})$.
\end{proof}

If $P$ is representable only over groups that are not elementary abelian, as for the Latin squares isotopic to $\Z_4$ and the other orbits marked ``exponent $4$ only'' in Table~\ref{g:tab:functional4}, Lemma~\ref{g:lem:trirep} gives Exact Triangle over $A^d$, whose non-elementary part has polynomial size and $\Theta(\log n)$ cyclic factors; Section~\ref{g:sec:triangle:p2} does not solve this (Open problem~\ref{g:prob:triexp}).

\begin{lemma}[gadgets give cubic hardness]\label{g:lem:trigadget}
If $P$ has an OV gadget of size $t$ (Definition~\ref{g:def:gadget}), then under OVH, for every $\delta>0$ there is $c$ such that $\mathrm{Tri}(P)$ with $d=c\log n$ has no algorithm in time $O(n^{3-\delta})$.
\end{lemma}

\begin{proof}
Let $\theta$ be the side assignment of the gadget and $F_i$ the maps of the proof of Lemma~\ref{g:lem:gadgethard}. Positions $1,2,3$ are the edges $(a,b)$, $(b,c)$, $(a,c)$ of a triangle, and any two of them share exactly one vertex.

\emph{Slice} ($\theta_i=\mathsf A$, $\theta_j=\mathsf B$, $\theta_k=\mathsf C$). Let $V$ be the part shared by the edges at positions $i$ and $j$, and $U$, $U'$ the parts at their other ends; the edge at position $k$ joins $U$ and $U'$. Given OV with $A_0,B_0\subseteq\{0,1\}^{d_0}$, $|A_0|,|B_0|\le N$, $d_0=c_0\log N$, put $n:=\lceil N^{2/3}\rceil+3\lceil N^{1/3}\rceil$. Cut $A_0$ and $B_0$ into blocks of $n$ vectors each, at most $N/n+1\le N^{1/3}+1$ blocks per list, filling the last block by repeating one of its vectors, which does not change the answer. There are at most $(N^{1/3}+1)^2\le n$ pairs of blocks; let the vertices of $V$ be the pairs $(\beta,\beta')$ of a block $\beta$ of $A_0$ and a block $\beta'$ of $B_0$ (repeated to fill $V$), and the vertices of $U$ and $U'$ the indices $1,\dots,n$. The edge at position $i$ between $v=(\beta,\beta')$ and $s\in U$ gets $F_i$ of the $s$-th vector of $\beta$, the edge at position $j$ between $v$ and $s'\in U'$ gets $F_j$ of the $s'$-th vector of $\beta'$, and every edge at position $k$ gets $F_k(0)$. By the argument of Lemma~\ref{g:lem:gadgethard}, the triangle $(s,v,s')$ satisfies $P$ in every coordinate if and only if these two vectors are orthogonal, and every pair in $A_0\times B_0$ occurs in some triangle. The instance has $n=O(N^{2/3})$ vertices per part and dimension $td_0\le\frac32tc_0\log n$, as $n\ge N^{2/3}$, and is built in $O(n^2td_0)$ time; an $O(n^{3-\delta})$ algorithm for $\mathrm{Tri}(P)$ with $c:=\frac32tc_0$ would solve OV in time $O(N^{2-2\delta/3})$.

\emph{Identification} (no position is $\mathsf C$, and one side is read at two positions $j,k$, the other at position $i$; by the symmetry of OV we may assume that $j,k$ read $\mathsf B$). The edges at positions $j$ and $k$ share a part $V$; the edge at position $i$ joins the other two parts $U$ and $U'$. Given OV with $|A_0|,|B_0|\le M$ and $d_0=c_0\log M$, let $n:=\lceil\sqrt M\rceil$, put the vectors of $A_0$ on the $n^2\ge M$ edges between $U$ and $U'$ (repeating one if needed; the edge gets $F_i(x)$), and cut $B_0$ into at most $n$ blocks of at most $n$ vectors. For each block, put its vectors $y_v$ on the vertices $v$ of $V$ and give both edges at $v$ at positions $j$ and $k$ the weights $F_j(y_v)$ and $F_k(y_v)$, so that both $\mathsf B$-positions read the same vector, as Definition~\ref{g:def:gadget} requires. The triangle $(u,u',v)$ satisfies $P$ in every coordinate if and only if the vector on $(u,u')$ is orthogonal to $y_v$. So OV on $M$ vectors is decided by at most $n$ instances of $\mathrm{Tri}(P)$ of dimension $tc_0\log M\le2tc_0\log n$, and an $O(n^{3-\delta})$ algorithm with $c:=2tc_0$ would solve it in time $O(n^{4-\delta})=O(M^{2-\delta/2})$.

In both cases OVH, applied with $2\delta/3$ respectively $\delta/2$, gives the claim.
\end{proof}

The all-edges version is at least as hard. Since brute force takes $O(n^3d)$ time, the relations with a gadget have $\mathrm{Tri}(P)$ of complexity $n^{3\pm o(1)}$ under OVH.

\begin{theorem}[the dichotomy at the cubic level]\label{g:thm:triD}
Let $P\subseteq\{0,1\}^3$, or $P=P_g$ for a map $g\colon\{0,1,2\}^2\to\{0,1,2\}$. Exactly one of the following holds.
\begin{enumerate}[label=\textup{(\roman*)}]
\item $P$ is representable, over $\Z$, $\F$ or $\mathbb F_3$ (the $10$ Boolean orbits marked REP in Table~\ref{g:tab:boolean}, and the $297$ maps of Table~\ref{g:tab:functional}), and $\mathrm{Tri}(P)$ is solved in time $n^{3-\eps\gamma+o(1)}$ given $\Hyp(\eps,\gamma)$ and in time $O(n^{2.99451})$ unconditionally.
\item $P$ has an OV gadget of size $1$, and under OVH, for every $\delta>0$ there is $c$ such that $\mathrm{Tri}(P)$ with $d=c\log n$ has no algorithm in time $O(n^{3-\delta})$.
\end{enumerate}
\end{theorem}

\begin{proof}
The classifications of Theorems~\ref{g:thm:D} and~\ref{g:thm:E} are statements about $P$ alone, so they apply unchanged; Lemma~\ref{g:lem:exclusive} gives the exclusion, Corollary~\ref{g:cor:trirepfast} gives~(i) (over $\mathbb F_3$ through Theorem~\ref{g:thm:triFp} with $p=3$), and Lemma~\ref{g:lem:trigadget} gives~(ii).
\end{proof}

The theorem rests on the computations of Section~\ref{g:sec:computer}, through Theorems~\ref{g:thm:D} and~\ref{g:thm:E}.

\begin{proposition}[the triangle version over four letters]\label{g:prop:triF}
\begin{enumerate}[label=\textup{(\alph*)}]
\item For a Latin square $P$ isotopic to $\F^2$, $\mathrm{Tri}(P)$ is Exact-XOR-Triangle in dimension $2d$, solved by Theorem~\ref{x:thm:xet}. For $P$ isotopic to $\Z_4$, $\mathrm{Tri}(P)$ and Exact Triangle over $\Z_4^d$ reduce to each other in $O(n^2d)$ time.
\item For the functional relations over four letters of Proposition~\ref{g:prop:F}(e): for the $33$ orbits representable over $\Z$, $\F$ or $\mathbb F_3$, $\mathrm{Tri}(P_g)$ is solved in time $n^{3-\eps\gamma+o(1)}$ given $\Hyp(\eps,\gamma)$ and in time $O(n^{2.99451})$ unconditionally; for the $9$ orbits representable over a group of exponent $4$ but over no $\Z^r$ and no $\Fp^r$, $\mathrm{Tri}(P_g)$ reduces to Exact Triangle over $\Z_4^{d'}$ with $d'=O(d)$; for the $167296$ orbits with a gadget, under OVH, for every $\delta>0$ there is $c$ such that $\mathrm{Tri}(P_g)$ with $d=c\log n$ has no algorithm in time $O(n^{3-\delta})$.
\end{enumerate}
\end{proposition}

\begin{proof}
(a) $P$ is representable over $\F^2$, respectively $\Z_4$, by the isotopy, so Lemma~\ref{g:lem:trirep} applies; Exact Triangle over $(\F^2)^d=\F^{2d}$ is Exact-XOR-Triangle. Conversely, an instance of Exact Triangle over $\Z_4^d$ becomes an instance of $\mathrm{Tri}(P)$ when every coordinate of the weights of the edges at position $i$ is replaced by its image under $\alpha_i^{-1}$; a coordinate triple of a triangle then lies in $P$ if and only if the three group elements sum to $0$. (b) The classification is that of Proposition~\ref{g:prop:F}(e). The first case is Corollary~\ref{g:cor:trirepfast}, the third Lemma~\ref{g:lem:trigadget}. In the second case Lemma~\ref{g:lem:universal}(c) gives a representation over a finite group $A$ of exponent dividing $4$, so $A\cong\Z_4^a\times\Z_2^b$, and Lemma~\ref{g:lem:trirep} reduces $\mathrm{Tri}(P_g)$ to Exact Triangle over $A^d$, which embeds into $\Z_4^{(a+b)d}$ by the injective homomorphism $x\mapsto2x$ from $\Z_2$ to $\Z_4$ on the last $b$ factors; Exact Triangle over $\Z_4^{d'}$ is open (Open problem~\ref{g:prob:triexp}).
\end{proof}

No Latin square has a gadget (Lemma~\ref{g:lem:latinnogadget}), so Lemma~\ref{g:lem:trigadget} gives no hardness for any of them, and for the Latin squares isotopic to $\Z_4$ and the nine orbits of~(b) that need exponent $4$ we know no truly subcubic algorithm either: Exact Triangle over $\Z_4^d$ is Open problem~\ref{g:prob:triexp}. For $L_5$, the four functional orbits of the third kind and the relations of Proposition~\ref{g:prop:F}(f) we know nothing beyond brute force in $O(n^3d)$ time, and no hardness. So for $\mathrm{Tri}(P)$ the relations whose status is open, with neither a truly subcubic algorithm nor cubic hardness known, already include the Latin squares isotopic to $\Z_4$, whereas for $\Prob(P)$ Theorem~\ref{g:thm:G} puts these on the algorithmic side. Over four letters, the functional relations whose status is open are, for $\mathrm{Tri}(P)$, the $13$ orbits of the third kind and of ``exponent $4$ only'' in Table~\ref{g:tab:functional4}, and for $\Prob(P)$ only the $4$ orbits of the third kind.

The hardness side cannot be extended to the Latin squares isotopic to $\Z_4$ by the same means. Lemma~\ref{g:lem:trigadget} is a deterministic reduction from Orthogonal Vectors, and for such a Latin square $P$, $\mathrm{Tri}(P)$ is Exact Triangle over $\Z_4^d$ (Proposition~\ref{g:prop:triF}(a)); by Proposition~\ref{g:prop:tricert} and the barrier after it, a deterministic reduction from Orthogonal Vectors to it would refute NSETH. For gadgets this follows also without NSETH, from Lemmas~\ref{g:lem:exclusive} and~\ref{g:lem:latinnogadget}.

\paragraph{Hypotheses.} The hard side needs only OVH, which the Strong Exponential Time Hypothesis implies; the cubic level comes from a polynomial blow-up of the OV instance ($N=n^{3/2}$ vectors for slices, $n^2$ against $n$ for identifications), not from a stronger hypothesis. In particular, the triangle version of $3$-OV ($P$ all but $111$, row 21 of Table~\ref{g:tab:boolean}) is cubic-hard under OVH, not only under the $3$-OV hypothesis. Hypotheses on 3SUM or APSP are not needed, and could not serve, since the preprint refutes the 3SUM and APSP conjectures.

\paragraph{Comparison with~\cite{BFK19}.} $\mathrm{Tri}(P)$ and the problems of Bringmann, Fischer and Künnemann are not directly comparable. Their properties are first-order properties of a graph, with one binary relation; in their vector problems the vectors sit on the quantified vertices $x_i$ (atoms $E(x_i,y)$), whereas $\mathrm{Tri}(P)$ puts a vector on every pair of quantified vertices, which is not a vector problem in their sense; an encoding with a vertex for every such pair would change the number of quantifiers and the input measure. Their measure is the number $m$ of edges and their baseline is $m^3$; ours is the number $n$ of vertices per part, with input size $\Theta(n^2d)$ and baseline $n^3$. So none of their bounds is a bound for $\mathrm{Tri}(P)$, and Theorem~\ref{g:thm:triD} neither refines nor contradicts their Theorem 3: their cubic-level dichotomy, for $\Prob(P)$ in the sparse model, makes only the $3$-OV orbit cubic-hard, while Theorem~\ref{g:thm:triD}, for the different problem $\mathrm{Tri}(P)$, makes all $12$ gadget orbits cubic-hard.

\subsection{What is known and what is new}\label{g:sec:triremarks}

\emph{Known.} We found no paper on Exact Triangle over $\Z_4^d$, over $\Z_{2^k}^d$ or over $\Fp^d$ with small $p$ (several targeted searches, not exhaustive). For the coordinatewise triangle problem, the closest work we found is the factored problems of Dalirrooyfard, Lincoln and Vassilevska Williams~\cite{DLW20} (Section~\ref{g:sec:coordremarks}): there edges carry factored labels, a set of strings per chunk, and a factored zero-triangle problem is $n^{3-o(1)}$-hard under each of the Strong Exponential Time Hypothesis, the 3SUM and the APSP hypotheses~\cite[Theorem~1.2]{DLW20}; these labels are more input than one letter per coordinate, so this hardness does not transfer to $\mathrm{Tri}(P)$, and Lemma~\ref{g:lem:trigadget} does not need it. We found nothing on a plain problem of triangles with orthogonal edge vectors (several targeted searches, not exhaustive).

\emph{New, to our knowledge.} In Sections~\ref{g:sec:triangle:el} and~\ref{g:sec:triangle:p2}, Theorem~\ref{g:thm:triFp} is a routine transfer of Theorem~\ref{x:thm:xet}, and Theorem~\ref{g:thm:triuniform} and Propositions~\ref{g:prop:triboundedrank}, \ref{g:prop:trismall} and~\ref{g:prop:trilight} combine it routinely with Part~\hyperref[m:partII]{II} and with the tools of Section~\ref{g:sec:groups}; what they add beyond them is the explanation of why the method of Theorem~\ref{g:thm:G} does not transfer (Remark~\ref{g:rem:trinogo}), which leaves Open problem~\ref{g:prob:triexp}, and two partial results on it: the reduction of Proposition~\ref{g:prop:tricopy}, which makes the carries edge-local by splitting vertices, and the certificates of Proposition~\ref{g:prop:tricert}, which adapt the zero-weight-triangle certificates of~\cite[Lemma~11]{CGIMPS16} to characters of $\Z_m^t$ and with their Theorem~2 explain why our hardness method stops short of $\Z_4^d$. The coordinatewise triangle problem (Section~\ref{g:sec:tri}) is routine on both sides once Theorems~\ref{g:thm:D}, \ref{g:thm:E}, \ref{x:thm:xet} and~\ref{g:thm:triFp} are available: the representable side is Exact Triangle over $\Z$, $\F^d$ or $\mathbb F_3^d$, and the hard side is the blow-up of an OV instance in Lemma~\ref{g:lem:trigadget}.

\section{Scripts and formal verification}\label{m:sec:status}

Two kinds of machine check accompany the paper: Python scripts, which are the ancillary files of the arXiv submission (Section~\ref{a:sec:anc}), and a formalisation in Lean~4 (Section~\ref{m:sec:lean}). Both are in the repository \url{https://github.com/Majdoddin/formal-math}: the scripts with their recorded outputs in the folder \texttt{scripts}, the Lean projects in the folders named below. The scripts also compute the values and exclusions that the text calls rate values (Remark~\ref{m:rem:exact}) or marks as numerical: numerical optima of proved closed-form bounds, the scores and exclusions of Section~\ref{m:sec:barriers} and Appendix~\ref{a:sec:overview}, and the numerical probe after Proposition~\ref{m:rem:border}. They are computed, not proved; the theorems are stated at explicit parameters.

\subsection{Ancillary scripts}\label{a:sec:anc}

The scripts listed in Table~\ref{tab:anc} are the ancillary files of this arXiv submission (folder \texttt{anc/}; the file \texttt{anc/README.txt} repeats the table with the command lines). They are Python~3 programs, run from inside \texttt{anc/} as \texttt{python3 FILE [MODE]}; five of them import \texttt{anc/profile\_model.py}, and \texttt{q2\_functional4.py} compiles two short C helpers shipped beside it, \texttt{enum\_functional.c} and \texttt{locrep.c}, with the system's \texttt{cc}. Each prints labelled lines, and the text cites a script with its mode or the label of the output line (for example ``output line A''). A script of kind P re-checks a statement that is proved in the text; one of kind C is part of a computer-assisted proof (Proposition~\ref{m:rem:border}, the ranks behind Remarks~\ref{a:rem:conditional} and~\ref{a:rem:koszul}, and the classifications of Section~\ref{g:sec:coord}, Section~\ref{g:sec:computer}); one of kind N computes a numerical value or heuristic evidence that is not proved.

\begin{table}[!htbp]
\centering\footnotesize
\renewcommand{\arraystretch}{1.1}\setlength{\tabcolsep}{3pt}
\begin{tabular}{@{}>{\raggedright}p{0.30\textwidth}>{\raggedright}p{0.38\textwidth}c>{\raggedright}p{0.12\textwidth}>{\raggedright\arraybackslash}p{0.10\textwidth}@{}}
file [mode] & supports & kind & needs & time\\\hline
\texttt{identity\_check.py} & Lemma~\ref{m:lem:identity}, Remark~\ref{m:rem:degree}, the two signs of Coppersmith's identity (Section~\ref{m:sec:tensor}) illegible in the scan & P & sympy & 1~s\\
\texttt{squared\_check.py} & Theorem~\ref{m:thm:correct}, Lemmas~\ref{m:lem:counts}, \ref{m:lem:boxes}, queries (Remark~\ref{m:rem:brute}) & P & numpy & 11~s\\
\texttt{omv512\_check.py} & the index identities of Proposition~\ref{m:rem:omv512} & P & -- & 1~s\\
\texttt{example\_squared.py} & the example of Section~\ref{m:sec:squared} & P & -- & 37~s\\
\texttt{explicit\_constants.py} & Corollary~\ref{m:cor:explicit} (P); Table~\ref{m:tab:rates}, lines E1, E2 (Table~\ref{tab:matrix}), E3 (Remark~\ref{m:rem:exact}) (N) & P, N & numpy, scipy & 6~s\\
\texttt{koszul\_exact.py} & Remark~\ref{a:rem:koszul} & C & -- & 1~s\\
\texttt{koszul\_modp.py} \texttt{base}\,|\,\texttt{excl}\,|\,\texttt{ctrl} & Remark~\ref{a:rem:conditional} (\texttt{base}), Remark~\ref{a:rem:koszul} (\texttt{excl}), controls & C & numpy & 7~s, 1~s, 1~s\\
\texttt{border\_apolarity.py} \texttt{G222 5}\,|\,\texttt{G221 5}\,|\,\texttt{G222 6} & Proposition~\ref{m:rem:border} (\texttt{G222 5}), controls & C & -- & 1~s, 1~s, 2~s\\
\texttt{border\_apolarity\_family.py} \texttt{S221 5}\,|\,\texttt{S222 5} & Proposition~\ref{m:rem:border}, controls & C & sympy & 6~s, 1~s\\
\texttt{border\_rank\_probe.py} & the probe after Proposition~\ref{m:rem:border} & N & numpy & 75~s\\
\texttt{exact\_ceiling.py} & the exact-count limits $0.1784$, $0.1292$, Section~\ref{m:sec:ceiling} & N & numpy, scipy & 12~s\\
\texttt{profile\_model.py} & the shared model; $0.003935$ (Schönhage $(3,3)$) & N & numpy & 2.5~min\\
\texttt{schoenhage\_scores.py} & Schönhage's family, $(3,3)$ largest & N & -- & 15~s\\
\texttt{border\_identity\_scores.py} & the family's scores; open shapes $\le0.00573$ & N & numpy & 25~min\\
\texttt{shared\_variable\_variants.py} & lines H ($0.005825$), A (Remark~\ref{m:rem:half}), K1 (one product per run) & N & numpy & 9~min\\
\texttt{pure\_inner\_cases.py} & the cases $R_0\ge2$, $kn\le30$ & N & numpy & 16~min\\
\texttt{vrana\_patterns.py} \texttt{enum} & Vrana's $464$ leaf counts & N & numpy & 6~min\\
\texttt{cw\_decompositions.py} \texttt{grouped}\,|\,\texttt{robust}\,|\,\texttt{wstyle}\,|\,\texttt{scan 2 4 8} & Coppersmith--Winograd decompositions & N & numpy & 14~min, 3~min, 42~s, 1~h\\
\texttt{thin\_exclusion.py} & Corollary~\ref{a:cor:145}: the shapes of the $145$ & C & -- & 1~s\\
\texttt{thin\_region.py} & the region of Proposition~\ref{a:prop:criterion}(iii) & N & -- & 10~min\\
\texttt{thin\_value.py} \texttt{squares}\,|\,\texttt{rect}\,|\,\texttt{sup} & what open thin shapes could be worth & N & -- & 1~s, 6~s, 1~s\\
\texttt{q2lib.py} & lattice tests with certificates, gadget search and orbits for the four scripts below & C & numpy & --\\
\texttt{q2\_boolean.py} & Theorem~\ref{g:thm:D}, Table~\ref{g:tab:boolean} & C & numpy & 1~s\\
\texttt{q2\_functional3.py} & Theorem~\ref{g:thm:E}, Table~\ref{g:tab:functional} & C & numpy & 38~s\\
\texttt{q2\_latin.py} & Proposition~\ref{g:prop:F}(b), (c) & C & numpy & 27~s\\
\texttt{q2\_functional4.py} \texttt{functional4}\,|\,\allowbreak\texttt{nonfunctional3}\,|\,\allowbreak\texttt{local4} & Proposition~\ref{g:prop:F}(e), Table~\ref{g:tab:functional4} (\texttt{functional4}); Proposition~\ref{g:prop:F}(f) (\texttt{nonfunctional3}); the counts before Open problem~\ref{g:prob:L5} (\texttt{local4}, N) & C, N & numpy, cc & 31~min, 19~s, 2.7~min\\
\texttt{enum\_functional.c}, \texttt{locrep.c} & C helpers of \texttt{q2\_functional4.py} & C, N & cc & --
\end{tabular}
\caption{The ancillary files. Kind: P re-checks a proved statement, C is part of a computer-assisted proof, N is a numerical value or heuristic evidence. ``--'' means the standard library only. Times are wall-clock times of one run on an 8-core desktop computer (\texttt{pure\_inner\_cases.py} uses $7$ processes, \texttt{vrana\_patterns.py enum} $8$, \texttt{q2\_functional3.py} and \texttt{q2\_functional4.py functional4} $4$); tested with Python 3.13, numpy 2.4, scipy 1.16 and sympy 1.14.}
\label{tab:anc}
\end{table}

\subsection{Lean formalisation}\label{m:sec:lean}

The mathematics of the paper outside Sections~\ref{x:sec:xor}--\ref{g:sec:triangle} was also formalised in Lean 4 with Mathlib (folder \texttt{thin-matrix-products}): the identities, the counts, the correctness lemmas and the arithmetic of every explicit constant and exponent are proved there from definitions that match the paper's, with no sorry and no axiom beyond Lean's three standard ones, except three correctness steps: the encoding into Set Disjointness in Theorem~\ref{s:thm:kpp} (only its parameter arithmetic is formalised), the composition of the steps of Corollary~\ref{e:cor:C} (each step is formalised) and the derivation of the output formula in Theorem~\ref{a:thm:shared}. Running times, the statements of the preprint and of the other cited works, and the border-rank arguments of Proposition~\ref{m:rem:border} and Appendix~\ref{a:sec:barriers} (Lemma~\ref{a:lem:subst} and the standard facts it uses; Mathlib has no theory of border rank), and the computer-assisted and numerical results are not formalised; the README of the formalisation lists every such item. Of Appendix~\ref{a:sec:full}, Lemmas~\ref{a:lem:link}, \ref{a:lem:types} (by a different proof, a tilted product measure and Chebyshev's inequality instead of the method of types, with an explicit rate), \ref{a:lem:forest} and~\ref{a:lem:rankthin}, Proposition~\ref{a:prop:criterion}, Corollary~\ref{a:cor:145} and Theorem~\ref{a:thm:full} for the preprint's run are formalised, and so is the preprint's Lemma~10, that the pruned recursion visits exactly $\Leaves(W)$, for every identity (the appendix applies it to every identity, and Part~\hyperref[m:partI]{I} on the $2L$ levels of a squared run); the squared-run case of Theorem~\ref{a:thm:full} and the numerical paragraphs are not. Three errors in a draft of this paper were found by the formalisation: the factor $2$ in Lemma~\ref{s:lem:shape}, $D\ge4$ in Corollary~\ref{x:cor:fg}(a), and $b\ge2$ in the last sentences of Theorem~\ref{a:thm:full} and Proposition~\ref{a:prop:criterion}, whose degenerate case $b=1$ has $D=1$ and a vacuous bound.

Of Section~\ref{x:sec:xor}, the following are formalised in a separate Lean project for that section (folder \texttt{xor-3xor}), with the same guarantee (no sorry and no axiom beyond the three standard ones): Lemmas~\ref{x:lem:select}, \ref{x:lem:greedy}, \ref{x:lem:wh}, \ref{x:lem:probes}, \ref{x:lem:counts}, \ref{x:lem:halving}, \ref{x:lem:correct} and~\ref{x:lem:cert}, the binary splitting in Lemma~\ref{x:lem:shape}, the correctness of the three algorithms (each answers YES exactly when a solution exists, and the all-edges version answers every pair correctly), the bound on the failed scans in the proof of Theorem~\ref{x:thm:xordet}, the parameter inequalities, and the values of the exponents of Corollary~\ref{x:cor:xor} (not the absorption of the logarithmic factors into them). There the counting algorithm is represented by the exact answers it must return, and the biased sets of Lemma~\ref{x:lem:aghp} are a hypothesis, which the construction formalised for the remark's route (below) satisfies but which the files of Section~\ref{x:sec:xor} do not import. Every call the three algorithms make is proved to be legal on the instance actually built, renumbered into $N$ vertices with isolated padding: it is an instance of $\Lop(N,D)$ with $D\le N^\varepsilon$ and $|W|$ within the bound of Lemma~\ref{x:lem:shape}, and the rows chosen from the answers of the calls are proved to be the greedy rows of Lemma~\ref{x:lem:greedy}. The cost of the calls is proved as a separate inequality and not composed with these proofs; the running times are not formalised.

The remark's route itself, which would give the entries marked~$\dagger$ in Table~\ref{tab:summary} and which Section~\ref{x:sec:xor} does not write out, is also formalised in the same repository, apart from its running time: the formalisation proves its correctness and its counts, as the analogues over~$\mathbb F_2$ of the preprint's Theorem~17, of its Theorem~21(a) and of the reduction of Chan and He~\cite{CH20}, gives an explicit construction of the biased sets of~\cite{AGHP92} with its bias bound, and proves the arithmetic of the exponent $1.99863$ and of the value $1.999125$ behind $1.9992$ (its rounding to $1.9992$ is in the unfinished running-time part). Of Sections~\ref{g:sec:groups}--\ref{g:sec:triangle}, the following are formalised in a third Lean project (folder \texttt{groups}), with the same guarantee: Lemmas~\ref{g:lem:onerow}, \ref{g:lem:greedy}, \ref{g:lem:fourier} (the identity over~$\mathbb C$), \ref{g:lem:instance}, \ref{g:lem:facts}, \ref{g:lem:survival} (except its last sentence), \ref{g:lem:light}, \ref{g:lem:partL} and \ref{g:lem:potential}--\ref{g:lem:Hrows}, Proposition~\ref{g:prop:C} (except its last sentence), Lemma~\ref{g:lem:powering} and the correctness part of Lemma~\ref{g:lem:onecall}; the correctness and the counts of Theorems~\ref{g:thm:A} and~\ref{g:thm:B}; for Theorem~\ref{g:thm:G}, that every solution lies in exactly one of the $7+(\lfloor\log n\rfloor+1)^3$ sub-instances, that each falls under Case~L or Case~H and is answered correctly with one detection call, and the bounds on the false positives; the correctness of Theorem~\ref{g:thm:triFp} in its detection and all-edges versions, with the halving of the incidences; Lemma~\ref{g:lem:repalg}, Theorems~\ref{g:thm:D} and~\ref{g:thm:E}, whose finite checks are certificates verified by Lean's kernel, and Proposition~\ref{g:prop:F}(b)--(d) with Lemmas~\ref{g:lem:gadgethard} and~\ref{g:lem:exclusive} for gadgets of every size. As for Section~\ref{x:sec:xor}, the detection and counting algorithms are represented by the exact answers they must return. Not formalised are the running times, the legality of the calls (the size of the instances and $D\le N^\varepsilon$), the arithmetic that turns $\varepsilon$ and $\gamma$ into the parameters, Proposition~\ref{g:prop:F}(a), (e) and~(f) and the counts of Tables~\ref{g:tab:boolean}, \ref{g:tab:functional} and~\ref{g:tab:functional4}, which rest on the computations of Section~\ref{g:sec:computer}, and the other statements of these sections.

\FloatBarrier
\appendix

\section{Details and proofs for Section~\ref{m:sec:barriers}}\label{a:sec:barriers}

\subsection{The class, the scores and the numerical exclusions}\label{a:sec:overview}

This subsection is the detailed version of Section~\ref{m:sec:barriers}; its notation is that of the first paragraph there. Its \emph{scores} are rate values of $\varepsilon\gamma$ (numerical) for the leaf count $\sum_\tau\min\{1,\rho\,s(\tau)\}$ of a run, which \texttt{anc/profile\_model.py} computes; here $s(\tau)$ is the number of selected output strings to which the leaf $\tau$ contributes and $\rho$ the density of the wanted ones. By Lemma~\ref{a:lem:link} in Appendix~\ref{a:sec:full} this count is, up to the factor $1-1/e$, the average of $|\Leaves(W)|$ over the sets $W$ of selected output strings of the given size, for every identity; for the tensor of Section~\ref{m:sec:tensor} and Schönhage's $(3,3)$ identity its rate is that of the exact leaf count of Remark~\ref{m:rem:exact}, which bounds the cost for every $W$ (the two counts agree order by order there).

\emph{The private-leaf condition} (Lemma~\ref{a:lem:private}). When $R_0\ge1$ this excludes $R_1\ge2$ for every inner length $q_0\le4^{c-1}$, and for $R_1\ge1$ it gives $R_0<\sqrt{q_0}$. That this forces $R_1=1$ is an argument from the ratios of the optima ($c-1\ge9$), not a proof: the outer outputs must have the sparsity of Schönhage's identity and of the tensor of Section~\ref{m:sec:tensor}. For $R_0=0$ the inequality is empty; that case is the Coppersmith--Winograd tensor below.

\emph{Exact identities} (proved bound, Proposition~\ref{a:prop:exact}; numerical scores). Without disjointness the bound fails: $z_{11}\mapsto z_{11}+z_0$ in Schönhage's identity gives an exact identity of effective inner length $R-k-n+1$, but it shares two variables, $x_1$ and $y_1$, which is unusable (Theorem~\ref{a:thm:shared}). Schönhage's identities attain the bound, and among those computed $(k,n)=(3,3)$ has the largest score, $\varepsilon\gamma=0.003935$ (numerical; ancillary files \texttt{anc/schoenhage\_scores.py} and \texttt{anc/profile\_model.py}).

\emph{Border identities.} The bound on shared variables (Theorem~\ref{a:thm:shared}) covers runs in which every output is selected; partial selections are excluded only heuristically, and that a single product per run gives no saving is computed (ancillary file \texttt{anc/shared\_variable\_variants.py}, output line K1). If the recursion's base case $S(1,n,q_0')$ has border rank at least $n+q_0'+1$ in general, as Koszul flattenings show for $n=2$, $q_0'\le6$ and for $q_0'=1$, $n\le6$, the family is optimal for every shape (Remark~\ref{a:rem:conditional}, a proved implication). The shapes $2\times4$ and $2\times5$ remain open, but even their hypothetical longest inner products would score at most $0.00573<0.005825$ (numerical; ancillary file \texttt{anc/border\_identity\_scores.py}). The family itself, $\langle k,1,n\rangle$ plus an inner product of length $(k-1)(n-1)+1$ in $kn+1$ terms, scores lower for larger $k,n$: $0.005489$ for $(2,3)$, $0.004771$ for $(3,3)$, $0.004720$ for $(2,4)$ (numerical; same file). So $k=n=2$, the tensor of Section~\ref{m:sec:tensor}, is the best member, as it is for Coppersmith's $\alpha$.

\emph{Other tensors (proved in this paper for Vrana's decomposition and, for every decomposition, for the full-product shapes of Appendix~\ref{a:sec:full}).} The Coppersmith--Winograd tensor~\cite{CW90} has an inner product of length $q$ from $q+2$ terms, but every outer output has a rank-two form and needs two terms; for its standard decomposition the pruned recursion visits at least $2^{m-1}$ times as many leaves as there are outputs, so there is no saving; other decompositions we tried give no saving either (numerical; ancillary file \texttt{anc/cw\_decompositions.py}). We have arguments for the standard and the W-style decompositions, which we do not include; the others are numerical, and a proof for all decompositions is not available. The partial product of Bini, Capovani, Romani and Lotti~\cite{BCRL79} has no thin products in one orientation, and in the others, with its original decomposition, the visited leaves exceed the outputs by a factor at least $\frac12(3/2)^{L-m}$ (an argument we do not include). Vrana's Example 4.10~\cite{Vra24}, a partial product with a border decomposition of length $5$ and two inner outputs per level, gives no saving with his decomposition: the pruned recursion computes at least $\sqrt D/2$ leaf products per wanted entry in every orientation (Proposition~\ref{a:prop:vrana}), and none of the support patterns that the flattening ranks allow for a five-term decomposition (the sets of letters of one factor that each term contains, which give $464$ distinct leaf counts) gives a saving (numerical; ancillary file \texttt{anc/vrana\_patterns.py}, mode \texttt{enum}). The $145$ decompositions of full products found by computer search~\cite{FBH+22,KM23,Nov25}\footnote{The decompositions are those of the AlphaTensor repository (\url{https://github.com/google-deepmind/alphatensor}, file \texttt{algorithms/factorizations\_r.npz}), of the flip-graph repository of Kauers and Moosbauer (\url{https://github.com/jakobmoosbauer/flips}, folder \texttt{solutions}, the 37 \texttt{.exp} files) and of the AlphaEvolve results (\url{https://github.com/google-deepmind/alphaevolve_results}, notebook \texttt{mathematical\_results.ipynb}); only the list of their shapes is taken from them.} (all rotations) compute a single product type per run, so their leaves are not shared between products of different types; no decomposition of any of their shapes, over any field and of any rank, gives a saving (Corollary~\ref{a:cor:145} in Appendix~\ref{a:sec:full}, which also maps how far beyond them the argument reaches and why it stops). Pan's trilinear aggregation~\cite{Pan78} is a direct sum with type selection, like Schönhage's identity, but every selected product is square, which gives no saving for every scheme (an argument we do not include).

\subsection{Setting}\label{a:sec:setting}

A base identity is a decomposition $\sum_{\lambda=1}^R\varphi_\lambda\psi_\lambda\chi_\lambda$, exact or border, of a trilinear form
\[
T=\sum_{i\le k,\,j\le n}x_iy_jz_{ij}+F\,z_0,
\]
where $\varphi_\lambda$, $\psi_\lambda$, $\chi_\lambda$ are linear forms in the left variables, the right variables and the outputs $z_{ij},z_0$, and $F$, the \emph{inner form}, is a bilinear form in the left and right variables. The left variables are the \emph{outer} $x_1,\dots,x_k$, which span a space $X$, and further \emph{inner} ones, and likewise on the right ($y_1,\dots,y_n$ spanning $Y$). The \emph{inner length} $q_0$ is the rank of $F$.

Let $\mathcal L(F)$ and $\mathcal R(F)$ be the left and right supports of $F$ (its images as a map from the dual of the right space to the left space, and conversely), and put $s_A:=\dim(X\cap\mathcal L(F))$ and $s_B:=\dim(Y\cap\mathcal R(F))$, the numbers of \emph{shared} variables on the two sides. A \emph{free change of basis} is a change of basis of the left space that maps $X$ to itself and one of the right space that maps $Y$ to itself, together with the change of the outer outputs $z_{ij}$ that keeps $\sum x_iy_jz_{ij}$ (the contragredient one); it fixes $z_0$, it does not change $q_0$, $s_A$ or $s_B$, and it preserves rank and border rank. A change that mixes $z_0$ with the $z_{ij}$ is not free.

By Lemma~\ref{a:lem:normal}, after a free change of basis $F$ is a sum of disjoint letter pairs, $F=\sum_{r\le q_0}u_rv_r$ with distinct left variables $u_r$ and distinct right variables $v_r$, of which exactly $s_A$ left and $s_B$ right letters are outer; for these coordinates an outer variable occurring in $F$ is \emph{shared}, and a pair $x_iy_j$ counts on both sides. This covers every identity of Section~\ref{m:sec:barriers} and Appendix~\ref{a:sec:overview}, among them Lemma~\ref{m:lem:identity} with $F=p_1y_1+p_2q$ ($s_A=0$, $s_B=1$). If $s_A=s_B=0$, then $F=\sum_{s\le q_0}p_sq_s$ and $T$ is $G(k,n,q_0):=\langle k,1,n\rangle\oplus\langle1,q_0,1\rangle$.

As in the preprint's Section 2.3, a run uses the identity at $L=cm$ levels; a selected output string has $z_0$ at a set $Q$ of exactly $m$ levels, its \emph{inner set}, and outer outputs elsewhere; a leaf is a string of $L$ terms, and it contributes to an output string $w$ if $\chi_{\tau_\ell}$ contains $w_\ell$ at every level $\ell$.

\subsection{The private-leaf inequality}\label{a:sec:private}

This is the inequality behind the private-leaf condition of Appendix~\ref{a:sec:overview}. Leaves that take a pure-inner term (output form a multiple of $z_0$) at every inner level of a wanted output and, at every other level, a term containing that output's outer letter and no other outer output, serve that output and no other; such leaves cannot be shared, so the pruned recursion computes their products one by one, and their number per wanted output must stay below $\sqrt D$ for a saving to be possible at all.

\begin{lemma}\label{a:lem:private}
Let $R_0\ge0$ terms be \emph{pure inner} (their output form is a nonzero multiple of $z_0$), and let every outer output $z_{ij}$ lie in at least $R_1$ terms whose output forms contain no other outer output. Then every selected output string $w$ has at least $R_0^mR_1^{L-m}$ leaves that contribute to $w$ and to no other selected output string. Consequently, if $|W|=N^2/\sqrt D$ entries are wanted, $D=q_0^m$, each being a selected output of one run, then the pruned recursion of the preprint's Section 2.4.2 computes the products at at least $N^2D^{-1/2}R_0^mR_1^{L-m}$ leaves, which is at least $N^2$ unless $R_0^2R_1^{2(c-1)}<q_0$.
\end{lemma}

\begin{proof}
Let $Q$ be the inner set of $w$. Take a leaf $\tau$ with a pure-inner term at every level of $Q$ and, at every other level $\ell$, a term whose output form contains $w_\ell$ and no other outer output; there are at least $R_0^mR_1^{L-m}$ of them. Each contributes to $w$. Let it contribute to a selected $w'$. At every level of $Q$ the output form of $\tau_\ell$ is a multiple of $z_0$, so $w'_\ell=z_0$ there; as $w'$ has $z_0$ at exactly $m$ levels, its inner set is $Q$. At a level $\ell\notin Q$, $w'_\ell$ is an outer output contained in the output form of $\tau_\ell$, whose only outer output is $w_\ell$; so $w'_\ell=w_\ell$, and $w'=w$. Hence the leaves found for distinct wanted outputs are distinct, and the pruned recursion, which computes the product at every leaf contributing to a wanted output, computes at least $|W|R_0^mR_1^{L-m}$ of them. This is below $N^2$ only if $R_0^mR_1^{L-m}<\sqrt D=q_0^{m/2}$, that is, raising to the power $2/m$ and using $L-m=(c-1)m$, only if $R_0^2R_1^{2(c-1)}<q_0$.
\end{proof}

\subsection{Exact identities with disjoint inner variables}\label{a:sec:exact}

This subsection proves the cap $q_0\le R-k-n$ of Appendix~\ref{a:sec:overview} for exact identities with $k,n\ge2$ whose inner variables are disjoint from the outer ones and whose error is harmless. The first paragraph defines harmless as a condition on the directions of the error monomials that guarantees the property of Schönhage's error the preprint's Lemma 9 relies on, that no product of level monomials with an error monomial among them reaches a selected output from selected inputs. The proof then goes through the three cases of error directions that the condition allows.

Give the variables $x_i$, $y_j$, $z_{ij}$ the class $0$ and the inner variables $p_s$, $q_s$, $z_0$ the class $1$. A monomial $abz$ of a trilinear form ($a$ left, $b$ right, $z$ output) has the \emph{direction} $(\mathrm{cl}(a)-\mathrm{cl}(z),\mathrm{cl}(b)-\mathrm{cl}(z))$, where $\mathrm{cl}$ is the class; the monomials of $G(k,n,q_0)$ have direction $(0,0)$. A product of level monomials that reaches a selected output string from selected input strings has directions summing to $(0,0)$, since the selected left, right and output strings have exactly $m$ inner letters each (the counting argument of the preprint's Lemma 9). We call the error $E$ of an exact identity $\sum_\lambda\varphi_\lambda\psi_\lambda\chi_\lambda=G(k,n,q_0)+E$ \emph{harmless} if the directions of the monomials of $E$ lie in one open half-plane of $\mathbb R^2$; by Gordan's theorem~\cite{Gor73} this says that no nonempty sum of error directions is $(0,0)$. The error of Schönhage's identity (the preprint's Lemma 6) is harmless (directions $(1,0)$, $(0,1)$, $(1,1)$). The possible nonzero directions are six: on an outer output, $p\,y$ gives $(1,0)$, $x\,q$ gives $(0,1)$ and $p\,q$ gives $(1,1)$; on $z_0$, $x\,q$ gives $(-1,0)$, $p\,y$ gives $(0,-1)$ and $x\,y$ gives $(-1,-1)$. Since $(0,0)$ lies in no open half-plane, a harmless error has no monomial of direction $(0,0)$: the coefficients of the monomials $x_iy_jz_{i'j'}$ and $p_sq_tz_0$ are those of $G$.

\begin{proposition}\label{a:prop:exact}
Let $k,n\ge2$ and let $\sum_{\lambda=1}^R\varphi_\lambda\psi_\lambda\chi_\lambda=G(k,n,q_0)+E$ be an exact identity with a harmless error $E$. Then $q_0\le R-k-n$.
\end{proposition}

\begin{proof}
The six directions lie at the angles $0,\frac\pi4,\frac\pi2,\pi,\frac{5\pi}4,\frac{3\pi}2$, and an open half-plane meets the unit circle in an open arc of length $\pi$. Four directions consecutive in this order span an angle of at least $\pi$, so the directions in an open half-plane are contained in three consecutive ones: $\{0,\frac\pi4,\frac\pi2\}$ (errors only on outer outputs), $\{\pi,\frac{5\pi}4,\frac{3\pi}2\}$ (errors only on $z_0$), $\{\frac\pi4,\frac\pi2,\pi\}$ or $\{\frac\pi2,\pi,\frac{5\pi}4\}$ (no monomial with $p$ and $y$), or $\{\frac{3\pi}2,0,\frac\pi4\}$ or $\{\frac{5\pi}4,\frac{3\pi}2,0\}$ (no monomial with $x$ and $q$). The last case is the mirror image of the third under exchanging left and right, so three cases remain. Write $T':=G(k,n,q_0)+E$, write $\varphi_\lambda=X_\lambda+P_\lambda$ with $X_\lambda$ in the span of the $x_i$ and $P_\lambda$ in that of the $p_s$, and $\psi_\lambda=Y_\lambda+Q_\lambda$ likewise.

\emph{Errors only on outer outputs.} Then the coefficient of $z_0$ in $T'$ is exactly $\sum_sp_sq_s$. Let $c_\lambda$ be the coefficient of $z_0$ in $\chi_\lambda$ and $\Lambda_0:=\{\lambda:c_\lambda\ne0\}$. Let $U$ be the matrix whose columns are the coefficient vectors of $c_\lambda\varphi_\lambda$, $\lambda\in\Lambda_0$, with the rows $U_x$ (coordinates $x_i$) above the rows $U_p$ (coordinates $p_s$), and $V$ the matrix of the $\psi_\lambda$ with rows $V_y$ above $V_q$. The coefficient of $z_0$ gives $UV^\top=\begin{pmatrix}0&0\\0&I_{q_0}\end{pmatrix}$. If a combination $\alpha V_y+\beta V_q$ of rows of $V$ vanishes, multiplying by $U_p^\top$ gives $\alpha\cdot0+\beta I_{q_0}=0$, so $\beta=0$; hence $\operatorname{rank}V=r_y+q_0$ with $r_y:=\operatorname{rank}V_y=\dim\operatorname{span}\{Y_\lambda:\lambda\in\Lambda_0\}$. The first block row says $U_xV^\top=0$: every row of $U_x$ is orthogonal to every row of $V$ in $\mathbb R^{\Lambda_0}$, so $r_x+r_y+q_0\le|\Lambda_0|$, where $r_x:=\operatorname{rank}U_x=\dim\operatorname{span}\{X_\lambda:\lambda\in\Lambda_0\}$. The monomials $x_iy_jz_{i'j'}$ of $T'$ are those of $\langle k,1,n\rangle=\sum x_iy_jz_{ij}$, so $\sum_\lambda X_\lambda Y_\lambda\chi^{\mathrm{out}}_\lambda=\langle k,1,n\rangle$, where $\chi^{\mathrm{out}}_\lambda$ is $\chi_\lambda$ without its $z_0$ part. Map the span of the $x_i$ onto its quotient by $\operatorname{span}\{X_\lambda:\lambda\in\Lambda_0\}$. The terms of $\Lambda_0$ vanish, and the other $R-|\Lambda_0|$ terms decompose the image of $\langle k,1,n\rangle$, whose output slices $\bar x_iy_j$ span a space of dimension $(k-r_x)n$; so $R-|\Lambda_0|\ge(k-r_x)n$, and likewise $R-|\Lambda_0|\ge(n-r_y)k$. Hence, using the average of the two bounds,
\begin{align*}
R&\ge q_0+r_x+r_y+\tfrac12\big((k-r_x)n+(n-r_y)k\big)=q_0+kn+r_x\big(1-\tfrac n2\big)+r_y\big(1-\tfrac k2\big)\\
&\ge q_0+kn+k\big(1-\tfrac n2\big)+n\big(1-\tfrac k2\big)=q_0+k+n,
\end{align*}
since $1-\frac n2\le0$, $r_x\le k$, $1-\frac k2\le0$ and $r_y\le n$.

\emph{No monomial with $p$ and $y$.} Then the errors on outer outputs are $x\,q$ and $p\,q$ monomials and those on $z_0$ are $x\,q$ and $x\,y$ monomials. For a left variable $v$ let $T'_v$ be the coefficient of $v$ in $T'$, a bilinear form in right variables and outputs; $T'_v=\sum_\lambda\varphi_\lambda(v)\psi_\lambda\chi_\lambda$, where $\varphi_\lambda(v)$ is the coefficient of $v$ in $\varphi_\lambda$. The $z_0$ part of $T'_{p_s}$ is $q_sz_0$, so the $T'_{p_s}$ are linearly independent, and therefore the $q_0\times R$ matrix $(\varphi_\lambda(p_s))$ has rank $q_0$; choose terms $\lambda_1,\dots,\lambda_{q_0}$ at which its columns form an invertible matrix. For each $a\le k$ choose numbers $c_{as}$ with $\varphi_{\lambda_t}(x_a)+\sum_sc_{as}\varphi_{\lambda_t}(p_s)=0$ for $t=1,\dots,q_0$. Then
\[
T'_{x_a}+\sum_sc_{as}T'_{p_s}=\sum_{\lambda\notin\{\lambda_1,\dots,\lambda_{q_0}\}}\Big(\varphi_\lambda(x_a)+\sum_sc_{as}\varphi_\lambda(p_s)\Big)\psi_\lambda\chi_\lambda .
\]
Let $\pi$ set the variables $q_s$ to $0$. Every monomial of $T'_{p_s}$ contains a $q$-variable, so $\pi T'_{p_s}=0$, and $\pi T'_{x_a}=\sum_jy_j(z_{aj}+\beta_{aj}z_0)$, where $\beta_{aj}$ is the coefficient of $x_ay_jz_0$ in $E$. So $\sum_ax_a\,\pi T'_{x_a}$ has a decomposition with $R-q_0$ terms. Its output slices with respect to the basis $z_{aj}+\beta_{aj}z_0$, $z_0$ of the outputs are the $kn$ independent forms $x_ay_j$, so its rank is at least $kn$, and $R\ge kn+q_0\ge q_0+k+n$, as $kn\ge k+n$ for $k,n\ge2$.

\emph{Errors only on $z_0$.} Then $T'=\sum x_iy_jz_{ij}+(\sum_sp_sq_s+x^\top B_1y+x^\top B_2q+p^\top B_3y)z_0$ for matrices $B_1,B_2,B_3$, writing $x,y,p,q$ for the vectors of variables. Since $(p+B_2^\top x)^\top(q+B_3y)=p^\top q+p^\top B_3y+x^\top B_2q+x^\top B_2B_3y$,
\[
T'=\sum_{i,j}x_iy_j\big(z_{ij}+(B_1-B_2B_3)_{ij}z_0\big)+\sum_sp'_sq'_sz_0,\qquad p':=p+B_2^\top x,\quad q':=q+B_3y,
\]
which is $G(k,n,q_0)$ after invertible changes of variables on the left ($p\mapsto p'$), on the right ($q\mapsto q'$) and of the outputs ($z_{ij}\mapsto z_{ij}+(B_1-B_2B_3)_{ij}z_0$). Such changes preserve the rank, so $G(k,n,q_0)$ has a decomposition with $R$ terms. $G(k,n,q_0)$ itself has no monomial with $p$ and $y$, so the previous case, applied to it with $E=0$, gives $R\ge kn+q_0\ge q_0+k+n$.
\end{proof}

\subsection{Border identities}\label{a:sec:border}

This subsection proves the border caps of Appendix~\ref{a:sec:overview}, Propositions~\ref{a:prop:borderG} and~\ref{a:prop:borderS}, by border substitution: a lower bound on the border rank of a tensor follows from lower bounds for the tensors obtained by killing one variable, and Lemma~\ref{a:lem:subst} says that the variable may be taken fixed by a connected solvable symmetry group, which keeps the quotients inside the families $G(k,n,q_0)$ and $S(k,n,q_0')$ so that an induction closes. Remarks~\ref{a:rem:conditional} and~\ref{a:rem:koszul} say what a stronger base case would give and what computer-calculated flattening ranks give for the shapes $(2,2)$, $(2,3)$ and $(3,3)$.

A border identity with $R$ terms for $T$ (with Bini's interpolation, $\sum_\lambda\varphi_\lambda(\epsilon)\psi_\lambda(\epsilon)\chi_\lambda(\epsilon)=\epsilon^hT+O(\epsilon^{h+1})$) shows that the border rank $\underline R(T)$ over $\mathbb C$ is at most $R$. We use two standard facts. First, if the slices of $T$ in one of its three factors are linearly independent, $b$ of them, then $\underline R(T)\ge b$, since the corresponding flattening has rank $b$, a sum of $r$ rank-one tensors has flattening rank at most $r$, and this is a closed condition. Second, border substitution~\cite{LM17}: for a tensor $T\in A\otimes B\otimes C$ and a line $\langle a\rangle\subseteq A$ let $T/a$ be the image of $T$ in $(A/\langle a\rangle)\otimes B\otimes C$.

\begin{lemma}[border substitution with a solvable symmetry group]\label{a:lem:subst}
Let $T\ne0$ have $\underline R(T)\le r$, and let $\mathcal B$ be a connected solvable linear algebraic group acting on $A$, $B$ and $C$ and fixing $T$. Then some $\mathcal B$-fixed line $\langle a\rangle\subseteq A$ has $\underline R(T/a)\le r-1$.
\end{lemma}

\begin{proof}
Let $T=\lim_{\epsilon\to0}T_\epsilon$ with $T_\epsilon=\sum_{\lambda\le r}a_\lambda(\epsilon)\otimes b_\lambda(\epsilon)\otimes c_\lambda(\epsilon)$. As $T\ne0$, some $a_\lambda(\epsilon)$ is nonzero for small $\epsilon\ne0$, say $a_1(\epsilon)$; along a sequence $\epsilon\to0$ its line converges to a line $\langle a\rangle$. Choose $g_\epsilon\in GL(A)$ with $g_\epsilon\to\mathrm{id}$ and $g_\epsilon a_1(\epsilon)\in\langle a\rangle$. Then $g_\epsilon T_\epsilon$ modulo $a$ has rank at most $r-1$ and converges to $T/a$, so $\underline R(T/a)\le r-1$. The set of lines $\langle a\rangle$ with $\underline R(T/a)\le r-1$ is therefore nonempty; it is closed in the projective space of $A$, since border rank at most $r-1$ is a closed condition (on an affine chart of the projective space of $A$, identify $A/\langle a\rangle$ with a fixed complement by a projection depending regularly on $a$), and it is stable under $\mathcal B$, which fixes $T$. By Borel's fixed point theorem~\cite[Section 21]{Hum75} the connected solvable group $\mathcal B$ has a fixed point in it.
\end{proof}

\begin{proposition}\label{a:prop:borderG}
For $k,n\ge1$ and $q_0\ge(k-1)(n-1)+1$, $\underline R(G(k,n,q_0))\ge q_0+k+n-1$. Hence a border identity with $R$ terms for $G(k,n,q_0)$ has $q_0\le R-k-n+1$.
\end{proposition}

\begin{proof}
For $k=1$ the right slices of $G(1,n,q_0)$, $x_1z_{1j}$ for $y_j$ and $p_sz_0$ for $q_s$, are $n+q_0$ independent forms, so $\underline R\ge n+q_0$. Let $k\ge2$. The group $\mathcal B$ of pairs of invertible upper-triangular matrices acting on $x_1,\dots,x_k$ (with the inverse transpose on the first index of the $z_{ij}$) and on $p_1,\dots,p_{q_0}$ (with the inverse transpose on the $q_s$) fixes $G(k,n,q_0)$ and is connected and solvable. Its diagonal part scales $x_i$ and $p_s$ by distinct characters, so a $\mathcal B$-fixed line of the left space is a coordinate line, and the upper-triangular part moves every $\langle x_i\rangle$ with $i\ge2$ and every $\langle p_s\rangle$ with $s\ge2$; so the fixed lines are $\langle x_1\rangle$ and $\langle p_1\rangle$. Since $G/x_1\cong G(k-1,n,q_0)$ and $G/p_1\cong G(k,n,q_0-1)$, Lemma~\ref{a:lem:subst} gives
\[
\underline R(G(k,n,q_0))\ge1+\min\big(\underline R(G(k-1,n,q_0)),\ \underline R(G(k,n,q_0-1))\big).
\]
We show by induction on $k+q_0$ that both terms of the minimum are at least $q_0+k+n-2$. As $q_0\ge(k-2)(n-1)+1$, the first is at least $q_0+(k-1)+n-1$ by induction. If $q_0-1\ge(k-1)(n-1)+1$, the second is at least $(q_0-1)+k+n-1$ by induction. Otherwise $q_0-1=(k-1)(n-1)$. If this is at least $1$, the output slices $x_iy_j$ and $\sum_sp_sq_s$ of $G(k,n,q_0-1)$ are $kn+1$ independent forms, and $kn+1=(q_0-1)+k+n$. If it is $0$, then $n=1$, $q_0=1$, and $G(k,1,0)$ has $k=q_0+k+n-2$ independent output slices. For the consequence: if $q_0\ge(k-1)(n-1)+1$ then $R\ge q_0+k+n-1$; otherwise $q_0\le(k-1)(n-1)=kn-k-n+1\le R-k-n+1$, since the $kn$ output slices $x_iy_j$ give $R\ge kn$.
\end{proof}

For one shared variable, let $S(k,n,q_0')$ be the tensor with inner form $\sum_{s\le q_0'}p_sq_s+c\,y_1$, where $c$ is a left inner variable and $y_1$ is shared; by Lemma~\ref{a:lem:normal}, up to a free change of basis and exchanging left and right, this is every tensor $T$ with $s_A+s_B=1$, and $G(k,n,q_0)$ is every $T$ with $s_A=s_B=0$. So Propositions~\ref{a:prop:borderG} and~\ref{a:prop:borderS} bound the inner length of every border identity with no, respectively one, shared variable, whatever the coordinates of its inner form. Its effective inner length is $q_0'+1$; the tensor of Part~\hyperref[m:partI]{I} is $S(2,2,1)$.

\begin{proposition}\label{a:prop:borderS}
For $k,n\ge1$ and $q_0'\ge(k-1)(n-1)+1$, $\underline R(S(k,n,q_0'))\ge q_0'+k+n-1$. Hence a border identity with $R$ terms for $S(k,n,q_0')$ has effective inner length $q_0'+1\le R-k-n+2$.
\end{proposition}

\begin{proof}
For $k=1$ the right slices, $x_1z_{11}+cz_0$ for $y_1$, $x_1z_{1j}$ for $y_j$, $j\ge2$, and $p_sz_0$ for $q_s$, are $n+q_0'$ independent forms. Let $k\ge2$. Let $\mathcal B$ consist of the upper-triangular changes of $x_1,\dots,x_k$ (inverse transpose on the first index of the $z_{ij}$; they do not touch $cy_1z_0$), the upper-triangular changes of $p_1,\dots,p_{q_0'}$ (inverse transpose on the $q_s$), and the scalings $c\mapsto vc$, $y_1\mapsto v^{-1}y_1$, $z_{i1}\mapsto vz_{i1}$ for all $i$. It fixes $S(k,n,q_0')$, is connected and solvable, and scales $x_i$, $p_s$, $c$ by distinct characters; as in the previous proof its fixed lines in the left space are $\langle x_1\rangle$, $\langle p_1\rangle$ and $\langle c\rangle$, with $S/x_1\cong S(k-1,n,q_0')$, $S/p_1\cong S(k,n,q_0'-1)$ and $S/c\cong G(k,n,q_0')$. So $\underline R(S(k,n,q_0'))$ is at least $1$ plus the least of three border ranks, and each is at least $q_0'+k+n-2$: $G(k,n,q_0')$ by Proposition~\ref{a:prop:borderG}; $S(k-1,n,q_0')$ by induction, as $q_0'\ge(k-2)(n-1)+1$; and $S(k,n,q_0'-1)$ by induction if $q_0'-1\ge(k-1)(n-1)+1$, and otherwise, when $q_0'-1=(k-1)(n-1)$, because its output slices $x_iy_j$ and $\sum_sp_sq_s+cy_1$ are $kn+1=(q_0'-1)+k+n$ independent forms. The consequence follows as in Proposition~\ref{a:prop:borderG}: otherwise $q_0'+1\le(k-1)(n-1)+1\le R-k-n+2$.
\end{proof}

\begin{remark}[a conditional improvement]\label{a:rem:conditional}
If $\underline R(S(1,n,q_0'))\ge n+q_0'+1$ for all $n,q_0'\ge1$, the same induction, with this base and with the outputs bound $kn+1=q_0'+k+n-1$ at $q_0'-1=(k-1)(n-1)$, gives $\underline R(S(k,n,q_0'))\ge q_0'+k+n$ for $q_0'\ge(k-1)(n-1)+1$, so one shared variable gives effective inner length at most $R-k-n+1$, as for disjoint inner variables, and Coppersmith's family is optimal for every shape with $kn+1$ terms. The base inequality is not proved here; Koszul flattenings give it for $n=2$, $q_0'\le6$ and for $q_0'=1$, $n\le6$ (computer calculation of ranks modulo a prime, which bound the rational ranks from below; ancillary file \texttt{anc/koszul\_modp.py}, mode \texttt{base}).
\end{remark}

\begin{remark}[the cases $(2,2)$, $(2,3)$, $(3,3)$]\label{a:rem:koszul}
Exact ranks of Koszul flattenings~\cite{LO13}, calculated by computer in rational arithmetic (ancillary file \texttt{anc/koszul\_exact.py}, which also reproduces the known border ranks $5$ and $7$ of $S(2,2,1)$ and $S(2,3,2)$; \texttt{anc/koszul\_modp.py}, mode \texttt{excl}, recomputes the four bounds modulo a prime), give $\underline R(S(2,2,2))\ge6$, $\underline R(S(2,3,3))\ge8$, $\underline R(S(2,3,5))\ge10$, and $\underline R(S(2,2,3))\ge7$ (rank $37>36$); with the three branches of the proof of Proposition~\ref{a:prop:borderS} ($G(3,3,5)\ge10$ by Proposition~\ref{a:prop:borderG}, $S(2,3,5)\ge10$, and $S(3,3,4)\ge10$ from its outputs) they give $\underline R(S(3,3,5))\ge11$. Likewise $\underline R(S(3,2,3))\ge8$, from the branches $G(3,2,3)\ge7$ (Proposition~\ref{a:prop:borderG}), $S(2,2,3)\ge7$ and $S(3,2,2)\ge7$ (its outputs); this is the shape $(2,3)$ with the shared variable among the two left variables, which $S(2,3,q_0')$ does not cover. Since $S(k,n,q_0'-1)=S(k,n,q_0')/p_{q_0'}$ and border rank does not increase under this map, these exclude with $kn+1$ terms any effective inner length beyond $(k-1)(n-1)+1$ for $(k,n)=(2,2),(2,3),(3,3)$. This part of Section~\ref{m:sec:barriers} and of Appendix~\ref{a:sec:overview} rests on these calculations.
\end{remark}

\paragraph{Proof of Proposition~\ref{m:rem:border}, lower bound.} For the lower bound we use weak border apolarity~\cite[Theorem 1.2 with Theorem 4.3]{BB21}, in the computational form of~\cite{CHL23}, which turns border rank at most $r$ into the existence of an ideal with prescribed dimensions, so that a finite computation can refute it: if the border rank $\underline R(T)$ is at most $r$ and a connected solvable group $\mathcal B$ fixes $T$, then the annihilator of $T$ in $\mathrm{Sym}(A^*)\otimes\mathrm{Sym}(B^*)\otimes\mathrm{Sym}(C^*)$ contains a $\mathcal B$-fixed multigraded ideal $I$ whose piece of each multidegree has codimension $\min(r,\dim)$ in that degree. Let $T=G(2,2,2)$ with $A=\langle x_1,x_2,p_1,p_2\rangle$, $B=\langle y_1,y_2,q_1,q_2\rangle$, $C=\langle z_{11},z_{12},z_{21},z_{22},z_0\rangle$, let $r=5$, and let $\mathcal B$ be generated by the diagonal maps fixing $T$, a $7$-dimensional torus, and the upper-triangular changes of $x_1,x_2$, of $y_1,y_2$ and of $p_1,p_2$, with the inverse transpose on the first, respectively second, index of the $z_{ij}$ and on $q_1,q_2$; the three triangular groups act on different variables and commute, and the torus normalises them, so $\mathcal B$ is connected and solvable. In multidegree $(1,1,0)$, $I$ has codimension $5$ in the $16$-dimensional $A^*\otimes B^*$ and lies in the $11$-dimensional annihilator there, so it equals that annihilator. In multidegrees $(1,0,1)$ and $(0,1,1)$ the annihilator is $16$-dimensional and $I$ is a $\mathcal B$-fixed hyperplane of it; every torus weight occurs there once, and exactly three hyperplanes in each degree are $\mathcal B$-fixed. For each of the $9$ pairs, the products $I_{110}C^*+I_{101}B^*+I_{011}A^*$ must lie in $I_{111}$, of dimension $80-5=75$, but they span $77$ or $78$ dimensions. So no such ideal exists, and $\underline R(T)\ge6$. The ranks are computed in exact rational arithmetic (ancillary file \texttt{anc/border\_apolarity.py}, under $2$~s); the same computation finds surviving candidates for Schönhage's $G(2,2,1)$ and for the tensor $S(2,2,1)$ of Lemma~\ref{m:lem:identity}, both of border rank $5$, and none for $S(2,2,2)$, of border rank at least $6$ (Remark~\ref{a:rem:koszul}; \texttt{anc/border\_apolarity\_family.py}).

A numerical probe agrees (ancillary file \texttt{anc/border\_rank\_probe.py}): minimising the squared distance from $G(2,2,2)$ to the sums of five rank-one tensors stops at a positive plateau with bounded parameters from every random start, as for two tensors whose border rank is known to exceed five, while for the tensor of Lemma~\ref{m:lem:identity} it tends to $0$ with growing parameters.

\subsection{At most one shared variable}\label{a:sec:shared}

This subsection proves Theorem~\ref{a:thm:shared}: in a run whose products all compute correct single outputs, at most one variable can be shared. Consider a run of a base identity whose inner form is a sum of disjoint letter pairs in the coordinates in which the run is read. (Lemma~\ref{a:lem:normal} does not remove this assumption here: a free change of basis mixes the outer outputs of a level among themselves, and the hypothesis of Theorem~\ref{a:thm:shared} is not invariant under it.) Let the run evaluate $T^{\otimes L}$ on a left array $a$ and a right array $b$ indexed by left and right strings, every entry $a[v]$ a linear form in the entries of a matrix $X$ and every entry $b[w]$ a linear form in those of $Y$, so that the errors do not reach the selected outputs (the preprint's selection with a harmless error, or Bini's interpolation). A selected output $o=(Q,I,J)$ has an inner set $Q$, a string $I$ of outer left letters and a string $J$ of outer right letters on the levels outside $Q$. Its \emph{inner indices} $\iota$ are the choices of one letter pair $u_rv_r$ of $F$ at every level of $Q$, $q_0^m=:D$ of them; $v_\iota$ is the left string equal to $I$ outside $Q$ and to the left letters of $\iota$ on $Q$, and $w_\iota$ the right string defined likewise from $J$. Reading the coefficient of the output string in $T^{\otimes L}$ level by level ($x_iy_j$ at an outer level, $F$ at an inner one), the run computes
\[
o=\sum_\iota a[v_\iota]\,b[w_\iota].
\]
Since the pairs are disjoint, the left letters of $\iota$ determine $\iota$, and so do the right letters. We say that $o$ is a \emph{correct single output} if there are an entry $(\rho,\sigma)$ and a set $\Theta$ of $D$ inner positions with $o=\sum_{t\in\Theta}X[\rho,t]\,Y[t,\sigma]$ as a bilinear form in the entries of $X$ and $Y$; $((\rho,\sigma),\Theta)$ is the \emph{entry and block} of $o$.

\begin{lemma}\label{a:lem:facts}
Let $o=\sum_\iota a[v_\iota]b[w_\iota]$ be a correct single output with entry and block $((\rho,\sigma),\Theta)$.
\begin{enumerate}[label=(\Alph*),nosep]
\item The forms $a[v_\iota]$ span $\operatorname{span}\{X[\rho,t]:t\in\Theta\}$ and the forms $b[w_\iota]$ span $\operatorname{span}\{Y[t,\sigma]:t\in\Theta\}$; in particular each of them is nonzero.
\item The forms $a[v_\iota]$ are linearly independent, and if $\sum_\iota a[v_\iota]\beta_\iota=\sum_\iota a[v_\iota]\beta'_\iota$ for linear forms $\beta_\iota,\beta'_\iota$ in the entries of $Y$, then $\beta_\iota=\beta'_\iota$ for every $\iota$.
\item The forms $b[w_\iota]$ are linearly independent.
\end{enumerate}
\end{lemma}

\begin{proof}
The bilinear form $\sum_{t\in\Theta}X[\rho,t]Y[t,\sigma]$ has rank $D$, since it pairs $D$ distinct variables of $X$ with $D$ distinct variables of $Y$. A sum $\sum_\iota\alpha_\iota\beta_\iota$ of $D$ products of linear forms has rank $D$ only if the $\alpha_\iota$ are linearly independent and the $\beta_\iota$ are linearly independent, which gives (C) and the first half of (B); and then the span of the $\beta_\iota$ is the image of the form as a map from the dual of the $X$-forms to the $Y$-forms, which for the right-hand side is $\operatorname{span}\{Y[t,\sigma]:t\in\Theta\}$, and likewise for the $\alpha_\iota$, which gives (A). For the second half of (B), choose for every $\kappa$ a point $e_\kappa$ of the space of $X$-entries with $a[v_\iota](e_\kappa)=[\iota=\kappa]$, which exists because the $a[v_\iota]$ are independent; evaluating both sides at $e_\kappa$ gives $\beta_\kappa=\beta'_\kappa$.
\end{proof}

\begin{theorem}\label{a:thm:shared}
Consider a run as above whose products are indexed by a family of inner sets of size $m$ with at least two members, and in which every output $(Q,I,J)$, $Q$ in the family and $I,J$ arbitrary, is a correct single output of a distinct entry and block.
\begin{enumerate}[label=(\alph*),nosep]
\item No two right variables are shared (and, by symmetry, no two left variables). This part does not use that the entries and blocks are distinct.
\item If, moreover, for every entry the blocks of the outputs that compute it are pairwise disjoint, then no left and right variable are both shared (in particular no pair $x_iy_j$ is a letter pair of $F$).
\end{enumerate}
So under the hypothesis of (b) at most one variable is shared, and a run of an identity with two shared variables has a single product.
\end{theorem}

\begin{proof}
Let $Q\ne Q'$ be two inner sets of the run and $\ell\in Q'\setminus Q$, which exists since $|Q|=|Q'|$.

(a) Let $y_1$ and $y_2$ be shared, in the letter pairs $c_1y_1$ and $c_2y_2$ of $F$. Let $w_1$ and $w_2$ be right strings that agree except at $\ell$, where $w_1$ reads $y_1$ and $w_2$ reads $y_2$, and that read $y_1$ at every other level of $(Q\setminus Q')\cup(Q'\setminus Q)$, the same right letter of a pair of $F$ at every level of $Q\cap Q'$, and the same outer letters elsewhere. In the product $Q'$ the strings agree outside $Q'$, where they read outer letters, and read right letters of pairs on $Q'$; so for every outer left string $I'$ they are the right strings $w_{\iota_1}$, $w_{\iota_2}$ of the output $o'=(Q',I',w_1|_{[L]\setminus Q'})$ at two distinct inner indices $\iota_1\ne\iota_2$. In the product $Q$ they read outer letters outside $Q$ (at $\ell$ they read $y_1$, $y_2$) and the same right letters of pairs on $Q$ (at $Q\setminus Q'$ the letter $y_1$ of the pair $c_1y_1$); so for every outer left string $I$ they are the right strings at the same inner index $\iota$ of the outputs $o_1=(Q,I,J_1)$ and $o_2=(Q,I,J_2)$, $J_i:=w_i|_{[L]\setminus Q}$, and $o_1$, $o_2$ have the same left strings $v_\kappa$ at every inner index $\kappa$. By Lemma~\ref{a:lem:facts}(A) for $o'$, $b[w_1]$ and $b[w_2]$ are nonzero forms in the column $\sigma'$ of $o'$; by (A) for $o_1$ and $o_2$, $b[w_i]$ is a form in the column $\sigma_i$ of $o_i$; hence $\sigma_1=\sigma'=\sigma_2$. The left strings of $o_1$ and $o_2$ coincide, so by (A) they have the same row $\rho$ and block $\Theta$, and $o_1=o_2$ as bilinear forms. Lemma~\ref{a:lem:facts}(B) then gives $b[w^{(1)}_\kappa]=b[w^{(2)}_\kappa]$ for every $\kappa$, where $w^{(i)}_\kappa$ are the right strings of $o_i$; at $\kappa=\iota$ this is $b[w_1]=b[w_2]$. But $b[w_1]$ and $b[w_2]$ are linearly independent by (C) for $o'$, a contradiction.

(b) Let $x_1$ be shared in the pair $x_1d$ and $y_1$ in the pair $cy_1$ of $F$ (if $x_1y_1$ is a pair, take $d=y_1$ and $c=x_1$). In the product $Q'$ take an output $o'=(Q',I',J')$ whose strings $I'$ and $J'$ read $x_1$ and $y_1$ at the levels of $Q\setminus Q'$, and two of its inner indices: one with the pair $x_1d$, and one with the pair $cy_1$, at every level of $Q'\setminus Q$, both with the same pairs on $Q\cap Q'$. Let $v$ be the left string of the first and $w$ the right string of the second. On $Q$, $v$ reads left letters of pairs ($x_1$ on $Q\setminus Q'$), and outside $Q$ it reads outer letters ($x_1$ on $Q'\setminus Q$, $I'$ elsewhere); so $v$ is a left string of the output $o_Q:=(Q,v|_{[L]\setminus Q},w|_{[L]\setminus Q})$ of the product $Q$, and by the same argument $w$ is a right string of $o_Q$. By Lemma~\ref{a:lem:facts}(A), $a[v]$ is a nonzero form in $\operatorname{span}\{X[\rho',t]:t\in\Theta'\}\cap\operatorname{span}\{X[\rho_Q,t]:t\in\Theta_Q\}$, so $\rho_Q=\rho'$ and $\Theta_Q\cap\Theta'\ne\emptyset$; and $b[w]$ gives $\sigma_Q=\sigma'$. So $o_Q$ and $o'$, which are distinct outputs ($Q\ne Q'$), compute the same entry with overlapping blocks. This contradicts the hypothesis of (b).
\end{proof}

The tilings of the preprint and of Part~\hyperref[m:partI]{I} satisfy the hypothesis of (b): the blocks used for one entry are the disjoint inner blocks of the products of the tile. The theorem concerns runs in which every output of every product is selected; for runs that select only part of the outputs it says only which triples of outputs cannot all be selected, and that no partial selection gains is heuristic (Appendix~\ref{a:sec:overview}).

\subsection{The normal form of the inner form}\label{a:sec:normal}

The lemma below is the normal form invoked in the setting, in Appendix~\ref{a:sec:overview} and in the paragraph before Proposition~\ref{a:prop:borderS}: a free change of basis, one that keeps the outer spaces $X$ and $Y$, turns every inner form into a sum of disjoint letter pairs with the numbers $s_A$, $s_B$ of shared letters unchanged, so that the border bounds proved for $G(k,n,q_0)$ and $S(k,n,q_0')$ apply, after exchanging left and right if needed for $S$, to every inner form with no, respectively one, shared variable. The paragraph after the proof says what the lemma is in matrix terms.

\begin{lemma}\label{a:lem:normal}
Let $F$ be a bilinear form of rank $q_0$ in the left and right variables, and let $X$, $Y$, $s_A$, $s_B$ be as in the setting. Then there are a basis of the left space whose first $k$ vectors are a basis of $X$ and a basis of the right space whose first $n$ vectors are a basis of $Y$ in which $F=\sum_{r=1}^{q_0}u_r\otimes v_r$ with distinct basis vectors $u_1,\dots,u_{q_0}$ and distinct basis vectors $v_1,\dots,v_{q_0}$; exactly $s_A$ of the $u_r$ lie in $X$ and exactly $s_B$ of the $v_r$ lie in $Y$. The change to these bases is a free change of basis.
\end{lemma}

\begin{proof}
Write $\mathcal{L}:=\mathcal{L}(F)$ and $\mathcal{R}:=\mathcal{R}(F)$; both have dimension $q_0$, and $F\in \mathcal{L}\otimes \mathcal{R}$. Contracting the first factor of $F$ with a linear form on $\mathcal{L}$ defines a linear map $\varphi$ from the dual $\mathcal{L}^*$ to $\mathcal{R}$, which is bijective since $F$ has rank $q_0$. Let $V:=\mathcal{L}\cap X$, of dimension $s_A$, let $W':=\varphi^{-1}(\mathcal{R}\cap Y)\subseteq \mathcal{L}^*$, of dimension $s_B$, and let $Z\subseteq \mathcal{L}$ be the annihilator of $W'$, of dimension $q_0-s_B$. Choose a basis of $V\cap Z$ and extend it to a basis of $V$ and to a basis of $Z$; the union of the two is a basis of $V+Z$, since $\dim(V+Z)=\dim V+\dim Z-\dim(V\cap Z)$; extend it to a basis $a_1,\dots,a_{q_0}$ of $\mathcal{L}$. So there are index sets $T_V$ and $T_Z$ such that $\{a_r:r\in T_V\}$ is a basis of $V$ and $\{a_r:r\in T_Z\}$ a basis of $Z$. Let $a_1^*,\dots,a_{q_0}^*$ be the dual basis of $\mathcal{L}^*$ and $b_r:=\varphi(a_r^*)$, a basis of $\mathcal{R}$. For every $\lambda\in \mathcal{L}^*$, contracting $\sum_ra_r\otimes b_r$ with $\lambda$ gives $\sum_r\lambda(a_r)\varphi(a_r^*)=\varphi\big(\sum_r\lambda(a_r)a_r^*\big)=\varphi(\lambda)$, so $F=\sum_ra_r\otimes b_r$. The annihilator of $Z=\operatorname{span}\{a_r:r\in T_Z\}$ is $\operatorname{span}\{a_r^*:r\notin T_Z\}$, and it is also $W'$; hence $\{b_r:r\notin T_Z\}$ is a basis of $\varphi(W')=\mathcal{R}\cap Y$, and the $b_r$ with $r\in T_Z$ are not in $Y$ (otherwise $\mathcal{R}\cap Y$ would have dimension above $q_0-|T_Z|=s_B$). Likewise the $a_r$ with $r\notin T_V$ are not in $X$, and moreover $\operatorname{span}\{a_r:r\notin T_V\}\cap X=0$: a vector in both lies in $\mathcal{L}\cap X=V=\operatorname{span}\{a_r:r\in T_V\}$, hence is $0$. So $\{a_r:r\in T_V\}$ extends to a basis of $X$, which together with $\{a_r:r\notin T_V\}$ is linearly independent and extends to a basis of the left space; on the right the same with $\{b_r:r\notin T_Z\}$ in $Y$. In these bases $u_r:=a_r$ and $v_r:=b_r$ are distinct basis vectors, $s_A=|T_V|$ of the $u_r$ lie in $X$ and $s_B=q_0-|T_Z|$ of the $v_r$ lie in $Y$. The first $k$ new left basis vectors span $X$, so the change of basis maps $X$ to itself, and likewise on the right; with the contragredient change of the $z_{ij}$ it is free.
\end{proof}

This is the normal form of a matrix under the row and column operations that preserve the flags $X\subseteq$ (left space) and $Y\subseteq$ (right space), stated for the two-step flags that the free changes preserve. In particular a tensor with $s_A=s_B=0$ is $G(k,n,q_0)$ after a free change of basis, and one with $s_A=0$, $s_B=1$ is $S(k,n,q_0-1)$, the outer right letter of the shared pair being renamed $y_1$ by a change of basis of $Y$.

\subsection{Vrana's partial product}\label{a:sec:vrana}

This subsection proves Proposition~\ref{a:prop:vrana} of Appendix~\ref{a:sec:overview}: with Vrana's five-term decomposition, every wanted output has $D$ leaves, and a leaf that takes $\tau_2$, the one term whose output factor contains both inner outputs, at $j$ levels serves at most $2^j$ read outputs; weighting each leaf by the inverse of this bound gives at least $\sqrt D/2$ leaf products per wanted output, so the pruning saves nothing in any orientation. The weights are what this count adds to the count of the private-leaf inequality.

Let $\Lambda:=[2]^3\setminus\{(1,1,1),(2,2,2)\}$ and $M_\Lambda:=\sum_{(i,j,k)\in\Lambda}x_{ij}y_{jk}z_{ki}$ \cite[Example 4.10]{Vra24}. Its border decomposition \cite[eq.~(42)]{Vra24} has the five terms
\begin{gather*}
\tau_1=x_{12}\,y_{12}\,z_{12},\qquad \tau_2=-(x_{12}+\epsilon^2x_{21})(y_{12}+\epsilon^2y_{21})(z_{12}+\epsilon^2z_{21}),\\
\tau_3=\epsilon^2x_{21}(y_{12}+\epsilon y_{11})(z_{12}+\epsilon z_{22}),\qquad \tau_4=\epsilon^2(x_{12}+\epsilon x_{22})\,y_{21}\,(z_{12}+\epsilon z_{11}),\qquad \\ \tau_5=\epsilon^2(x_{12}+\epsilon x_{11})(y_{12}+\epsilon y_{22})\,z_{21},
\end{gather*}
with $\tau_1+\dots+\tau_5=\epsilon^3M_\Lambda+O(\epsilon^4)$ (expand). Read with the $x$-variables as outputs, $M_\Lambda$ has the outer outputs $x_{11}=y_{12}z_{21}$ and $x_{22}=y_{21}z_{12}$ and the inner outputs $x_{12}=y_{21}z_{11}+y_{22}z_{21}$ and $x_{21}=y_{11}z_{12}+y_{12}z_{22}$, each an inner product of length $2$. A run at $L$ levels reads output strings with an inner output at exactly $m$ levels, its inner set, so that each read output is an inner product of length $D=2^m$; a leaf (a string of $L$ terms) contributes to an output string $w$ if the $x$-factor of its term at every level $\ell$ contains $w_\ell$. The substitution $x_{ij}\mapsto y_{ij}\mapsto z_{ij}\mapsto x_{ij}$ maps $M_\Lambda$ to itself, since $\Lambda$ is invariant under $(i,j,k)\mapsto(j,k,i)$, and maps the set of five terms to itself ($\tau_3\mapsto\tau_4\mapsto\tau_5\mapsto\tau_3$, and $\tau_1$, $\tau_2$ fixed), so everything below holds in the $y$- and $z$-orientations as well.

\begin{proposition}\label{a:prop:vrana}
For every $L$, $m$ and every set $W$ of read output strings, the pruned recursion of the preprint's Section 2.4.2 with decomposition~(42) computes the products at at least $|W|\,2^{m/2-1}=|W|\sqrt D/2$ leaves. So with $|W|=N^2/\sqrt D$ wanted entries it computes at least $N^2/2$ leaf products, whatever $c=L/m$: there is no saving, in any of the three orientations.
\end{proposition}

\begin{proof}
The $x$-factors are $x_{12}$ ($\tau_1$), $x_{12}+\epsilon^2x_{21}$ ($\tau_2$), $x_{21}$ ($\tau_3$), $x_{12}+\epsilon x_{22}$ ($\tau_4$) and $x_{12}+\epsilon x_{11}$ ($\tau_5$). So $\tau_1,\tau_2,\tau_3$ contain no outer output, the outer output $x_{11}$ lies only in $\tau_5$ and $x_{22}$ only in $\tau_4$. For $w\in W$ with inner set $Q$ let $\mathcal T(w)$ be the set of leaves with, at every $\ell\in Q$, a term among $\tau_1,\tau_2,\tau_3$ containing $w_\ell$ ($\tau_1$ or $\tau_2$ for $x_{12}$, $\tau_2$ or $\tau_3$ for $x_{21}$), and at every $\ell\notin Q$ the term $\tau_5$ if $w_\ell=x_{11}$ and $\tau_4$ if $w_\ell=x_{22}$. They contribute to $w$, and there are $2^m$ of them; for $0\le j\le m$, exactly $\binom mj$ of them have $\tau_2$ at $j$ levels. Let $\tau\in\mathcal T(w)$ contribute to a read $w'$. At every level of $Q$ the $x$-factor of $\tau_\ell$ contains only inner outputs, so the inner set of $w'$ contains $Q$, and as both have $m$ levels it is $Q$. At $\ell\notin Q$, $w'_\ell$ is then an outer output in the $x$-factor of $\tau_\ell$, whose only outer output is $w_\ell$. At $\ell\in Q$, $w'_\ell$ is $x_{12}$ if $\tau_\ell=\tau_1$, $x_{21}$ if $\tau_\ell=\tau_3$, and one of the two if $\tau_\ell=\tau_2$. So $\tau$ lies in $\mathcal T(w')$ for at most $s(\tau):=2^{j(\tau)}$ read outputs $w'$, where $j(\tau)$ is its number of levels with $\tau_2$. The pruned recursion computes the product at every leaf that contributes to a wanted output, so at least
\begin{align*}
\Big|\bigcup_{w\in W}\mathcal T(w)\Big|&\ge\sum_{w\in W}\sum_{\tau\in\mathcal T(w)}\frac1{s(\tau)}\ge\sum_{w\in W}\ \sum_{j\le m/2}\binom mj2^{-j}\\
&\ge|W|\,2^{-m/2}\sum_{j\le m/2}\binom mj\ge|W|\,2^{m/2-1}
\end{align*}
of them, where the first inequality holds because each leaf of the union is counted at most $s(\tau)$ times with weight $1/s(\tau)$, and the last because $\sum_{j\le m/2}\binom mj\ge2^{m-1}$. With $|W|=N^2/\sqrt D=N^22^{-m/2}$ this is at least $N^2/2$. The other orientations follow by the cyclic symmetry.
\end{proof}

\subsection{Thin full products}\label{a:sec:full}

A full product used as a base is given by its shape $\langle o_1,b,o_2\rangle$, three positive integers, and by a decomposition of it, of which each shape has infinitely many over every infinite field. A computer search tries finitely many decompositions, so it cannot show that a shape is unusable. The results below remove the decompositions from the question: the forest bound and the rank bound (Lemmas~\ref{a:lem:forest} and~\ref{a:lem:rankthin}) bound the output supports of every decomposition, over every field and of every rank, in terms of $o_1$, $b$, $o_2$ alone, and Proposition~\ref{a:prop:criterion} turns this into a condition on the three integers that excludes the shape. The question becomes one about a discrete set of triples, and most of it is settled at once: every shape whose smaller outer side is at least $2$ and at most $b$, and all the shapes of the $145$ decompositions found by computer search, for every decomposition of these shapes and not only the ones found (Corollary~\ref{a:cor:145}). What remains are the thin shapes that fail all three cases of the proposition; for each $b$ they include every shape whose smaller side is large enough, since each of the three cases of the proposition bounds the smaller side in terms of $b$, so there are infinitely many, and numerically the smallest by number of outputs is $\langle34,2,35\rangle$, with $1190$ outputs against $144$ for the largest thin shape among the $145$. This proves what Appendix~\ref{a:sec:overview} states about the $145$. The leaf count is first linked to the cost of the pruned recursion (Lemmas~\ref{a:lem:link} and~\ref{a:lem:types}, Theorem~\ref{a:thm:full}); two bounds on the output supports of a decomposition (Lemmas~\ref{a:lem:forest} and~\ref{a:lem:rankthin}) then give a criterion on the shape alone (Proposition~\ref{a:prop:criterion}) that covers the $145$ (Corollary~\ref{a:cor:145}). The last paragraphs, on the region covered beyond them, on the value of the shapes left open, and on why the argument stops, are partly numerical and say so.

Throughout, $\sum_{l=1}^R u_l\otimes v_l\otimes w_l$ is a decomposition over a field of the matrix multiplication tensor
\[
\langle o_1,b,o_2\rangle=\sum_{i\le o_1,\,j\le b,\,c\le o_2}x_{ij}\,y_{jc}\,z_{ic},\qquad o_1,b,o_2\ge1,
\]
read with $w$ as the output factor; by the cyclic symmetry of the tensor, every rotation of every decomposition of a full product has this form. Terms with $w_l=0$ are dropped. In this subsection $\nu:=o_1o_2$ is the number of outputs, $Z_l$ is the set of outputs $z_{ic}$ with a nonzero coefficient in $w_l$, $S_l:=|Z_l|$, and $\kappa:=\min(o_1,o_2)$; the shape is \emph{thin} if $b<\kappa$, and $\Gamma:=\min_{t\in[0,1]}\big(\ln\sum_lS_l^t-\frac t2\ln b\big)-\ln \nu$.

\paragraph{The run.} Used as a base identity at $L$ levels with every output string selected, the decomposition computes in one run the single product $\langle o_1^L,b^L,o_2^L\rangle$, so $D=b^L$. An $N\times D$ by $D\times N$ product, with $N$ a common multiple of $o_1^L$ and $o_2^L$, is tiled by $N^2/\nu^L$ runs whose selected output strings are in bijection with the entries of the product. A leaf $\tau$, a string of $L$ terms, contributes to the $s(\tau):=\prod_\ell S_{\tau_\ell}$ output strings of its own run whose letters lie in the output supports of its terms. The pruned recursion of the preprint's Section 2.4.2 is defined for every identity (its step (2) forms the set passed to a term from the sets of the outputs the term contributes to), and the proof of the preprint's Lemma 10 applies verbatim: for a set $W$ of wanted entries it visits exactly $\Leaves(W)$, the leaves contributing to an output string of $W$, and computes one product at each of them. For $b\ge2$, so that $D$ grows with $L$, we say that the decomposition \emph{gives a saving} if for some $\gamma>0$ the pruned recursion computes at most $N^2/D^\gamma$ leaf products for every $N$ and every set $W$ of at most $N^2/\sqrt D$ entries; the leaf products are the cost of the preprint's method (its Section 2.4.4), so this is what a base identity must give.

\begin{lemma}\label{a:lem:link}
Let a finite set of leaves contribute to the elements of a finite nonempty set $E$, leaf $\tau$ to $s(\tau)$ of them, let $0\le w\le|E|$, $\rho:=w/|E|$ and $V:=\sum_\tau\min\{1,\rho\,s(\tau)\}$. The average of $|\Leaves(W)|$ over the sets $W\subseteq E$ of exactly $w$ elements lies between $(1-1/e)V$ and $V$. In particular some $W$ with $|W|=w$ has $|\Leaves(W)|\ge(1-1/e)V$.
\end{lemma}

\begin{proof}
A leaf $\tau$ is missed by $W$ exactly when $W$ avoids its $s=s(\tau)$ elements, which $\binom{|E|-s}w$ of the $\binom{|E|}w$ sets do. The ratio is $\prod_{i<s}\frac{|E|-w-i}{|E|-i}\le(1-\rho)^s\le e^{-\rho s}$, so $\tau$ lies in $\Leaves(W)$ for a fraction at least $1-e^{-\rho s}\ge(1-1/e)\min\{1,\rho s\}$ of the sets, the last step because $1-e^{-x}$ is concave and increasing and equals $1-1/e$ at $x=1$. By the union bound the fraction is also at most $\min\{1,\rho s\}$. Sum over the leaves.
\end{proof}

So $V$, the leaf count $\sum_\tau\min\{1,\rho\,s(\tau)\}$, is the average cost of the pruned recursion over the wanted sets of a given size, up to the factor $1-1/e$; Lemma~\ref{a:lem:link} holds for every identity, not only for full products.

\begin{lemma}\label{a:lem:types}
Let $R\ge1$ and $S_1,\dots,S_R\ge1$ be integers, $\beta>0$, $F(t):=\ln\sum_lS_l^t-t\ln\beta$, $F^*:=\min_{t\in[0,1]}F(t)$ and $V_L:=\sum_{\tau\in[R]^L}\min\{1,\beta^{-L}\prod_\ell S_{\tau_\ell}\}$. Then $\frac1L\ln V_L\to F^*$ as $L\to\infty$.
\end{lemma}

\begin{proof}
For every $t\in[0,1]$, $\min\{1,y\}\le y^t$ gives $V_L\le e^{LF(t)}$, so $\limsup\frac1L\ln V_L\le F^*$. For the lower bound group the leaves by their type $k=(k_1,\dots,k_R)$, the numbers of levels with each term. The $\binom L{k}\ge(L+1)^{-R}e^{LH(k/L)}$ leaves of type $k$ (\cite[Theorem 11.1.3]{CT06}; $H$ is the entropy in nats) all have the same product $\prod_\ell S_{\tau_\ell}=e^{L\langle k/L,\ln S\rangle}$, so $\frac1L\ln V_L\ge g(k/L)-\frac RL\ln(L+1)$ with $g(p):=H(p)+\min\{0,\langle p,\ln S\rangle-\ln\beta\}$. As $g$ is continuous on the simplex and the types $k/L$ are dense in it, $\liminf\frac1L\ln V_L\ge g(p)$ for every $p$. Let $p_t\propto(S_l^t)_l$, so that $H(p_t)=\ln\sum_lS_l^t-t\langle p_t,\ln S\rangle$ and $F'(t)=\langle p_t,\ln S\rangle-\ln\beta$, and let $t^*$ minimise the convex function $F$ on $[0,1]$. If $0<t^*<1$, then $F'(t^*)=0$ and $g(p_{t^*})=H(p_{t^*})=F(t^*)$. If $t^*=0$, then $F'(0)\ge0$ and $g(p_0)=H(p_0)=\ln R=F(0)$. If $t^*=1$, then $F'(1)\le0$ and $g(p_1)=H(p_1)+\langle p_1,\ln S\rangle-\ln\beta=F(1)$. In each case $g(p_{t^*})=F^*$.
\end{proof}

\begin{theorem}\label{a:thm:full}
If $\sum_lS_l^t\ge \nu\,b^{t/2}$ for every $t\in[0,1]$, that is, if $\Gamma\ge0$, then for every $L$ and every $N$ as above (a multiple of $\nu^L$ for the squared run) some set $W$ of exactly $\lceil N^2/\sqrt D\rceil$ entries makes the pruned recursion, in the preprint's run and in the squared run of Section~\ref{m:sec:squared} with this base, compute at least $(1-1/e)N^2e^{-\delta_LL}$ leaf products, where $\delta_L\to0$ depends only on the decomposition. So, if $b\ge2$, there is no $\gamma>0$ for which it computes at most $N^2/D^\gamma$ leaf products for every such $W$: the decomposition gives no saving.
\end{theorem}

\begin{proof}
Apply Lemma~\ref{a:lem:link} with $E$ the $N^2$ entries, the leaves of all runs, and $w=\lceil N^2/\sqrt D\rceil$, so $\rho\ge b^{-L/2}$. Each of the $N^2/\nu^L$ runs contributes $\sum_{\tau\in[R]^L}\min\{1,\rho\,s(\tau)\}\ge V_L$ with $\beta=\sqrt b$, and by Lemma~\ref{a:lem:types} $V_L\ge e^{L(F^*-\delta_L)}$ with $\delta_L\to0$. The hypothesis says $F^*\ge\ln \nu$, so $V\ge(N^2/\nu^L)\,\nu^Le^{-\delta_LL}$, and Lemma~\ref{a:lem:link} gives $W$. For the squared run the leaves are strings of $2L$ terms with the same output factors (the second half exchanges only left and right), the read output strings are in bijection with the entries of a tile, $D=b^{2L}$ and $\rho\ge b^{-L}$, so the count is $V_{2L}$ over $N^2/\nu^{2L}$ runs, which gives the bound with $2\delta_{2L}$ in place of $\delta_L$.
\end{proof}

Conversely, if $\Gamma<0$, the upper halves of both lemmas bound the average of $|\Leaves(W)|$ over the sets $W$ of this size by about $N^2e^{L\Gamma}$, so the pruned recursion saves on most sets $W$, though not necessarily on every $W$ as the matrix theorem requires; $\Gamma\ge0$ is thus exactly the condition that it saves nothing on average.

\begin{lemma}[forest bound]\label{a:lem:forest}
Call a set $F$ of outputs a \emph{forest} if the bipartite graph on the $o_1$ rows and $o_2$ columns with an edge $\{i,c\}$ for each $z_{ic}\in F$ has no cycle.
\begin{enumerate}[label=(\alph*),nosep]
\item For every forest $F$, at least $b|F|$ terms have $Z_l\cap F\ne\emptyset$.
\item $\sum_l\min\{S_l,f\}\ge b\nu$, where $f:=\nu/(o_1+o_2-1)$.
\end{enumerate}
\end{lemma}

\begin{proof}
(a) Let $T|_F:=\sum_{z_{ic}\in F}\sum_jx_{ij}y_{jc}z_{ic}$. Projecting every $w_l$ onto the outputs in $F$ gives a decomposition of $T|_F$ whose nonzero terms are those with $Z_l\cap F\ne\emptyset$, so it suffices to show that every decomposition of $T|_F$ has at least $b|F|$ terms. We induct on $|F|$; let $F\ne\emptyset$ and take a decomposition $\sum_{l\le R'}u'_l\otimes v'_l\otimes w'_l$ of $T|_F$. A nonempty forest has a vertex of degree one, say the column $c$ with the single edge $z_{ic}$ (a row is symmetric, with $x$ and $y$ exchanged). The slices of $T|_F$ at $y_{1c},\dots,y_{bc}$ are $x_{ij}\otimes z_{ic}$, $j\le b$, which are linearly independent, and the slice at $y_{jc}$ is $\sum_lv'_l[y_{jc}]\,u'_l\otimes w'_l$; so the $b\times R'$ matrix $(v'_l[y_{jc}])$ has rank $b$, and some set $J$ of $b$ terms gives an invertible minor. Then the span of $\{v'_l:l\in J\}$ is a complement of the span $B'$ of the right variables $y_{jc'}$, $c'\ne c$; let $\pi$ be the projection onto $B'$ along it. Applying $\pi$ to the right factors kills the terms of $J$, fixes the summands of $T|_F$ at outputs other than $z_{ic}$ (they contain no $y_{jc}$, as $c$ has degree one), and maps the summands at $z_{ic}$ to some tensor $\sum_jx_{ij}\otimes\pi(y_{jc})\otimes z_{ic}$; projecting the output factors onto $F\setminus\{z_{ic}\}$ then removes these. This gives a decomposition of $T|_{F\setminus\{z_{ic}\}}$ with $R'-b$ terms, and $F\setminus\{z_{ic}\}$ is again a forest, so $R'-b\ge b(|F|-1)$ by induction.

(b) The group $G:=\mathbb Z_{o_1}\times\mathbb Z_{o_2}$ acts on the outputs by $z_{ic}\mapsto z_{i+a,c+d}$, simply transitively, and by automorphisms of the bipartite graph. Let $F_0$ be the spanning tree consisting of the first row and the first column, with $o_1+o_2-1$ outputs. Its $\nu$ translates $gF_0$ are forests, and each output lies in exactly $o_1+o_2-1$ of them. Summing (a) over $g\in G$ gives $\sum_l|\{g:Z_l\cap gF_0\ne\emptyset\}|\ge b\,\nu\,(o_1+o_2-1)$, and each summand on the left is at most $\min\{\nu,S_l(o_1+o_2-1)\}$; divide by $o_1+o_2-1$.
\end{proof}

With $F$ a transversal (one output in each column of the longer side) instead of a tree, the same argument gives $\sum_l\min\{S_l,\kappa\}\ge b\nu$; the trees halve the cap for square shapes, since $f\approx\kappa/2$ there.

\begin{lemma}[rank bound]\label{a:lem:rankthin}
Every decomposition of $\langle o_1,b,o_2\rangle$ has $R\ge \nu+(b-1)\max(o_1,o_2)$ terms, over every field.
\end{lemma}

\begin{proof}
For $b=1$ the slices at the $\nu$ outputs, $x_{i1}\otimes y_{1c}$, are linearly independent, so $R\ge \nu$. For $b\ge2$, the slices at $x_{1b},\dots,x_{o_1b}$ are $\sum_cy_{bc}\otimes z_{ic}$, linearly independent, so as in the proof of Lemma~\ref{a:lem:forest} some $o_1$ terms have left factors spanning a complement of the span of the $x_{ij}$, $j<b$. Projecting the left factors onto that span along them kills these terms, and then setting $y_{bc}=0$ for every $c$ removes the summands with $j=b$ and leaves $\langle o_1,b-1,o_2\rangle$. So $R\ge o_1+R(\langle o_1,b-1,o_2\rangle)$, and in the same way, with the slices at $y_{b1},\dots,y_{bo_2}$ and then $x_{ib}=0$, $R\ge o_2+R(\langle o_1,b-1,o_2\rangle)$. Induct on $b$, using the larger of the two steps each time.
\end{proof}

\begin{proposition}[criterion]\label{a:prop:criterion}
Let $o_1,o_2\ge2$, so that $f>1$, and put $r:=1+(b-1)/\kappa$ and $c_t:=(f^t-1)/(f-1)$. Every decomposition of $\langle o_1,b,o_2\rangle$ satisfies $\sum_lS_l^t\ge \nu\,b^{t/2}$ for every $t\in[0,1]$, and so, if $b\ge2$, gives no saving by Theorem~\ref{a:thm:full}, in each of the following cases:
\begin{enumerate}[label=(\roman*),nosep]
\item $f\le b$;
\item $f^{2(b-1)}\ge b^{f-1}$;
\item $(1-c_t)\,r+c_t\,b\ge b^{t/2}$ for every $t\in[0,1]$.
\end{enumerate}
Case (i) holds whenever $\kappa\le b$, and case (ii) whenever $\kappa^{2(b-1)}\ge b^{\kappa-1}$.
\end{proposition}

\begin{proof}
For an integer $s\ge1$ and $t\in[0,1]$, $s^t\ge(1-c_t)+c_t\min\{s,f\}$: for $s\le f$ because $x^t$ is concave and the right side is its chord through $(1,1)$ and $(f,f^t)$, and for $s>f$ because $s^t\ge f^t$. As $0\le c_t\le1$, summing over the terms and using $R\ge r\nu$ (Lemma~\ref{a:lem:rankthin}, since $\max(o_1,o_2)=\nu/\kappa$) and Lemma~\ref{a:lem:forest}(b) gives $\sum_lS_l^t\ge(1-c_t)r\nu+c_tb\nu$, which is case (iii). In case (ii) with $f\ge b$, put $\lambda:=(b-1)/(f-1)\le1$; then $r\ge1$ and the weighted inequality of arithmetic and geometric means give $(1-c_t)r+c_tb\ge1+(b-1)c_t=(1-\lambda)+\lambda f^t\ge f^{\lambda t}\ge b^{t/2}$, the last step being $2(b-1)\ln f\ge(f-1)\ln b$. Case (ii) with $f<b$ is case (i). In case (i), $s^t\ge\min\{s,f\}^t\ge\min\{s,f\}f^{t-1}$, so $\sum_lS_l^t\ge f^{t-1}\,b\nu\ge b^{t-1}\,b\nu\ge \nu\,b^{t/2}$ by Lemma~\ref{a:lem:forest}(b). Finally $f\le\kappa$ (with $\kappa=o_1$, this is $o_2\le o_1+o_2-1$), so $\kappa\le b$ gives (i); and $\varphi(x):=2(b-1)\ln x-(x-1)\ln b$ is concave with $\varphi(1)=0$, so $\varphi(\kappa)\ge0$ gives $\varphi(f)\ge0$, which is (ii).
\end{proof}

\begin{corollary}\label{a:cor:145}
No decomposition, over any field and of any rank, of the shape of any rotation of any of the $145$ decompositions of full products found by computer search~\cite{FBH+22,KM23,Nov25} gives a saving, in the preprint's run or in the squared run.
\end{corollary}

\begin{proof}
Every dimension of the $145$ shapes is at least $2$, as Proposition~\ref{a:prop:criterion} requires. The $435$ rotations of the $145$ decompositions comprise $96$ thin ones, with $81$ distinct shapes up to $\langle12,11,12\rangle$, and $339$ with $\kappa\le b$. For the $81$ thin shapes the integer inequality $\kappa^{2(b-1)}\ge b^{\kappa-1}$ holds, with equality only for $\langle9,3,11\rangle$, by an exact check (ancillary file \texttt{anc/thin\_exclusion.py}, which lists the shapes). Apply Proposition~\ref{a:prop:criterion} and Theorem~\ref{a:thm:full}.
\end{proof}

The decompositions themselves are not needed: only the list of their shapes enters, read once from the files named in Appendix~\ref{a:sec:overview}. The same script checks that case (ii), which is exact in integers for rational $f$, covers every thin shape with all dimensions at most $12$.

\paragraph{How far the criterion reaches (numerical).} Case (iii) covers, by interval subdivision in $t$ in floating point with a relative margin of $10^{-12}$ (ancillary file \texttt{anc/thin\_region.py}), every $\langle o_1,b,o_2\rangle$ with $b<o_1\le K_{\rm all}$ and any $o_2\ge o_1$, and every square $\langle k,b,k\rangle$ with $k\le K_{\rm sq}$, where $(K_{\rm all},K_{\rm sq})=(23,34)$, $(31,46)$, $(38,58)$, $(46,69)$ for $b=2,3,4,5$, and about $7b+10$ and $11b+12$ up to $b=12$; the case $o_2\to\infty$ is the worst, because the bound only weakens as $f$ grows towards $o_1$. The smallest shapes left open thus grow linearly in $b$: $\langle34,2,35\rangle$, $\langle46,3,48\rangle$, $\langle58,4,60\rangle$, $\langle70,5,70\rangle$, with $1190$, $2208$, $3480$ and $4900$ outputs, against $144$ for the largest thin shape among the $145$; elongated shapes are open from $\langle24,2,418\rangle$ and $\langle32,3,786\rangle$ on. Shapes just beyond this region would not beat our rate even if they saved: by the same two bounds, a shape can reach at most $\varepsilon\gamma=-\Gamma/\ln\max(o_1,o_2)$ (a run needs $N\ge\max(o_1,o_2)^L$ for $D=b^L$), and this stays below the $0.005825$ of Remark~\ref{m:rem:exact} for squares with $k\le50$, $61$, $73$ when $b=2,3,4$, and for all shapes with $\kappa\le41$, $47$ when $b=2,3$ (numerical, $t$ on a grid and $o_2$ sampled; ancillary file \texttt{anc/thin\_value.py}). For larger shapes the two bounds allow more, up to $0.0195$ for $b=2$ and above $0.07$ for larger $b$, so they do not show that the open region is worthless.

\paragraph{Why the argument stops.} The two bounds, and the further constraints of this kind that we know, do not exclude every thin shape. Lemmas~\ref{a:lem:forest} and~\ref{a:lem:rankthin} are linear in the numbers of terms of each support size, with caps that grow with $\kappa$, and the rank bounds we know for thin shapes, Lemma~\ref{a:lem:rankthin} among them, exceed $\nu$ only by $O(b\kappa)$. For $\langle k,2,k\rangle$ with $k$ even, take $\nu$ supports of size $1$ and $6k$ of size $k/2$. Then $R=\nu+6k\ge \nu+k$, $\sum_l\min\{S_l,f\}=\nu+3k^2=4\nu\ge2\nu$, and further proved constraints on the terms, which we do not include, hold as well; but $\sum_lS_l^t=\nu+6k(k/2)^t$, which at $k=1000$ and $t=0.3$ is $1.039\nu<2^{0.15}\nu=1.110\nu$, so $\Gamma<0$. Whether a decomposition with these supports exists is not known. Excluding every thin shape would need either a rank bound $R(\langle k,b,k\rangle)\ge(1+c)\nu$ with $c>0$ independent of $k$, or a bound on how many terms of support one a term of large support can replace. One form of the latter: for the slices $B_{ic}:=\sum_jx_{ij}\otimes y_{jc}$ (each of rank $b$) and any rank-one $P_{ic}$, the $B_{ic}-P_{ic}$ span a space of dimension at least $c'\nu$. This asks how far a family of matrices stays from its rank-one perturbations, a question of the kind studied as matrix rigidity~\cite{Val77}, where strong explicit bounds are a long-standing open problem. Here we can show only that the dimension can be as small as $3\nu/4$ (one dependency per $2\times2$ block of outputs), and we do not know whether it is always at least $c'\nu$.

\end{document}